\documentclass[fleqn,10pt]{SelfArx}
\usepackage[english]{babel}
\usepackage{float}
\usepackage{stfloats}
\usepackage{textcomp}
\usepackage{gensymb}
\usepackage{mathtools}
\usepackage{array}
\DeclarePairedDelimiter\ceil{\lceil}{\rceil}
\DeclarePairedDelimiter\floor{\lfloor}{\rfloor}

\definecolor{color1}{RGB}{0,80,80} 
\definecolor{color2}{RGB}{59,182,179} 
\definecolor{color3}{RGB}{255,255,255} 
\definecolor{color4}{RGB}{10,30,30} 

\usepackage{hyperref} 

\hypersetup{
hidelinks,	colorlinks,	breaklinks=true,
urlcolor=color4,	citecolor=color4,
linkcolor=color4,	bookmarksopen=false,
pdftitle={The JPEG~XL Image Coding System: History, Features, Coding Tools, Design Rationale, and Future},
pdfauthor={Jon Sneyers et al},
}

\JournalInfo{arXiv:2506.05987 [cs.MM]}
\Archive{June 13th, 2025, CC BY-SA 4.0}

\PaperTitle{The JPEG~XL Image Coding System: History, Features, Coding Tools, Design Rationale, and Future}
\PaperMainTitle{The JPEG~XL Image Coding System}
\PaperSubTitle{History, Features, Coding Tools, Design Rationale, and Future}

\Authors{Jon~Sneyers\textsuperscript{1†}*,
Jyrki~Alakuijala\textsuperscript{2†},
Luca~Versari\textsuperscript{2†},
Zoltán~Szabadka\textsuperscript{2†},
Sami~Boukortt\textsuperscript{2},
Amnon~Cohen-Tidhar\textsuperscript{1},
Moritz~Firsching\textsuperscript{2},
Evgenii~Kliuchnikov\textsuperscript{2},
Tal~Lev-Ami\textsuperscript{1},
Eric~Portis\textsuperscript{1},
Thomas~Richter\textsuperscript{3},
Osamu~Watanabe\textsuperscript{4}
}

\affiliation{\textsuperscript{1}\textit{
Cloudinary Research, Petah Tikva, Israel}}
\affiliation{\textsuperscript{2}\textit{
Google Research, Z\"urich, Switzerland}}
\affiliation{\textsuperscript{3}\textit{
Moving Picture Technologies, Fraunhofer IIS, Erlangen, Germany}}
\affiliation{\textsuperscript{4}\textit{
Department of Electronic Systems Engineering, Takushoku University, Tokyo, Japan}}
\affiliation{\textbf{\textsuperscript{†} Equal contribution}\hfill * Corresponding author: jon@cloudinary.com}

\newcommand{\keywordname}{Keywords} 

\Abstract{
JPEG~XL is a new image coding system offering state-of-the-art compression performance, lossless JPEG recompression, and advanced features. It aims to replace JPEG, PNG, GIF, and other formats with a single universal codec. This \ThisName{} provides an overview of JPEG~XL, including its history, design rationale, coding tools, and future potential. It can be used as a companion document to the standard (ISO/IEC 18181), or as a standalone \ThisName{} to better understand JPEG~XL, either at a high level or in considerable technical detail.
}

\begin{document}

\maketitle
\tableofcontents

\thispagestyle{empty} 

\section*{Introduction}
\addcontentsline{toc}{section}{Introduction}

JPEG~XL \cite{18181-1} is a practical, efficient, and royalty-free image codec designed for a variety of use cases \cite{jxl-use-cases}, including professional and consumer photography, medical and scientific imagery, digital art, websites, and social media. It can be used across the workflow, from capturing and authoring to storage, archiving, interchange, and web delivery.

This \ThisName{} provides a complete overview of JPEG~XL, including its history, design rationale, coding tools, and future potential. It can be used as a companion document to the standard (ISO/IEC 18181), or as a standalone \ThisName{} to better understand JPEG~XL, either at a high level or in considerable technical detail.

\paragraph{Disclaimer.}
This \ThisName{} is not a substitute for
the JPEG~XL standard ISO/IEC 18181 \cite{18181-1,18181-2}.
Although it goes into considerable depth, it is not at all a complete specification.
In case there are inconsistencies between this \ThisName{} and the standard, the only authoritative specification is provided by the most recent edition of the standard.
Should there be any such discrepancies, please inform the corresponding author.

\begin{figure*}[b]\centering
\includegraphics[width=\linewidth]{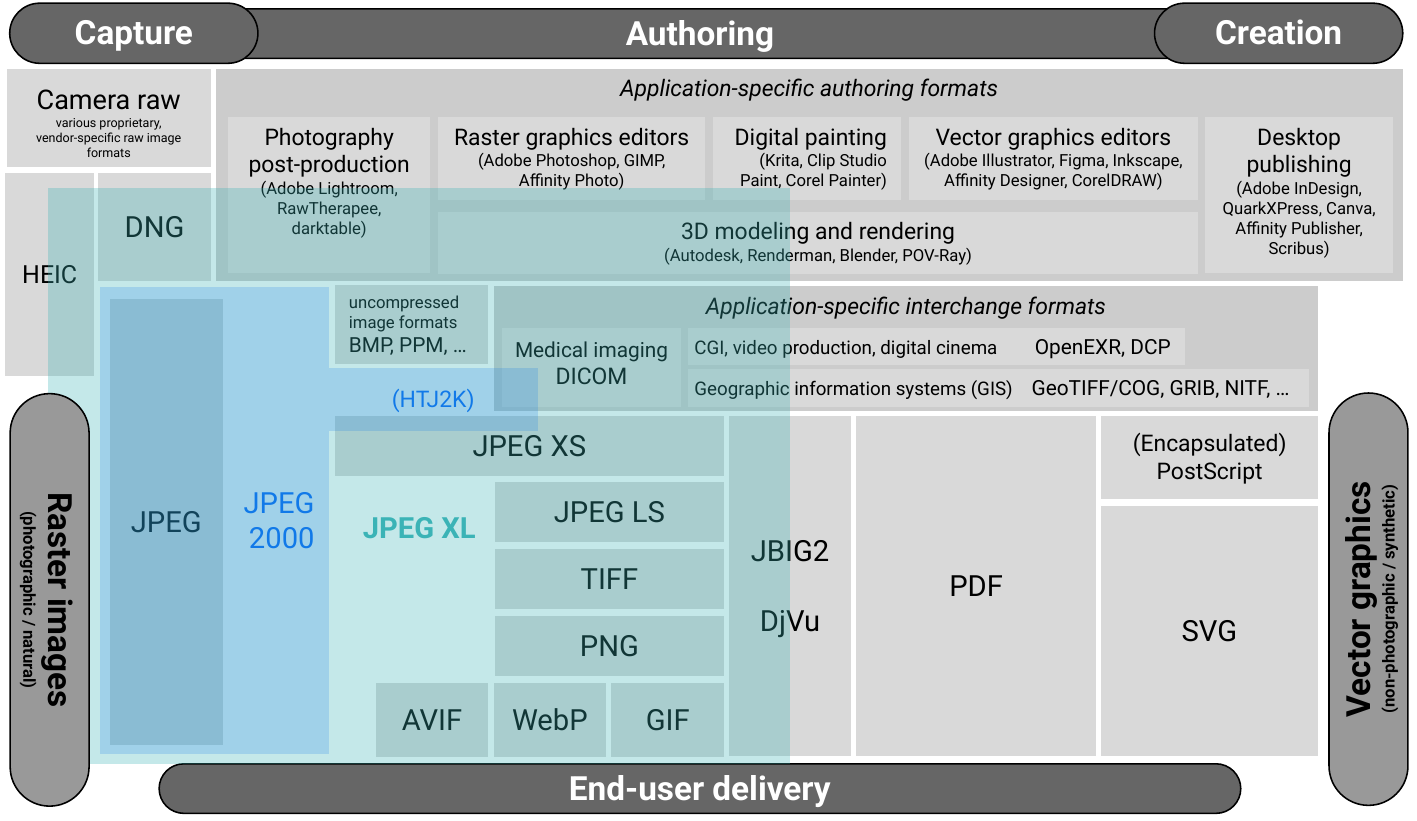}
\caption[Scope and positioning]{Schematic illustration of the scope of the JPEG~XL standard and its positioning with respect to other image formats.
The horizontal axis depicts a spectrum from purely photographic raster images (left) to purely synthetic vector graphics (right), with a range in between that contains mixed content and non-photographic raster images.
The vertical axis depicts various stages of the digital asset lifecycle, from capturing or creation (top) to end-user delivery (bottom), with a range in between that includes production (mezzanine codecs), interchange, storage, and archiving.}
\label{fig:scope}
\end{figure*}

\subsection*{TL;DR overview of JPEG~XL}
\phantomsection
\addcontentsline{toc}{subsection}{TL;DR overview of JPEG~XL}

JPEG~XL was designed to match or outperform all previous raster image formats in terms of both functionality and compression.
Figure~\ref{fig:scope} illustrates the scope of JPEG~XL relative to other image formats. 
In terms of functionality, JPEG~XL can do everything that JPEG, PNG, GIF, TIFF, BMP, WebP, AVIF, and HEIC can do, and more.
Effectively, it eliminates the need to select the right image format for a specific task.
It is also suitable as a payload codec in various application-specific formats, including DNG, DICOM, and GIS formats.

In terms of compression, JPEG~XL offers state-of-the-art performance, both lossless and lossy across a broad quality range, cutting storage and bandwidth costs in half compared to JPEG.

A unique feature of JPEG~XL is its ability to losslessly recompress existing JPEG images, reducing file sizes --- and thus transmission and storage costs --- by about 20\% while allowing byte-for-byte exact reconstruction of the original JPEG file. This is possible because the coding tools of JPEG~XL are a superset of the tools available in JPEG.
This helps to preserve our digital heritage, since no generation loss (additional compression artifacts caused by lossy transcoding) is introduced in the transition process when migrating from JPEG to JPEG~XL.

Some of the key benefits of JPEG~XL include:
\begin{description}
\item[improved color accuracy and fidelity:] wide gamut, high dynamic range with no limit on the precision;
\item[additional channels:] support for alpha transparency, depth maps, CMYK, spot colors, selection masks, multispectral images, \ldots
\item[layered images, animations,] and multi-page images;
\item[substantially improved compression:] saving about 50\%, both for lossy and lossless image compression;
\item[lossless JPEG recompression,] saving about 20\% when migrating existing images, completely risk-free;
\item[advanced progressive decoding,] for a great user experience even under challenging network conditions;
\item[wide range of trade-offs] between quality, speed, and file size, with particular focus on reliable and consistent high-fidelity compression;
\item[various content:] works great for both photographs and other types of raster images such as digital art, screenshots, illustrations, computer-generated imagery;
\item[fast encoding] and decoding, making good use of modern multi-core CPUs; no special hardware needed;
\item[works across the workflow,] from authoring and interchange to storage and end-user delivery;
\item[royalty-free, open source,] high-quality, production-ready reference software is available.
\end{description}

The flexible design of the codestream syntax allows one to balance the following three main aspects, obtaining trade-offs depending on the requirements of the application:
\begin{description}
\item[image fidelity:] from `web quality' to high quality, visually or mathematically lossless at any precision needed;
\item[compression ratio:] typically 2:1 to 6:1 for lossless compression, 20:1 to 50:1 for lossy compression;
\item[speed] of the encoding and/or decoding process.
\end{description}

\subsection*{Structure of this \ThisName{}}
\phantomsection
\addcontentsline{toc}{subsection}{Structure of this \ThisName{}}

This \ThisName{} is organized as follows:
\begin{itemize}
\item \SectionName{}~\ref{history} reviews the historical context and notable precursors to JPEG~XL.
\item \SectionName{}~\ref{overview} provides a high-level overview of the different parts of the standard, the overall architecture of the codec, and a brief description of the main features of the JPEG~XL codestream and file format.
\item \SectionName{}s~\ref{headers}-\ref{progressive} contain a detailed technical description of the various codestream elements and coding tools, as well as motivations for the various design choices.
The topics discussed include:
\begin{itemize}
\item header syntax (\SectionName{}~\ref{headers}),
\item color spaces (\SectionName{}~\ref{color}),
\item the two main coding modes, Modular (\SectionName{}~\ref{modular}) and VarDCT (\SectionName{}~\ref{vardct}),
\item specific coding tools related to image features and restoration filters (\SectionName{}~\ref{features-filters}),
\item entropy coding (\SectionName{}~\ref{entropy}), and
\item bitstream ordering (\SectionName{}~\ref{progressive}).
\end{itemize}
\item \SectionName{}~\ref{performance} investigates the lossless and lossy compression performance compared to other codecs.
\item \SectionName{}~\ref{future} concludes with a discussion of adoption, application domains, and potential future revisions or extensions of the standard.
\end{itemize}

The more technical \SectionName{}s~\ref{headers} to \ref{progressive} are to a large extent covering independent aspects of the codec, so they can be read in any order.
For a high-level overview, the combination of \SectionName{}s~\ref{overview}, \ref{performance}, and \ref{adoption_applications} is a good starting point.

Compared to earlier JPEG~XL codec overviews \cite{jpegxl2019,jxl-white-paper}, this \ThisName{} is substantially more detailed and up to date.


\section{History and related work}
\label{history}
\begin{figure}\centering
\includegraphics[width=1\linewidth]{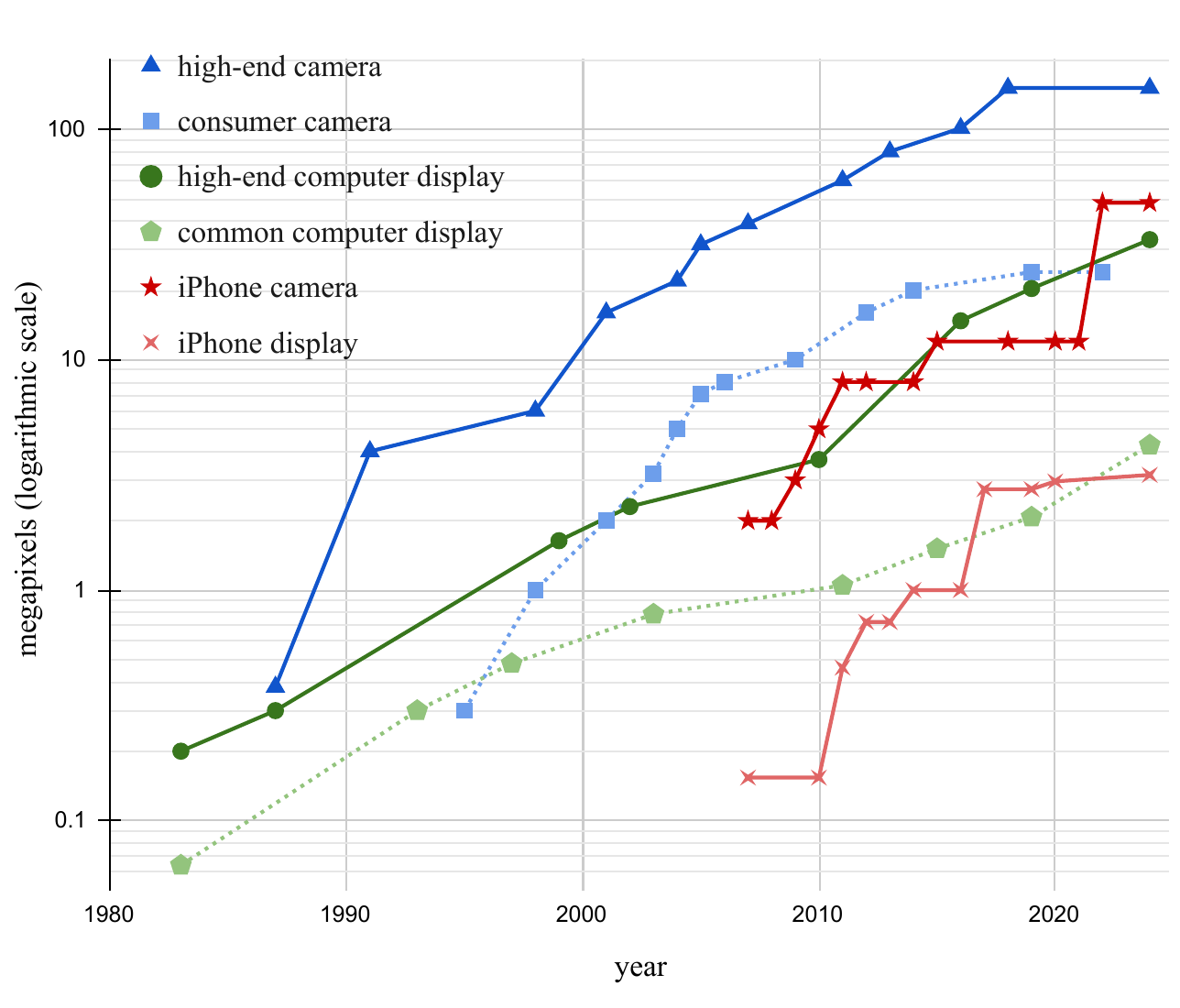}
\caption{Evolution of camera and display resolutions.}
\label{fig:megapixels}
\end{figure}

The history of digital image capturing and display technology is intertwined with the history of image (and video) codecs and file formats.
Resolution and color precision have improved greatly over the years: from $640 \times 200$ monochrome (1-bit) or $320 \times 200$ 4-color palette (2-bit) CGA graphics in the early 1980s,
256-color (8-bit palette) at VGA resolution ($640 \times 480$) in the late 1980s,
`high color' (15 or 16-bit, RGB555 or RGB565) at SVGA resolution ($800 \times 600$) in the early 1990s,
`true color' (24-bit, 8-bit per RGB component) at XGA resolution ($1024 \times 768$) in the late 1990s,
digital cameras quickly overtaking analog (film) photography in the early 2000s and the ensuing megapixel race (see also Figure~\ref{fig:megapixels}),
full HD ($1920 \times 1080$) resolution becoming common in the 2010s,
to, at the time of writing, 
`deep color' (10-bit per component) wide color gamut and high dynamic range displays at 4K or higher resolution being readily available in consumer devices.

Advances in image fidelity inevitably imply an increase in image data: an uncompressed full screen image on a CGA display corresponds to 16 kilobytes, while an uncompressed 4K HDR image is 31 \emph{mega}bytes, almost 2000 times the size of a CGA image.
The first commercially available digital SLR camera (the Kodak DCS) produced 1.3 megapixel images at 1.3 megabytes per uncompressed image (8-bit grayscale or Bayer data); currently the most popular digital camera is the iPhone, which produces 48 megapixel images at 182 megabytes per uncompressed image (10-bit RGB).

Both lossless and lossy compression have played an important role in many use cases of digital images; in particular for saving storage costs (contributing greatly to making digital photography economically feasible) and for reducing bandwidth consumption and speeding up image transfer for websites and mobile apps.

\subsection{Precursors}
\begin{figure*}\centering
\includegraphics[width=1\linewidth]{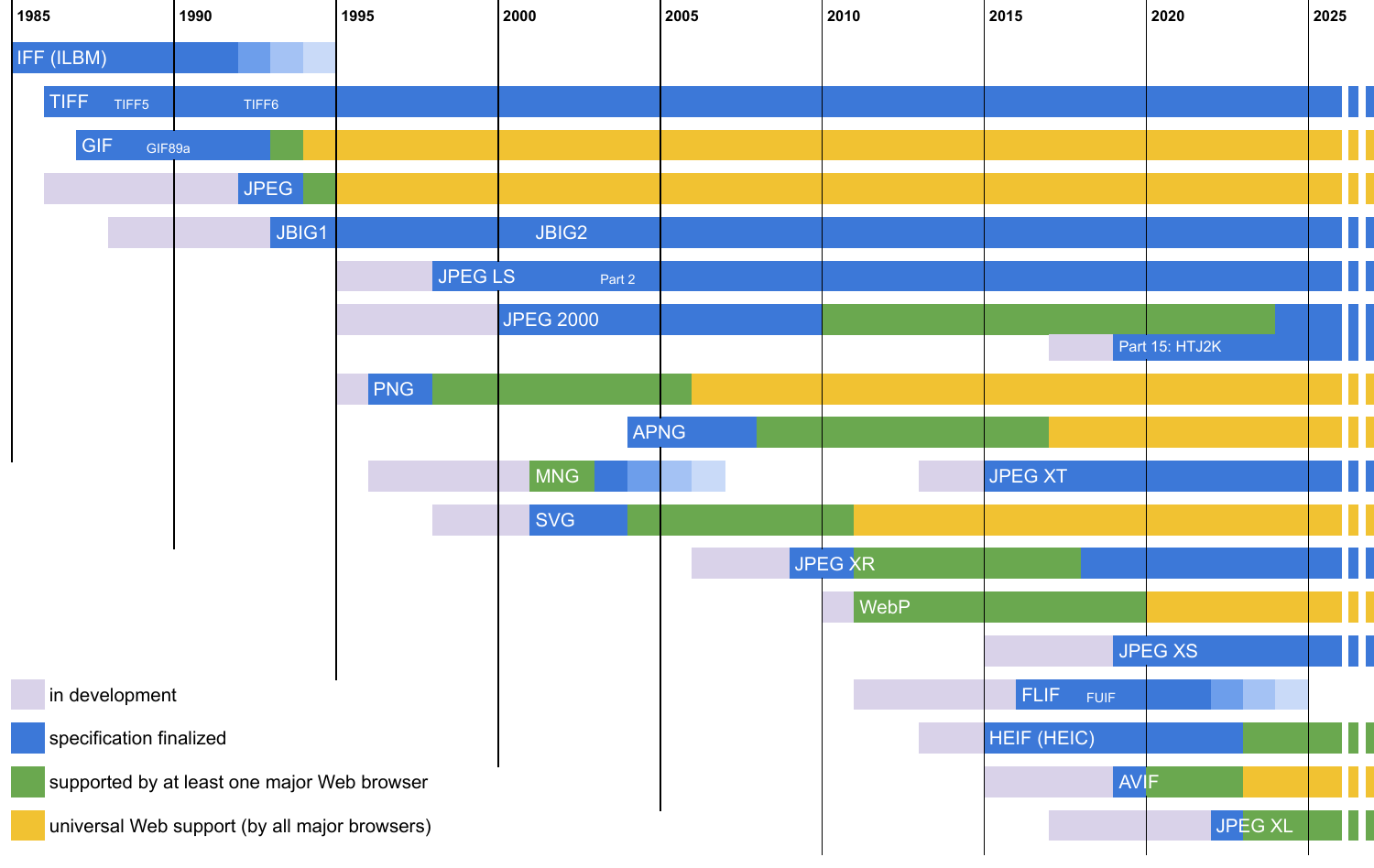}
\caption{Timeline of image formats and their adoption on the Web.}
\label{fig:timeline}
\end{figure*}

It is not within the scope of this \ThisName{} to give a detailed and exhaustive overview of the various image codecs and related work that preceded JPEG~XL.
However we will discuss the most notable precursors, focusing on lessons that can be drawn from them, and key elements that have been either preserved or changed in JPEG~XL.
Figure~\ref{fig:timeline} gives an overview of image codecs on a timeline.

\subsubsection[IFF]{IFF (Interchange File Format)}
The IFF format, introduced in 1985 by Electronic Arts and Commodore, can be considered the `grandmother of container formats'.
It consists of chunks identified by a four-byte type identifier (consisting of human-readable ASCII characters) and a big-endian 32-bit size field determining the length of the chunk payload. This mechanism allows implementations to skip any unknown chunks, allowing new chunk types to be added in a backwards compatible way.
Microsoft's RIFF container was inspired by IFF, and so was Apple's QuickTime container format, which would become the basis for ISOBMFF (e.g. MP4 and HEIF).
Also arguably the PNG chunk structure was inspired by IFF.

The Interleaved Bitmap (ILBM) format, popular on the Commodore Amiga platform, was based on IFF. It offered palette images with up to 256 colors (with later extensions to 24 bit RGB), with a simple optional compression based on a run-length encoding (RLE) scheme known as PackBits.

\subsubsection[GIF]{GIF (Graphics Interchange Format)}
Introduced in 1987 by CompuServe, GIF was the first major interchange format for digital images \cite{gif_etc}. It only supports palettes of up to 256 colors, which was also a limitation of contemporary display technology, so it was considered a `lossless' image format.
GIF is based on Lempel-Ziv-Welch (LZW) entropy coding, which was a substantial improvement over simple Run Length Encoding (RLE) or no compression at all. In particular for non-photographic images, it brought significant compression gains.
A revision in 1989 brought simple alpha transparency (bi-level only) and support for animation. In the early days of the world wide web, GIF was supported by the first graphical web browsers.

\paragraph{Composite images.}
GIF introduced the notion of an image canvas on which frames with potentially smaller dimensions are positioned in order to create a composed image. This mechanism also exists in JPEG~2000 (\ref{jpeg2000}, APNG and MNG (\ref{png}), HEIF \ref{heif}, and JPEG~XL. In animated GIF (or JPEG~XL), this mechanism can be used as a simple inter-frame coding tool. The frame durations can vary from one frame to the other in both GIF and JPEG~XL.


\paragraph{Interlacing.}
Another idea pioneered by GIF was the notion of progressive decoding: a preview of the image can be shown while the image is still being transferred, using partial image data to obtain a lower-resolution version of the full image. In GIF this is accomplished using a simple 4-pass row interlacing scheme: first every 8th row is coded, then every missing 4th row, every missing even row, and finally all odd rows.
The `Venetian blinds' style of progressive rendering this produces is rather primitive compared to what would later be possible in JPEG and JPEG~XL, but in the days of dial-up internet access, it was nevertheless a very much appreciated feature.

\subsubsection[JPEG]{JPEG (Joint Photographic Experts Group)}
The single most successful image codec, by far, is the JPEG codec \cite{jpeg}. Standardized in 1992 and somewhat confusingly named after the committee that created it, it was the first \emph{lossy} image codec. For photographic images, JPEG quickly became the codec of choice, since it allowed visually lossless compression at bit rates an order of magnitude lower than in any other image format. Digital cameras would not have been economically feasible without JPEG. Web browsers quickly added support for JPEG and it would soon become the most ubiquitous image format ever.

\paragraph{FOSS implementation.}
Part of JPEG's success was due to the fact that its baseline mode was royalty-free, and that free and open source software (FOSS) was available early on: the libjpeg library by the Independent JPEG Group (IJG), initially mainly authored by Tom Lane, was first released already in 1991. It however also lead to a \emph{de facto} standard that was only a subset of the full 1992 standard: patent-encumbered elements like arithmetic coding, and modes that had less practical relevance in the 1990s (like 12-bit precision, hierarchical and lossless mode) were simply not supported in most deployments, so they did not become part of the \emph{de facto} JPEG format.
Nevertheless, even with the limitations of the \emph{de facto} JPEG format, it has still become an extremely popular and widely adopted image format, to the point that even three decades after its introduction, it is still the most widely used image format.

One of the important lessons drawn from the history of JPEG, was that having a high-quality FOSS implementation available early on, which fully implements the codec, is crucial for gaining traction and adoption and to avoid creating a \emph{de facto} standard that is only a subset of the full standard.
Patent-encumbered technology was avoided and significant effort was spent on ensuring that the reference implementation (libjxl) was effectively `ready to use' for early adopters: no licensing obstacles, a high quality of implementation, including an encoder with good perceptual optimization.

\paragraph{Legacy.}
There are trillions of JPEG images in existence.
A key element in the design of JPEG~XL was to ensure that JPEG~XL would become a superset of the \emph{de facto} JPEG format, in the sense that existing JPEG image data can also be represented in JPEG~XL. JPEG is based on the Discrete Cosine Transform (DCT) applied on $8 \times 8$ blocks of sample values, quantized according to an arbitrary quantization table, and typically applied to YCbCr components where the chroma components (Cb, Cr) are optionally subsampled. It is possible to take an existing JPEG image and store the exact same quantized DCT coefficients in a JPEG~XL image, allowing a lossless transcoding. 

The entropy coding used in JPEG is Huffman coding, which has the advantage of being fast and simple, but it can be improved. 
Arithmetic coding is part of the JPEG standard and it brings better compression, though it is not part of the \emph{de facto} JPEG format so it cannot be used in a fully interoperable way.
In JPEG~XL, Huffman coding is still an option (which can be useful in case very fast encoding is required) but there are also better options available, including Asymmetric Numeral Systems (ANS) and an optional use of LZ77.
Additionally, JPEG~XL adds sophisticated context modeling options, which further improves entropy coding results. 
Since the entropy coding step in both JPEG and JPEG~XL is itself lossless, existing JPEGs can be converted to JPEG~XL while taking advantage of the improved entropy coding options, reducing the file size by roughly 20\% (see also Figure~\ref{fig:lossless_jpeg_results} in \SectionName{}~\ref{lossless}).

\paragraph{Parallelism.}
JPEG was designed for the single-threaded processors of its time and the opportunities for parallel decoding are limited. Due to the design of the JPEG~XL's data format, which enables the use of both instruction-level and multi-threaded parallelism, JPEG~XL exhibits superior encoding and decoding speed compared to the otherwise lighter and simpler JPEG standard on multi-core processors. This efficiency improvement is also observed when using losslessly recompressed JPEG source data.

\paragraph{Progressive decoding.}
Although both JPEG and JPEG~XL support progressive decoding, their underlying methodologies differ. In legacy JPEG, progressive mode is optional; consequently, the majority of existing files decode sequentially with top-to-bottom rendering. By contrast, JPEG~XL’s VarDCT mode forces the encoder to provide a baseline level of progression. The authors contend that mandating this feature will incentivize developers to implement progressive rendering more broadly across applications such as image viewers, thumbnail generators, and web browsers. To balance utility with data overhead, an 8x8 subsampled progression was selected as this guaranteed baseline. Furthermore, JPEG~XL gives encoders greater flexibility beyond this requirement, allowing for custom refinement orders---such as center-first---which are not possible in legacy JPEG.

\subsubsection[PNG]{PNG (Portable Network Graphics)}
\label{png}
When CompuServe decided to use LZW compression in the design of GIF, they were not aware that it was patented by a company that would later become Unisys. 
In 1994, on December 24th, Unisys announced that they would enforce the patent and require fees to be paid by commercial on-line information service providers.
This lead to widespread condemnation and the development of a royalty-free replacement format called PNG \cite{png} --- officially ``Portable Network Graphic'', but it is 
pronounced ``ping'' and also considered to be a recursive acronym meaning ``PING Is Not GIF''.

PNG is based on Deflate compression, which is the combination of LZ77 sliding window dictionary coding with Huffman coding; it is the same method used in the ZIP and gzip file formats.
Compared to GIF, PNG brought substantial improvements:
besides palettes, it also supports true color (8-bit and even 16-bit per component).
Instead of the simple bi-level transparency of GIF (either fully opaque or fully transparent), it supports semitransparent pixels, allowing smooth anti-aliased edges in the transparency mask as well as modeling shadows and translucent regions.

Another improvement over GIF is the option to use a 7-pass interlacing scheme (known as Adam7), which results in better progressive previews than in GIF.
It does however come at a cost in compression performance.

PNG became a popular interchange format since it offers lossless compression and has the color precision and alpha transparency support required for authoring workflows.
Also on the web it quickly became universally supported and surpassed GIF (but not JPEG) in terms of usage.

\paragraph{Animation.}
However, PNG did not initially support animation. There was disagreement on how to extend PNG with animation support. One proposal was MNG (Multiple-image Network Graphics) which is rather complex and includes the option to use JPEG payloads, also known as JNG. The other proposal was APNG (Animated PNG) which is much simpler.
Eventually APNG won, but it would take until 2017 before all major browsers supported it.

Because PNG lacked animation support for nearly two decades and the patents on GIF had already expired by 2004, PNG did not fully displace GIF in common usage.
While for still images, GIF had been rendered obsolete by the combination of JPEG and PNG, it remains the only widely supported animated image format, to the point that ``GIF'' has become a synonym for ``short looping animation'' regardless of the actual codec that is used.

\paragraph{Universality.}
While PNG can be used in lossy ways (e.g. by reducing the colors and using a palette instead of preserving the original colors), it is considered to be and was designed specifically as a lossless image format.
Although the JPEG standard did have a lossless mode, it did not become part of the \emph{de facto} JPEG format, so ``JPEG'' is to this day associated with inherently lossy compression.

In other words, by the end of the 1990s, three different image formats had emerged that were universally supported and that each had their specific role: JPEG for lossy photographic images, PNG for lossless and for non-photographic images (for which the DCT is not very effective), and GIF for animations.

In the design of JPEG~XL, one of the goals was to create a universal, \emph{general-purpose} format that could replace not just JPEG but also PNG and GIF. A lesson learned from PNG was that if the new format does not fully cover all of the functionality of the old format it intends to replace, the old format will inevitably remain in use too.
Although PNG has been a very successful image format, it failed to render GIF obsolete since it lacks support for animation.
Something similar happened with lossless WebP (see \ref{webp}), which failed to make PNG obsolete since it lacks support for high bit depth and large image dimensions.
In JPEG~XL, covering at least all the functionality available in JPEG, PNG, and GIF,
without any exceptions, was considered an important target.
This would make it possible to use a single format for all types of images.

\subsubsection[TIFF]{TIFF (Tag Image File Format)}
Predating even GIF, the first version of TIFF was released in 1986 by Aldus Corporation, which would later be acquired by Adobe.
TIFF \cite{tiffspec,tiff} is a rather complicated and versatile format that can contain many different payload codecs and image data types.
Over the years, many extensions of TIFF have been introduced, including many proprietary ones.

Baseline TIFF (Part 1 of TIFF 6.0) offers limited options in terms of compression: only uncompressed and the simple PackBits RLE scheme are supported.
Better compression options, including LZW, Deflate, JPEG, JBIG, JPEG~2000, JBIG2,
and even JPEG~XL are available in various extensions of TIFF.
Few if any applications support all extensions of TIFF though, so in practice, if interoperability is desired, one has to stick to the baseline format.

\paragraph{Authoring workflows.}
However, the flexibility of TIFF also makes it a widely used format in authoring workflows and the publishing industry.
Even after the introduction of JPEG and PNG, until JPEG~2000 was introduced, TIFF remained the only interoperable image format that can represent CMYK images losslessly, store multi-layer or multi-page images, or represent multi-spectral images like satellite imagery.

One of the advantages of TIFF for authoring workflows is that it can be encoded quickly. This is mostly due to the simplicity of its (baseline) compression, but also because it supports tiles. This is important when a large image needs to be saved frequently, e.g. during editing.

It was a goal in the design of JPEG~XL to have most (if not all) of the features of TIFF, including CMYK support, layered and multi-page images, no limitations on the number of components, and fast encode options.
This rich feature set makes JPEG~XL a feasible replacement for TIFF as an interchange format in authoring workflows.

\subsubsection{JPEG 2000}
\label{jpeg2000}
JPEG 2000 \cite{j2k-overview,j2k-suite}, named for its year of standardization, was developed as the successor to the original JPEG format. However, its gains in compression performance over JPEG were only modest, especially within the typical quality range for which JPEG was commonly used. As the conclusion of \cite{j2k-overview} notes:
``JPEG-2000 is unlikely to replace JPEG in low complexity applications at bitrates in the range where JPEG performs well. However, for applications requiring either higher quality or lower bitrates, or any of the features provided, JPEG-2000 should be a welcome standard.''

\paragraph{Flexibility.}
The main advantage JPEG~2000 brings, is not in compression performance but in terms of functionality and flexibility. It offers a lot of control over the bitstream ordering: an image can be split into \emph{tiles}, which are split into \emph{precincts} which results in \emph{packets} for every quality \emph{layer}. The packets can be almost arbitrarily reordered within a codestream. This leads to great flexibility, though also to considerable decoder implementation complications.

In JPEG~XL, the design of the bitstream ordering syntax tries to find a sweet spot between complexity and flexibility. The tiles, precincts, packets and layers in JPEG 2000 terminology correspond respectively to frames, groups, sections and passes in JPEG~XL terminology.
Compared to JPEG~2000, there is somewhat less flexibility in JPEG~XL in the way the bitstream is ordered: while sections can be permuted arbitrarily, some aspects like group dimensions are largely fixed. The main trade-off aimed for in the JPEG~XL design, is to avoid adding flexibility if it comes at a substantial computational cost for decoding, especially if it offers no or hardly any additional functionality (besides potential encoder convenience).

\paragraph{Adoption.}
JPEG~2000 gained traction in specific niches such as medical imagery (DICOM), digital cinema (DCP), satellite imagery and meteorological data (GIS).
It is also one of the payload codecs that can be used to encode raster images in PDF (since PDF version 1.5).
However it did not become as ubiquitous as JPEG.
One plausible factor is that, for typical consumer use cases, JPEG~2000’s compression gains were not seen as large enough to justify switching from JPEG, which was already `good enough' in many workflows.
Another reason was a certain nervousness about potential patent issues. Even though the baseline JPEG~2000 codec (Part 1 of the standard) was designed to be royalty-free, recent experiences with Unisys' claims on GIF, followed by a company called Forgent Networks that made patent claims on JPEG (which would eventually be invalidated, but that took until 2006), lead to a high level of \emph{fear, uncertainty, and doubt} when it came to adopting new codecs, especially highly complicated ones like JPEG~2000.

Finally, possibly the biggest obstacle to widespread adoption of JPEG~2000 was the lack of a high-quality free and open source implementation for well over a decade after the codec was introduced.
The reference software JasPer was available from the start, and was released with an open source license in 2004, but it was designed for reference, not performance, making it rather slow and impractical.
The Java-based JJ20000 implementation was in a similar situation. 
This caused JPEG~2000 to get a reputation of being too computationally demanding.
Commercial implementations such as Kakadu are fast, but not free.
The OpenJPEG library is both free and reasonably fast, but it took until 2014 before it was sufficiently mature to be fully conformant with Part 1 of the standard.

To avoid such obstacles for the adoption of JPEG~XL, the development of the standard went hand in hand with the development of a fast and free implementation, libjxl.

\subsubsection[LS]{JPEG LS}
In a 1993 addition to the JPEG standard, a lossless mode was added. It did not become part of the \emph{de facto} standard and is generally not supported by JPEG decoders, although it does get used as a payload codec in the DICOM and DNG file formats.

JPEG LS \cite{jpegls}, introduced in 1999, was designed to improve upon lossless JPEG while keeping the complexity low. Its compression performance is substantially better than that of lossless JPEG and comparable to that of lossless JPEG~2000, while it is simpler and faster than JPEG~2000.

However, for non-photographic images such as screenshots, charts, logos, illustrations, and pixel art, the compression performance of PNG is better than that of both JPEG~LS and JPEG~2000. For this type of image content, the dictionary-based entropy coding (and the option of using a color palette) in PNG brings a clear advantage.
This probably explains why JPEG~LS saw limited adoption in scenarios where PNG was already entrenched.

JPEG~XL brings substantial lossless compression improvements over PNG regardless of the type of image content (photographic or not). Unlike earlier JPEG codecs, it also includes coding tools specifically aimed at non-photographic images.

\subsubsection[JBIG2]{JBIG2 (Joint Bi-level Image Experts Group)}
When in the 1970s and 1980s fax machines, photocopiers and printers transitioned from analog to digital, at least initially the focus was mainly on bi-level images (1-bit, black and white).
The \emph{Comité Consultatif International Téléphonique et Télégraphique} (CCITT, now known as ITU-T) created standards for fax machines known as Group 3 (ITU-T T.4) and Group 4 (ITU-T T.6). These are based on a combination of Run Length Encoding with Huffman coding, 
where lines are predicted by the line above.

JBIG1 was introduced in 1993 as an improvement over Group 4 compression.
In 2000, JBIG2 was released as a further improvement over JBIG1.
It remains the most advanced standardized format for bi-level images, and is part of the PDF standard since PDF version 1.4.

One of the key coding tools of JBIG2 is pattern matching and substitution (PM\&S), with the option of doing soft pattern matching (SPM) with refinement data.
Effectively this is a generalization of LZ77 dictionary coding to allow referencing 2-dimensional patterns rather than just 1-dimensional strings.
Since refinement data is optional, it is possible to apply JBIG2 in a lossy way by allowing `close enough' patterns to be used as a match.

\paragraph{Fidelity.}
There was some controversy around JBIG2 when in 2013 it was reported that Xerox Workcentre photocopiers were producing substitution errors where scanned documents such as construction blueprints would have altered numbers, e.g. replacing a ``6'' by an ``8''. This turned out to be caused by applying lossy JBIG2 compression.
Unlike the artifacts of overly aggressive JPEG compression, where there would be obvious DCT noise or blurring, these artifacts of excessive JBIG2 compression are not at all obvious.
This has caused government institutions in Germany and Switzerland to recommend not to use JBIG2.

In JPEG~XL, the `Patches' coding tool (see \SectionName{}~\ref{patches}) can be seen as a variant of JBIG2's pattern matching. While in theory it is possible to create an encoder that causes substitution errors like the ones that caused controversy for JBIG2, it is hard to do so by accident in JPEG~XL: the Patches are added on top of the image data, which still provides `refinement data' that would have to be explicitly removed by an encoder in order to get substitution errors.
Moreover, in the libjxl encoder, if Patches are used at all, they are used only in a very conservative way which cannot lead to substitution errors even if the refinement data would be removed.
In general, an overall design goal for JPEG~XL was to allow lossy encoding with a consistently high image fidelity.

\paragraph{Security.}
More recently, in September 2021, there was again bad press for JBIG2 when the `ForcedEntry' security exploit (CVE-2021-30860) was discovered to have been used to target political dissidents and human rights activists.
This was a ``zero-click'' exploit in iOS that allowed compromising phones without requiring any user interaction, simply by sending a malicious message.
It consists of a file with a \verb|.gif| filename extension that actually contains a PDF with a JBIG2 payload. By pretending it was a GIF file, it would automatically be sent to the ImageIO library for decoding, which would then happily parse the PDF file and decode the JBIG2 image inside.
The combination of an integer overflow bug in the implementation of the JBIG2 decoder with an extremely expressive bitstream syntax that makes JBIG2 Turing complete, allowed the malicious actors to effectively use JBIG2 as a scripting language that could be used to circumvent Apple's ``BlastDoor'' sandbox and install spyware.

In the design of JPEG~XL, while the bitstream is very expressive, anything that would cause it to become Turing complete was avoided.
While implementation bugs can always cause security issues, during libjxl development, a lot of effort has been spent already (and keeps being spent) on minimizing the risk, using static and dynamic code analysis tools, automated testing and fuzzing techniques.

\subsubsection{WebP}
\label{webp}
In 2010, Google announced the WebP \cite{webp} format as a replacement for JPEG on the web.
It is based on an intra-only subset of the VP8 video codec, which was released as a royalty-free format after Google acquired On2 Technologies, the company that had designed VP8. WebP lacks the progressive decoding feature of JPEG, a limitation inherited from its video codec origins where such a feature is largely irrelevant. While not truly progressive, WebP's libwebp library does support incremental decoding, rendering the image from top to bottom as the data arrives.

In 2012, following the promises made in the initial launch and the following public feedback on lossy WebP, Jyrki Alakuijala developed a new image compression format to add lossless and transparency coding capabilities to WebP. This new codec consistently outperforms PNG in compression. Key innovations in the Lossless WebP were the \emph{entropy image} concept --- a two-dimensional entropy coding map that was a simpler and more rigid version of what later evolved into \emph{context maps} in Brotli and JPEG~XL, the \emph{Select} nonlinear predictor, and two-dimensional distance codes later also used in JPEG~XL modular mode. Thanks to its good browser support, lossless WebP is currently the top choice for web delivery of logos, screen content, and other images that benefit from pixel-based compression. 

With the addition of lossless compression and transparency, WebP was ready to replace JPEG, PNG, and GIF on the web. Because of some of the limitations inherited from VP8, such as having only support for 8-bit limited-range YCbCr with 4:2:0 chroma subsampling, 8-bit limited-range RGBA, and a maximum resolution of $16383 \times 16383$ pixels, WebP cannot replace formats such as PNG or TIFF in professional authoring workflows.

\paragraph{Scope.}
As the name suggests, WebP was designed specifically for the Web.
This explains why certain bitstream limitations --- in particular, forced chroma subsampling, the limitation to 8-bit precision, and limits on the image dimensions --- were considered acceptable or even desirable.
It also explains why there was relatively little focused effort on software support for WebP outside web browsers. For example, while Chrome has supported WebP since early 2011 (i.e., even before the format was finalized), it took until 2018 for GIMP to add WebP support, until 2022 for Adobe Photoshop to add support, and until 2023 for Microsoft Windows Photos to add support.
This situation has led to frustrating end-user experiences where an image saved from a web page could not be opened in any other program than the browser.

By contrast, JPEG~XL explicitly targets  a wide range of use cases \cite{jxl-use-cases}, rather than focusing on a single primary application.
Nevertheless, web delivery was considered to be one of the major application domains during the design of JPEG~XL, which led to some specific bitstream features like minimal header overhead and progressive decoding that are especially useful for images on the web.

\paragraph{MozJPEG.}
Google actively promoted WebP from the beginning, among others by suggesting to web developers using its PageSpeed Insights tool or Lighthouse audits to replace JPEG and PNG images by WebP.
However Mozilla (Firefox) and Apple (Safari) were not convinced that WebP actually does bring a `big-enough improvement' over well-optimized JPEG.
To demonstrate their point, Mozilla started developing a new JPEG encoder called MozJPEG \cite{mozjpeg} with better compression performance than the commonly used libjpeg-turbo encoder, at the cost of a slower (but still acceptable) encoding speed.
Due to the format's popularity, eventually both Firefox and Safari added WebP support --- in 2019 and 2020, respectively.

\subsubsection[FLIF]{FLIF (Free Lossless Image Format)}
Introduced in 2015, the Free Lossless Image Format \cite{flif} was created by Jon Sneyers and Pieter Wuille to improve the compression performance of PNG.
The bitstream syntax includes a signaled chain of transforms, such as the YCoCg-R color transform and a Palette transform --- this mechanism would eventually evolve into JPEG~XL's Modular mode.
Besides the usual `scanline order' (row by row, from left to right in each row), FLIF also includes an interlaced sample ordering, referred to as `Adam-$\infty$' interlacing, which generalizes the Adam7 interlacing of PNG.
In contrast to PNG, the interlacing scans are not coded independently, but samples from previous scans are used in the prediction and context model for the current sample.
This makes it possible to refer to already-decoded neighboring samples at 7 of the 8 adjacent positions, as opposed to only 4 out of 8 (W, NW, N, NE) when using the scanline order.

A careful ordering of the interlacing scans that prioritizes Y over Co and Cg, in combination with an interpolation algorithm to fill the missing samples, causes the progressive previews to look substantially better than in interlaced PNG.
It also allows to use FLIF in a lossy way by simply truncating a file.

The main innovation in FLIF is the use of a Meta-Adaptive context model, which consists of a signaled decision tree (called the MA tree). This is a binary tree where each node tests the value of a local `property' against a threshold value, with one branch for the `above the threshold' case and another for the `below or equal' case.
Properties include the sample values at the current location in other, already-decoded components, differences between neighboring already-decoded samples, and the predicted value of the current sample.
The property indices and threshold values are signaled, leading to a context model that can itself be adapted by an encoder to the image contents.

\paragraph{Speed.}
The high compression density of FLIF is a direct result of its core components: the interlacing method, a highly conditional context model, and a complex variant of CABAC for entropy coding. However, these components constrain its decoding speed. The intricate data dependencies introduced by the interlacing and context modeling inhibit effective parallel processing and lead to poor memory locality. Furthermore, the frequent use of computationally expensive branching instructions, integral to its compression strategy, creates a performance bottleneck during decoding.
Decoding speed is a crucial element for image compression, particularly for the web, where slow decoding can nullify the gains of a faster file transfer. When the total time to display an image to the user is not reduced, the main purpose of data compression is defeated. These factors as well as non-competitive lossy image quality likely explain why FLIF saw very limited adoption and was never widely implemented in web browsers.

While JPEG~XL's Modular mode was heavily inspired by FLIF, it was redesigned with an emphasis on high-speed decoding. To this end, during JPEG~XL development phase, Luca Versari replaced several of FLIF's core components with more performant alternatives. The entropy coding transitions from a complex, branch-heavy CABAC variant to a higher-throughput ANS implementation with static probabilities, requiring the most heavy computation to be done at encoding time only. Furthermore, FLIF's intricate interlacing method was eliminated in favor of an optional Squeeze transform, which offers the dual benefits of higher-quality progressive decoding and an increase of locality in memory access patterns during decoding.

\subsubsection{Brotli}
In 2013, Jyrki Alakuijala and Zoltán Szabadka introduced Brotli, a novel general-purpose data compression format that offers superior compression compared to the established Deflate (gzip) method, originally developed for W3C's Web Open Font Format 2 (WOFF2). Brotli uses prefix coding and --- similar to LZ4 and Zstandard --- replaces duplicate data with quadruplet metadata of [copy-length, copy-offset, insert-length, insert-string]. Brotli distinguishes itself from other fast general-purpose data compressors through the incorporation of 2\textsuperscript{nd}-order context modeling, a predefined dictionary, a move-to-front transform for block switches, and joint probability encoding for copy-length and insert-length. Approved in 2016 by the Internet Engineering Task Force (IETF) \cite{brotli}, it became widely adopted as a content encoding for compressed HTTP transfer.

\paragraph{Context map.}
One of the key innovations in Brotli is the introduction of a \emph{context map}, which allows the use of a relatively large context model without the overhead of signaling a large amount of prefix codes.
This concept was also used in JPEG~XL, where it is used in both VarDCT and Modular mode. In combination with MA trees, the resulting data structure is effectively a directed acyclic graph (DAG).
It is a crucial ingredient to allow combining a large, meta-adaptive context model with static probability distributions.

\begin{figure*}\centering \small
\begin{tabular}{llll}
libjpeg-turbo -quality 50 & & libjpeg-turbo -quality 10 & \\
\multicolumn{2}{l}{\includegraphics[width=0.47\linewidth]{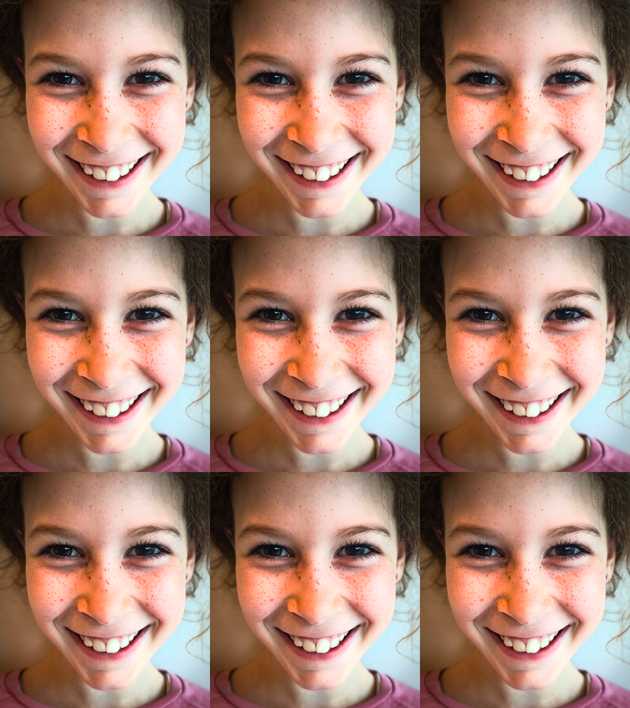}} &
\multicolumn{2}{l}{\includegraphics[width=0.47\linewidth]{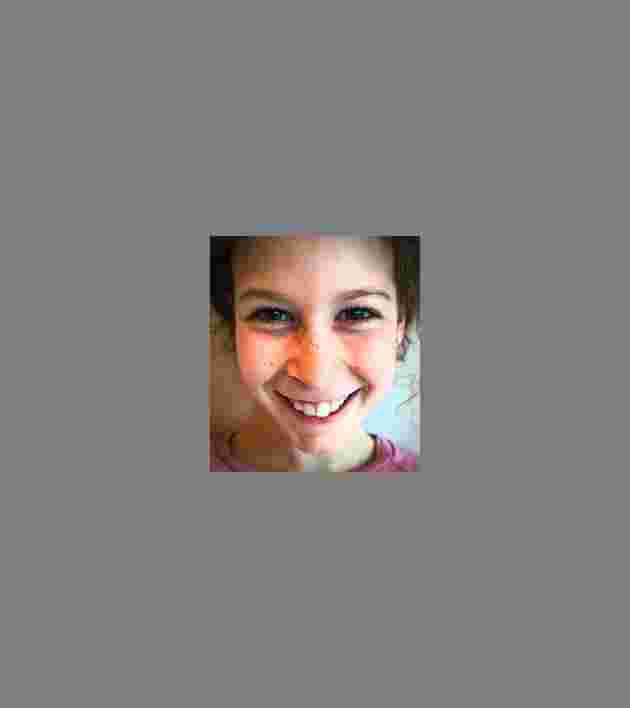}} \\
PSNR-Y: 36.192 & 
SSIM: 0.9435 & 
PSNR-Y: {\bf 37.755} &
SSIM: {\bf 0.9797} \\
MS-SSIM: 0.9931 &
NLPD: 0.1009 &
MS-SSIM: {\bf 0.9946} &
NLPD: {\bf 0.0770} \\
VMAF: {\bf 92.496} & 
Butteraugli p-norm: {\bf 1.598} & 
VMAF: 79.307 &
Butteraugli p-norm: 4.281\\
SSIMULACRA 2.1: {\bf 70.75} & 
CVVDP: {\bf 9.5512} & 
SSIMULACRA 2.1: 49.63 &
CVVDP: 9.2752\\
\end{tabular}
\caption[Non-correlation of  $L^2$ metrics with perceptual quality due to homogeneous backgrounds.]{Objective metrics can give deceiving results 
when an image has an `easy' background. The best (highest quality) metric score is indicated in {\bf bold}.
The image on the left has a better visual quality than the image on the right, which has a large amount of padding causing inflated scores for metrics based on an $L^2$ norm. This issue is of practical relevance for subjects captured against naturally uniform backgrounds, such as a bird in flight or a portrait against a wall. 
}
\label{fig:metric-norm}
\end{figure*}

\subsubsection[Butteraugli, XYB]{Butteraugli and XYB Color Space}

The core technologies of the Pik image format (and later JPEG~XL VarDCT mode) originated from the creators of Brotli and WebP lossless. Pivoting to lossy compression, their first innovations included the Butteraugli metric and XYB color space by Jyrki Alakuijala, the Brunsli JPEG recompressor by Zoltán Szabadka, and the Guetzli encoder, a joint project led by Alakuijala.
A significant outcome of this research was Butteraugli, a psychovisual distortion metric developed in 2016 to facilitate the optimization of high-fidelity image encoders. Butteraugli is based on a computational model of the human visual system and introduces the XYB color space, which is designed to approximate the photoreceptor response of the human retina. The effectiveness of this color space led to its subsequent adoption for internal data representation in the Pik, Jpegli, and lossy JPEG~XL codecs.

Unlike simpler metrics, Butteraugli creates a detailed heatmap of perceptual differences by considering complex color and masking effects. It then uses a distinct aggregation method to provide a single, reliable score. This made it an ideal tool for guiding optimization decisions in the development of high-quality encoders like Guetzli, Pik, and JPEG~XL.

To evaluate image quality, metrics traditionally average the error using an $L^1$ norm (mean absolute error) or $L^2$ norm (PSNR). However, this can be misleading. For example, in a product photo with a large white background, these metrics might report a high score even if the product itself is poorly compressed, because the ``easy'' background skews the average. Butteraugli avoids this pitfall by focusing on the worst errors. It uses higher norms, such as the max norm ($L^\infty$) or a specialized "3-norm" (a mixture of  $L^3$, $L^6$, and $L^{12}$ norms), to ensure that the most significant visual flaws heavily influence the final score, leading to more reliable encoder decisions.

\paragraph{Improving JPEG1 coding.}
In 2017, Jyrki Alakuijala invented a novel variable dead-zone quantization technique that enabled adaptive quantization within traditional lossy compressors. This method was showcased in the Guetzli encoder, guided iteratively by Butteraugli measurements. Guetzli demonstrated significant compression gains and provided valuable user feedback for subsequent designs.
The Guetzli JPEG encoder, too slow for widespread use, serves as a proof-of-concept that focusing on the highest quality range can lead to substantial advancements in JPEG compression techniques \cite{guetzli-study}.
During the JPEG~XL development work Alakuijala refined such optimizations to be made heuristically at practical coding speeds, and found a way to add more than 8 bits of color dynamics in the old 8-bit JPEG formalism. Zoltán Szabadka integrated all these ideas into a new JPEG Encoder/Decoder pair, opensourced as Jpegli \cite{jpegli-study}, a fast JPEG codec with perceptual optimizations.
It outperforms libjpeg-turbo in terms of quality.

\subsubsection{Brunsli}
Numerous systems have been developed for the lossless recompression of JPEG images, including the proprietary software StuffIt by Alladin Systems and open-source projects like PackJPG \cite{packjpg}, which later influenced the Lepton format. The increasing proliferation of digital photography and the growing dimensions of image files have intensified the need for efficient JPEG storage solutions, driving the development of these specialized recompressors. Zoltán Szabadka and Jyrki Alakuijala introduced Brunsli, a computationally efficient JPEG recompression tool which had been initially integrated into the Pik (2017) image format and was subsequently released as a standalone utility (2019).

Specialized lossless JPEG recompression tools can typically achieve file size reductions of approximately 20\%, a significant improvement over the 1--3\% gains, if any, offered by applying general-purpose compressors to a JPEG file. 

Although not an initial requirement by the JPEG Committee for the JPEG~XL standard, the inclusion of JPEG recompression functionality was strongly advocated by key contributors Jyrki Alakuijala and Jon Sneyers. They successfully persuaded the committee of its significance as a critical feature for adoption. Initially, this capability was realized by integrating the Brunsli recompression algorithm directly into the JPEG~XL framework. Luca Versari subsequently refined the JPEG~XL specification to a more integrated solution where JPEG recompression is incorporated into the format's existing VarDCT (Variable-blocksize DCT) mode.

This design evolution streamlined the process, requiring only a small, optional payload of ``JPEG bitstream reconstruction data'' (\verb|jbrd|) to be stored separately. The image is decodable without this data, and this supplementary data is exclusively required for the bit-exact reconstruction of the original JPEG file. The  (\verb|jbrd|) data block includes metadata such as the entropy codes, the progressive scan script, the application of restart markers, and the specific values of padding bits.

The Brunsli algorithm influenced the design of the Discrete Cosine Transform (DCT) coefficient coding mechanisms employed in the PIK format and its successor, JPEG~XL. Several innovative techniques originating from Brunsli were adopted to enhance compression efficiency. Notably, in the coding of AC coefficients, Brunsli introduced an explicit counter for the number of non-zero values within a block, a method that supplants the traditional End-of-Block (EOB) symbol used in baseline JPEG. Furthermore, it moved beyond the static zig-zag scan by enabling custom coefficient orderings for more adaptive entropy coding. Finally, the mandated availability of 8x8 progressive decoding, a technique originating from Brunsli, was also incorporated into the design principles of these subsequent formats.

\begin{figure*}\centering
\includegraphics[width=1\linewidth]{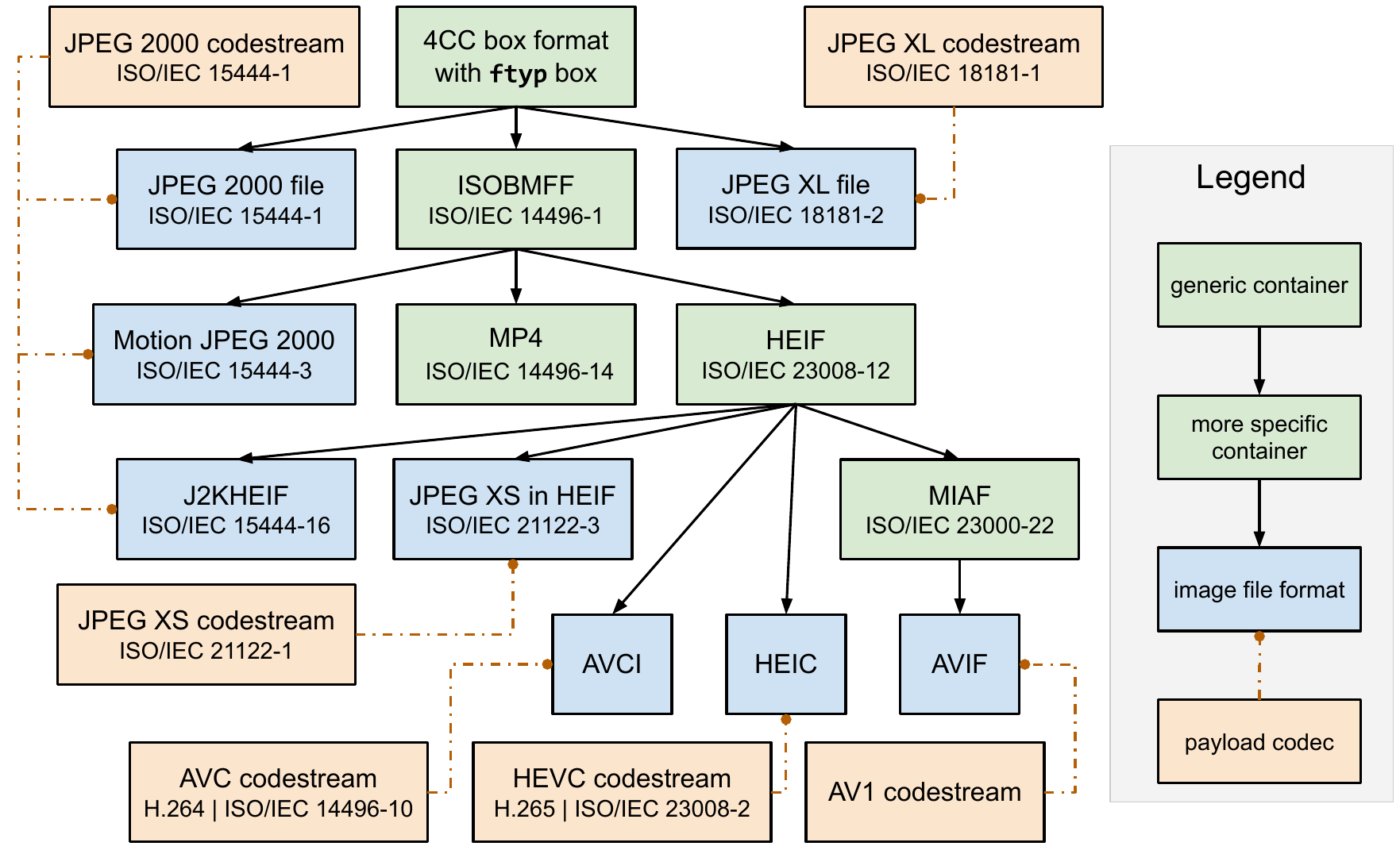}
\caption[Containers, payload codecs, and image file formats]{Overview of various containers, payload codecs, and image file formats.}
\label{fig:containers}
\end{figure*}

\subsubsection[HEIF]{HEIF (High Efficiency Image File Format)}
\label{heif}
Still image codec improvements had seemingly stalled after JPEG~2000. While JPEG~XR \cite{jpegxr} was introduced in 2009, it positioned itself between JPEG and JPEG~2000 in terms of compression performance, complexity, and features, and did not get traction outside of the Microsoft Windows platform.

Video codec research, on the other hand, was very active. While initially focusing on better inter-frame coding tools, advances were also made in intra-frame coding. In particular for low-bitrate compression, various coding tools such as deblocking filters were developed that can reduce visible artifacts and produce more appealing images than low-quality JPEGs.
WebP was the first still image format to leverage a video codec (VP8).

In 2013 MPEG introduced HEVC (H.265) \cite{hevc}, the successor to AVC (H.264). FFmpeg developer Frabice Bellard subsequently defined the still image format BPG (Better Portable Graphics) in 2014, as a container for intra-only HEVC payloads with optional metadata and a separately coded alpha channel.
Meanwhile, Nokia researchers created a different, more generic image container format called HEIF \cite{heif}
which became an MPEG standard (ISO/IEC 23008-12) in 2015.

The HEIF container is based on the ISO base media file format (ISOBMFF, ISO/IEC 14496-12), which is also the basis on which the MP4 container was built, as well as Motion JPEG~2000.
In a sense, HEIF bridges the feature gap between video codecs and still image formats, by adding constructs to overcome the limitations of video codecs and especially hardware implementations of video codecs.
For example, HEIF defines container-level tiling constructs that can be used to compose a larger image out of separately coded tiles, allowing the use of a hardware HEVC encoder with resolution limits.
It also defines a way to add alpha transparency as a separately coded single-component image, as well as syntax for embedding ICC profiles and metadata, and instructions for image cropping and orientation.

HEIF containers can have various payload codecs, including JPEG and AVC, but are most commonly used with a HEVC payload. The resulting files are called HEIC.
Apple adopted HEIC in 2017 and made it the default capture format for the iPhone camera.

In 2019, a constrained subset of HEIF was defined, called MIAF (ISO/IEC 23000-22), in an effort to reduce the complexity of implementing all of the format options available in HEIF.
Figure~\ref{fig:containers} illustrates the relationships between various container formats, payload codecs and image file formats.

\paragraph{Patents.}
Both the HEIF container and the HEVC payload codec are claimed to be covered by patents. Nokia does grant a royalty-free license for its patents related to HEIF, but only for non-commercial purposes.
For the HEVC codec itself, the patent licensing situation is complicated, to say the least.
Thousands of patents are claimed to be essential for implementing HEVC, and three patent pools were created:
MPEG Licensing Administration (MPEG LA, now merged into the Via Licensing Alliance),
HEVC Advance, and Velos Media.
Moreover not all companies claiming essential patents are part of any of the pools.

This situation has led to a fairly limited adoption of HEIC as an image codec; it did not really get traction outside the Apple ecosystem.

\subsubsection[AVIF]{AVIF (AV1 Image File Format)}
The AOMedia Video 1 (AV1) codec \cite{av1} was created in 2018 by the Alliance for Open Media, an industry consortium formed with the aim of developing open and royalty-free formats for multimedia delivery.
Introduced in 2019, the AVIF image format is based on the MIAF subset of the HEIF container, with an AV1 payload.

\subsection{JPEG~XL development}

In September 2016, at the IEEE ICIP conference in Phoenix, Arizona, the JPEG committee organized the Image Compression Grand Challenge. 
HEVC (the payload codec of HEIC)
and Daala (one of the precursors of AV1 and AVIF)
were judged to be the best lossy codecs at this time.
The Free Lossless Image Format (FLIF) was the best lossless codec.

Overall, the conclusion of the Grand Challenge was that significant improvements in compression efficiency could be obtained, which led to the creation of a new ad hoc group within the JPEG committee ("Ad Hoc Group on next generation image compression standard") in January 2017.
It was initially chaired by Jan De Cock (Netflix) and David Taubman (University of New South Wales).

\subsubsection{Call for Proposals}

In October 2017, a Draft Call for Proposals
\footnote{\href{https://jpeg.org/downloads/jpegxl/jpegxl-draft_cfp.pdf}{jpeg.org/downloads/jpegxl/jpegxl-draft\_cfp.pdf}} for a next-generation image coding standard (JPEG~XL)
was published, followed by a Final Call
\footnote{\href{https://jpeg.org/downloads/jpegxl/jpegxl-cfp.pdf}{jpeg.org/downloads/jpegxl/jpegxl-cfp.pdf}} in April 2018.

The call for proposals mentioned the aim to develop a standard for image coding that offers substantially better compression efficiency than existing image formats (e.g. $>60\%$ over JPEG), along with features desirable for web distribution and efficient compression of high-quality images, including higher resolution, higher bit depth, higher dynamic range and wider color gamut coding. 

Seven proposals were submitted in response to this call.
They were evaluated \cite{akyazi2019assessment} by the JPEG committee and presented and discussed at the October 2018 JPEG meeting in Vancouver, Canada.

\subsubsection{AVIF, FUIF, PIK}

The selection process for the JPEG~XL involved three primary proposals that were considered the most promising:

\begin{description}
\item[AVIF,] from the Alliance for Open Media (AOM), based on the AV1 video codec;
\item[PIK,] from the Google Research compression team;
\item[FUIF,] from Cloudinary, an evolution of FLIF.
\end{description}
The trajectory for JPEG~XL's development was significantly altered in early 2019. The JPEG committee had initially selected AVIF as the foundational codec in October 2018, but that proposal was later withdrawn. When the committee invited the PIK team to lead the effort, its lead Jyrki Alakuijala advocated for a new more collaborative strategy. Citing the strengths of the FUIF approach and knowing the inherent difficulties of building upon an unfamiliar codec, he proposed an unusual approach of full, day-one integration of the PIK and FUIF platforms instead of incrementally bringing features of FUIF into PIK. The committee accepted this recommendation, formally deciding in January 2019 to merge the two proposals to create the foundation for JPEG~XL.

\subsubsection{Collaborative phase}

The integration of disparate codec architectures into a unified framework presents a significant technical challenge. The precursor to JPEG~XL, known as PIK, exemplified this complexity by initially functioning as a container for three distinct subcodecs. The first was a lossy mode, which drew its perceptual modeling principles from Guetzli. The second was a lossless mode, architecturally derived from lossless WebP but enhanced with a self-correcting predictor developed by Alexander Rhatushnyak. The third was a JPEG recompression mode, which was functionally equivalent to Brunsli.

An early proposal for a combined codec further compounded this heterogeneity by incorporating a fourth subcodec, FUIF (Free Universal Image Format). These constituent codecs operated with near-total independence, each employing its own entropy coding schemes and internal data representations. For instance, the lossy PIK mode operated on three-component, 32-bit floating-point values; the lossless PIK mode handled up to four components of 8-bit or 16-bit unsigned integers; and FUIF was designed for an arbitrary number of components represented as 32-bit integers.

Luca Versari and Jon Sneyers led the collaborative effort to meaningfully unify these four distinct subcodecs into a single, cohesive architecture. The lossless mode of PIK and the FUIF codec were merged to form what is now the Modular mode of JPEG~XL. Concurrently, the lossy PIK mode evolved into JPEG~XL's VarDCT mode. In the initial design of the lossy mode, various auxiliary images—such as the low-frequency (LF) image, signaling for DCT block selection, adaptive quantization maps, chroma-from-luma predictors, Edge-Preserving Filter (EPF) sharpness modulation, and Patches—were handled through ad-hoc signaling methods.

Ultimately, the representation and signaling of these auxiliary images, along with alpha transparency and depth maps (which were generalized as 'extra channels'), were consolidated. They are now all encoded as Modular sub-bitstreams within the main bitstream. This architectural unification significantly simplified the overall codec design while simultaneously enhancing the expressivity and compression efficiency of the JPEG~XL format.


A significant simplification of the bitstream was achieved by eliminating the `Brunsli mode' for JPEG recompression. This involved two key changes: factoring out the JPEG bitstream reconstruction data (\verb|jbrd|) and generalizing VarDCT to allow for YCbCr components with chroma subsampling.

In order to enable parallel decoding and region-of-interest decoding, tiling was introduced in both VarDCT and Modular mode.
Additionally, progressive passes were added, allowing both VarDCT and Modular image data (with the Squeeze transform) to be encoded in a combined progressive bitstream.

Originally the FUIF codec used a range coder (a variant of CABAC) as the entropy coder while PIK used several different entropy coders including two variants of ANS and Huffman coding, as well as Brotli.
At one point there were six different entropy coding methods used in the combined codec.
These were then unified into a single entropy coding scheme that generalizes all the different methods: all symbols were eventually encoded using the HybridUint representation, which uses either prefix coding or ANS coding for the entropy-coded tokens plus a variable number of raw bits (for the high-entropy least significant bits of large-amplitude symbols), with the option of using LZ77 (distance, length) pairs.

Like FLIF, FUIF uses a meta-adaptive context model in which an MA tree is signaled and used to define an image-dependent dynamic context model.
Like Brotli, PIK uses a static large context model with a context map that clusters the contexts so only a small set of histograms needs to be signaled.
Both ideas were combined in JPEG~XL's Modular mode: effectively the addition of context clustering turns the MA tree into a directed acyclic graph (DAG) since leaf nodes with similar distributions can be merged.

While FLIF and FUIF use the MA tree only for context modeling and the predictor is signaled separately (and coarsely),
in JPEG~XL the MA tree also signals the predictor to be used.
The set of predictors was extended significantly, as well as the set of properties that can be used in the decision nodes of MA trees.

Coding tools were tested with an ablation study and some of them were eventually dropped. For example, Dot modeling was investigated as a specific coding tool for small ellipse-shaped image elements such as stars or planets in a night sky, which are hard for DCT-based compression. But it was removed after it was found that it did not perform substantially better than the alternative of using small Patches.

This development process took about two years.
By the end of 2020, the bitstream was considered frozen and libjxl version 0.2 was released.
The Final Draft International Standard (FDIS) was submitted in January 2021 for Part 1 (core codestream) and in April 2021 for Part 2 (file format).
The first editions of the ISO/IEC 18181 standard were published in October 2021 (for part 2) and in March 2022 (for part 1).

\section{High-level overview}
\label{overview}

The goal of JPEG~XL is rather ambitious: it aims to be a royalty-free general-purpose image codec that can become a universal interchange format for all use cases of still images, ranging from image capturing and authoring workflows to web delivery.
In terms of both features and compression, it aims to unify and replace the various existing image formats, including JPEG, JPEG 2000, JPEG XR, PNG, GIF, WebP, AVIF, HEIC, TIFF, and OpenEXR.

In contrast to WebP and AVIF, which focus on the specific use case of web delivery at `web quality', JPEG~XL targets a wide range of use cases. This includes lossless and very high-quality compression, as required in workflows where subsequent editing is expected.
It also targets printing (e.g. CMYK and spot colors), scientific and medical applications (e.g. multi-spectral and high-precision imagery), large images, and various trade-offs between encoding effort and compression.

\subsection{Parts of the ISO/IEC 18181 standard}

The JPEG~XL format is defined in ISO/IEC 18181. This standard consists of four parts:

\begin{itemize}
\item   18181-1: Core codestream \cite{18181-1}
\item   18181-2: File format \cite{18181-2}
\item   18181-3: Conformance testing
\item   18181-4: Reference software
\end{itemize}

\subsubsection{Codestream}

The core codestream contains all the data necessary to decode and display a still image (possibly with multiple layers) or animation.
Besides the pixel data itself, this also includes basic metadata like image dimensions, color space information, orientation, upsampling, frame or layer blending, etc.
It is worth noting that in this respect JPEG~XL deviates from traditional JPEG designs where the codestream only reconstructs sample values on a regular sampling grid, bare any interpretation. JPEG and JPEG~2000 follow this design, where --- for example --- color space information is provided only at file format level. In JPEG~XL however, the codestream signals not just the sample values but also their colorimetric interpretation (see also \SectionName{}~\ref{color}) and any other metadata needed to properly display the image.

\subsubsection{File format}

The JPEG~XL file format can take two forms:

\begin{description}
\item[`raw' codestream:]
In this case, only the image data itself is stored, and no additional metadata can be included. Such a file starts with the bytes \verb|0xFF0A| (the JPEG marker for ``start of JPEG~XL codestream'').
\item [ISOBMFF-style container:]
This is a box-based container format that includes a JPEG~XL codestream box (\verb|jxlc|), and can optionally include other boxes with additional information, such as Exif metadata. In this case, the file starts with
the bytes \verb|0x0000000C| \verb|4A584C20| \verb|0D0A870A|.
\end{description}

\subsubsection{Conformance}

Part 3 of the standard defines precision bounds and test cases for conforming
decoders, to verify that they implement all coding tools correctly and accurately.
Tolerances are defined in terms of a series of reference decoded images with thresholds on the peak error (maximum absolute difference, per-sample) and on the root mean square error (RMSE).
Two conformance levels are currently defined, corresponding to Level 5 and Level 10 of the Main profile (see \SectionName{}~\ref{profiles}).

The conformance test bitstreams and evaluation scripts are available on
GitHub\footnote{\href{https://github.com/libjxl/conformance}{github.com/libjxl/conformance}}.

\subsubsection{Reference software}

The reference implementation of JPEG~XL is called libjxl.
It is available on GitHub\footnote{\href{https://github.com/libjxl/libjxl}{github.com/libjxl/libjxl}}.
The software copyright license is the 3-clause BSD license, a permissive open source license.

\subsection{Codec architecture}

\begin{figure*}\centering
	\includegraphics[width=\linewidth]{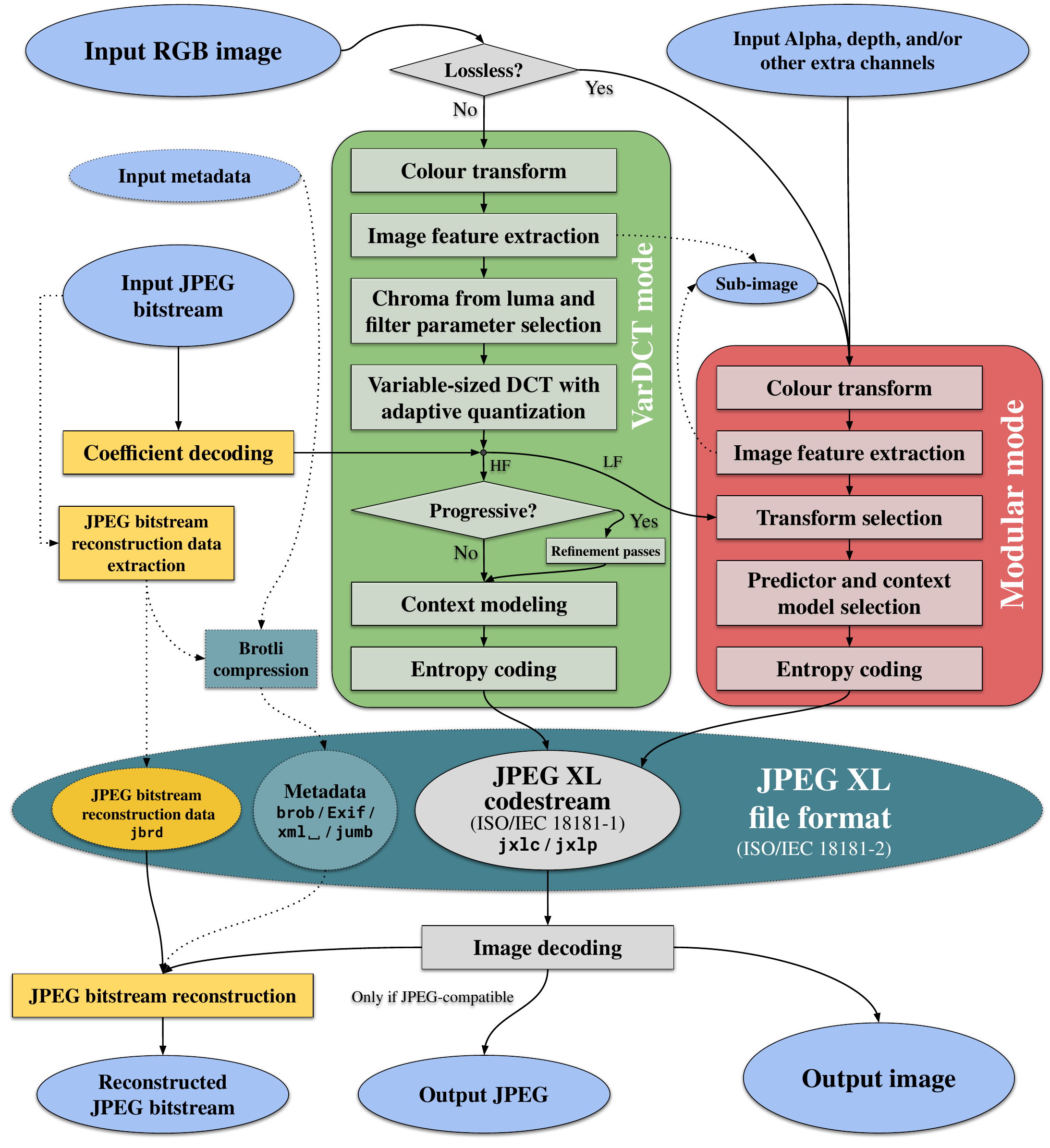}
	\caption{Overview of the JPEG~XL codec architecture.}
	\label{fig:overview}
\end{figure*}

JPEG~XL makes a clear separation between image data and metadata.
Everything that is needed to correctly display an image (or animation) is considered to be image data, and is part of the core codestream.
This includes elements that have traditionally been considered `metadata', such as ICC profiles and Exif orientation, since those are essential to display the image correctly.
Any other kind of metadata (such as copyright declarations or GPS coordinates) is considered not part of the image data and not part of the core codestream.
The goal is to reduce the ambiguity and potential for incorrect implementations that can be caused by having a `black box' codestream that only contains numerical pixel data, leaving it to applications to figure out how to correctly interpret the data (i.e. apply color transforms, upsampling, orientation, blending, cropping, etc.).
By including this functionality in the codestream itself, the decoder can provide output in a normalized way by default (e.g. in RGBA, orientation already applied, frames blended and coalesced), simplifying the task of correctly rendering images from the application point of view by handling error-prone subtasks like frame blending or applying image orientation in the decode library rather than leaving them to be implemented at the application level.
Knowing the colorimetric interpretation of the sample data also allows a lossy encoder to make better decisions and to produce more consistent results for a given quality setting.

The remaining metadata, e.g. Exif or XMP, can be stored in the container format, but it does not influence the main image rendering (although it could describe alternative rendering options).
In the case of Exif orientation, this field has to be ignored by applications, since the orientation in the codestream always takes precedence (and will already have been applied transparently by the decoder). This also means that stripping metadata can be done without affecting the displayed image.

Figure~\ref{fig:overview} gives a high-level overview of the overall JPEG~XL codec architecture.
The top part of this diagram shows the encoder side; the decoder side is at the bottom and mostly omitted for brevity since it consists of the same components, just applied in reversed order: e.g. entropy decoding, then applying inverse DCT, rendering image features, undoing the color transform.


\subsection{Codestream features}

The codestream contains one image, which has specific dimensions (up to $2^{30} \times 2^{30}$) and components: the main image data has either one (grayscale) or three (color) components, and there can be up to 4096 additional components, known as `extra channels'.
The image dimensions define the size (in pixels) of a rectangular canvas.
An image consists of one or more frames that are rendered on this canvas. Frames can represent animation frames, layers of a composite still image, or individual pages of a multi-page image. Each frame contains pixel data (samples for each component) that is encoded either using VarDCT mode or Modular mode.
The data is subdivided into groups, which can be coded independently, allowing parallelization and region-of-interest decoding.

\subsubsection{Frames}

A JPEG~XL codestream contains one or more frames. In the case of animation, these frames have a duration and can be looped (infinitely or a specific number of times).
Zero-duration frames are possible and represent different layers of the image.
Maximum-duration frames indicate `page breaks' in a multi-page image.

Frames have a blend mode (Replace, Add, Alpha-blend, Multiply, etc.) and they can use any previous frame as a base.
They can be smaller than the image canvas, in which case the pixels outside the crop are copied from the base frame.
Frames can be positioned at an arbitrary offset from the image canvas. This offset can also be negative and frames can also be larger than the image canvas, in which case parts of the frame will be invisible and only the intersection with the image canvas will be shown.
It is also possible to have invisible `ReferenceOnly' frames that are not rendered, but that can still be used as a base for frame blending or as a source for Patches.

By default, decoders will blend and coalesce frames, producing only a single output frame when there are subsequent zero-duration frames.
As a result, all output frames are of the same size (the size of the image canvas) and have either no duration (in case of a still image) or a non-zero duration (in case of animation).
This causes layered images to be shown in viewing applications as a single merged image, cropped to the image canvas, while authoring tools can still decode the individual layers.
It also simplifies the implementation of animation viewers since viewers do not need to implement frame blending at the application level.

\subsubsection{Pixel data}

Every frame contains pixel data coded in one of two modes:

\begin{description}
\item[VarDCT mode:]
In this mode, variable-sized DCT transforms are applied
and the image data is encoded in the form of DCT coefficients. This mode is
always lossy.
In case the JPEG~XL codestream was created by transcoding a JPEG encoded image to the JPEG~XL syntax, only a subset of the JPEG~XL features is used, for example only the 8x8 DCT is applied. Such a JPEG~XL codestream allows to recover the original JPEG image it was created from. As JPEG~XL provides improved entropy coding methods, the JPEG~XL codestream will, in general, be of smaller size, even though it represents exactly the same image.

\item[Modular mode:]
In this mode, only integer arithmetic is used, which
enables lossless compression. However, this mode can also be used for lossy
compression. Multiple transforms can be used to improve compression or to
obtain other desirable effects: reversible color transforms (RCTs),
(delta) palette transforms, and a modified non-linear Haar transform
called Squeeze, which facilitates (but does not require) lossy compression
and enables progressive decoding.
\end{description}

These coding modes are not mutually exclusive. The three main color components (or the grayscale component) are coded in one of the two modes; any additional components (known as `extra channels') such as alpha transparency are always coded in Modular mode. Moreover, internally, the VarDCT mode uses Modular sub-bitstreams for
various auxiliary images: the `LF image' (a 1:8 downscaled version
of the image that contains the DC coefficients of DCT8x8 and low-frequency
coefficients of the larger DCT transforms), and weights for adaptive quantization.

In both modes, optional, separately coded additional `image features'
are rendered on top of the decoded image:

\begin{description}
\item[Patches:]
rectangles from a previously decoded frame (typically a ReferenceOnly frame)
can be blended using one of the blend modes on top of the current frame.
This allows the encoder to identify repeating patterns (such as letters of
text) and encode them only once, using patches to insert the pattern in
multiple spots. These patterns are encoded in a previous frame, making
it possible to combine Modular-encoded pixels with a VarDCT-encoded frame or
vice versa. Luca Versari designed the patches subsystem.

\item[Splines:]
centripetal Catmull-Rom splines can be encoded, with a color
and a thickness that can vary along the arclength of the curve.
Although the current encoders do not use this bitstream feature yet, we
anticipate that it can be useful to complement DCT-encoded data, since
thin lines are hard to represent faithfully using the DCT. Sami Boukortt and Jyrki Alakuijala co-designed the splines subsystem.
For example, individual strands of hair, thin branches of a tree, or line art could be coded more effectively using splines.

\item[Noise:]
luma-modulated synthetic noise can be added to an image, e.g.
to emulate photon noise, in a way that avoids poor compression due to
high frequency DCT coefficients. Jyrki Alakuijala designed the photon noise subsystem.
The JPEG~XL photon noise subsystem is different from more complex `film grain synthesis' tools, since the photon noise operates through a Laplacian filter. This makes it impossible for the photon noise subsystem to introduce lower frequency features or accidental clusters of noise. In physics-based measurements the shot noise is related to the square root of the signal strength. However, given psychovisual compression this can mean that the noise intensity is highly non-linear. In JPEG~XL we solve this by providing a small lookup-table to express the intensity dependence of noise.  
In AVIF and MPEG standards similar features are called `film grain synthesis'. In JPEG~XL no effort to model film grain was made, but the noise model is just more simple shot noise. Both film grain synthesis and photon noise model help compression by making flat or smooth image regions to appear more natural.
\end{description}

Finally, both modes can also optionally apply two filtering methods to
the decoded image, which both have the goal of reducing block artifacts
and ringing:

\begin{description}
\item[Gabor-like transform (`Gaborish'):]
a small (3x3) blur that gets applied across block and group boundaries, reducing blockiness. The encoder applies the inverse sharpening transform before encoding, effectively getting the benefits of lapped transforms without the disadvantages.

\item[Edge-preserving filter (`EPF'):]
similar to a bilateral filter, this smoothing filter avoids blurring edges while reducing ringing.
The strength of this filter is signaled locally.
Compared to the smoothing filters typically available in MPEG standards and other video codecs like AV1, the EPF of JPEG~XL is computationally somewhat more expensive, has a more subtle smoothing effect, and can be locally adjusted in a more fine-grained way.
\end{description}

\subsubsection{Groups}

In both modes (Modular and VarDCT), the frame data is signaled as
a sequence of groups, that is, rectangular image tiles. These groups can be decoded independently,
and the frame header contains a table of contents (TOC) with bitstream
offsets for the start of each group. This enables parallel decoding,
and also partial decoding of a region of interest or a progressive preview.

In VarDCT mode, all groups have dimensions $256 \times 256$, where groups at the right and bottom edge are potentially clipped to the image dimension. First the LF image is encoded, also in
256x256 groups (corresponding to $2048 \times 2048$ pixels, since this data
corresponds to the 1:8 image). This means there is always a basic
progressive preview available in VarDCT mode.

Optionally, the LF image can be encoded separately in a (hidden)
LF frame, which can itself recursively be encoded in VarDCT mode
and have its own LF frame. This optional pyramidal scheme makes it possible to represent huge
images while still having an overall preview that can be efficiently
decoded.

Then the HF groups are encoded, corresponding to the remaining AC
coefficients. The HF groups can be encoded in multiple passes for
more progressive refinement steps; the coefficients of all passes
are added. Unlike JPEG progressive scan scripts, JPEG~XL allows
signaling any amount of detail in any part of the image in any pass since there is no forced spectral selection or selection of bit planes: in any pass it is possible to revise all coefficients. It is for example possible to encode the most salient regions in full detail in the first pass, with only coarse detail for the other regions.

In Modular mode, groups can have dimensions $128 \times 128$, $256 \times 256$, $512 \times 512$ or $1024 \times 1024$.
If the Squeeze transform was used, the data will
be split in three parts: the Global groups (the top of the Laplacian
pyramid that fits in a single group), the LF groups (the middle part
of the Laplacian pyramid that corresponds to the data needed to
reconstruct the 1:8 image) and the HF groups (the base of the Laplacian
pyramid), where the HF groups are again possibly encoded in multiple
passes (up to three: one for the 1:4 image, one for the 1:2 image,
and one for the 1:1 image).

In case of a VarDCT image with extra channels (e.g. alpha), the
VarDCT groups and the Modular groups are stored together in the corresponding bitstream sections. This allows for progressive previews and efficient cropped decoding of all the channels.

The default group order is to encode the LF and HF groups in
scanline order (top to bottom, left to right), but this order
can be permuted arbitrarily. This allows, for example, a center-first
ordering or a saliency-based ordering, causing the bitstream
to prioritize progressive refinements in a different way.

\subsection{File format features}

Besides the image data itself (stored in the \verb|jxlc| codestream box),
the optional container format allows storing various types of additional information.

\subsubsection{Metadata}

Three types of metadata can be included in a JPEG~XL container:

\begin{itemize}
\item Exif (\verb|Exif| box) \cite{exif}
\item XMP (\verb|xml␣| box) \cite{xmp}
\item JUMBF (\verb|jumb| box) \cite{jumbf}
\end{itemize}

This metadata can contain information about the image, such as copyright
notices, GPS coordinates, camera settings, etc.
If it contains rendering-impacting information (such as Exif orientation),
the information in the codestream takes precedence.
JUMBF metadata can include JPEG Trust \cite{jpegtrust}
media authenticity annotations \cite{jumbf_trust}.

\subsubsection{brob}

The container allows the above metadata to be stored either uncompressed
(e.g. plaintext XML in the case of XMP) or by Brotli-compression \cite{brotli}.
In the latter case, the box type is \verb|brob| (Brotli-compressed Box) and
the first four bytes of the box contents define the actual box type
(e.g. \verb|xml␣|) it represents.

\subsubsection{JPEG recompression}

JPEG~XL can losslessly recompress existing JPEG files.
The general design philosophy still applies in this case:
all the image data is stored in the codestream box, including the DCT
coefficients of the original JPEG image and possibly some of the `metadata' that is embedded through Application segments (APP markers) in JPEG but is considered part of the image data in JPEG~XL, such as an ICC profile or the
Exif orientation field. The remaining Exif metadata, which does not have an impact on the image display, is stored in the \verb|Exif| box, not in the codestream.

In order to allow bit-identical reconstruction of the original JPEG file
(not just the image but the actual file), additional information is needed,
since the same image data can be encoded in multiple ways as a JPEG file.
For example, the exact Huffman codes that were used, restart markers, the JPEG bitstream layout (e.g. sequential or progressive), or even the contents of padding bits to keep Huffman-coded segments byte-aligned: this is all information that is not part of the image data itself, but is still needed to reconstruct the original JPEG bitstream exactly.
The \verb|jbrd| box (JPEG Bitstream Reconstruction Data) contains this information.
Typically it is relatively small. Using the image data from the codestream,
the JPEG bitstream reconstruction data, and possibly other metadata boxes
that were present in the JPEG file (Exif/XMP/JUMBF), the exact original
JPEG file can be reconstructed.

This \verb|jbrd| box is not needed to display a recompressed JPEG image; it is only needed to bit-exactly reconstruct the original JPEG file.

\subsubsection{Frame index}

The container can optionally store a \verb|jxli| box, which contains an index
of offsets to keyframes of a JPEG~XL animation. It is not needed to display
the animation, but it does facilitate efficient seeking.
Keyframes are frames that can be used as an independent starting point for decoding, that is, none of the earlier frames before a keyframe are referenced by the keyframe or by any subsequent frame.

\subsubsection{Partial codestream}

The codestream can optionally be split into multiple \verb|jxlp| boxes;
semantically, this is equivalent to a single \verb|jxlc| box that contains the
concatenation of all partial codestream boxes.
This makes it possible to create a file that starts with
the data needed for a progressive preview of the image, followed by
metadata, followed by the remaining image data.

\subsubsection{Profiles and levels}
\label{profiles}
Profiles and levels structure the interoperability space between implementations. While a profile defines the subset of coding tools a decoder needs to support in order to decode an image, levels constrain the quantitative parameters of the codec such as image dimensions or component precision. Levels are typically inclusive, i.e. a higher level decoder is able to decode a file conforming to a lower profile.
Currently only one profile is defined for JPEG~XL: the ``Main Profile''.
It includes all coding tools.
Two levels are defined: level 5 and level 10.

Level 5 is designed for end-user image delivery, including web browsers and mobile apps.
It does not allow CMYK, restricts the number of extra channels to 4 and the total number of pixels per displayed frame to $2^{28}$ (268 megapixels). It also limits the precision to allow for an implementation with 16-bit intermediate buffers and puts bounds on the amount and size of splines, as well as the size of the MA trees used in entropy coding.

Level 10 is very permissive and is designed to accommodate a broad range of use cases including authoring workflows. It does not impose strong limitations compared to the implicit limits imposed by the signaling syntax. The maximum dimensions in this level are $2^{40}$ pixels (1099 gigapixels) and the maximum number of extra channels is 256. The limitations of this level are mostly aimed at allowing decoders to do some `sanity checking' and to detect malicious or corrupt bitstreams, rather than imposing bounds that are limiting in practice.

If not signaled otherwise, a JPEG~XL codestream is assumed to conform to Level 5 of the Main Profile. In particular, this is the case for a JPEG~XL file consisting of a `raw' codestream (without the ISOBMFF-style container).
Conformance to a specific (profile and) level can be indicated using the \verb|jxll| box.

\subsubsection{Gain maps}

The third edition of ISO/IEC 18181-2 adds an additional \verb|jhgm| box that can store a HDR gain map as specified in ISO 21496‐1. Since JPEG~XL can represent HDR images directly, the main use case for this box is not to provide an SDR-to-HDR gain map (although it can also be used for that), but an inverse HDR-to-SDR gain map that effectively represents a custom local tone mapping. This allows artistic control over the rendition of an HDR image on SDR displays or on HDR displays that do not support the entire dynamic range needed for the image.

\section{Image and frame headers}
\label{headers}

One of the design goals for JPEG~XL was to minimize the overhead of headers. Compared to a typical photographic image, the size in bytes of the header data seems almost too insignificant to worry about. But for a small icon or thumbnail used on a website, obligatory headers can in fact get larger than the image data itself.

Many header fields in JPEG~XL have a variable length and are signaled only conditionally. The use of default values in common cases allows for a very concise header.
For example, the header of an image that represents a small $120 \times 80$ thumbnail can be as small as 4 bytes: 2 bytes for the codestream signature (\verb|0xFF0A|), followed by 9 bits to signal the image dimensions (one bit that says ``the height is divisible by 8 and small'', then five bits for the height divided by 8, then three bits to say ``the aspect ratio is 3:2 so the width is 120''), followed by 1 bit that says ``default image header: it's an 8-bit sRGB still image without extra channels, encoded in XYB, nothing special'', then another bit that says ``use the default XYB and upsampling weights'', and finally one more bit that says ``default frame header: it's a regular frame, VarDCT-encoded, nothing special''.

\begin{figure*}\centering
	\includegraphics[width=0.96\linewidth]{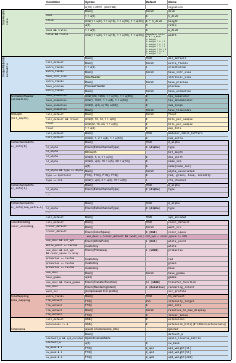}
	\caption[Image header syntax]{Image header syntax. Note that most fields are signaled conditionally and have a variable length. 
    }
	\label{fig:image_header}
\end{figure*}
\subsection{Image header}
The image header at the very start of a JPEG~XL codestream contains the basic information about an image or animation, as well as the header fields that are global for the entire image, that is, they are the same for all frames.
Other header information is signaled on a per-frame basis in the frame headers.

The image header always starts with the two-byte sequence \verb|0xFF0A| which is also known as the signature, magic, or `start of JPEG~XL codestream' marker. It can be used to identify JPEG~XL files or codestreams.

Figure~\ref{fig:image_header} gives an overview of the image header. The notational conventions are as follows: u($n$) denotes $n$ bits read in little-endian order and interpreted as an unsigned integer; 
    Bool() is a synonym for u(1);
    F16() denotes 16 bits interpreted as a half float (binary16);
    U32($a_0,a_1,a_2,a_3$) denotes first $k =$ u(2) followed by $a_k$;
    Enum() is short for U32(0, 1, 2 + u(4), 18 + u(6));
    and U64() is short for U32(0, 1 + u(4), 17 + u(8), longU64())
    where longU64() is defined as follows:
\begin{verbatim}
v = u(12); s = 12;
while (u(1) == 1) {
  if (s == 60) {
    v += u(4) << s; break;
  }
  v += u(8) << s; s += 8;
}
return v;
\end{verbatim}

\subsubsection{Dimensions}
The width and height of the image is signaled using a SizeHeader representation, which allows signaling image dimensions of up to $2^{30} \times 2^{30}$. The height is signaled first, and if the width can be derived from the height through one of seven predefined aspect ratios (1:1, 6:5, 4:3, 3:2, 16:9, 5:4, 2:1) then only three more bits are needed. There is also a special case for ``small images with multiple-of-8 dimensions'', allowing image dimensions up to $256 \times 256$ to be represented in just 14 bits (or even just 9 if the aspect ratio is one of the predefined ones).
Dimensions up to $8192 \times 8192$ can be represented in at most 30 bits (17 bits if the aspect ratio is a predefined one).

Besides the main image dimensions, which are always signaled, there are two additional dimensions that can optionally be signaled: the ``intrinsic size'' and the dimensions of an optional preview frame.
The intrinsic size indicates the recommended display dimensions of the image, in CSS reference pixels where the width of 1 pixel corresponds to a visual angle of about 0.0213 degrees, as defined in Section 4.3.2 of \cite{css}. 
By default the intrinsic size is identical to the image dimensions but it may be different. This can be useful to represent images with non-square pixels, or to annotate `Retina' images (with device-pixel-ratio 2) as having intrinsic dimensions that are half the actual image dimensions. In contexts such as web browsing, the viewing application is expected to display the image by default based on the intrinsic size, resizing the image if needed.
In other image formats, this information can be signaled indirectly based on a creative interpretation of specific Exif metadata \footnote{See \href{https://github.com/eeeps/exif-intrinsic-sizing-explainer}{github.com/eeeps/exif-intrinsic-sizing-explainer}.} but in JPEG~XL it can be signaled directly.

\subsubsection{Orientation}
In JPEG images produced by a camera, it is common to use Exif metadata to indicate the orientation of the image, besides other metadata that does not influence the rendering of the image, such as the camera model and settings or the date and location of the picture.
Applications are assumed to apply the corresponding orientation correction when displaying the image.
In JPEG~XL, the orientation information is stored in the image header and the libjxl decoder will automatically apply the necessary correction, so applications will effortlessly display the image correctly.
The possible values of this field are the same as in Exif:
1	(none),
2	(flip horizontally),
3	(rotate 180 degrees),
4	(flip vertically),
5	(transpose, i.e. rotate 90 clockwise then flip horizontally),
6	(rotate 90 degrees clockwise),
7	(flip horizontally then rotate 90 clockwise),
8	(rotate 90 degrees counterclockwise).

\subsubsection{Animation}
\label{header_animation}
One bit in the image header determines if the codestream represents an animation or a still image.
In both cases, there can be multiple frames, but the semantics are different: in the case of an animation, the different frames are shown sequentially (either as in a video, or as in a multi-page document), while in the case of a still image, the frames represent layers, which are blended over one another to form a single merged composite still image.

In the animation case, the image header defines the duration of a `tick', which will be the unit used to define frame durations.
It also defines whether the animation should be looped forever, played once, or repeated a specific number of times. Finally there is also the option of storing SMPTE timecodes for every frame.

The total number of frames is not signaled in the image header; instead, every frame header indicates whether it is the last frame or not. This allows streaming encoders to emit a valid codestream without knowing in advance how many frames there will be.

Frames with a duration of zero (which is the duration of all frames in the non-animated case) are to be blended on the canvas and the intermediate result is not displayed: only the frames with a non-zero duration (or the final merged image, in the non-animated case) are displayed.

\subsubsection{Other fields}
The image header also signals the color space information, including the bit depth and information about extra channels besides the color channels (or grayscale channel). \SectionName{}~\ref{color} describes this part of the image header.

Images or individual extra channels can optionally be stored in a subsampled way, in which case they will be upsampled at the end of the decoding process. The upsampling procedure (\SectionName{}~\ref{upsampling}) is parameterized, and custom weights can be signaled in the image header.

Finally there is a mechanism for future extensions to the image header that is designed to allow graceful degradation, i.e. decoders ignore header extensions they do not understand. Currently no extensions are defined; if in the future there will be any extensions, they will be designed in a way that allows at least some extent of forward compatibility. That is, a decoder that does not know about the extension can still decode the image and produce a reasonable result. 

\begin{figure*}\centering
	\includegraphics[width=\linewidth]{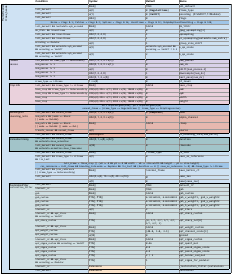}
	\caption{Frame header syntax.
    }
	\label{fig:frame_header}
\end{figure*}
\subsection{Frame header}
An image consists of one or more frames. Figure~\ref{fig:frame_header} gives an overview of the frame header syntax.

Frames can be visible or invisible. Invisible frames are not conceptually part of the decoded image, but play an auxiliary role in reconstructing it, that is, they are referenced by visible frames by means of Patches (\SectionName{}~\ref{patches}), frame blending (\SectionName{}~\ref{blending}), or LF frames (\SectionName{}~\ref{lf-frame}).

There can also optionally be a preview frame if its presence and dimensions are signaled in the image header. In this case, the very first frame in the codestream represents a preview (typically a lower resolution / lower quality version of the image, or a representative frame or title frame of an animation). To decode the full image, the preview frame can be skipped.

\subsubsection{Frame types}
There are four frame types, signaled as a two-bit number in the frame header:

\begin{description}
\item[RegularFrame (0):]
This is the default, `normal' frame type. Frames of this type are part of the decoded image or animation sequence.
\item[LFFrame (1):]
Frames of this type represent the low-frequency (LF) information of a future frame, i.e. an 8x downscaled version of that frame. They are not part of the decoded sequence of frames.
\item[ReferenceOnly (2):]
Frames of this type are invisible and not part of the decoded sequence of frames. They will only be used as a source of Patches or as source frames for blending.
\item[SkipProgressive (3):]
This frame type is similar to RegularFrame: SkipProgressive frames are visible frames that are part of the decoded image. However when showing a progressive preview of the decoded image, frames of this type do not result in an update of the preview. 
\end{description}

The distinction between RegularFrame and SkipProgressive allows layered images to either have progressive previews at each layer or to only show the image when the final layer is available.

If the frame uses the LF information from a previously decoded LF frame, the frame header signals this. LF frames can recursively have their own LF frames at up to four levels: the first level LF frame is 8x downsampled, the second level is 64x downsampled, the third level is 512x downsampled and the fourth level is 4096x downsampled. This allows representing huge images while starting the codestream with a small LF frame that can be used as a thumbnail.

If the decoded frame has to be stored for future reference (which will always be the case for ReferenceOnly frames, but can also be the case for other frame types), this is signaled in the frame header. There are four `slots' for storing decoded frames (not including the implicit slots for LF frames).
Frames can be stored either before applying the inverse color transform (i.e. as XYB components), or after (i.e. as RGB components). Patch references are only allowed to frames in the internal color space used for coding the image (so before applying the inverse color transform), while frame blending happens in the image color space.

\subsubsection{Mode}
There are two encoding modes: Modular (\SectionName{}~\ref{modular}) and VarDCT (\SectionName{}~\ref{vardct}). Since the mode is signaled in the frame header, each frame can use different modes. For example, it can be useful to encode parts of the image using a ReferenceOnly Modular frame, which then is referenced via Patches from the main RegularFrame which may be VarDCT encoded.

Regardless of the mode, the frame header signals an (extensible) set of flags that indicate the presence of the additional coding tools described in \SectionName{}~\ref{features-filters}.

In VarDCT mode, if the image is XYB encoded (see \SectionName{}~\ref{xyb}), the frame header also signals two global weights to adjust the quantization of the X and B components.

In Modular mode, the frame header signals the group size to be used ($128 \times 128$, $256 \times 256$, $512 \times 512$ or $1024 \times 1024$). In VarDCT mode, the group size is always $256 \times 256$. The reason for this is that Modular-encoded images can benefit from using a larger (and occasionally a smaller) group size, in order to reduce the number of poorly-predicted pixels at the first row and first column of a group, or to allow a more effective use of RCTs and Palette transforms.
However in VarDCT mode the effect of the group size on compression performance is negligible so a fixed group size is used to reduce implementation complexity.

\subsubsection{Upsampling}
The main color channels and the extra channels can optionally be downsampled 2x, 4x, or 8x (in both dimensions). The frame header signals the upsampling factor to be applied to the main channels and to each of the extra channels.

In case the image is not XYB-encoded, the frame header can indicate that the YCbCr color transform was applied. In that case, the chroma channels can optionally be subsampled 2x (in one or both dimensions). The reason for this is to be able to represent JPEG images, which typically store YCbCr image data with 4:4:4, 4:2:2, or 4:2:0 subsampling.
Subsampled chroma samples are centered w.r.t. the corresponding $2 \times 1$ or $2 \times 2$ luma samples, as is the convention in JPEG, not co-sited (top-left aligned) like in Blu-ray video, or left aligned (horizontally cosited, vertically centered) like in MPEG-4.

\subsubsection{Crop}
\label{header_crop}
Optionally a frame can have dimensions that are different from the image dimensions, that is, the frame is cropped. A frame header bit indicates whether this is the case, and if so, the frame dimensions are signaled. It can be bigger or smaller than the image dimensions.
For cropped frames that are not ReferenceOnly, the frame header also signals an $(x_0,y_0)$ offset to position the frame on the image canvas. The offset can be negative. Parts of the frame that do not overlap with the image canvas are not displayed in the decoded image, i.e. a decoder will by default show frames cropped to the image canvas. Authoring tools can decode the full uncropped layers though, e.g. to allow recropping an image.

Layers or animation frames only update the region of the image canvas that corresponds to the crop.

\subsubsection{Other fields}
The frame header signals which (if any) previously decoded and stored frame to use as the source (`background') for frame blending, and which blend mode to use.
This is detailed in \SectionName{}~\ref{blending}.

In case of animations, the frame header also contains the frame duration and optionally a SMPTE timecode. \SectionName{}~\ref{animation} describes this in more detail.

The image data of a frame can be signaled in multiple passes, in order to obtain more fine-grained progressive (or partial) decoding steps. The frame header signals the number of passes and what information will be available after each pass.
\SectionName{}~\ref{passes} provides more details on this.

Frames can optionally have a name of up to 1071 bytes, which is interpreted as a UTF-8 encoded string. This can be useful to identify image layers in authoring tools, or to label some frames of an animation.

\section{Color spaces and transforms}
\label{color}
Traditionally, image and video codecs have focused on representing the sample values themselves, i.e. one 2D array or a sequence of 2D arrays of numerical data. In order to interpret this data as a raster image, the color space of the sample values has to be known, either by signaling it or by assumption or convention.

\subsection{Color space signaling}
In most image formats, color spaces are either not signaled at all (e.g. GIF), or signaled only optionally. As a result, the correct interpretation can be ambiguous, although the general consensus is now that untagged images are to be assumed to be in the sRGB color space.

\subsubsection{Design philosophy}
In JPEG~XL, images always have a fully defined color space, i.e. it is always
unambiguous how to colorimetrically interpret the pixel values. There are two options (and the image header signals which of these two options is used):

\begin{itemize}
\item   Pixel data is in a specified (non-XYB) color space, and the decoder will produce
a pixel buffer in this color space plus an ICC profile that describes that
color space. Mathematically lossless encoding can only use this option.
\item   Pixel data is in the XYB color space, which is an absolute color space.
In this case, the decoder can produce a pixel buffer directly in any desired
display space like sRGB, Display-P3 or Rec.2100 PQ.
\end{itemize}

The image header always contains color space signaling; however, its meaning depends on which of the above two options were used:

\begin{itemize}
\item In the first case (non-XYB), the signaled color space (and bit depth) defines how to interpret the pixel data.

\item In the second case (XYB), the pixel data is always in the XYB color space, and the signaled color space (and bit depth) is merely a \emph{suggestion} of an output color space to represent the image in when storing a decoded image, i.e. it is the RGB color space the original image was in, that has a sufficiently wide gamut and a suitable transfer curve to represent the image data with high fidelity using the suggested bit depth. However, to render the image on a display, it is fine to ignore the signaled color space and to instead directly convert the XYB pixel data to the display color space. The signaled space would be used mainly when converting a JPEG~XL image to another image format such as PNG.
\end{itemize}

Color spaces can be signaled in two ways in JPEG~XL: `Enum'-style ColorEncoding, and compressed ICC profiles.

\subsubsection{ColorEncoding}
This is a very compact representation that covers most or all of the common RGB color spaces. The libjxl decoder can convert XYB to any of these color spaces without requiring an external color management library.
The representation is inspired by H.273 CICP \cite{h273}, i.e. a concise way of identifying a color space by means of enumerated fields.
However, not all of the field values defined in H.273 are included --- the more esoteric ones that are rarely if ever used in practice were excluded. Also new field values were added to allow defining custom values for specific fields.

The fields of the ColorEncoding representation are as follows:
\begin{description}
\item[all\_default (Bool):]
If true, then the color space is sRGB. This means that in the common case of sRGB images, the color space signaling requires only a single bit.
\item[want\_icc (Bool):]
If true, then the ColorEncoding can still contain informative Primaries to describe the color gamut, but the actual color space is encoded using an ICC profile.
\item[ColorSpace (Enum):]
Possible values are 0 for RGB, 1 for Gray, 2 for XYB, and 3 for Unknown.
\item[WhitePoint (Enum):]
CIE xy chromaticity coordinates of the white point.
Possible values are 1 for CIE Standard Illuminant D65 (0.3127, 0.3290), 
10 for CIE Standard Illuminant E (1/3, 1/3),
11 for DCI-P3 from SMPTE ST 428-1 (0.314, 0.351),
and 2 for a custom white point given by two signed integers with an implicit denominator of $10^6$.
\item[Primaries (Enum):]
CIE xy chromaticity coordinates of the red, green, blue primary colors.
Possible values are 1 for the sRGB primaries (0.64, 0.33; 0.30, 0.60; 0.15, 0.06),
9 for the ITU-R BT.2100-2 primaries (0.708, 0.292; 0.170, 0.797; 0.131, 0.046),
11 for the DCI-P3 primaries from SMPTE ST 428-1 (0.680, 0.320; 0.265, 0.690; 0.150, 0.060),
and 2 for custom primaries given by six signed integers with an implicit denominator of $10^6$.
\item[TransferFunction (Enum):]
Either a pure gamma function with an exponent signaled as an integer number with an implicit denominator of $10^7$, or one of the following transfer functions:
1 for ITU-R BT.709-6,
8 for linear (a shorthand for gamma 1),
13 for the sRGB transfer function (IEC 61966-2-1),
16 for PQ (ITU-R BT.2100-2),
17 for DCI (SMPTE ST 428-1),
and 18 for HLG (ITU-R BT.2100-2).
\end{description}

All of the commonly used RGB color spaces can be represented in this way.

\subsubsection{ICC}
Arbitrary ICC profiles can also be used, including CMYK ones. The ICC profile data gets compressed. In this case, external color management software (e.g. lcms2 or skcms) has to be used for color conversions.
There is no restriction to a particular version of ICC; the normative reference  in the JPEG~XL standard to the ICC standard (ISO 15076-1) is not dated, so future versions of ICC can automatically be used in JPEG~XL.

ICC profiles are compressed using a special-purpose compression scheme. Arbitrary byte sequences can be encoded using this scheme, but it has a specific context model and reconstruction procedure that is optimized to efficiently represent ICC profiles. For example, there are predefined strings for common ICC tag codes such as  \verb|rTRC|, \verb|rXYZ|, \verb|cprt|, \verb|wtpt|, \verb|bkpt|, \verb|rXYZ|, \verb|gXYZ|, \verb|bXYZ|, \verb|kXYZ|, \verb|rTRC|, \verb|gTRC|, \verb|bTRC|, \verb|kTRC|, \verb|chad|, \verb|desc|, \verb|chrm|, \verb|dmnd|, \verb|dmdd|, \verb|lumi|.
Additionally, for tables of numerical data that may be stored in an ICC profile, predictive coding is applied.
This consists of applying $n$-th order prediction on the resulting element sequence to predict element $x_i$ from up to three previous elements $x_{i-n}$ to $x_{i-1}$, as follows:
\begin{itemize}
\item if $n=1$, the prediction is $x_{i-1}$;
\item if $n=2$, the prediction is $2 x_{i-1} - x_{i-2}$;
\item if $n=3$, the prediction is $3 x_{i-1} - 3 x_{i-2} +x_{i-3}$;
\end{itemize}
This prediction process happens at the byte level, but it is taking into account that the width of the table elements can be 1, 2 or 4 bytes using optional shuffling operations to reorder the bytes by significance.
For large tables like 3D look-up tables, this predictive coding scheme can be very effective in improving compression.

\subsubsection{Bit depth}
The color space assumes a nominal range of $[0,1]$ for the sample values. Values outside this range can also be represented and correspond to out-of-gamut colors.
Two types of sample representation are supported:
\begin{description}
\item[Integer:]
In this case, the sample data is given as integer numbers with an implicit denominator of $2^n - 1$, where $n$ is the bit depth. While the nominal range is $[0,2^n-1]$, it is possible to represent values outside of this range, including negative values.
The maximum bit depth is 31 in this case, to allow using buffers of type \verb|int32_t|.

\item[Floating-point:]
In this case, the sample data is interpreted as floating-point values (ISO/IEC/IEEE 60559). Any floating point type that fits within a single-precision (32-bit) \verb|float| can be used, i.e. the number of exponent bits is at most 8 and the number of mantissa bits is at most 23. This allows representing both types of half-floats currently in common use: binary16 (5 exponent bits, 10 mantissa bits) and bfloat16 (8 exponent bits, 7 mantissa bits).
\end{description}

In the XYB case, the signaled bit depth is merely a suggestion on what representation to use for a decoded image, assuming that integers will be clamped to the nominal range.
In the non-XYB case however (typically only used for lossless compression), the bit depth does determine how to interpret the sample data.

JPEG~XL can losslessly represent 32-bit floating point data. For integer data, in principle up to 31-bit unsigned integer or 32-bit signed integer data can be stored losslessly, but the reference implementation libjxl is limited to an integer precision of 24-bit since it internally uses a 32-bit float representation in most of the encoding and decoding implementation.

\subsubsection{HDR}
The \verb|intensity_target| (signaled in the image header) indicates an upper bound on the intensity (in nits, i.e. candela / m\textsuperscript{2}).
For SDR images, the default value of 255 nits is used.
Other information relevant for HDR tone mapping can also be signaled in the image header.

\subsection{Color transforms}
Color transforms are seen as a coding tool in JPEG~XL, not as something to be handled at the application level.
These transforms can be applied at three different levels:
\begin{itemize}
\item Image: the choice between XYB and non-XYB is signaled in the image header, and applies globally to the entire image, i.e. to all frames.
\item Frame: the choice between RGB and YCbCr (and the choice of chroma subsampling) is signaled in the frame header. This allows for example taking an existing JPEG image, adding a PNG logo as an alpha-blended overlay to it, and representing the result as a single composite still image in JPEG~XL which contains a YCbCr frame and an RGBA frame.
\item Group: in Modular mode, reversible color transforms (RCTs) such as YCoCg can be signaled at the frame level but also at the group level.
\end{itemize}

\subsubsection{XYB}
\label{xyb}

The XYB color space is derived from perceptual models \cite{cam} based on the absorbance spectra of the long (L), medium (M), and short (S) wavelength cone cells of the human retina. Structurally analogous to opponent-process color spaces like CIELAB, XYB utilizes three axes: a luminance channel (Y) representing black to white, a red-green opponent channel (X), and a blue-yellow opponent channel (B').

The development of the XYB color space was motivated by a key observation made by Jyrki Alakuijala during the development of the Butteraugli perceptual metric. It was noted that the contribution of the S-cone receptors to perceived luminance is scale-dependent. At fine, high-frequency scales, such as those corresponding to individual pixels, the influence of the S-cone response on luminance perception is practically negligible. However, at larger angular scales, such as the 2-degree standard observer defined by the CIE, the S-cone contribution to luminance is substantial.

This discrepancy presented an opportunity for a more efficient color representation optimized for digital images. By designing a color model that accounts for the diminished role of the S-cone signal in high-frequency luminance information, it becomes possible to encode color data more compactly. Alakuijala subsequently formalized this insight into the XYB color space, which is engineered to align more closely with the spatial characteristics of human color perception, thereby enabling more effective compression.

\begin{figure}\centering
	\includegraphics[width=\linewidth]{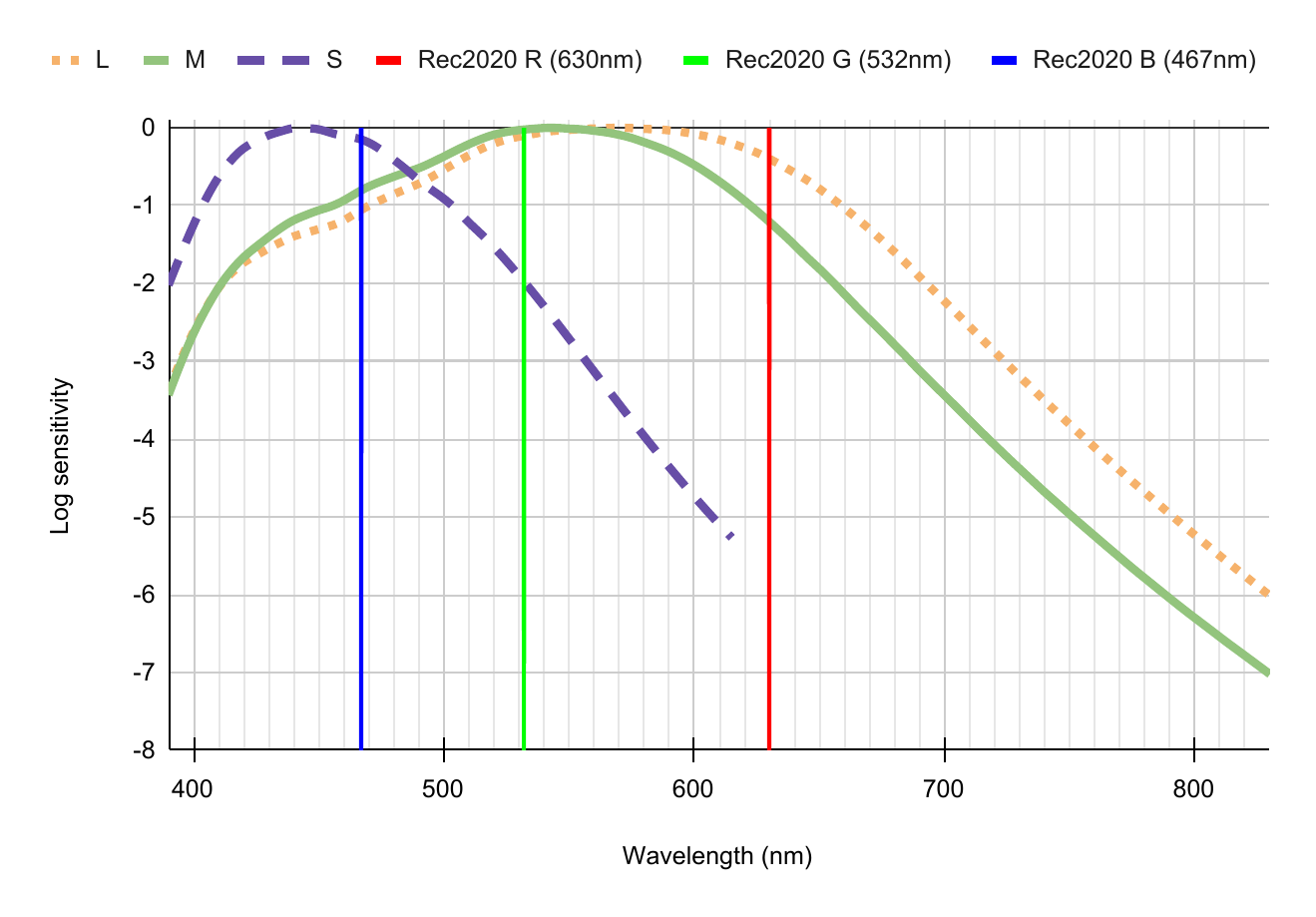}
	\caption{Spectral sensitivity of human cone cells \cite{Stockman-Sharpe}.}
	\label{fig:cone-sensitivity}
\end{figure}

\begin{figure}\centering
	\includegraphics[width=\linewidth]{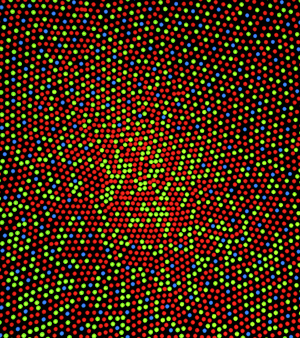}
	\caption{Illustration of the distribution of cone cells in the fovea, from \cite{why_is_color}.}
	\label{fig:cone-distribution}
\end{figure}

\paragraph{RGB to XYB conversion.}
Starting from RGB sample values in some RGB color space, the forward XYB transform is defined as follows.
First the RGB values are converted to $(R_l, G_l, B_l)$ which are relative to
the sRGB color primaries and D65 white point, with a linear transfer function and a nominal range of $[0,1]$ for colors within the sRGB gamut and within a standard dynamic range, though out-of-range values are allowed in case of wide gamut and/or high dynamic range.
In particular, the value $(1,1,1)$ corresponds to white light at an intensity of \verb|intensity_target| nits, which is 255 nits by default but for HDR images higher values like 1000 or 10000 nits can be used.
From this normalized absolute RGB space (``linear sRGB''), the conversion to $XYB$ is as follows. First, $(R_l, G_l, B_l)$ samples are converted to $(L_m, M_m, S_m)$:

\begin{flalign}
&b =	0.00379307325527544933  &\nonumber\\ 
&L_m =	0.3	R_l	+ 0.622 G_l	+ 0.078	B_l	+ b  &\\ 
&M_m =	0.23 R_l + 0.692 G_l + 0.078 B_l + b  &\nonumber\\
&S_m =	0.2434227 R_l +	0.2047674 G_l + 0.5518099 B_l + b  &\nonumber
\end{flalign}

Compared to the Hunt-Pointer-Estévez definition of LMS \cite{LMS-HPE}, the
$L_m M_m S_m$ space is less spectrally sharp. It more explicitly models that the peak spectral sensitivities of L and M cones occur at relatively nearby wavelengths, as shown in Figure~\ref{fig:cone-sensitivity}.
Both L and M peak at a color between green and yellow; L more towards yellow and M more towards green.
L cones peak at 570.2nm (lime green), while M cones peak at 542.8nm (chartreuse green).

\begin{figure*}\centering
	\includegraphics[width=\linewidth]{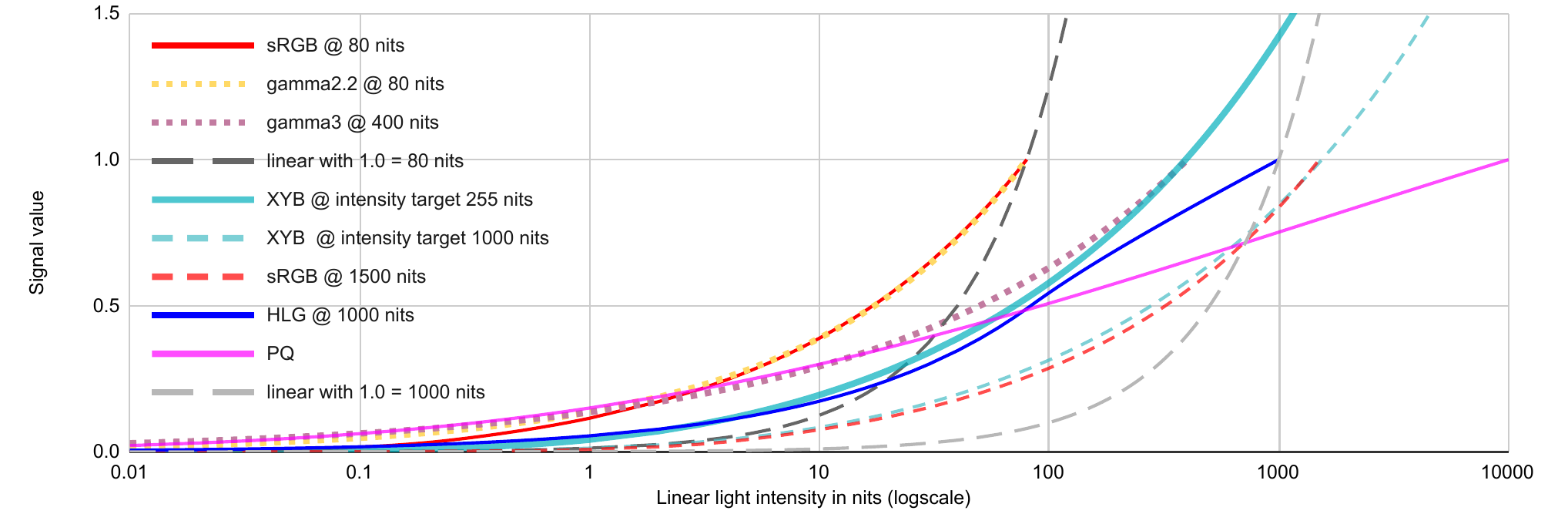}
	\includegraphics[width=\linewidth]{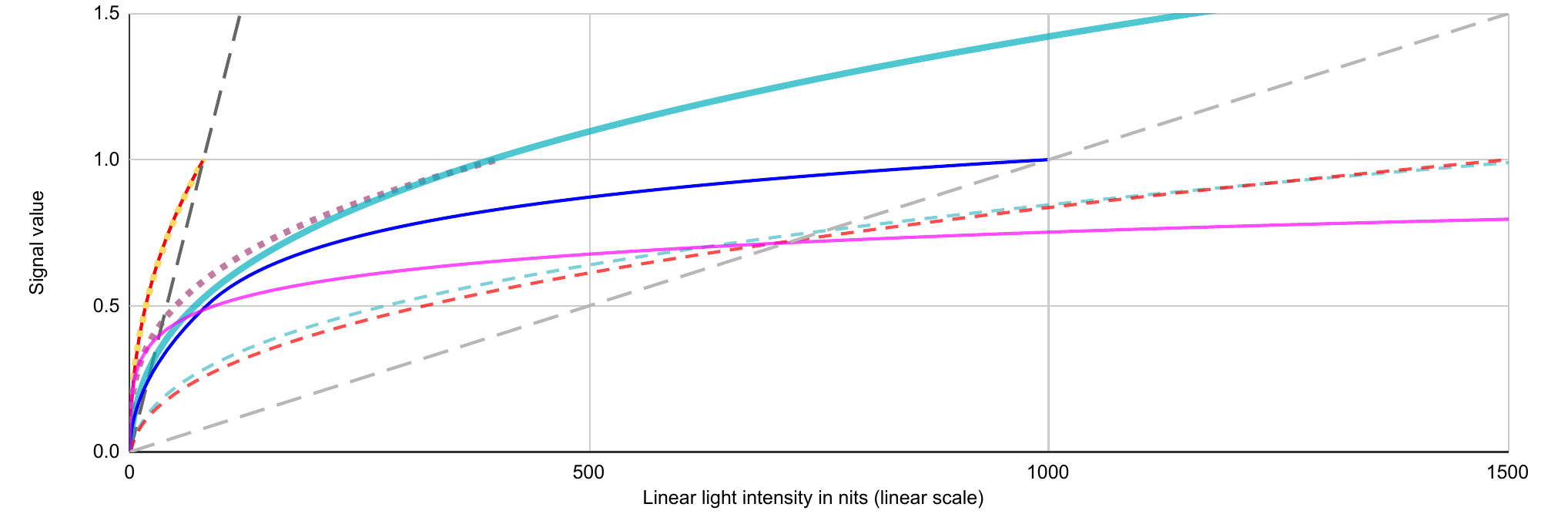}
	\caption{Comparison between various transfer functions.}
	\label{fig:tf}
\end{figure*}


The $L_m M_m S_m$ values are then `gamma compressed':

\begin{flalign}
&L_g = \sqrt[3]{L_m} - \sqrt[3]{b} \\
&M_g = \sqrt[3]{M_m} - \sqrt[3]{b} &&\nonumber\\
&S_g = \sqrt[3]{S_m} - \sqrt[3]{b} &\nonumber
\end{flalign}

The $b$ bias term in the above equations models spontaneous opsin activation.
The cube root in the forward transform translates to a cube in the inverse transform. This reduces the computational complexity of the decoder.

Figure~\ref{fig:tf} shows how the biased cubic transfer function used in XYB compares to other transfer functions like sRGB, PQ and HLG.
Note that unlike PQ and HLG, which are designed to be used with a fixed $[0,1]$ signal range, the range of $L_m M_m S_m$ values is positive but not necessarily bounded by 1, though the values will be smaller than 0.85 if the intensities do not exceed \verb|intensity_target|.

Finally the $XYB$ values are defined as follows:

\begin{flalign}
&X = (L_g - M_g)/2 &\\
&Y = (L_g + M_g)/2 &\nonumber\\
&B = S_g &\nonumber
\end{flalign}

While $B$ corresponds to just the S cone response, in practice in JPEG~XL it will rarely be used directly. In VarDCT mode, `chroma from luma' is typically used to subtract $Y$ from $B$, and in Modular mode, the `integerization' of XYB is defined in a way that does the same thing. So in practice, the actual color space that is used is $XYB'$, with
$B' = B - Y$.

For grayscale (achromatic colors) where $R_l = G_l = B_l$, it will be the case that $L_m = M_m = S_m$, so $X = B' = 0$.

\paragraph{Motivation.}
The lightness component $Y$ is based only on the responses of L and M cones, ignoring the contribution of the S cones. The reason is illustrated in Figure~\ref{fig:cone-distribution}: in the fovea centralis, which is the part of the retina responsible for sharp central vision, the density of S cones is very low.
Effectively this means that after frequency transforms, the high-frequency signal corresponding to the $B$ (or rather, $B'$) component can be quantized more aggressively.

\begin{figure}\centering
	\includegraphics[width=\linewidth]{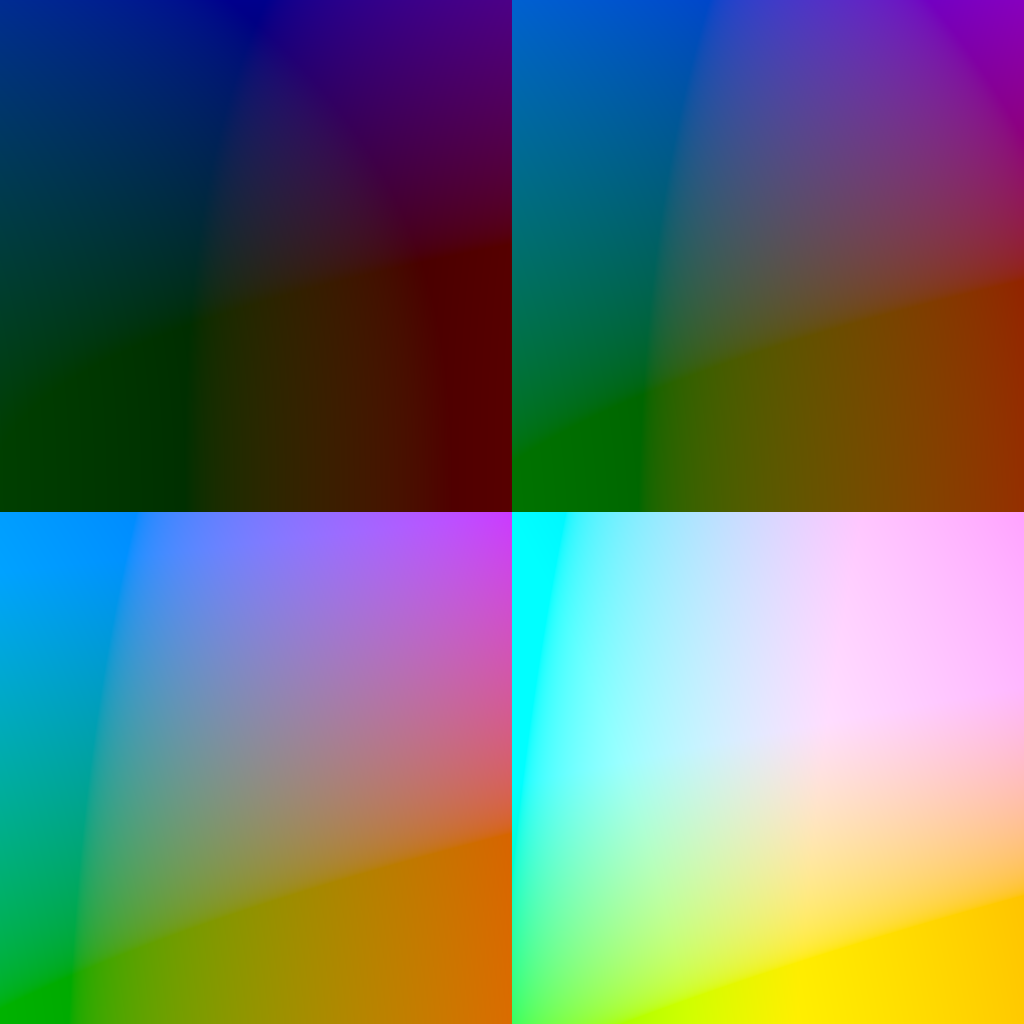}
	\caption[Visualization of the XYB' color space]{Visualization of the XYB' color space: slices of the XB' plane at four values of Y.}
	\label{fig:xyb}
\end{figure}

\begin{figure}\centering
	\includegraphics[width=\linewidth]{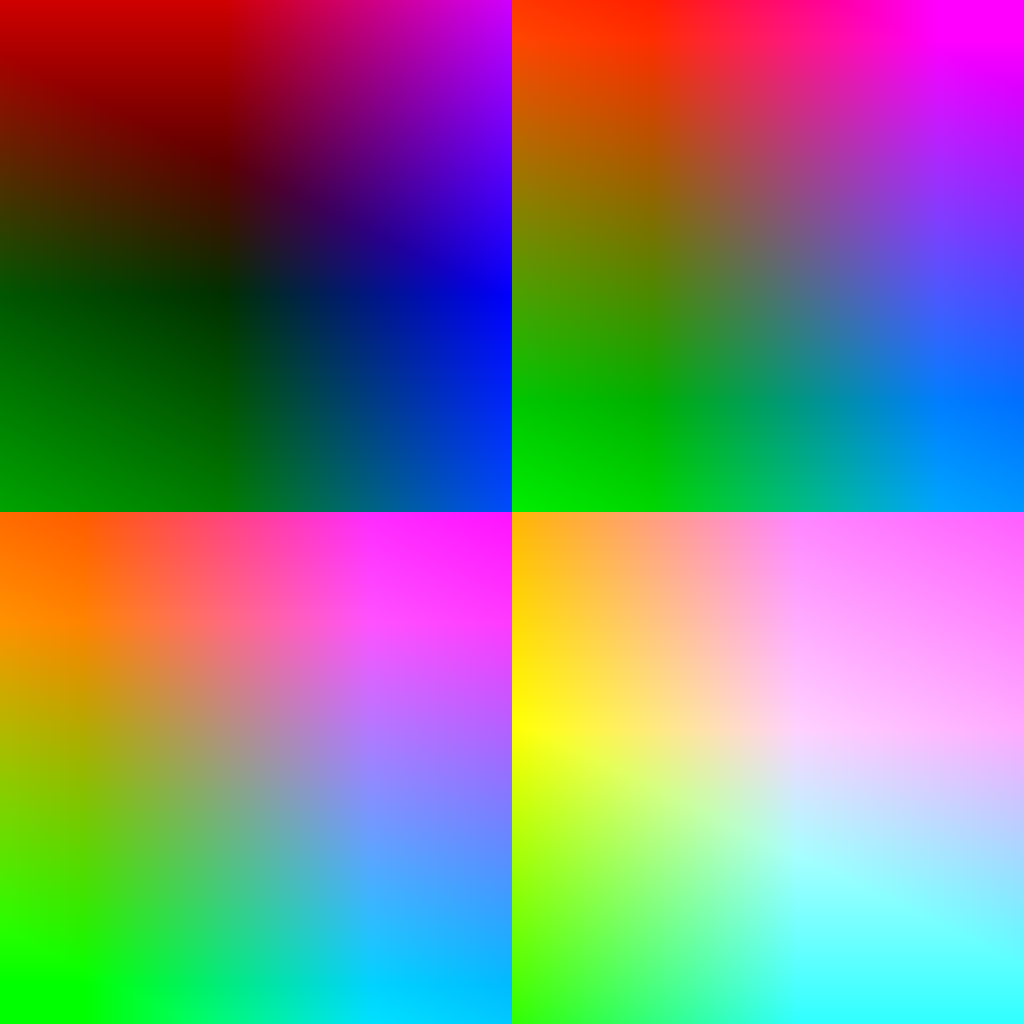}
	\caption[Visualization of the YCbCr color space]{Visualization of the YCbCr color space: slices of the CbCr plane at four values of Y.}
	\label{fig:ycbcr}
\end{figure}

\begin{figure*}
\centering
	\includegraphics[width=\linewidth]{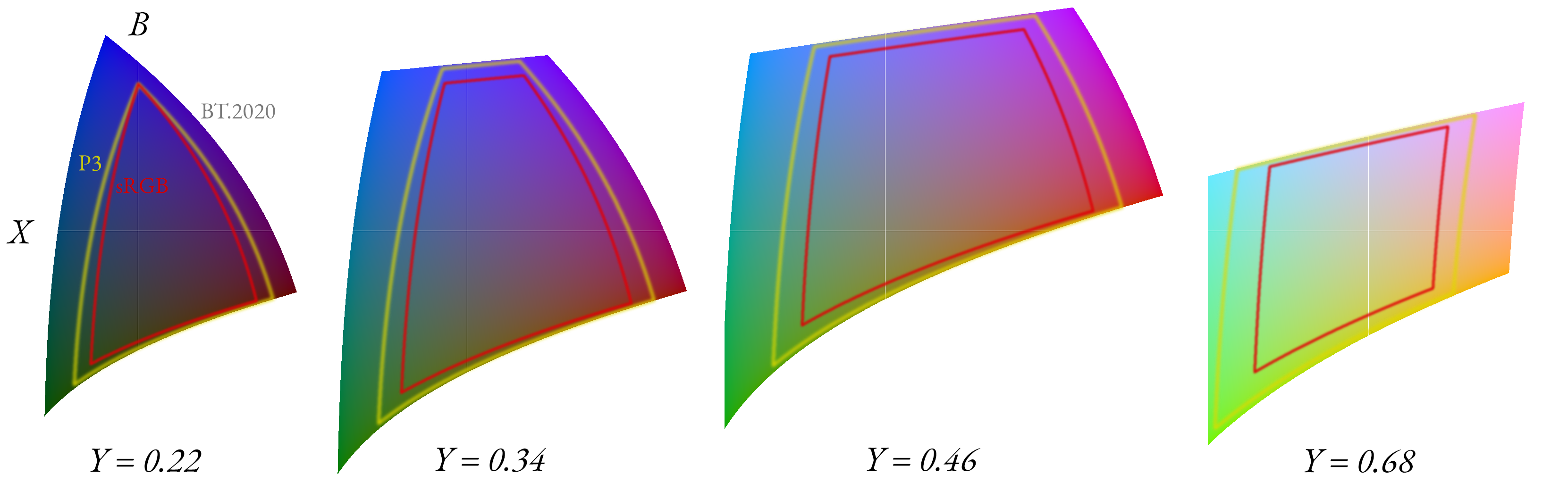}
	\caption{Slices of the XB' plane at four values of Y, with the color gamuts of sRGB, P3, and BT.2020 indicated.}
	\label{fig:xyb-gamuts}
\end{figure*}

The XYB' color space is illustrated in Figure~\ref{fig:xyb}. 
Compared to YCC color spaces commonly used in image compression like YCbCr and YCoCg, the main differences are:
\begin{itemize}
\item XYB' is perceptually more uniform and better models the human visual system than YCC spaces. Compare e.g. Figures~\ref{fig:xyb}~and~\ref{fig:ycbcr}.

\item While YCC spaces are relative to the RGB space they are derived from, XYB' is an absolute color space: wider gamut translates to a larger range for the X and B' components, higher dynamic range translates to a larger range for the Y component. Figure~\ref{fig:xyb-gamuts} shows how the gamuts of sRGB, P3, and BT.2020 correspond to increasingly larger subvolumes of the XYB' space.
\end{itemize}

A key innovation in the JPEG~XL approach is to use a single absolute color space as the internal color space when doing lossy compression.
No matter which color space the original input image uses, an encoder will convert it to XYB. Only for lossless compression and for JPEG recompression, the XYB space is not used.

The main advantage of always using the same color space for encoding, is that an encoder can apply perceptual optimization much more accurately and effectively, since it `knows' the exact colorimetric interpretation of the input data it is encoding, and thus the visual impact of the choices it makes.
Using a perceptually uniform color space as the internal space makes it easier to keep the visual fidelity balanced across different regions of the image, rather than it being modulated by the effects of perceptually less uniform color spaces such as sRGB or YCbCr-transformed sRGB.

Compared to perceptual color spaces that have been proposed in color appearance models like CIELAB and CIECAM02 and color distance metrics like CIE $\Delta E*$ \cite{CIEDE2000}, the main difference is that these models are typically based on experiments where the color stimuli (e.g. physical samples or rectangles on a screen) occupy a visual angle of $2\degree$ (or more),
while XYB' was intended to be used for pixel-sized colors.
The CSS pixel unit nominally corresponds to a visual angle of $0.0213\degree$, and current displays typically have a device pixel ratio (DPR) of 2 or even 3, which corresponds to a pixel size of about $0.01\degree$, i.e. more than two orders of magnitude smaller than the stimuli typically used to determine color appearance models and color distance metrics.
This implies that typical color distance metrics are including parifoveal vision, while for pixel-sized details, foveal vision dominates and the contribution of S cones plays a smaller role.

\subsubsection{YCbCr}

In the non-XYB case, the frame header signals whether or not the YCbCr transform was applied to the RGB (or CMY) sample data. The exact variant of YCbCr used in JPEG~XL is defined in such a way that it is identical to the way it is implemented in JPEG (or rather, JFIF and Exif, since the JPEG standard itself does not specify any color transforms):

\begin{flalign}
&R = Y + \frac{128}{255} + 1.402 Cr &\\
&G = Y + \frac{128}{255} - 0.344136 Cb - 0.714136 Cr &\nonumber\\
&B = Y + \frac{128}{255} + 1.772 Cb &\nonumber
\end{flalign}

The main reason this color transform is included in JPEG~XL is because it is required for lossless JPEG recompression. When encoding an image from pixels, the recommended color space to use is XYB, since it is based on a better model of the human visual system and is thus more perceptually uniform and more suitable for perceptual optimization than YCbCr.
Figure~\ref{fig:ycbcr} visualizes the YCbCr color space (for the common case of it being applied to RGB samples in the sRGB color space) by showing the CbCr plane at four different constant Y values. If the space were perceptually uniform, each of these planes would look evenly bright. Compared to the XYB' color space (Figure~\ref{fig:xyb}), it is obvious that the perceptual uniformity of YCbCr is somewhat lacking. This can be explained by sRGB itself not being perceptually uniform, and a simple linear matrix multiplication like YCbCr does not suffice to correct that.

\subsubsection{Chroma subsampling}
\label{chroma_subsampling}
When YCbCr is used, the frame header also signals the use of chroma subsampling.
Only subsampling by a factor of two is allowed in JPEG~XL, either horizontally or vertically or both. This means 4:4:4, 4:2:2, 4:4:0, and 4:2:0 subsampling can be expressed, but not e.g. 4:1:1 or 3x subsampling, which are possible in JPEG but usage is very rare since no major JPEG encoder implementations produce such images.
The reason for this limitation is to have a balance between implementation complexity and coverage of existing JPEG images. Not allowing chroma subsampling at all would simplify JPEG~XL implementations but it would cause many existing JPEG images to not be representable losslessly in JPEG~XL. Allowing arbitrary chroma subsampling factors would only have a tiny impact on coverage but a substantial impact on implementation complexity.

Chroma subsampling is not a useful coding tool in JPEG~XL since effectively the same result can be obtained by zeroing all DCT coefficients except those in the top-left quadrant and reordering the coefficients such that those coefficients are at the end. With the entropy coding used in JPEG~XL, it is possible to do this with very low signaling overhead. But generally it is better not to remove high-frequency information in such a blunt and global way.

The details of how chroma subsampling influences the image padding (to ensure an integer number of blocks for all components) are defined to correspond with JPEG.

\subsection{Extra channels}
Besides the `main image channels' --- either one Grayscale channel, or three color channels ---
there can be up to 4096 `extra channels' in a JPEG~XL image.
The image header signals the number of extra channels, and for each extra channel the following information is signaled:
\begin{description}
\item[d\_alpha (Bool):]
This field signals a particularly common set of defaults. If true, the extra channel is of type Alpha (unassociated), 8-bit, at 1:1 resolution and without a name, and no more fields are signaled for this channel.
\item[type (Enum):]
Determines the semantics of the extra channel. The various types are described below.
\item[bit\_depth (Bundle):]
Indicates the bit depth and whether it is integer or floating-point (and if floating-point, how many exponent bits there are).
\item[dim\_shift (integer):]
The extra channel is subsampled (in both dimensions) by a factor of $2^\textrm{dim\_shift}$.
In particular for channels encoding depth or thermal (infrared) information, this optional subsampling can be useful.
\item[name (string):]
Optionally, extra channels can have a name (UTF-8 encoded) with a maximum length of 1071 bytes.
This name can be used to add a human-readable description, or it can use an application-specific naming convention, in particular for the generic extra channel types (NonOptional and Optional).
\item[Type-dependent information.]
Depending on the type of extra channel, additional fields may be signaled that are specific to that type.
\end{description}

\subsubsection{Alpha}
The most common extra channel type is used for alpha transparency.
By convention, alpha sample values of 0 indicate full transparency,
while 1 indicates full opacity. Out-of-range alpha values are allowed, and are interpreted differently depending on the blending information signaled in the frame header (see \SectionName{}~\ref{blending}) or as part of the Patches data (see \SectionName{}~\ref{patches}): either they are clamped to the $[0,1]$ range before blending, or the blending formulas are applied with unclamped values.

Alpha channels can be either associated (also known as `premultiplied') or unassociated; this is signaled as an additional boolean field in the extra channel information of type Alpha.
Associated alpha (as in e.g. the OpenEXR image format) is more expressive than unassociated alpha (as in e.g. the PNG image format) since it allows modeling layers with light-emitting but transparent pixels, such as a candle flames or reflections in glass. Such pixels would correspond to an alpha value of zero and color components that are non-zero.

There can be multiple alpha channels. The first alpha channel is considered to be the `main' alpha channel that is used to render the image.

\subsubsection{Depth}
This extra channel contains depth information, i.e. the (estimated) distance from the camera at each position in the image. This data can be used to separate the foreground from the background in an image (e.g. in order to apply an artificial bokeh effect in post-production), or to aid in the reconstruction of a 3D scene from one or more 2D images.

The convention used in JPEG~XL is that higher sample values in the depth channel indicate a larger distance from the camera. The exact scale to be used is not standardized; if needed, applications can use a channel naming convention to disambiguate if needed.

\subsubsection{Thermal}
The extra channel type `Thermal' is used for infrared thermography images. Higher sample values indicate a warmer temperature. Again, the exact scale is not standardized and applications can use a channel naming convention to disambiguate if needed.

\subsubsection{CFA}
In digital cameras, image data is typically captured using a Color Filter Array (CFA), also known as a Bayer filter mosaic.
This data can be stored using extra channels of type `CFA'.
A type-specific integer cfa\_channel is signaled in the extra channel information, which is an index used to identify the subpixel color and position within the filter mosaic.

\subsubsection{Black (K)}
For CMYK images, the Key component (denoting black ink) is represented as an extra channel.
The use of a CMYK ICC profile is obligatory for CMYK images.
The RGB channels are in this case interpreted a Cyan, Magenta, Yellow.

In the case of CMYK, the sample value 0 indicates `full ink' while the sample value 1 indicates `no ink'. This convention is common for CMYK images since the color model is subtractive, so the invariant remains that all-zero sample values are black and all-one sample values are white, just like in RGB.

Since VarDCT is `hard-coded' for three color components, and all extra channels are Modular-encoded, it is not possible to apply lossless JPEG recompression to CMYK JPEG images. In principle the CMY components of the JPEG image data could be represented losslessly, but the K component would have to be decoded to sample values and stored in that way, which would be inefficient and not fully lossless (as the DCT coefficients are lost).
Having to store the K component separately in an extra channel, and thus not being able to apply lossless JPEG recompression to CMYK JPEG images, is an intentional limitation of JPEG~XL. The rationale is as follows: for most use cases where CMYK is desirable, lossless compression is more suitable than lossy compression --- that is, if lossy compression is acceptable, then generally it will also be acceptable to represent the image in a (sufficiently wide gamut) RGB space. In particular, the K component often contains rasterized text, for which DCT-based compression does not perform very well. Thus, requiring Modular mode when compressing CMYK data was deemed to be an acceptable limitation.
On the upside, the fixed number of components in VarDCT helps to reduce the implementation complexity that would arise from dealing with a variable number of components.

\subsubsection{SpotColor}
Besides Alpha and Black (K) channels, SpotColor channels are the only extra channels that have a standardized rendering impact.
These are monochrome channels with a specific color and solidity. They are effectively rendered as if there is an additional layer of constant color, with the SpotColor channel as the alpha channel. Solidity is an additional multiplicative factor in the alpha blending.

These channels can be used to model non-process inks in offset printing, including metallic or fluorescent inks, or the scratchable area on a scratchcard.
The color to be used for rendering the image on a display is signaled as part of the extra channel information, and it can be a color that is out-of-gamut with regard to the image color space. In any case this color may only be an approximation, since spot color channels are particularly useful to represent image data that cannot be reproduced with regular display or printing technology. The details of the spot color definition (such as a reference to a Pantone\textregistered{} color) can be stored using application-specific channel naming conventions.

\subsubsection{SelectionMask}
A selection mask indicates a region of interest, e.g. a part of the image that can be `selected' (or masked) in an image editor in order to selectively apply further processing steps.
The sample value zero means that the pixel is not part of the selection; sample value one means that the pixel is fully-selected. Intermediate values signify partial selection and are useful for fuzzy or `feathered' masks.

\subsubsection{NonOptional}
Since more extra channel types than the ones described above may be needed, a generic extra channel type `NonOptional' is available. If this type is used, the channel name follows a naming convention that identifies the interpretation of this channel.
If an application does not know how to handle this type of data, it indicates that it cannot correctly render the image (e.g. it refuses to load the image, or shows a warning).
In this sense, the data is ``not optional''; it is necessary to understand the semantics of this channel to render the image correctly. For example, the render-impacting data encoded in Alpha, Black, and SpotColor channels would be appropriate to store in a NonOptional channel if those standard channel types had not been defined.

\subsubsection{Optional}
For generic extra channels that can be safely ignored by applications that do not know how to handle them, the `Optional' type can be used. This is meant for data that has no impact on the (default) rendering of an image. For example, data encoded in Depth, Thermal, and SelectionMask channels would be appropriate to store in an Optional channel if those standard channel types had not been defined.

\section{Modular sub-bitstream}
\label{modular}
Most of the types of image data in JPEG~XL --- essentially everything except the metadata and the HF VarDCT coefficients --- is encoded using Modular sub-bitstreams.
An image frame can be encoded using only Modular sub-bitstreams for all image data (`Modular mode'), or it can be encoded using VarDCT, which still depends on Modular sub-bitstreams for much of its signaling, with the major exception of the HF data (which is the bulk of the image data) which has a dedicated coding scheme.
Part of the reason why it is called ``Modular'' is that this encoding mode is effectively used as a `submodule' in the overall codec.

The Modular sub-bitstream can encode a set of 2D arrays of integer numbers, called channels.
These channels can correspond to color components and extra channels, LF coefficients, color palettes, quantization tables, block type selections, adaptive quantization weights, chroma from luma multipliers, local filter strengths, or the result of any of these after applying one or more modular transforms.

The number of channels, their dimensions, and their semantics are not signaled explicitly in the Modular sub-bitstream, but implicitly derived from the image and frame header information and the specific section of the codestream in which the Modular sub-bitstream is `invoked'.
For example, for a losslessly compressed RGBA image, a single modular sub-bitstream could correspond to 4 channels, each $256 \times 256$ (the default Modular group size), with semantics Red, Green, Blue, Alpha.

\subsection{Modular transforms}
When using Modular encoding, a chain of transforms may be applied to the image data before encoding it. These transforms are signaled, and decoders apply the corresponding inverse transforms when decoding the image.
The flexibility or `modularity' obtained by having a signaled transform chain is the other reason for the name of the Modular coding mode.

For example, one transform could be an RCT transform to convert RGBA to YCoCgA; a subsequent transform could apply a Palette transform to the two chroma channels (say there are only 100 unique combinations of Co and Cg used in the image). The encoded data would then contain, in order:
1) a $100 \times 2$ `metachannel' that represents the palette of (Co,Cg) colors,
2) a $256 \times 256$ luma (Y) channel, 
3) a $256 \times 256$ index channel with samples in range $[0,99]$ that refer to the palette, and finally,
4) a $256 \times 256$ alpha channel.

\begin{figure}\centering
	\includegraphics[width=\linewidth]{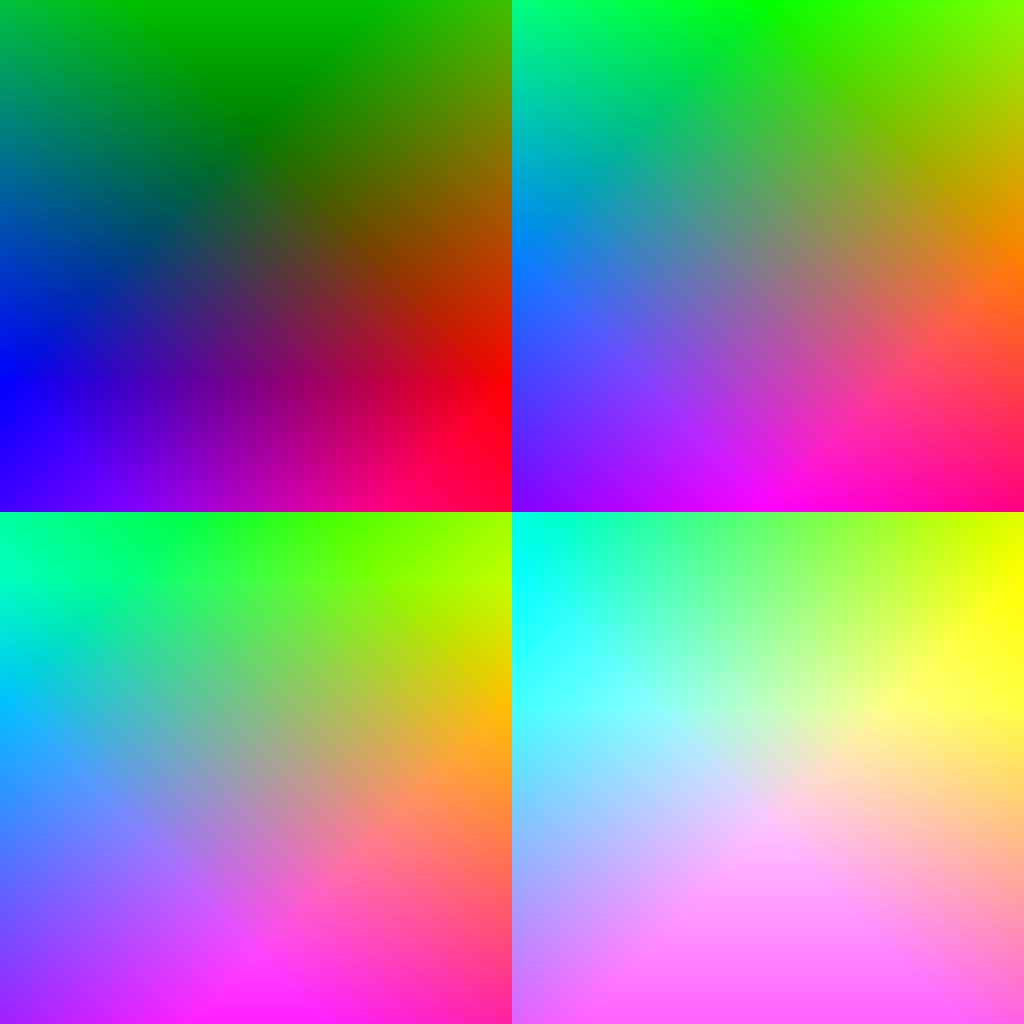}
	\caption[Visualization of the YCoCg-R color space]{Visualization of the YCoCg-R color space: slices of the CoCg plane at four values of Y.}
	\label{fig:ycocg}
\end{figure}
\begin{figure*}\centering
\begin{tabular}{rccc}
\includegraphics[width=0.22\linewidth]{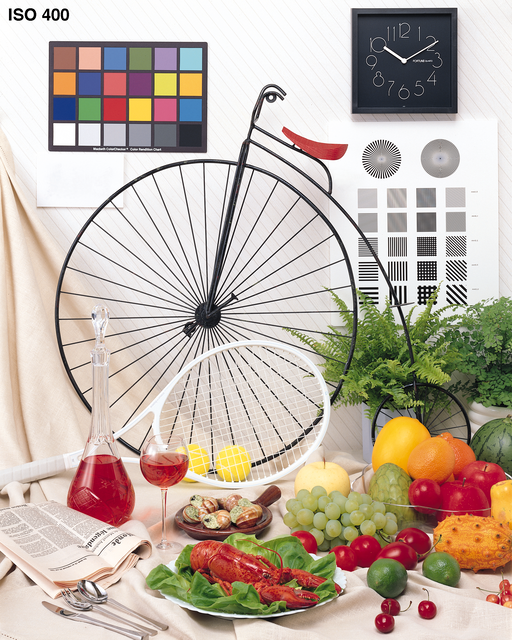} &
\includegraphics[width=0.22\linewidth]{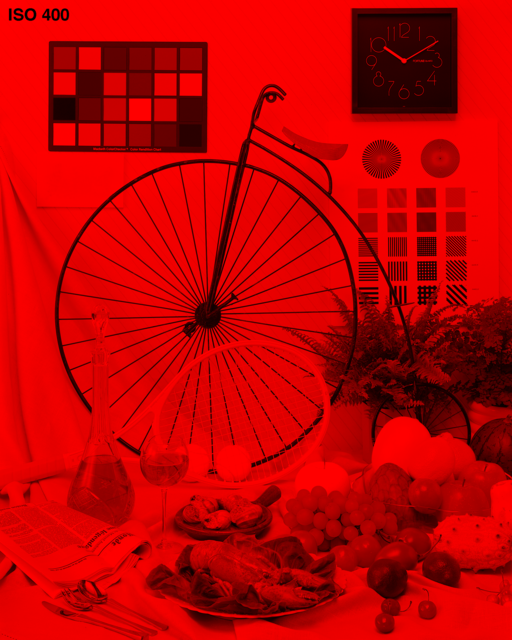} &
\includegraphics[width=0.22\linewidth]{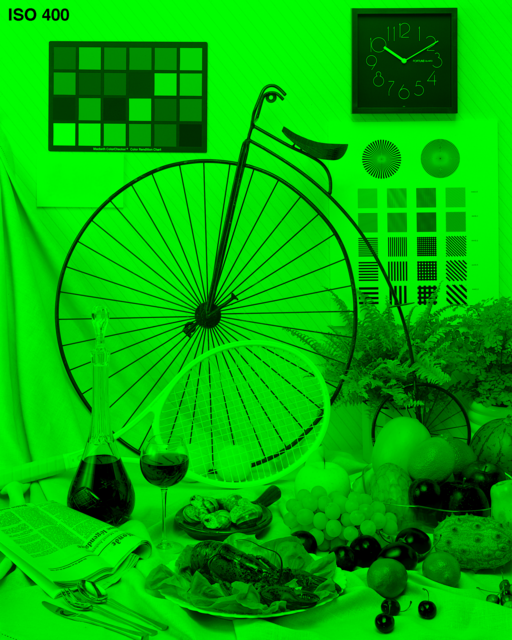} &
\includegraphics[width=0.22\linewidth]{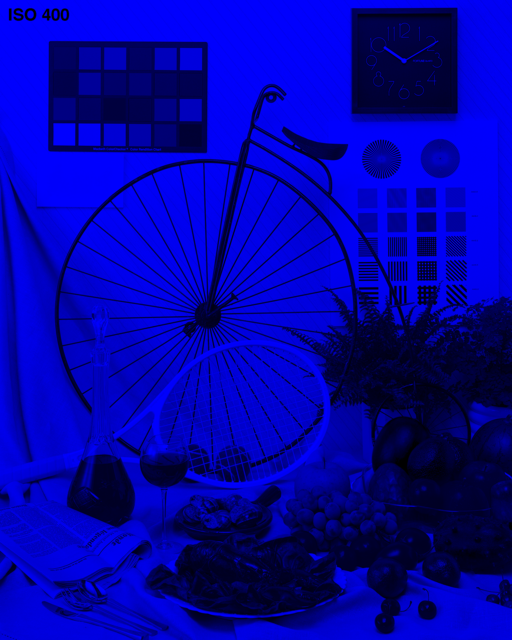} \\
\multicolumn{1}{l}{Color image \hfill RGB:} & R & G & B \\
YCoCg-R: & \includegraphics[width=0.22\linewidth]{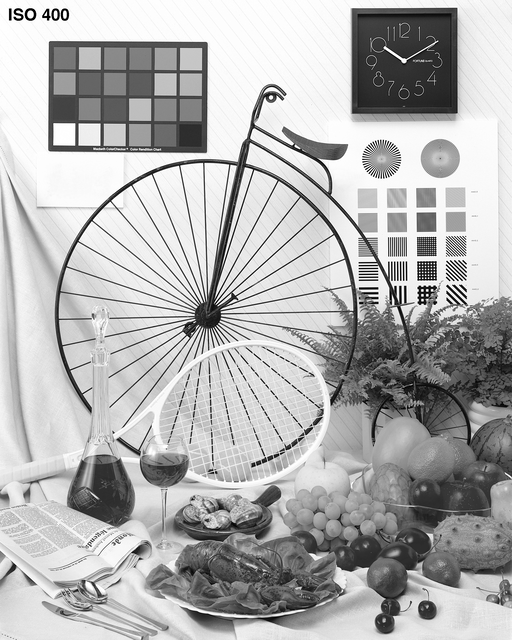} &
\includegraphics[width=0.22\linewidth]{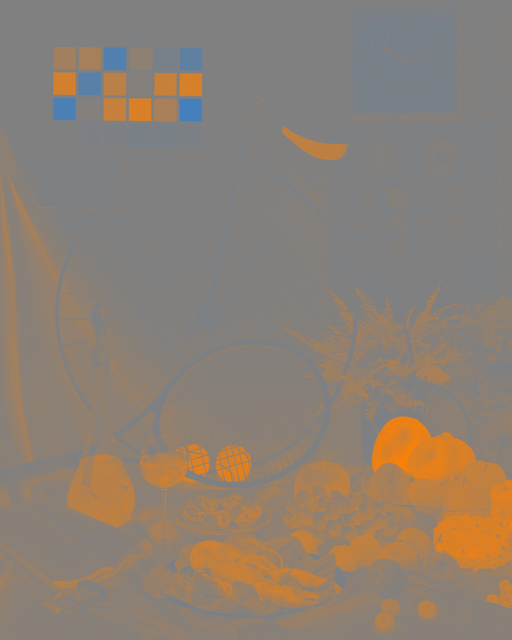} &
\includegraphics[width=0.22\linewidth]{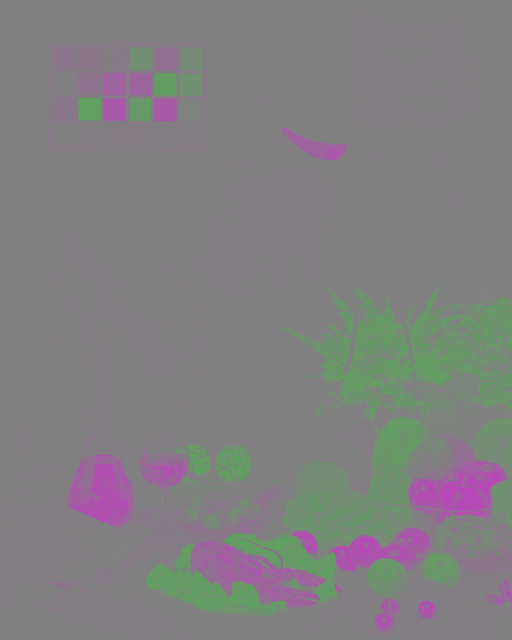} \\
& Y & Co & Cg \\
XYB: & \includegraphics[width=0.22\linewidth]{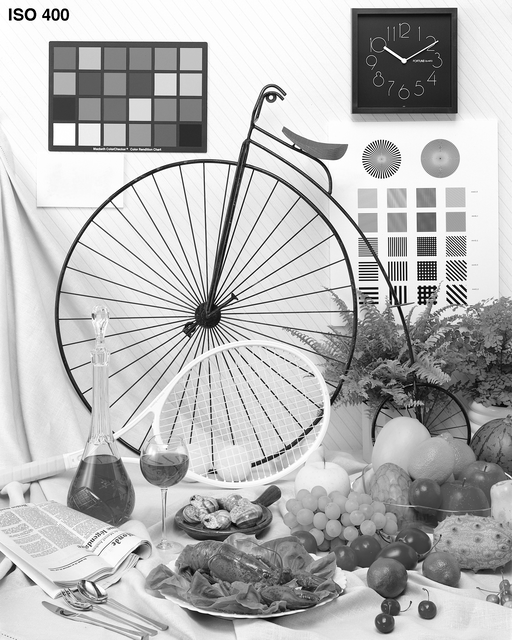} &
\includegraphics[width=0.22\linewidth]{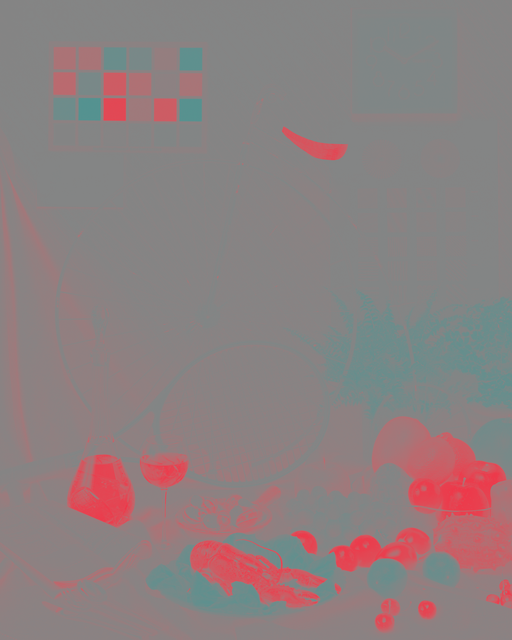} &
\includegraphics[width=0.22\linewidth]{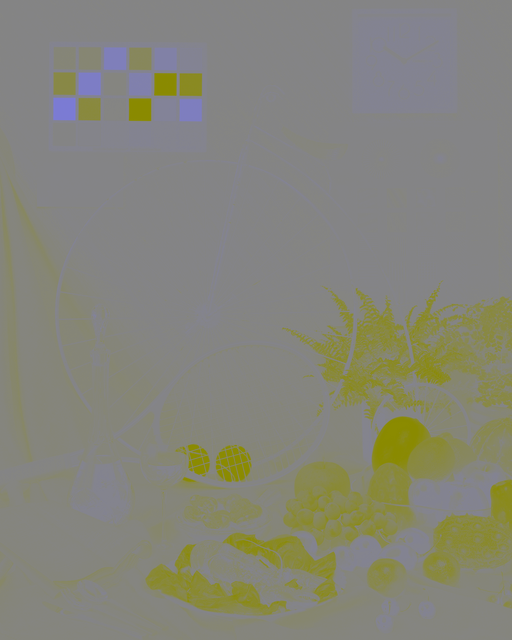} \\
& Y & X & B' \\
\end{tabular}
\caption[Decomposition of an example image into color components]{Decomposition of an example image into color components.
Top: color image (left) and RGB components, which are obviously highly correlated.
Middle: YCoCg-R, a decorrelation transform suitable for lossless compression.
Bottom: XYB, a perceptually motivated color decorrelation transform suitable for lossy compression.}
\label{fig:components_comparison}
\end{figure*}
\subsubsection{RCT}
Reversible Color Transforms can be useful to de-correlate color channels (or correlated extra channels). This transform takes any three consecutive channels with identical dimensions, permutes them in an arbitrary way, and then applies one of seven reversible transforms on them.
In total there are $42 = 7 \times 3!$ different RCTs.
The RCT transform has a parameter begin\_c which is the index of the first of the three channels on which the transform is applied, and a parameter rct\_type which identifies which RCT to apply. For example, if the initial channels are R,G,B,A and begin\_c is zero, then the transform is applied to the R,G,B channels; if begin\_c is 1, then it is applied to the G,B,A channels.

Starting from three channels RGB, rct\_type divided by 7 (rounded down) indicates the permutation: 0 means no-op (RGB), 1 corresponds to GBR, 2 to BRG, 3 to RBG, 4 to GRB, and 5 to BGR.
The remainder of the division (rct\_type \% 7) indicates which transform to apply after the permutation:
\begin{description}
\item[0 (no-op):] $(A,B,C) \longmapsto (A,B,C)$ 
\item[1:] $(A,B,C) \longmapsto (A,B,C-A)$
\item[2:] $(A,B,C) \longmapsto (A,B-A,C)$
\item[3:] $(A,B,C) \longmapsto (A,B-A,C-A)$
\item[4:] $(A,B,C) \longmapsto (A,B-\floor{\frac{A+C}{2}},C)$
\item[5:] $(A,B,C) \longmapsto (A,B-\floor{\frac{A+C}{2}},C-A)$
\item[6 (YCoCg-R):] $(A,B,C) \longmapsto (t + \floor{\frac{B-t}{2}}, A-C, B-t)$\\ 
where $t = C + \floor{\frac{A-C}{2}}$
\end{description}

For example, rct\_type 10 first permutes RGB to GBR and then applies transform number 3 so the overall effect is $(R,G,B) \longmapsto (G, B-G, R-G)$, which is also known as the `SubtractGreen' transform.

The YCoCg-R transform maps $(R,G,B)$ approximately to $(\frac{R+2G+B}{4}, R-B, G - \frac{R+B}{2})$, that is, the luma component Y corresponds to a $(0.25, 0.5, 0.25)$ weighted sum of RGB, while Co corresponds to an `orange-blue' chroma component and Cg to a `green-purple' chroma component.

Figure~\ref{fig:ycocg} visualizes the YCoCg-R color space and Figure~\ref{fig:components_comparison} illustrates it on an example image, compared to RGB and XYB decompositions.
Similar to YCbCr, YoCoCg-R is not as perceptually uniform as XYB', but unlike YCbCr and XYB', it has the advantage of being reversible in integer arithmetic while only requiring two additional precision bits: if the RGB samples are $n$-bit, then after applying YCoCg-R, the resulting Y values are still $n$-bit while the Co and Cg values each require $n+1$ bits as their range expands from $[0,2^n-1]$ to $[-2^n+1,2^n-1]$.

The no-op transform can be used to simply permute channels to change their order.
This can be useful in case `PrevChannel' properties are used in the MA tree that defines the context model and predictors (see \SectionName{}~\ref{modular_properties}), i.e. it may be beneficial for compression.
Note that multiple RCT transforms can be chained, which allows arbitrary permutations to be performed on an arbitrary number of channels.

\begin{figure*}\centering
	\includegraphics[width=0.48\linewidth]{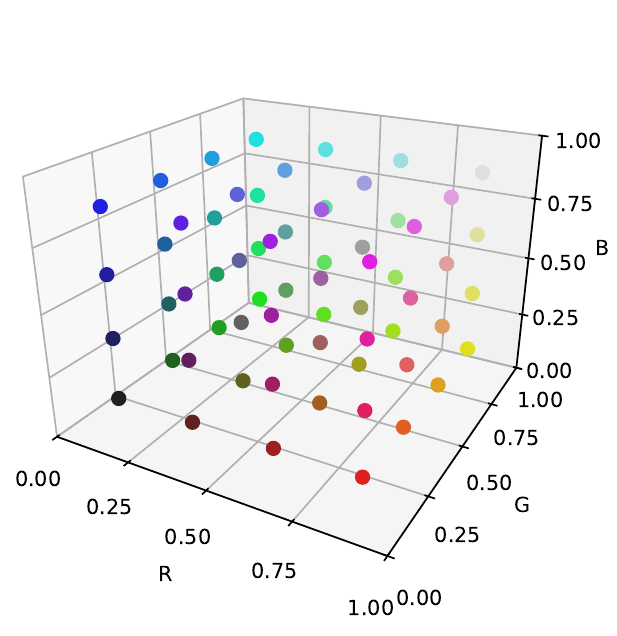}
	\includegraphics[width=0.48\linewidth]{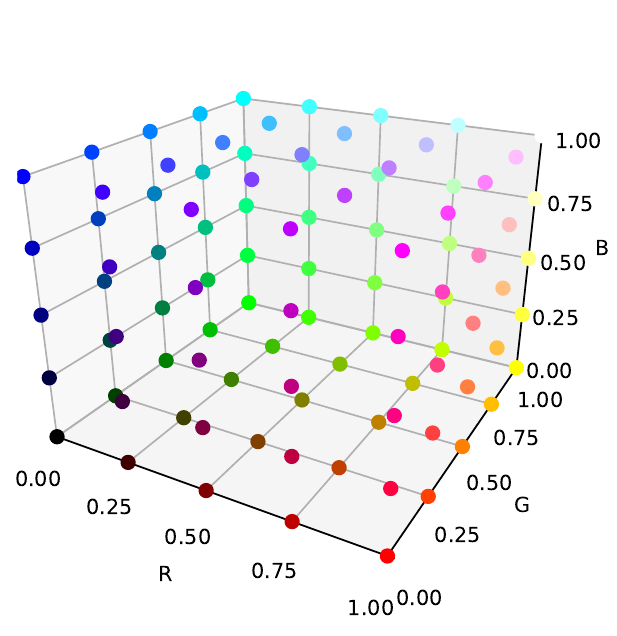}
	\caption[The implicit palette seen as two color cubes]{Left: the first 64 colors of the implicit palette, forming a $4 \times 4 \times 4$ cube in the RGB space of the image (here assumed to be sRGB). Right: the next 125 colors of the implicit palette, forming a $5 \times 5 \times 5$ cube (inner $3 \times 3 \times 3$ colors omitted for clarity).}
	\label{fig:implicit_palette_cubes}
\end{figure*}

\begin{figure}\centering
	\includegraphics[height=0.5\textheight]{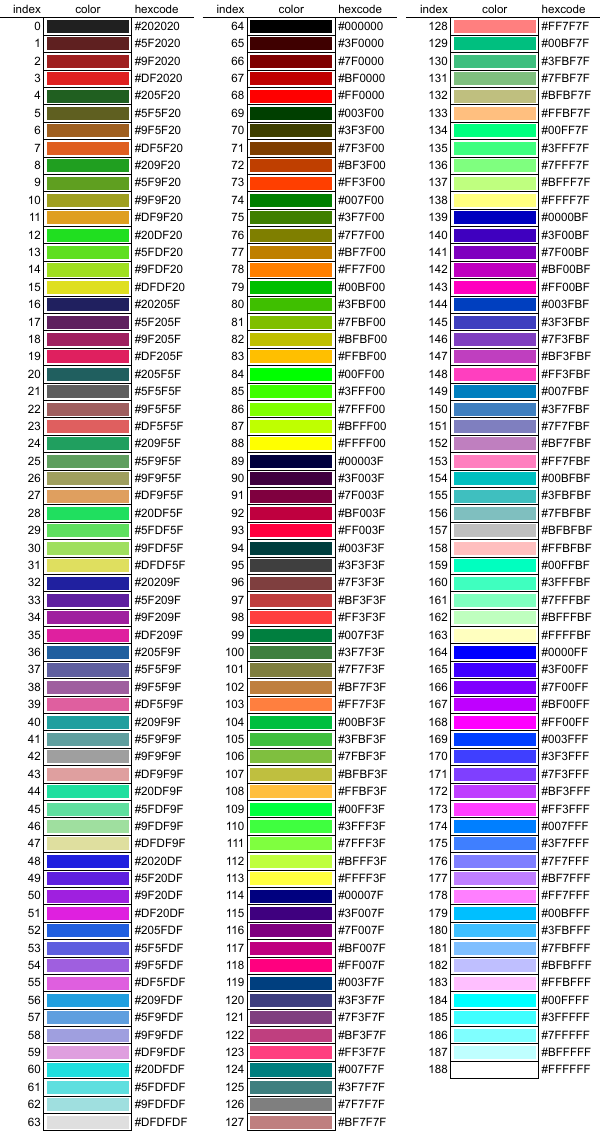}
	\caption{The implicit palette, in case of 8-bit sRGB.}
	\label{fig:implicit_palette}
\end{figure}

\begin{figure}\centering
	\includegraphics[width=\linewidth]{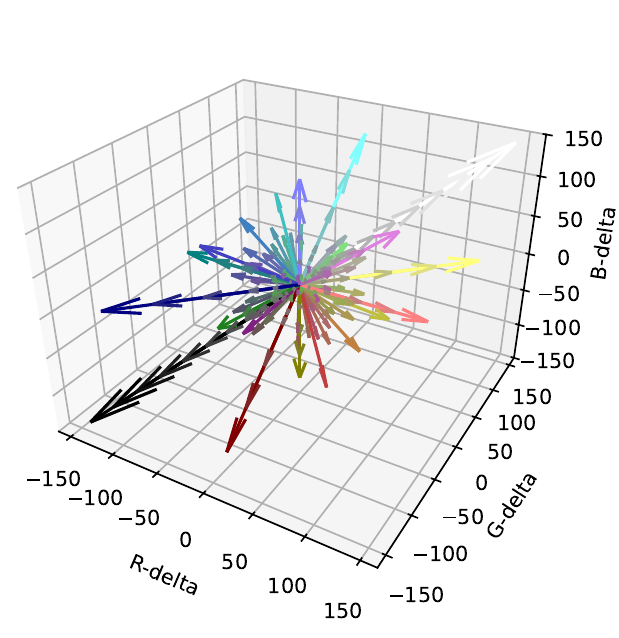}
	\caption[The implicit delta palette entries]{The implicit delta palette entries. Each delta entry corresponds to a difference vector that gets added to the predicted color; the colors shown here correspond to the resulting color if the predicted color is gray.}
	\label{fig:implicit_delta_palette}
\end{figure}
\begin{figure}\centering
	\includegraphics[width=\linewidth]{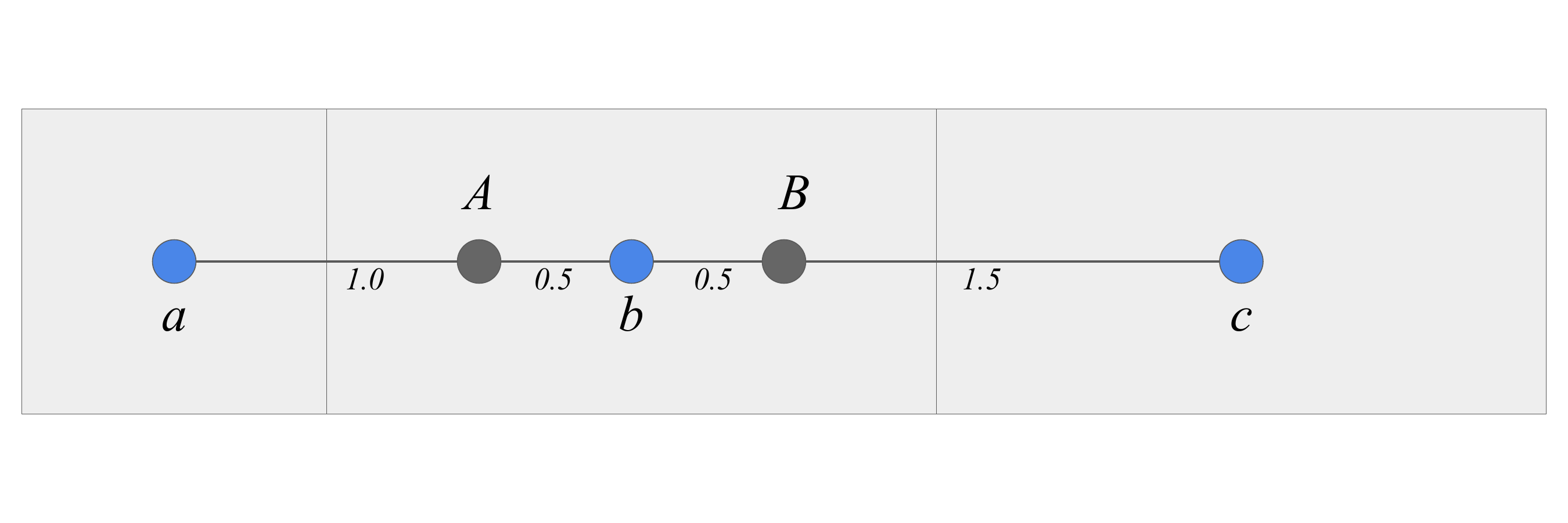}
	\caption[Sample positions related to the tendency term in Squeeze]{The sample positions in the linear interpolation conditionally used in the tendency term. }
	\label{fig:squeeze-interpolation}
\end{figure}

\subsubsection{(Delta)Palette}
The Palette transform in JPEG~XL's Modular encoding is a generalization of the color palette concept as it is available in GIF, PNG, and lossless WebP.
In these older image formats, a palette can have at most 256 distinct colors (i.e. the indices fit within a byte) and the colors have 3 or 4 components (RGB or RGBA).
JPEG~XL's Palette transform takes an arbitrary number $k$ of consecutive channels with identical dimensions, and replaces them with a single index channel of the same dimensions. An additional palette `metachannel' of dimensions $n \times k$ is added, where $n$ is the number of colors (which can be larger than 256). The maximum number of colors is limited to 70911 by the transform description syntax. This palette channel lists the colors as $n$ tuples of $k$ sample values.

\paragraph{Channel palette.}
In particular, the Palette transform can be applied to a single channel, which can be useful to reduce the range of the encoded indices. For example, an image could nominally have a bit depth of 16-bit, but contain only 3000 distinct sample values. In that case, entropy coding will be substantially improved when the main data is in $[0,2999]$ range rather than scattered throughout the $[0,65535]$ range.

\paragraph{Implicit palette.}
Additionally, a set of 189 predefined implicit colors is added to the palette, corresponding to a $4 \times 4 \times 4$ cube interleaved within a $5 \times 5 \times 5$ cube, both uniformly sampling the RGB volume. These colors do not need to be signaled explicitly. Whenever indices outside the range (i.e. indices $\geq n$) are used, they refer to this implicit palette.
Figure~\ref{fig:implicit_palette_cubes} illustrates how the implicit palette is defined.
The color values are scaled according to the bit depth of the image and they are relative to the RGB color space defined in the image header.
Figure~\ref{fig:implicit_palette} lists the colors.

Thanks to this implicit palette, the Palette transform can be useful even for $n = 0$, i.e. without any explicitly signaled palette colors.

\paragraph{Delta palette.}
Optionally, the first $d$ of the $n$ palette colors can be interpreted as `delta entries'.
This means that the signaled `color' values do not represent a fixed color, but rather as a difference vector that gets applied to the \emph{predicted} sample values for the current pixel position. Both the number $d$ and the predictor choice are signaled.
For example, if interpreted as a delta entry, the `color' $(0,0,0)$ does not correspond to black, but to the predicted sample values based on the reconstructed neighboring samples, without modification.
As another example, a delta entry $(-5,0,3)$ means that for the first component 5 is subtracted from the predicted value, for the second component the prediction is used as is, and for the third component 3 is added to the predicted value.

While the WebP format also incorporates a delta palette feature, its implementation is more constrained. Within the WebP lossless specification, a given palette must consist entirely of either delta values or absolute color values; a hybrid palette containing both types of entries is not permissible within that formalism. When utilized in conjunction with a West (or Left) predictor, the delta palette mechanism exhibits a functional analogy to the Hold-And-Modify (HAM) display mode developed for the Commodore Amiga \cite{amiga} in the late 1980s. This predictive coding technique operates by taking the color value of the adjacent pixel to the west (left) and applying a delta modification from the palette to derive the color of the current pixel. This "hold-and-modify" principle is conceptually similar to how the HAM mode would hold the value of a previous pixel and modify one of its color channels. In the empirical evaluations conducted to guide the development of the Delta Palette mode, optimal compression performance was observed when employing the AvgAll-predictor in conjunction with error dithering. It is noteworthy that the optimal choice of predictor appears to be context-dependent. For instance, experimental implementation of Delta Palette coding within the WebP lossless specification yielded superior performance when utilizing the AvgW+N-predictor. This suggests that the ideal predictive strategy for delta-encoded palettes can vary based on the specific architectural details of the codec.

\paragraph{Implicit delta palette.}
Finally there are also implicit delta palette entries. If the index channel contains negative indices, they refer to one of the 143 implicit delta entries.
Figure~\ref{fig:implicit_delta_palette} illustrates the set of implicit delta entries.

Delta palette entries substantially enhance the expressivity of lossy palette-based approaches like `\verb|pngquant|'. With traditional palette coding, the number of colors is limited, which necessitates the use of dithering methods to create a more satisfactory result when the image contains slow gradients (since without dithering, the color banding would be very noticeable). However, dithering tends to counteract some of the compression gains that can be made by using palette coding.
Delta palette coding allows reproducing gradients much more accurately by using delta entries to leverage a predictor, which in many cases can predict smooth gradients well.

\begin{figure*}\centering
	\includegraphics[width=\linewidth]{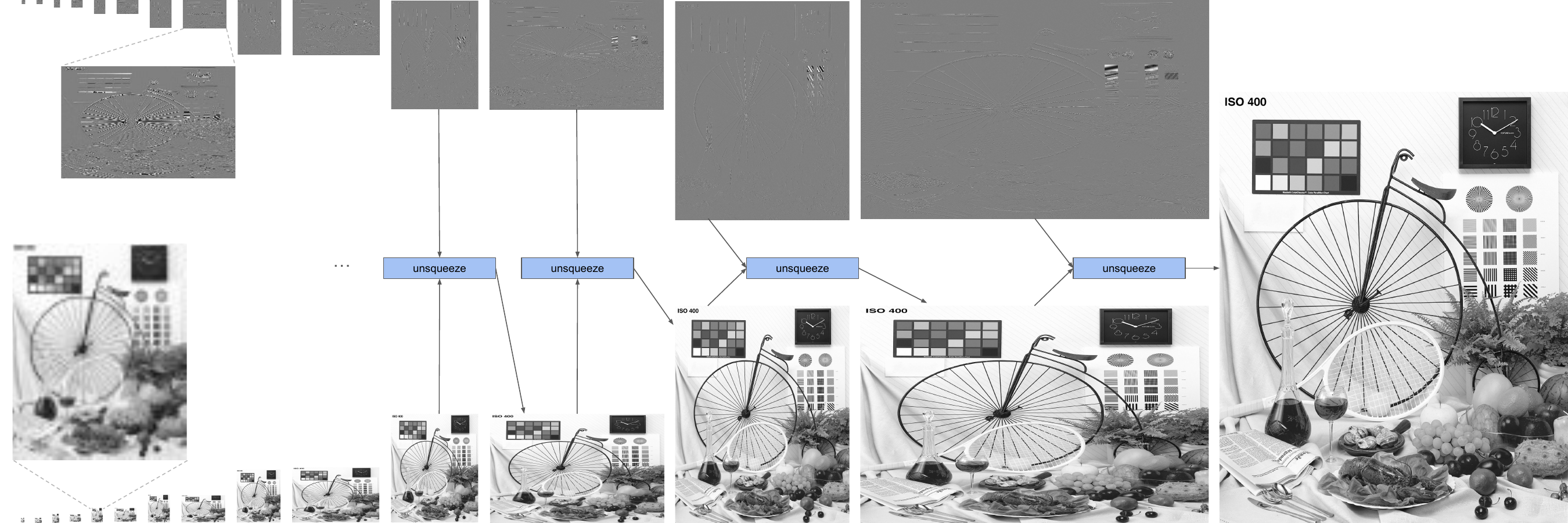}
	\caption[Result of the forward Squeeze transform with default parameters]{Top: result of the forward Squeeze transform with default parameters: a series of residual channels of increasing dimensions. Bottom: intermediate results and final result of applying the inverse Squeeze transform.}
	\label{fig:squeeze}
\end{figure*}

\begin{figure*}\centering
\includegraphics[width=0.8\linewidth]{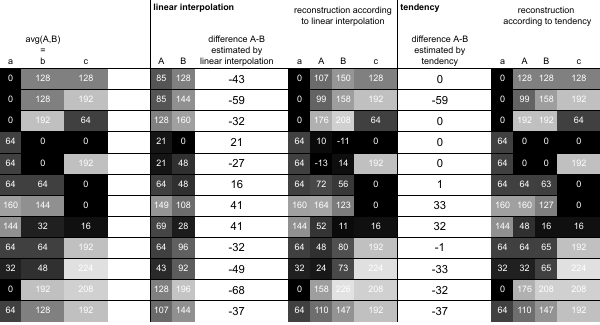}
	\caption[Examples illustrating the nonlinear tendency term]{Examples illustrating the difference between unconditional linear interpolation and the nonlinear tendency term.}
	\label{fig:tendency-table}
\end{figure*}

\begin{figure*}\centering
	\includegraphics[width=\linewidth]{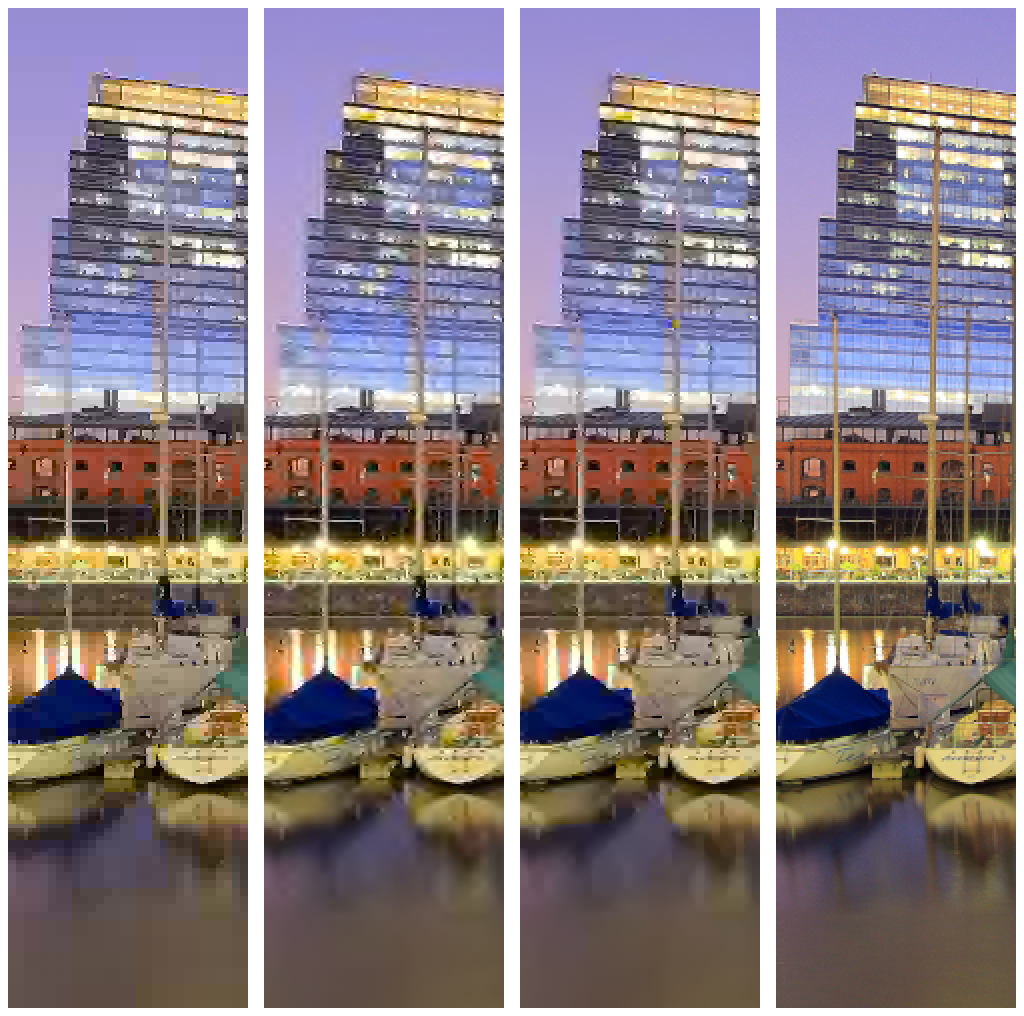}
	\caption[Visual comparison of alternatives to the tendency term]{Crop of an image compressed to 0.4 bpp using the Squeeze transform
 with quantization of the residuals. From left to right: 1) no tendency term at all,
 2) an unconditional linear interpolation tendency term,
 3) the nonlinear tendency term used in JPEG~XL,
 4) original image, for reference.}
	\label{fig:squeeze-comparison}
\end{figure*}

\subsubsection{Squeeze}
The Squeeze transform consists of a series of Squeeze steps applied in an order that is either signaled explicitly or corresponds to a predefined default order.
In each step, a channel gets replaced by two new channels that both have one of their dimensions halved, i.e. either the horizontal or the vertical dimension is divided by two.
One of these new channels will contain a downscaled version of the image (`squeezed', since the aspect ratio will change), while the other contains residuals that allow reconstructing the full-resolution image when combined with the downscaled image.
Figure~\ref{fig:squeeze} illustrates the result of the default Squeeze parameters, which repeatedly apply Squeeze steps (in alternating directions) to the squeezed channel (but not to the residual channels).

The downscaled image is obtained by averaging, i.e. for every $2 \times 1$ or $1 \times 2$ block of sample values $A$ and $B$, the value is $(A+B)/2$; if this is not an integer number, then it is rounded up if $A>B$ and down otherwise. This rounding rule mitigates the bias that would occur from using fixed rounding, which would either slightly darken or brighten the downscaled image. It also ensures that $A$ and $B$ can be easily reconstructed at decode time: adding $\floor{(A-B)/2}$ to the properly rounded average will result in $A$.

\paragraph{Tendency term.}
The residuals are not simply $A-B$, but the value of $A-B$ can be derived from the residuals. Using just the differences as residuals would result in a pixelated image if the high-frequency residuals get quantized. Instead, a `tendency term' is subtracted from the residuals, which is based on the previously reconstructed nearest neighboring sample $a$ (on the left in case of a horizontal squeeze step, above in case of a vertical squeeze step) and on the sample values $b$ and $c$ in the downscaled image at the current and next (right or below) positions. 
It can be seen as a conditional, constrained (and thus non-linear) linear interpolation: if this sequence $a,b,c$ of samples (where $b = (A+B)/2$) is monotonically increasing ($a \leq b \leq c$) or decreasing ($a \geq b \geq c$), then the tendency term will correspond to the interpolated difference, constrained to a value that respects the monotonicity (i.e. $a \leq A \leq B \leq c$, or $a \geq A \geq B \geq c$); if however the sequence is not like that --- i.e. $b$ is a local maximum ($a < b > c$) or minimum ($a > b < c$) --- then the tendency term is zero.

Figure~\ref{fig:squeeze-interpolation} illustrates the sample positions.
Given $a$ and $b$, the value of $A$ can be interpolated as $(a+2b)/3$.
Given $b$ and $c$, the value of $B$ can be interpolated as $(3b+c)/4$.
So the difference $A-B$ can be estimated to be $(4a - 3c - b) / 12$.
However, as shown in Figure~\ref{fig:tendency-table}, using this estimate unconditionally can lead to overshoot/undershoot, which can cause ringing artifacts (if the high-frequency residuals are quantized).
For this reason, the tendency term is defined in a nonlinear way which ensures that the estimated difference $A-B$ will always keep the reconstructed $A$ and $B$ values within the range $[\min(a,b,c),\max(a,b,c)]$, making overshoot/undershoot impossible.

The advantage of the nonlinearity in the tendency term is illustrated in Figure~\ref{fig:squeeze-comparison}. Here an image was compressed aggressively to 0.4 bpp using two variants of Squeeze (one without any tendency term, the other with unconditional linear interpolation) and with Squeeze as it is defined in JPEG~XL. In the first variant, the pixelation or `blockiness' is obvious. In the second variant, pixelation is avoided but there are ringing artifacts at the strong edges. With the nonlinear tendency term, artifacts are more balanced and both pixelation and ringing can be mitigated.

\subsection{Channel coding}
The (possibly transformed) image data is coded in a planar, channel-by-channel way, not in an interleaved (pixel-by-pixel) way.
Each channel is traversed in the usual scanline order (row by row, from left to right in each row).
However the data encoded in a single modular sub-bitstream typically corresponds only to a $256 \times 256$ region of the image, corresponding to a single group (see \SectionName{}~\ref{groups}).

For each sample value to be coded, a decision tree called the MA tree is traversed, which determines the predictor to use, the multiplier and offset to apply to the residual, and the context to use for entropy coding.
The following pseudocode describes the decoding process:

{\footnotesize
\begin{verbatim}
for (i = 0; i < channel.size(); i++) {
  for (y = 0; y < channel[i].height; y++) {
    for (x = 0; x < channel[i].width; x++) {
      props = GetProperties(i, x, y);
      node = MA_tree_root;
      while (node.is_decision_node()) {
        if (props[node.property] > node.value)
             node = node.left_child;        
        else node = node.right_child;
      }
      diff = DecodeHybridUint(node.context);
      diff = UnpackSigned(diff);
      diff *= node.multiplier;
      diff += node.offset;
      guess = Predict(node.predictor, i, x, y);
      channel[i].sample(x, y) = guess + diff;
    }
  }
}
\end{verbatim}
}

The array of local properties ({\small \verb|props|}) is described in \SectionName{}~\ref{modular_properties}.
These properties are referenced by decision nodes in the MA trees (\SectionName{}~\ref{modular_trees}).
Entropy coding ({\small \verb|DecodeHybridUint|}) will be discussed later, in \SectionName{}~\ref{entropy}.
The various predictors are described in \SectionName{}s~\ref{modular_predictors} and \ref{modular_weighted}.

\begin{figure}\centering
\small
\begin{tabular}{l|l}
Index & Property value \\
\hline
0 & $i$ (channel index) \\
1 & stream index \\
2 &	$y$ (row number)\\
3 &	$x$ (column number)\\
4 &	$|N|$\\
5 &	$|W|$\\
6 &	$N$\\
7 &	$W$\\
8 &	$W -$ (previous value of property 9)\\
9 & $W + N - NW$ \\
10 & $W - NW$ \\
11 & $NW - N$ \\
12 & $N - NE$ \\
13 & $N - NN$ \\
14 & $W - WW$ \\
15 & max\_error (see \SectionName{}~\ref{modular_weighted})\\
16 & $|$PrevChannel$|$\\
17 & PrevChannel\\
18 & $|$PrevChannelErr$|$\\
19 & PrevChannelErr\\
20 & $|$Prev2Channel$|$\\
21 & Prev2Channel\\
22 & $|$Prev2ChannelErr$|$\\
23 & Prev2ChannelErr\\
23 & $|$Prev3Channel$|$\\
\ldots & \ldots\\
\end{tabular}
\caption{Properties that can be referenced in an MA tree.}
\label{fig:properties}
\end{figure}

\subsubsection{Properties}
\label{modular_properties}
The following names are used for neighboring already-coded sample values:

\begin{center}
\begin{tabular}{|c|c|c|c|c|}
\hline
 & &		$NN$ & & \\		
\hline
 & $NW$ & $N$ & $NE$ &	$NEE$\\
\hline
$WW$	& $W$ & $C$	 & & \\
\hline
\end{tabular}
\end{center}

To handle edge cases: for the very first sample, all neighbors are assumed to be zero; otherwise, if $W$ is missing $N$ is used instead, if $N$, $NW$, or $WW$ are missing $W$ is used instead, if $NE$ or $NN$ are missing, $N$ is used instead, and if $NEE$ is missing, $NE$ is used instead.

Figure~\ref{fig:properties} lists all the local properties that can be referenced in an MA tree decision node.

Some of these properties are static within the inner coding loop, which means that in an optimized implementation, it can be worthwhile to specialize the tree to avoid unnecessary branching on tests involving these properties.
For example, the stream index (property 1) is a unique identifier for each modular sub-bitstream and remains constant during the entire sub-bitstream; the channel index i (property 0) and the row number y (property 2) are also unchanged during the inner loop.

The two nearest (available) neighbors, N and W, can be referenced either directly (properties 6 and 7) or using their absolute value instead (properties 4 and 5). If the sample values are all positive, this is of course redundant, but in many cases the sample values are signed and centered on zero, for example when the channel corresponds to a chroma channel (in XYB or if an RCT was used) or when it corresponds to Squeeze residuals.
In these cases it can be useful in an MA tree to distinguish between low-amplitude and high-amplitude values, which can be done with a single test against the absolute value.

Property 9 is the unclamped `gradient' predictor, which is a simple yet effective prediction for the current pixel. Property 8 then corresponds to the `prediction miss' at the previous position.

Properties 10 to 14 are identical to the neighbor differences used in the context model of FFV1 \cite{ffv1}. One difference though is that in FFV1 these differences are quantized while here they are used as is.

The value of max\_error (property 15) is related to the self-correcting predictor (\SectionName{}~\ref{modular_weighted}).

The remaining properties refer to previously coded channels; in libjxl these properties are not used by default since it causes a substantial degradation in memory locality and thus speed (for both encoding and decoding), though these properties can often be useful to improve compression.
PrevChannel is the sample value at the same position as the current sample, but in the previous already-coded channel with the same dimensions as the current channel.
PrevChannelErr corresponds to the prediction miss w.r.t. the clamped gradient predictor.
Prev2Channel is the same as PrevChannel but for the channel that came before the previous one, and so on.

In this way, any previously coded channel (with the same dimensions) can be used as part of the context model. This also implies that the channel ordering does matter, since there can only be references to already-coded channels.
For that reason, the RCT transform allows doing transforms that simply permute the channels --- if not for these `previous channel' properties, such permute-only RCTs would serve no purpose.

\begin{figure}\centering
\setlength{\tabcolsep}{3pt}
\begin{tabular}{l|l|l}
 & Name & Predictor value \\
\hline
0 &	Zero & $0$ \\
1 &	West & $W$ \\
2 &	North & $N$ \\
3 &	AvgW+N & $(W + N) / 2$ \\
4 &	Select & \footnotesize if $|N - NW| < |W - NW|$, then $W$, otherwise $N$ \\
5 &	Gradient & \footnotesize$\min(\max(W + N - NW, \min(W, N)), \max(W, N))$ \\
6 &	Weighted	& $p''$, see \SectionName{}~\ref{modular_weighted} \\
7 &	NorthEast & $NE$ \\
8 &	NorthWest & $NW$ \\
9 &	WestWest & $WW$ \\
10 & AvgW+NW & $(W + NW) / 2$ \\
11 & AvgN+NW & $(N + NW) / 2$ \\
12 & AvgN+NE & $(N + NE) / 2$ \\
13 & AvgAll & $\frac{6 N - 2 NN + 7 W + WW + NEE + 3 NE + 8}{16}$\\
\end{tabular}
\caption[Predictors available in Modular coding]{Predictors available in Modular coding. Division is integer division, rounding towards zero.}
\label{fig:predictors}
\end{figure}
\subsubsection{MA trees}
\label{modular_trees}
A meta-adaptive (MA) tree is a binary tree that is interpreted as a decision tree to determine both the context and predictor.
In a modular sub-bitstream, either a `local MA tree' is signaled in the beginning of the sub-bitstream, or a previously coded `global MA tree' is reused.
An encoder can use either approach: in principle a single global tree is just as expressive as having many local trees, since the stream index is one of the properties that can be referenced.
Thus, different local trees can be represented as subtrees of a global tree that starts with decision nodes that create different branches for each stream index.
However, an encoder may want to start producing output before it has received all input, or process different modular sub-bitstreams in parallel. In these cases, local trees can be more convenient and efficient.

MA trees work as follows.
Decision nodes (all of the inner nodes) are all of the same form: ``is the value of property number $i$ larger than threshold $v$?''. If the answer is yes, the left branch is followed, otherwise the right branch is followed.
In terms of signaling, the tree is traversed in a breadth-first way; for decision nodes only the property index $i$ and threshold $v$ have to be coded explicitly and the pointers to both branches can be left implicit.

Leaf nodes contain the following information:
\begin{description}
\item[context:]
The index of the context to be used for entropy coding.
This is not signaled explicitly but simply corresponds to the enumeration order of the leaf nodes. The context map (\SectionName{}~\ref{context_map}) then remaps these indices, so after remapping different leaves can share the same context.

\item[predictor:]
Which predictor to use. Figure~\ref{fig:predictors} lists the 14 predictors that are available.

\item[multiplier:]
A constant that is multiplied with the residual before it gets added to (or in an encoder, subtracted from) the predicted value.
These multipliers act like a quantization factor and can be used in this way by an encoder.

\item[offset:]
An additive constant that further modifies the final version. This mechanism can be useful to apply a global shift or bias to the distribution of residuals in a given context, which may improve opportunities for sharing contexts between different leaf nodes and thus reduce histogram signaling overhead.
\end{description}

\begin{figure*}\centering
\setlength{\tabcolsep}{2.5pt}
\renewcommand{\arraystretch}{1.3}
\begin{tabular}{ccccc}
Zero & West & North & AvgW+N & Select \\
\includegraphics[width=0.19\linewidth]{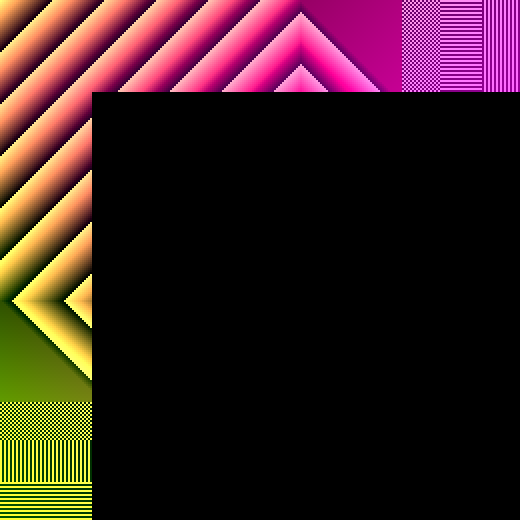} &
\includegraphics[width=0.19\linewidth]{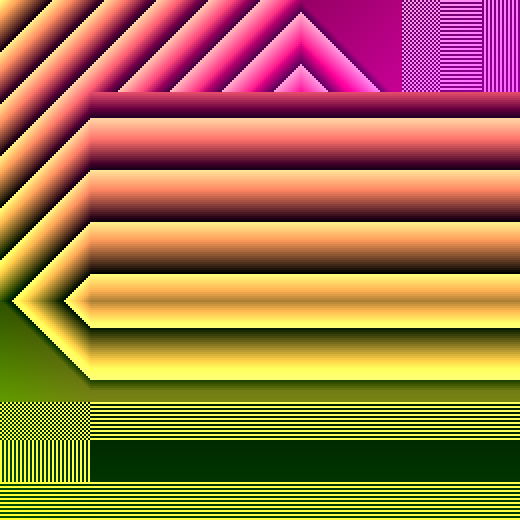} &
\includegraphics[width=0.19\linewidth]{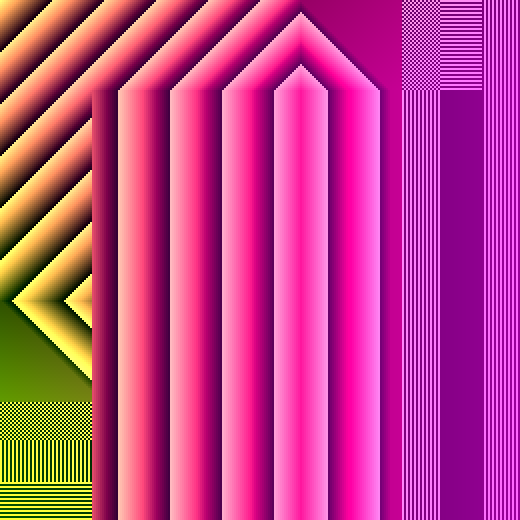} &
\includegraphics[width=0.19\linewidth]{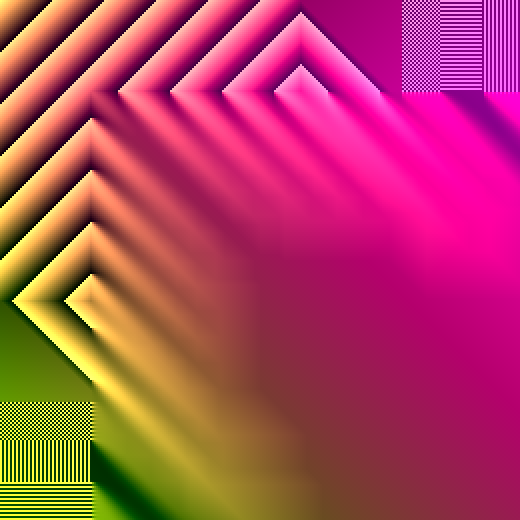} &
\includegraphics[width=0.19\linewidth]{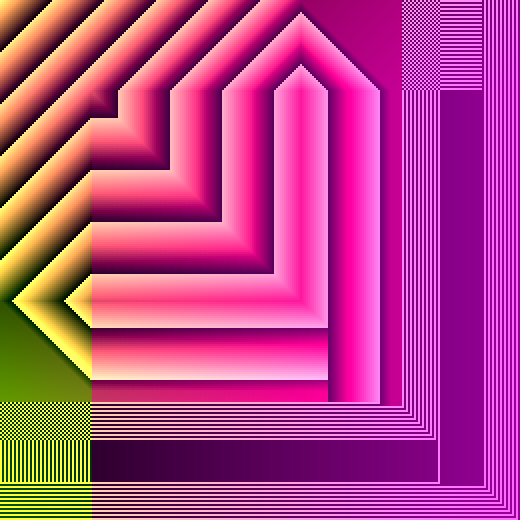} \\
Gradient & Weighted & NorthEast & NorthWest & WestWest \\
\includegraphics[width=0.19\linewidth]{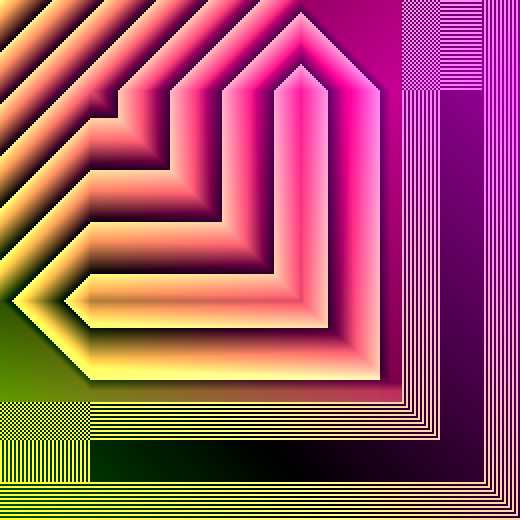} &
\includegraphics[width=0.19\linewidth]{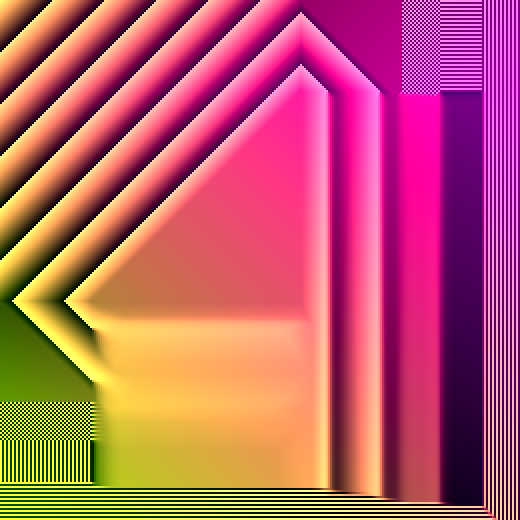} &
\includegraphics[width=0.19\linewidth]{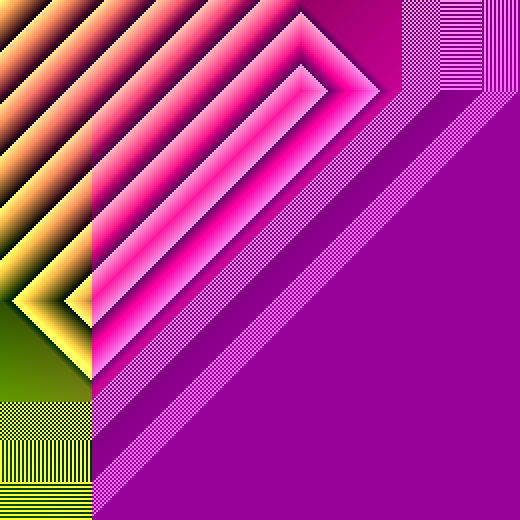} &
\includegraphics[width=0.19\linewidth]{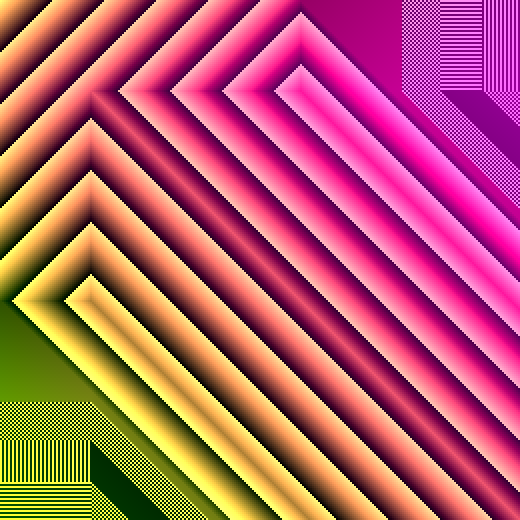} &
\includegraphics[width=0.19\linewidth]{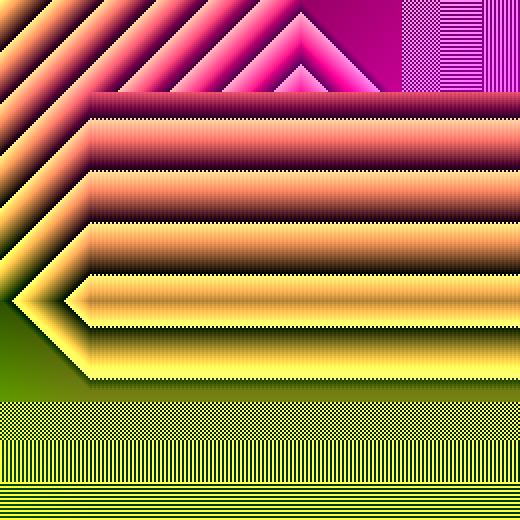} \\
AvgW+NW & AvgN+NW & AvgN+NE & AvgAll & \\
\includegraphics[width=0.19\linewidth]{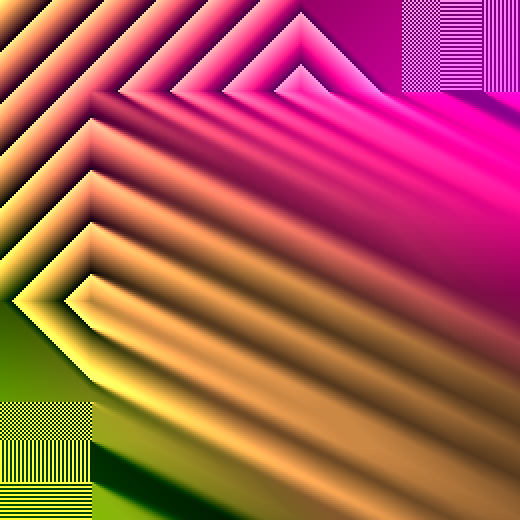} &
\includegraphics[width=0.19\linewidth]{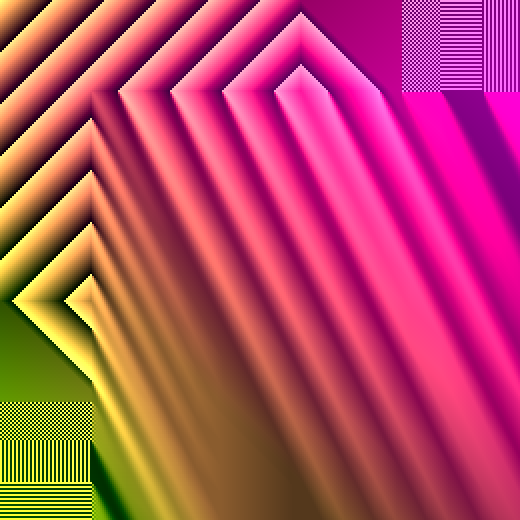} &
\includegraphics[width=0.19\linewidth]{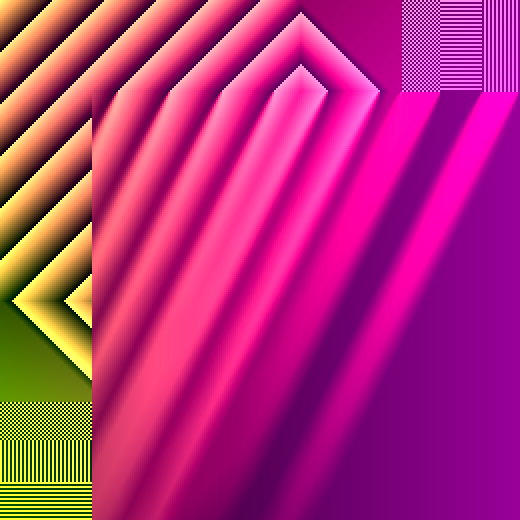} &
\includegraphics[width=0.19\linewidth]{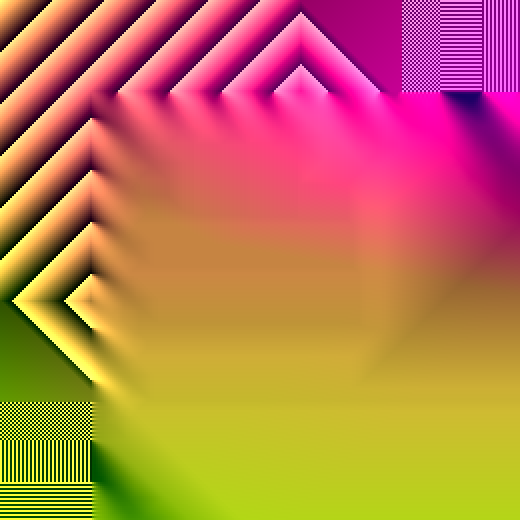} & \\
\end{tabular}
\caption[Example of how the various predictors continue a pattern]{Example of how the various predictors would continue an initial image pattern if all prediction residuals would be zero.
\href{https://jxl-art.lucaversari.it/?zcode=fZAxb8MgEIV3fsWbI0XBVhxVHTo0SycPSSVmy5D6JBssfFXovy_YccniLtwD3r374J1Ym5E7vAhFOtbyJMWHoa-OZ3k5f-IkBN3wgzccq6TCog47jN5oatl5sIM2g7MT-4bNK3YHsYeac4yGFMl9o74HdwZkianp4d19QmM1Wtd_D3bCnSLB5AaDsWE23qacOLGNE4sYWMf1gVJUcpWlzLLM8vgk57Y6ShlTrob_allVCXTe15g71KatVsv81aGwL7JDimfUdLd-1oM1ZNaQWUNmDZl1G2I9UQvt1qMWXPkP7i8}{(link to jxl-art)}}
\label{fig:predictors-example}
\end{figure*}

\begin{figure*}\centering
\setlength{\tabcolsep}{1pt}
\renewcommand{\arraystretch}{0.5}
\small
\begin{tabular}{ccccc}
\emph{Image} & \emph{MA tree} & \emph{Contexts} & \emph{Predictors} & \emph{Residuals}\\
\hline
\includegraphics[width=0.12\linewidth]{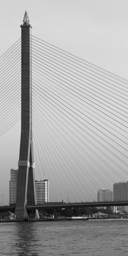} &
\multicolumn{1}{b{0.5\linewidth}}{
\begin{center}
Effort 2: 20,048 bytes\\
(uncompressed: 32,768 bytes)\\
\vspace{1em}
\includegraphics[width=0.98\linewidth]{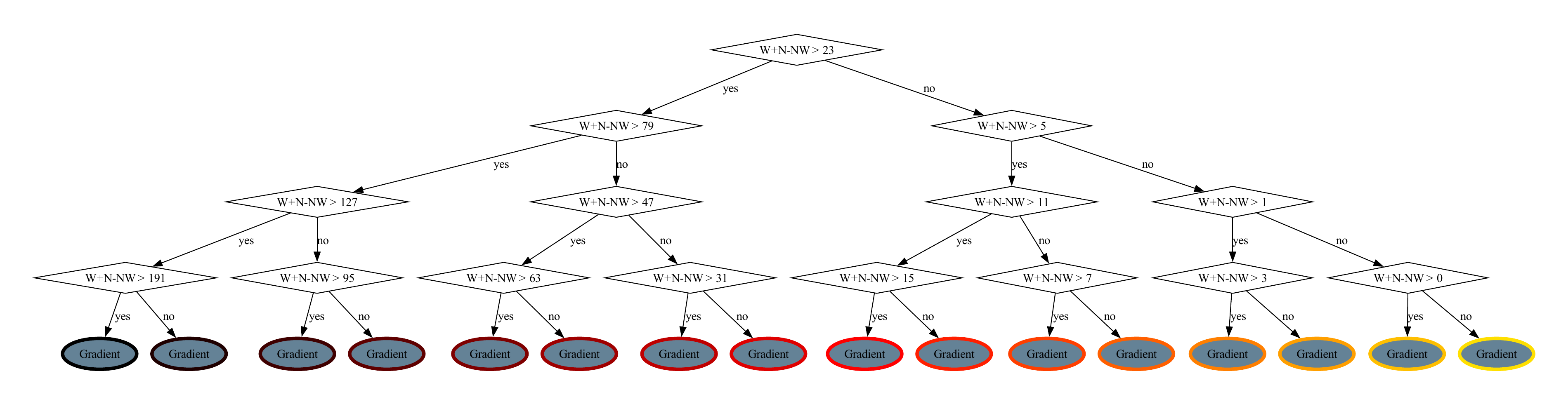}
\end{center} 
}
&
\includegraphics[width=0.12\linewidth]{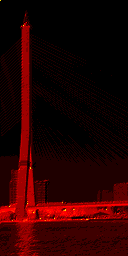} &
\includegraphics[width=0.12\linewidth]{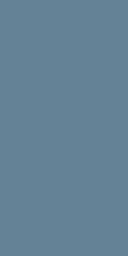} &
\includegraphics[width=0.12\linewidth]{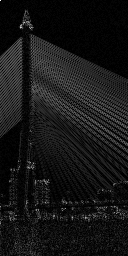} \\
\hline
\multicolumn{2}{b{0.6\linewidth}}{
\begin{center}
Effort 3: 17,216 bytes\\
\vspace{1em}
\includegraphics[width=\linewidth]{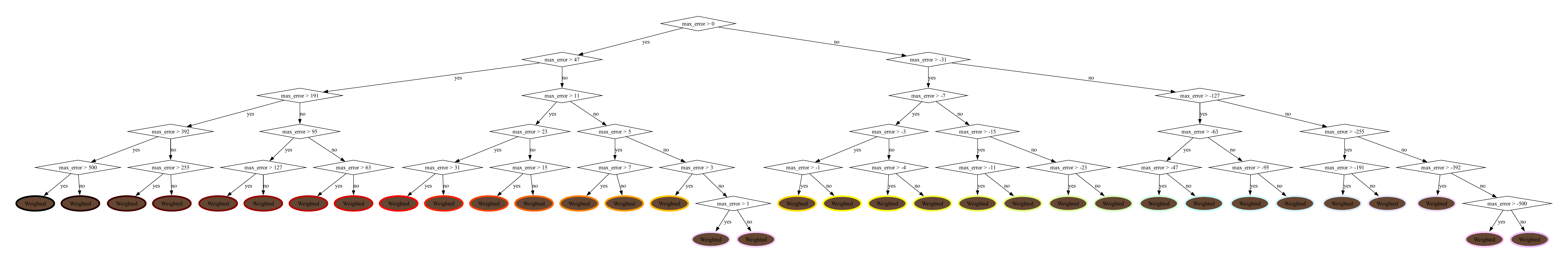}
\end{center}}&
\includegraphics[width=0.12\linewidth]{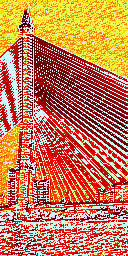} &
\includegraphics[width=0.12\linewidth]{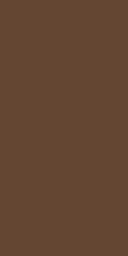} &
\includegraphics[width=0.12\linewidth]{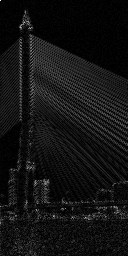} \\
\hline
\multicolumn{2}{b{0.6\linewidth}}{
\begin{center}
Effort 4: 16,625 bytes\\
\vspace{1em}
\includegraphics[width=\linewidth]{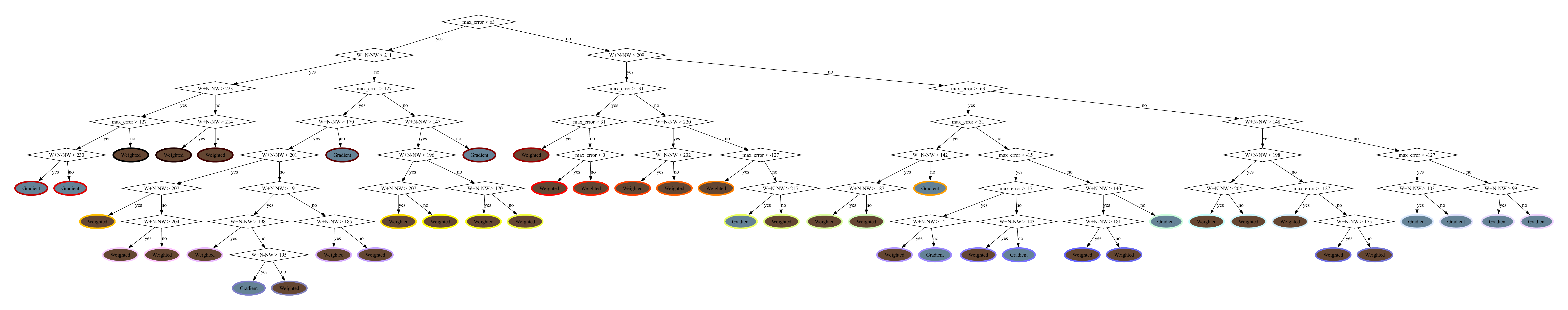}
\end{center}}&
\includegraphics[width=0.12\linewidth]{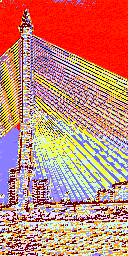} &
\includegraphics[width=0.12\linewidth]{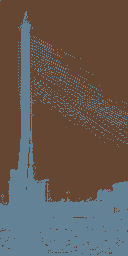} &
\includegraphics[width=0.12\linewidth]{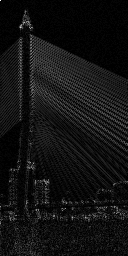} \\
\hline
\multicolumn{2}{b{0.6\linewidth}}{
\begin{center}
Effort 7: 14,945 bytes\\
\includegraphics[width=0.87\linewidth]{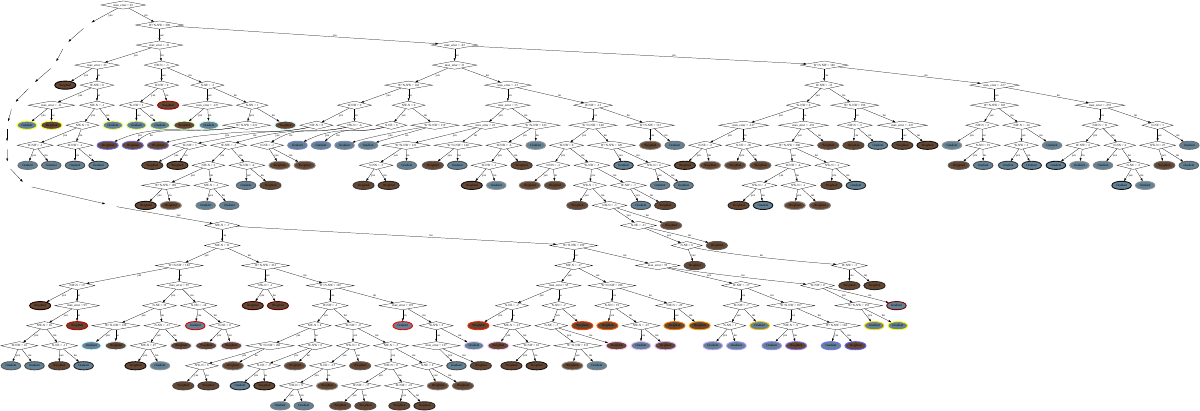}
\end{center}}&
\includegraphics[width=0.12\linewidth]{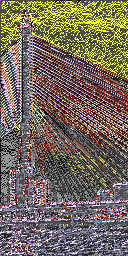} &
\includegraphics[width=0.12\linewidth]{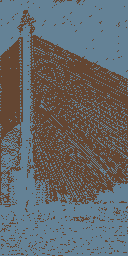} &
\includegraphics[width=0.12\linewidth]{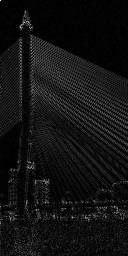} \\
\hline
\multicolumn{2}{b{0.6\linewidth}}{
\begin{center}
Effort 11: 12,789 bytes\\
\vspace{0.3em}
\includegraphics[width=\linewidth]{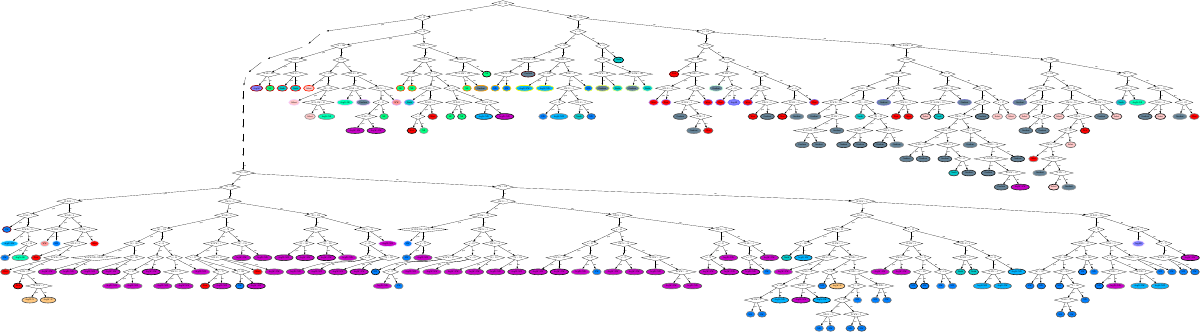}
\end{center}}&
\includegraphics[width=0.12\linewidth]{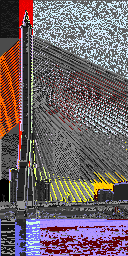} &
\includegraphics[width=0.12\linewidth]{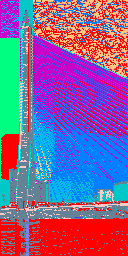} &
\includegraphics[width=0.12\linewidth]{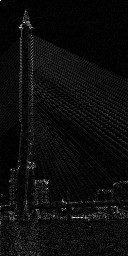}\\
\end{tabular}
\raggedleft \noindent
\includegraphics[width=0.9\linewidth]{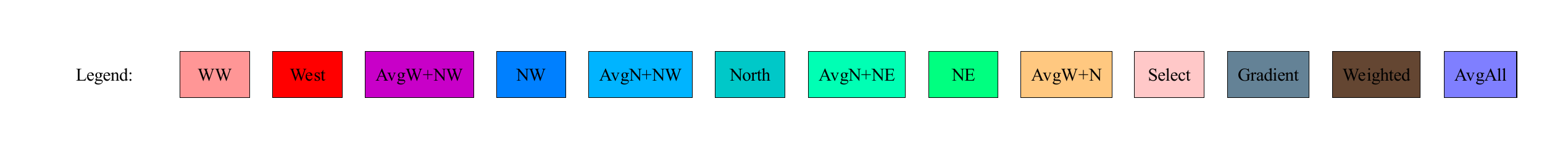}

\caption[Modular encoding at different libjxl effort settings]{Modular encoding at different libjxl effort settings. Efforts 2 and 3 use a fixed predictor and a fixed MA tree; efforts 4 and 7 use two different predictors and a constructed MA tree; effort 11 uses all possible predictors.
The leaf nodes of MA tree trees are filled with a color corresponding to the predictor and have a border corresponding to the context.
}
\label{fig:modular_example}
\end{figure*}

\subsubsection{Predictors}
\label{modular_predictors}
The different predictors are listed in Figure~\ref{fig:predictors}.
An example of how these predictors behave, if they were to be used to `fill in' a large image region, is given in Figure~\ref{fig:predictors-example}.

Figure~\ref{fig:modular_example} illustrates the overall Modular channel coding process: an MA tree defines both the context and the predictor;  better prediction leads to lower amplitude residuals; better context modeling leads to lower overall entropy.
For the same image, five different examples of MA trees are given in Figure~\ref{fig:modular_example}, corresponding to different encode effort settings.

The first four (Zero, West, North, AvgW+N) are simple and identical (at least in the 8-bit case) to filters available in PNG (None, Sub, Up, Average). The West predictor is also what JPEG uses for DC prediction and often used in TIFF.

The Select predictor originates from lossless WebP, and can be seen as a simplified yet improved version of the Paeth filter of PNG. One way to interpret is to select between $W$ and $N$ depending on which of them is closest to the (unclamped) gradient predictor $W+N-NW$, using $N$ in case of a tie.
In terms of compression results, this predictor tends to outperform Paeth, possibly since it always returns a direct neighbor rather than also having $NW$ as a third option.
It can also be implemented more efficiently.

The (clamped) Gradient predictor is also used in FFV1 and non-interlaced FLIF, as well as in lossless JPEG and WebP (albeit without clamping or with different clamping). It has the nice property that in case the local neighborhood corresponds to a smooth linear gradient in any direction, the prediction is correct.
While the Select predictor is mostly useful for non-photographic images with a low number of distinct colors, the Gradient predictor is more generally applicable and also works well for images containing smooth color changes.
The clamping is equivalent to returning the median value of ${W+N-NW, W, N}$; this is how it is defined in FLIF and FFV1, where it is called the median predictor.
This clamping ensures that the predicted values remain within the original range. In lossless WebP (which only handles 8-bit sample values), values are simply clamped to the $[0,255]$ range.

The Weighted (or Self-correcting) predictor is rather complicated. It works particularly well for photographic images. It will be explained in the next section.

The next three predictors --- NorthEast, NorthWest, and WestWest --- are less useful as general-purpose predictors that can be used throughout an image; after all, they correspond to further-away neighbors while nearer neighbors could be used instead.
However, in combination with a meta-adaptive context model, they can become useful.
The MA tree can restrict the use of these predictors to specific cases where they work well, e.g. in regions where the predominant local image features are diagonal ($NW$, $NE$) or where dithering patterns or other `dotted' pixel-level details are common ($WW$).

Finally there are three simple averaging predictors that can be seen as diagonal predictions at intermediate angles, and the ``AvgAll'' predictor that is based on a larger set of six neighboring pixels, including relatively distant samples at a sample grid distance of 2 ($NN$ and $WW$) and even $\sqrt{5}$ ($NEE$).
The AvgAll predictor was designed specifically for use in combination with the Delta Palette transform.

\subsubsection{Self-correcting predictor}
\label{modular_weighted}

Predictor number 6 is known as the ``weighted'' or ``self-correcting'' predictor.
Unlike the other predictors, it requires maintaining a state (beyond just accessing the values of neighboring already-decoded samples). The state that it requires is proportional to the row length. This state contains the prediction errors (the differences between predicted values and the actual values) for previously-coded samples, both on the current row (to the left of the current sample) and on the row above.

This predictor is based on four different sub-predictors. The final prediction it returns is a weighted sum of those sub-predictors, where the weights depend on the local past performance of each sub-predictor.
For this reason the absolute value of the prediction error is maintained separately for each sub-predictor. Also the signed error relative to the final weighted prediction is maintained, which is used to feed back some of the prediction error into the sub-predictors themselves, which causes them to become `self-correcting'.

The sub-predictors, weighted prediction, and absolute or signed prediction errors are computed with 3 bits of additional precision, i.e. all sample values are first multiplied by 8 ({\small \verb|sample << 3|}), and the final prediction result is divided by 8 again, rounding to the nearest integer, rounding half down ({\small \verb|(prediction + 3) >> 3|}).
For simplicity of notation, we use $X'$ to denote $8X$ and $X''$ to denote $\floor*{\frac{X + 3}{8}}$.

Still using the neighboring sample names as in \SectionName{}~\ref{modular_properties}, we use $p$ to denote the prediction at the current position and $e = p - C'$ to denote the (signed) prediction error at the current position; a notation like $e_N$ is then used to denote the prediction error at the corresponding position, i.e. $e_N = p_N - N'$.\\
The four sub-predictors $s_0, s_1, s_2, s_3$ are defined as follows:

\begin{flalign}
&s_0 = W' + NE' - N'&\\
&s_1 = N' - \floor*{\frac{w_1 (e_W + e_N + e_{NE})}{32}}&\\
&s_2 = W' - \floor*{\frac{w_2 (e_W + e_N + e_{NW})}{32}}&\\
&s_3 = N' - \floor*{\frac{
u + w_6 (NN'-N') + w_7 (NW' - W')}{32}}&\\
& \mathrm{where\ } u = w_3 e_{NW} + w_4 e_N + w_5 e_{NE}  \nonumber
\end{flalign}

The weights $w_1, \ldots, w_7$ are signaled as part of the header of a modular sub-bitstream, with default values (cheaply signaled in a single bit) being $w_1 = 16, w_2 = 10, w_3 = w_4 = w_5 = 7, w_6 = w_7 = 0$, but arbitrary 5-bit values can be used instead.

For each sub-predictor $s_i$, a weight $\alpha_i$ is computed that is inversely proportional to the sum of the absolute errors of the sub-predictor at neighboring positions $W$, $WW$, $NW$, $N$, and $NE$. This computation also involves a global 4-bit multiplier that can be different for each sub-predictor and is also signaled in the header of the modular sub-bitstream; this can be used to globally adjust the balance between the four sub-predictors.

The final prediction $p$ is computed as a weighted sum
\begin{equation}
p \approx \frac{\sum_{i=0}^3 \alpha_i s_i}{\sum_{i=0}^3 \alpha_i}
\end{equation}
where the use of actual integer division operations (four divisions to compute the values of $\alpha_i$, one division for the weighted sum) is avoided, since that would make it hard to implement this predictor in an efficient way --- integer division is typically one of the slowest CPU instructions.
Instead this computation is done in an approximate way that can be implemented without any actual integer divisions, using only arithmetic shifts and a small look-up table.

For the purpose of channel coding, the prediction that is used is $p''$, i.e. the additional 3 precision bits are removed again. But for the purpose of computing prediction errors to feed them back into the sub-predictors, the additional precision is used.

\section{VarDCT}
\label{vardct}

VarDCT mode is designed for lossy compression.
While JPEG uses a fixed block size ($8 \times 8$) to apply the discrete cosine transform, and then applies a fixed quantization step to the coefficients,
in JPEG~XL both the block sizes and quantization are \emph{variable}, hence the name VarDCT.

\subsection{Block sizes and types}

\begin{figure*}\centering
	\includegraphics[width=1\linewidth]{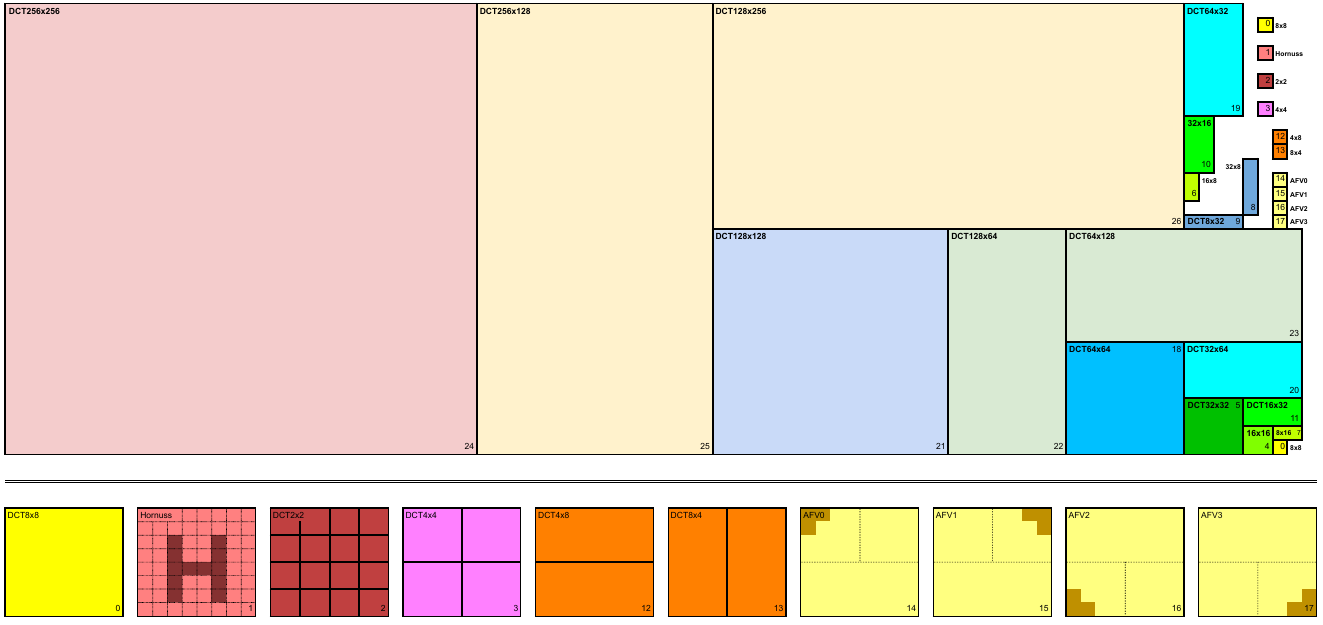}
	\caption[The block sizes and types available in VarDCT mode]{The block sizes and types available in VarDCT mode. The number in the bottom-right corner is the index of the block type, used to signal the image segmentation / block type selection. If two block types have the same color, they are rotated variants of the same transform, which share quantization tables and coefficient orderings.}
	\label{fig:blocktypes}
\end{figure*}

Figure~\ref{fig:blocktypes} gives an overview of the different block sizes and types available in VarDCT mode. Note that the transform naming uses the `rows x columns' convention (which is different from the usual `width x height' convention to denote image dimensions).

There are 10 transforms that work on an $8 \times 8$ block of pixels (they are shown in more detail in the bottom of Figure\ref{fig:blocktypes}), 5 larger square DCT transforms (from $16 \times 16$ to $256 \times 256$), and 12 rectangular DCT transforms:
5 horizontal ones with a 2:1 aspect ratio, 5 vertical ones with a 1:2 aspect ratio,
and the DCT8x32 and DCT32x8 transforms with 4:1 and 1:4 aspect ratios.

Among the $8 \times 8$ transforms, one is the DCT8x8 transform as is used in JPEG. Others can be seen as a further segmentation into smaller blocks, such as the DCT4x8 transform which is a $8 \times 8$ transform that composed of two horizontal rectangular DCT transforms.

The Hornuss transform is a direct coding of sample values in $4 \times 4$ chunks, relative to the average sample value in each chunk.
The AFV transforms are composed of a horizontal DCT4x8, a square DCT4x4, and a variant of DCT4x4 with one `corner cut'. Effectively three pixels in a corner (the corner pixel itself and its two nearest neighbors) are coded separately while a DCT transform is applied covering the remaining 13 pixels of the $4 \times 4$ block.

\begin{figure*}\centering
	\includegraphics[width=0.32\linewidth]{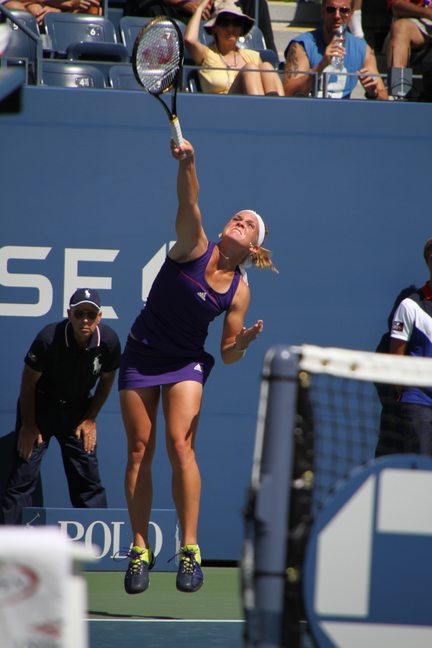}
	\includegraphics[width=0.32\linewidth]{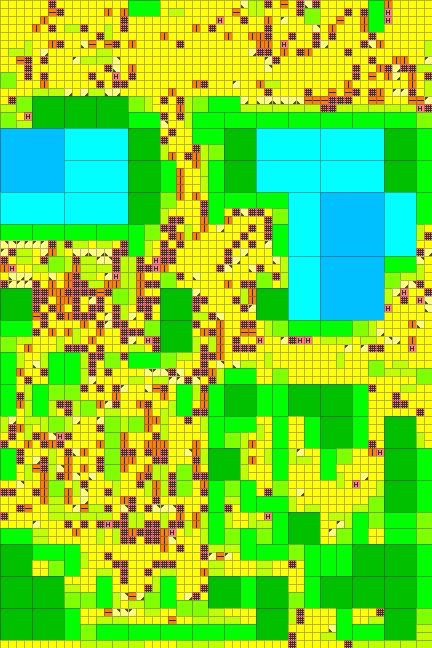}
	\includegraphics[width=0.32\linewidth]{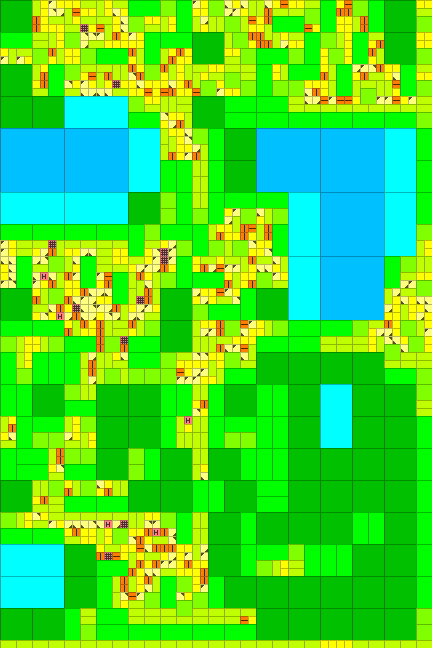}
	\caption[Example segmentations into various block types]{Example image (left) and two possible ways to segment it into various block types, corresponding to choices made by the libjxl v0.11 encoder. The middle image corresponds to a relatively high fidelity setting (distance 1), the image on the right to a lower fidelity setting (distance 4).}
	\label{fig:ac_strategy}
\end{figure*}
\begin{figure*}\centering
\setlength{\tabcolsep}{3pt}
\begin{tabular}{ccc}
& & JPEG~XL, 361 bytes \\
JPEG, 790 bytes & JPEG~XL, 361 bytes & no adaptive LF smoothing\\
\includegraphics[width=0.31\linewidth]{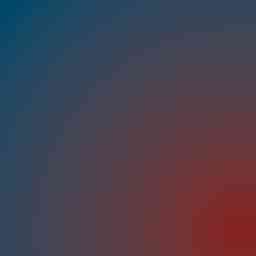} &
\includegraphics[width=0.31\linewidth]{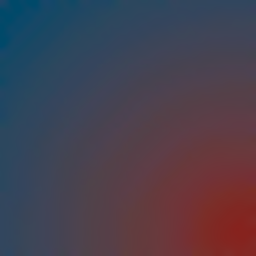} &
\includegraphics[width=0.31\linewidth]{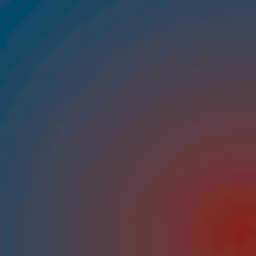} \\
AVIF, 366 bytes & HEIC, 631 bytes& WebP, 370 bytes\\
\includegraphics[width=0.31\linewidth]{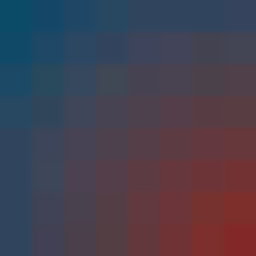} &
\includegraphics[width=0.31\linewidth]{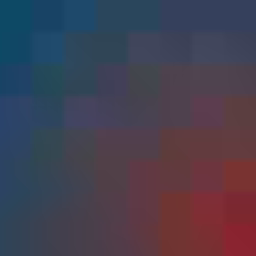} &
\includegraphics[width=0.31\linewidth]{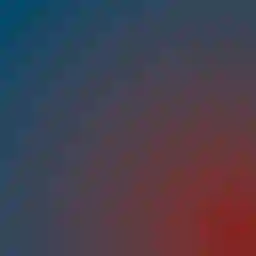} \\
\end{tabular}
\caption[Color banding artifacts in different codecs]{Color banding artifacts in slow gradients encoded at a low quality setting in different codecs.}
\label{fig:gradient}
\end{figure*}
\subsection{LF image}
In JPEG, after applying the DCT8x8 transform, the lowest frequency coefficient (also known as the DC coefficient) corresponds to the average of all the samples of the $8 \times 8$ block.
In other words, the image corresponding to just the DC coefficients is effectively a 1:8 downscaled version of the image.
When using progressive JPEG, this DC image will be the first preview.

In JPEG~XL, even though there are different block types and block sizes available, the coding is still based on a 1:8 downscaled version of the image, called the LF image.
For the DCT8x8 transform, this image contains just the DC coefficient; however for larger transforms, it also contains additional information that can be converted into additional low-frequency coefficients.

For example, the DCT8x16 transform is a horizontal rectangular transform covering two $8 \times 8$ blocks. In this case, the 1:8 image does not directly represent the DC coefficient of the $16 \times 8$ block; instead it contains the average value in each $8 \times 8$ block. From those two average values, it is then possible to reconstruct the two lowest frequency coefficients of the DCT8x16 transform, that is, the DC and the `horizontal gradient' coefficient.

Smaller transforms, such as DCT4x8, are defined in such a way that they always cover an $8 \times 8$ region and have a lowest-frequency coefficient covering the whole block. That is, the DC coefficients of their smaller components are combined into a single LF coefficient plus additional low-frequency coefficients from which the DC coefficients can be reconstructed.

There are two ways to code the LF image: directly, or by referencing a previously coded LF frame. In the case of direct coding, Modular sub-bitstreams are used to code the LF image in groups of $256 \times 256$ LF coefficients, which correspond to $2048 \times 2048$ regions of the frame.
In the other case, the entire LF image is coded as a separate LF frame at 1:8 resolution, which can itself be encoded in either Modular or VarDCT mode (in which case it can have its own second-level LF frame at 1:64 resolution, up to four levels).

\paragraph{Adaptive LF smoothing.}
When the LF image is coded directly, its sample values are quantized to integers.
Optionally, it can be signaled that an adaptive LF smoothing step has to be applied by a decoder after dequantization.
This step has the effect of smoothing the LF image in regions corresponding to slow gradients --- regions where differences between adjacent LF samples are similar to the quantization bucket size --- while not affecting the LF at all in other regions.
This helps to avoid color banding artifacts even when the LF quantization is relatively coarse and would otherwise cause noticeable blocking or color banding.
Figure~\ref{fig:gradient} illustrates how even at a very low quality setting (distance 12), JPEG~XL can preserve smooth gradients without glaring banding or blocking artifacts, partially thanks to adaptive LF smoothing.

\subsection{HF metadata}
\label{hf_metadata}
Together with the LF groups, a number of control fields are signaled to define the segmentation into blocks, weights for adaptive quantization, chroma from luma multipliers, and filter strengths.

This `HF metadata' is bundled per region corresponding to an LF group (i.e., a $2048 \times 2048$ frame region) to allow better entropy coding and achieve a lower signaling overhead compared to what would be the case if this data were to be signaled as part of the HF groups (i.e. $256 \times 256$ frame regions).
These control fields are encoded using a Modular sub-bitstream.

\subsubsection{XFromY, BFromY}
\label{xbfromy}
The first two channels of the Modular-encoded HF metadata are at 1:64 resolution, so $32 \times 32$ (or smaller, at the right and bottom edges of the frame or for frames smaller than $2048 \times 2048$).
They represent the XFromY and BFromY multipliers that are used in `Chroma from luma' (\SectionName{}~\ref{cfl}).

\subsubsection{DctSelect}
Figure~\ref{fig:ac_strategy} shows two examples of possible block segmentations for an image.

Since the block sizes can vary considerably, it would be wasteful to represent the segmentation as a 2D image at 1:8 resolution, since then larger blocks would cover multiple sample positions in that image. While it would be possible to define a more sophisticated variant of Modular coding that skips such implied sample positions (any position not in the top-left corner of a block), it would be a substantial complication with little added compression gain over the approach that was chosen instead.

The map of block types is `flattened' into a 1D array by traversing the blocks in scanline order, only recording each block when it is first encountered, i.e. at its top-left location.
The total number of blocks used in the current LF group (nb\_blocks) is signaled explicitly; it can be at most 65536 (if all blocks have the smallest size, $8 \times 8$) but typically the actual number will be lower.

Any segmentation is allowed that satisfies the following constraints: the whole frame region is covered by blocks, none of the blocks overlap, and none of the blocks cross a boundary between HF groups.

The 1D array of block types is coded as one long row of a nb\_blocks $\times 2$ channel that is the third channel of the HF metadata. The `sample values' in this row are in the range $[0,26]$.
In case of JPEG recompression, where only DCT8x8 is used, this row will be all zeroes, which will have a negligible signaling overhead.

\subsubsection{HfMul}
\label{hfmul}
The second row of this nb\_blocks $\times 2$ channel contains weights that influence the quantization of the HF coefficients of each block.
More precisely: if the sample value in this channel is $x$ for a given block, then all the quantization factors are multiplied by
$\frac{2^{16}}{g (1 + x)}$ , where $g$ is a signaled global scaling constant.

In case of JPEG recompression, where only fixed quantization is possible, it suffices to set $g = 2^{16}$ and to signal all zeroes also for this row.

\subsubsection{Sharpness}
\label{sharpness}
Finally, the fourth channel of the HF metadata is a 1:8 resolution image that signals the local strength of the edge-preserving filter (see \SectionName{}~\ref{epf}).
The sample values in this image indicate the amplitude of smoothing, in the integer range $[0,7]$. By default the value $i$ gets mapped to $\sigma = i/7$, but an arbitrary lookup table can be signaled to implement a different mapping.

\subsection{HF coding}
In VarDCT mode, the bulk of the image data consists of high-frequency (HF) coefficients, which represent all of the detail that is not present in the 1:8 resolution LF image.

\begin{figure*}\centering
	\includegraphics[width=\linewidth]{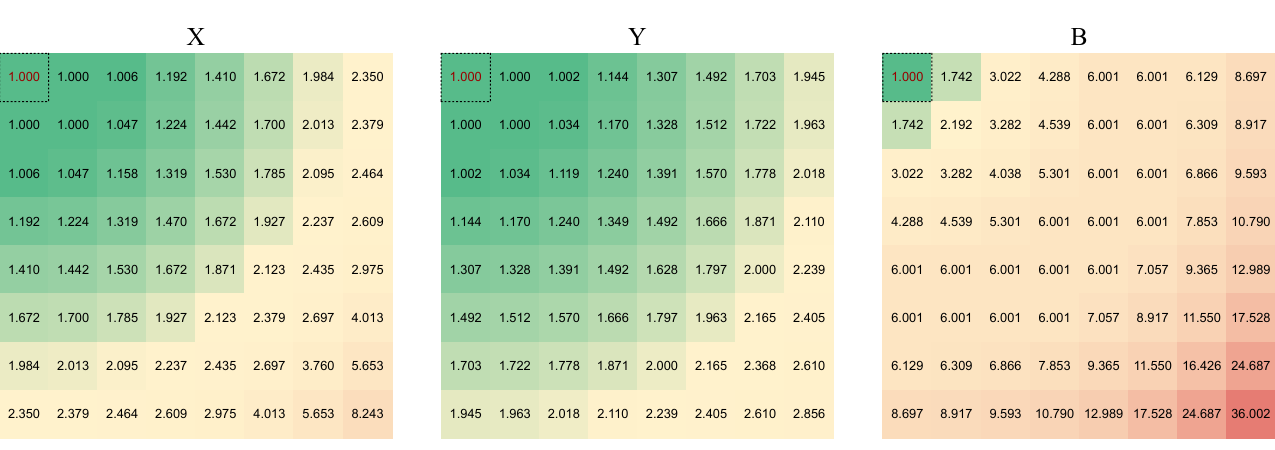}
	\caption[Default quantization tables for DCT8x8]{Default quantization tables for DCT8x8. In this figure, the factors are rescaled per component to be relative to the DC factor, since the differences in the scales of the X,Y,B components would otherwise make it hard to compare them. The coloring goes from green (finest quantization) to yellow to red (coarsest quantization).}
	\label{fig:default_quant8x8}
\end{figure*}
\begin{figure}\centering
	\includegraphics[width=1\linewidth]{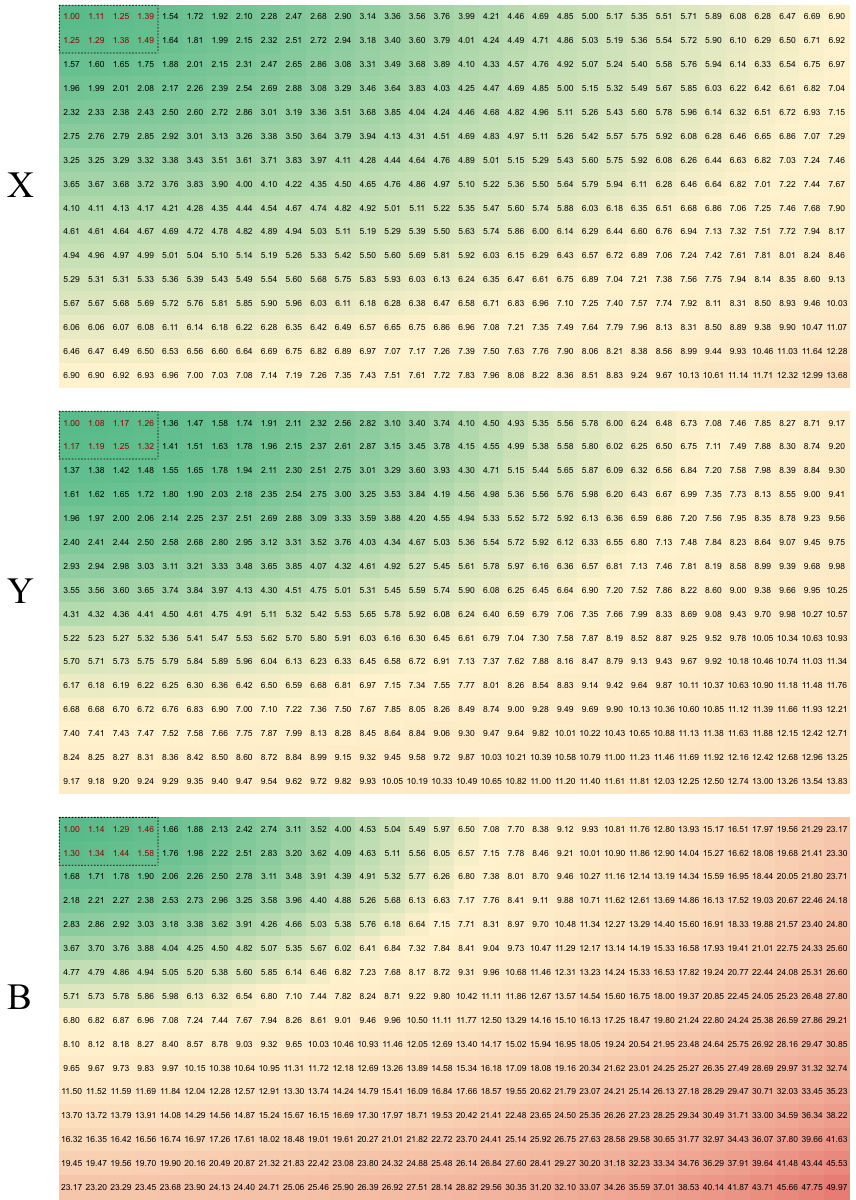}
	\caption{Default quantization tables for DCT32x16.}
	\label{fig:default_quant32x16}
\end{figure}

\subsubsection{Quantization tables}
The high-frequency DCT coefficients are quantized, just like in JPEG, using a quantization table that assigns a quantization factor to each coefficient position. Different quantization tables can be used for each component; in JPEG it is common to use two tables: one for Y and one for both chroma components (Cb and Cr). In JPEG~XL, each of the three XYB components uses a different quantization table.

The quantization tables in JPEG consist of integer numbers which are stored uncompressed, using either one or two bytes per coefficient.
In JPEG~XL, the quantization factors are not restricted to integers, allowing more fine-grained quality settings and more precise spectral allocation of bit rate. The quantization tables themselves are coded in a compressed way.

Explicit uncompressed quantization tables would have a prohibitive signaling overhead: in JPEG, typically the quantization tables require 134 bytes (two quantization tables of 64 bytes each, plus DQT header overhead), but that is just for the DCT8x8. In the case of larger transforms, the size of the quantization table also grows: for DCT256x256, the quantization table would require 65536 bytes when using 1 byte per factor!
For this reason, compression is used for coding the quantization tables, and explicit tables can generally be avoided.

Before coding the actual quantization tables, first a global scaling factor is signaled that will be multiplied to all the quantization factors.
There are three tiers of explicitness in JPEG~XL's coding of quantization tables; different options can be used for different transform types:
\begin{description}
\item[Explicit.]
The `RAW' coding mode for quantization tables consists of first signaling a half-float denominator, and then a modular sub-bitstream that represents a 3-channel image with the same dimensions as the quantization table. Each channel corresponds to the quantization table for one of the components. 
\item[Parametrized.]
In this case, quantization tables are computed using an algorithm that is part of the specification and that takes a small number of parameters as input in order to produce a quantization matrix.
\item[Default.]
The parameters for the parametrized coding mode have default values, which allow signaling quantization tables very concisely.
\end{description}

For lossless JPEG recompression, the RAW coding mode has to be used.
In other cases, libjxl simply uses the default tables, only scaling them using the global scaling factor.
This means that the overhead of signaling quantization tables is very small.
As an example, in Figures~\ref{fig:default_quant8x8}~and~\ref{fig:default_quant32x16} the default quantization tables for DCT8x8 and DCT32x16 are shown. Note that the top-left corner of the quantization table corresponds to LF coefficients which are coded in the 1:8 LF image, so they are irrelevant for HF coding.
Note that the default quantization tables for the B component are substantially steeper --- the highest-frequency coefficients get quantized much more coarsely than the lower-frequency coefficients --- than the tables for the other two components.
This difference is perceptually motivated, as explained in \SectionName{}~\ref{xyb}.

The quantization tables are signaled globally per frame.

\subsubsection{Coefficient reordering}
\label{coef_ordering}
The order in which HF coefficients are traversed during coding matters.
Especially in relatively smooth regions of an image, the highest frequency coefficients will tend to get quantized to zero, leading to longer runs of zeroes if the coefficients are ordered from low to high frequency.
For this reason, in JPEG, the coefficients are traversed in a zig-zag order.

JPEG~XL also uses a (generalized) zig-zag ordering by default, but it allows signaling an arbitrary coefficient order.
In combination with multiple progressive passes, this makes it possible to adapt the ordering to the specific pass: e.g. for a pass corresponding to a 1:2 resolution preview, all coefficients that are not in the top-left quadrant are zero, so by using an ordering that puts them at the end, they can effectively be skipped without signaling overhead.

Coefficient reordering permutations are signaled using a Lehmer code. The same signaling mechanism is used to signal table of contents (TOC) permutations, as described in \SectionName{}~\ref{lehmer}.

\subsubsection{NonZeros}
The coefficient coding scheme used in JPEG allows representing a series of consecutive zeroes with a single code, and in particular it has a special EOB (``end-of-block'') code to indicate that all remaining coefficients of the current block are zeroes.
The approach used in JPEG~XL is somewhat different.

The number of non-zero coefficients in each block is signaled explicitly. It is predicted as the average of the corresponding numbers in the blocks above and to the left, with an adjustment to handle variable-sized blocks: in case of blocks larger than $8 \times 8$, the number of non-zero coefficients is distributed evenly over the $8 \times 8$ sub-blocks. This prediction is only used for context modeling; the signaled numbers are not prediction residuals but absolute counts.

Coefficients are traversed in the order that was signaled, counting the number of non-zero coefficients coded so far. Since the total number of non-zero coefficients is signaled in advance, there is no need for an EOB code: the coding loop simply ends immediately after coding the last non-zero coefficient.

\subsubsection{Context model}
\label{context_hf}

The context model used in the entropy coding of HF coefficients is quite sophisticated.
\paragraph{Block context.}
For every block, there is a `block context' that depends on:
\begin{itemize}
\item the component index (0,1,2);
\item the block type;
\item the adaptive quantization weight corresponding to the block, quantized according to a signaled sequence of thresholds;
\item the quantized LF values for each component, corresponding to the block or its top-left sub-block, further quantized according to signaled thresholds.
This only applies when the LF is coded directly. It does not apply when an LF frame is used.
\end{itemize}

This leads to at most 2496 different initial block context indices, which are then further reduced to at most 16 different block contexts using a block context map that is signaled globally per frame.

\paragraph{Non-zero context.}
The context used to code the number of non-zero coefficients depends on the block context and a predicted value. For the purpose of computing this predicted value, the non-zero counts of previously-decoded blocks that cover more than one $8 \times 8$ block are distributed proportionally over the covered sub-blocks, that is, a non-zero count of $k$ in a block that covers $N$ sub-blocks (so it corresponds to $64N$ sample values) leads to assigning the value $\ceil{\frac{k}{N}}$ to each of these $N$ sub-blocks.
The predicted number of non-zeroes is computed as $\ceil{\frac{W+N}{2}}$, where $W$ and $N$ are the number of non-zeroes in the (sub-)blocks to the left and above the current position.
The prediction cannot be larger than 64, regardless of the size of the block types involved; in case of block types larger than $8 \times 8$, it is not a prediction for the actual non-zero count but only for the scaled-down value ($\ceil{\frac{k}{N}}$) that will be assigned to each sub-block.
The context to be used is a combination of the block context and the prediction; there is a distinct context for predictions between 0 and 7, and an additional context for every pair of consecutive predicted values between 8 and 64.
There are 37 contexts per block context, so there are at most 592 contexts in total for coding non-zero counts (since there are at most 16 distinct block contexts).

\paragraph{Coefficient context.}
The context used for the actual HF coefficients depends on:
\begin{itemize}
\item the block context;
\item the coefficient position (after reordering), quantized to 31 distinct values, with finer granularity at the beginning than at the end;
\item the number of non-zero coefficients left, quantized to 8 distinct levels;
\item a boolean that indicates ``the previous coefficient was non-zero''.
\end{itemize}

The coefficient position and number of remaining non-zeroes are combined into 229 different contexts.
Not all of the $8 \times 31 = 248$ combinations are possible since the position limits the number of remaining coefficients, which is an upper bound for the number of remaining non-zeroes.
Together with the boolean, there are 458 contexts per block context, so at most 7328 contexts in total for coding the actual coefficients.

Overall, for HF coding, there are $37+458=495$ contexts per block context, or at most 7920 contexts in total.
However, each $256 \times 256$ HF group (and in case of multiple progressive passes, also each pass) can use its own set of contexts, so the total number of contexts is multiplied by a number that is at most equal to the number of groups multiplied by the number of passes.
Different groups can use the same set of contexts; if the characteristics of the image content are similar across groups, it can be useful to reduce the signaling overhead by sharing contexts between groups.

For each HF pass in a frame, the contexts are clustered using a context map (see \SectionName{}~\ref{context_map}) into at most 255 different distributions.

\begin{figure}\centering
	\includegraphics[width=0.9\linewidth]{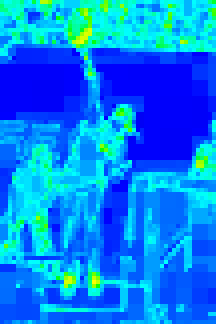}
	\caption[Heatmap showing adaptive quantization weights]{Heatmap showing adaptive quantization weights, for the middle image of Figure~\ref{fig:ac_strategy}.}
	\label{fig:quant_heatmap}
\end{figure}

\subsubsection{Adaptive quantization}
The HfMul weights that are signaled as part of the HF metadata 
(see \SectionName{}~\ref{hfmul}) can be used to locally modulate the coarseness of quantization of the HF coefficients.

In JPEG, the same quantization factors are used uniformly across the entire image. This implies that the only way to avoid artifacts in a particularly `hard' region of the image, is to globally increase the image quality.
By allowing adaptive quantization, JPEG~XL encoders can ensure a more consistent image quality.
Figure~\ref{fig:quant_heatmap} shows an example assignment of adaptive quantization weights.

The adjustment of quantization granularity doesn't just affect the precision of the HF coefficients. It also sets the baseline modulation of the edge-preserving filter (EPF, see \SectionName{}~\ref{epf}). Regions with coarser quantization are smoothed more aggressively, though the EPF strength can be additionally modulated using the Sharpness weights (see \SectionName{}~\ref{sharpness}).
Having the base EPF strength proportional to the width of the quantization buckets does allow reducing the signaling overhead for the Sharpness weights since even a simple constant Sharpness will produce meaningful EPF strength modulation.

\subsubsection{Chroma from luma}
\label{cfl}
Chroma information is locally often somewhat correlated to luma information.
The chroma from luma mechanism of JPEG~XL is quite simple: multipliers XFromY and BFromY are signaled for each $64 \times 64$ region, and in the (unquantized) HF coefficients of the X and B components, the value of the corresponding Y coefficient, multiplied by these multipliers, is subtracted before encoding (and added after decoding).

Since Chroma from Luma operates in the DCT domain and only on the HF coefficients (with the LF coefficients signaled separately as part of the LF image), this approach effectively corresponds to applying an arbitrary linear decorrelation in the pixel domain.
In the pixel domain, it can be seen as a prediction of the form $C = \alpha L + \beta$ as in \cite{av1cfl}, where the additive term $\beta$ (a level shift) is provided by the LF coefficients and the multiplier $\alpha$ is signaled through XFromY and BFromY in the HF metadata (see \SectionName{}~\ref{xbfromy}).

\section{Features, filters, and upsampling}
\label{features-filters}
Regardless of the main coding mode (Modular or VarDCT), image frames can store auxiliary data that is applied on top of the main pixel data.
There are three kinds of such auxiliary data: specific image features which are rendered over the initial image, restoration filters that can be used to reduce blocking or other compression artifacts, and upsampling algorithms that can lift image data coded at a lower resolution to a higher resolution.

\subsection{Image features}
\label{image_features}
Some kinds of image features are hard for the main coding tools (DCT and Modular) to encode efficiently. JPEG~XL offers three dedicated coding tools which provide more-effective representations of these `hard' features.

\subsubsection{Splines}
Thin high-contrast curved lines are challenging for DCT-based approaches: they require fine-grained quantization of all coefficients including the highest frequency ones, which translates to high bit rates after entropy coding. Alternatively, with a too-aggressive quantization, either the lines will get blurred (or even disappear, in extreme cases), or there will be noticeable ringing or DCT noise artifacts.

For this reason, a dedicated coding tool was added that can represent such lines directly in the form of centripetal Catmull-Rom splines \cite{splines}.
These splines are defined using a sequence of control points which are coded using a double delta encoding: the starting point and the number of points are signaled explictly and the subsequent control points are given by signaling the horizontal and vertical difference between the relative positions of the control points.

The resulting line traces an interpolated curve that passes through each of the control points. It can potentially self-intersect.
Along the arclength of this path, the color and thickness of the line can vary.
Both the color components and the Sigma parameter (which controls thickness) are signaled as entropy-coded one-dimensional DCT32 coefficients: the first coefficient represents the base color and thickness, while the remaining coefficients allow modulation of these values along the path.

Splines are rendered by adding the (signed) color sample values to the underlying background image.
The Sigma parameter can be interpreted as the variation of a Gaussian function that modulates the amplitude according to the distance from the path. This allows representing lines that can be thinner than a pixel or that have an arbitrary thickness that does not need to be an integer in pixel units, with proper anti-aliasing.
The thickness parameter is allowed to be negative, which causes the color sample values to be subtracted instead of added to the color sample values of the underlying image.

\begin{figure*}\centering
\setlength{\tabcolsep}{1pt}
\begin{tabular}{cc}
\includegraphics[width=0.65\linewidth]{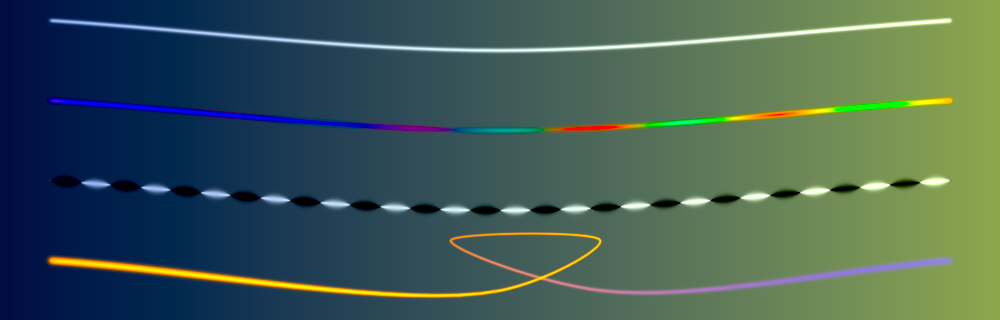} &
\includegraphics[width=0.31\linewidth]{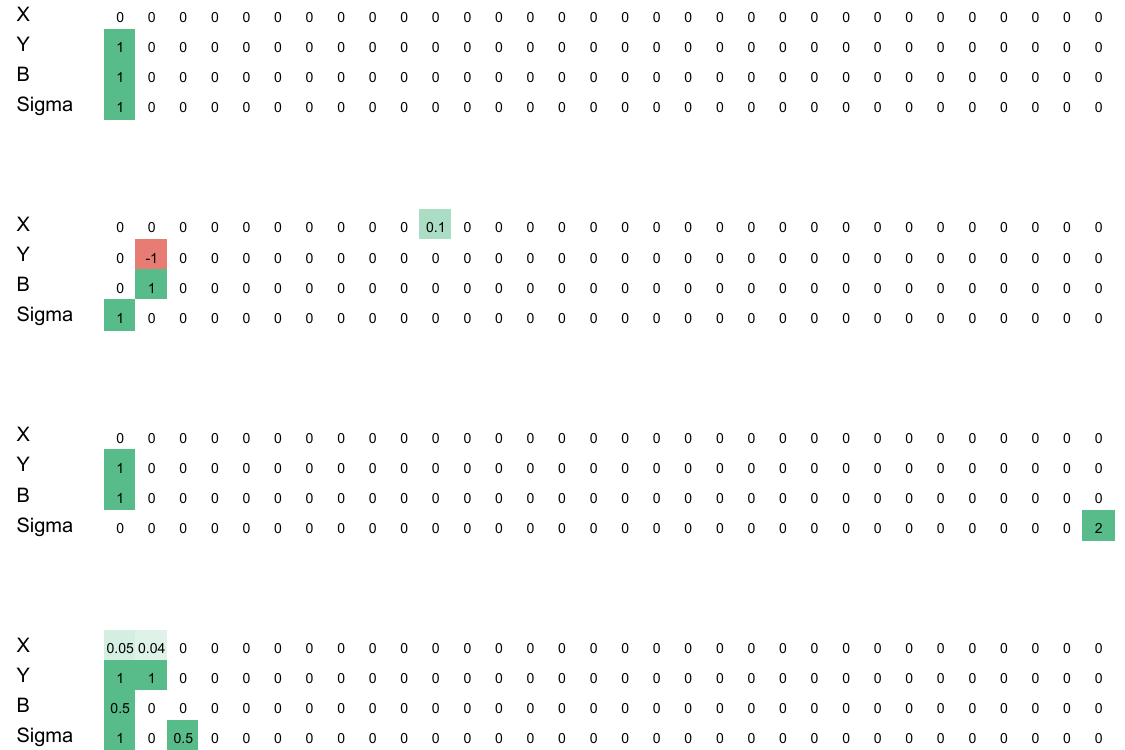}
\end{tabular}
\caption[Four example splines and the corresponding 1D DCT coefficients]{Four example splines and the corresponding 1D DCT coefficients representing the evolution of the X,Y,B components and Sigma (thickness) along the arclength.
\href{https://jxl-art.lucaversari.it/?zcode=1ZPNCsIwDMfveYq8wMa_3Tq2iwdB8L7D9OymG4h42MHHN21zUBBRKsgoNB9N0_RHstuvqZv6eWQDgLbDdBpnLiyI2ut5ugwETlpkFn7fga3fwQ7URGtz6ZXOW0r5F29LhiypVElg_g1KeiiQMkVE5e0PWS2oI9LqtwFUHUFZo6Dql6ByOOkilInfTeyr3P0CeEqWMISVImtAlZclqHRRVupvNO4BJU1HPvBKcqhivHILnow7GToR7TDLdEO1zALPQVZP_MEd}{(link to jxl-art)}}
\label{fig:splines-example}
\end{figure*}

Figure~\ref{fig:splines-example} shows a few example splines as well as the 1D DCT coefficients that describe the color and thickness.

Currently no encoder implements spline detection.
However, manual experiments indicate that substantial compression performance gains could be made by extracting splines and coding them separately.

\begin{figure*}\centering
\setlength{\tabcolsep}{2pt}
\begin{tabular}{ccc}
\includegraphics[width=0.4\linewidth]{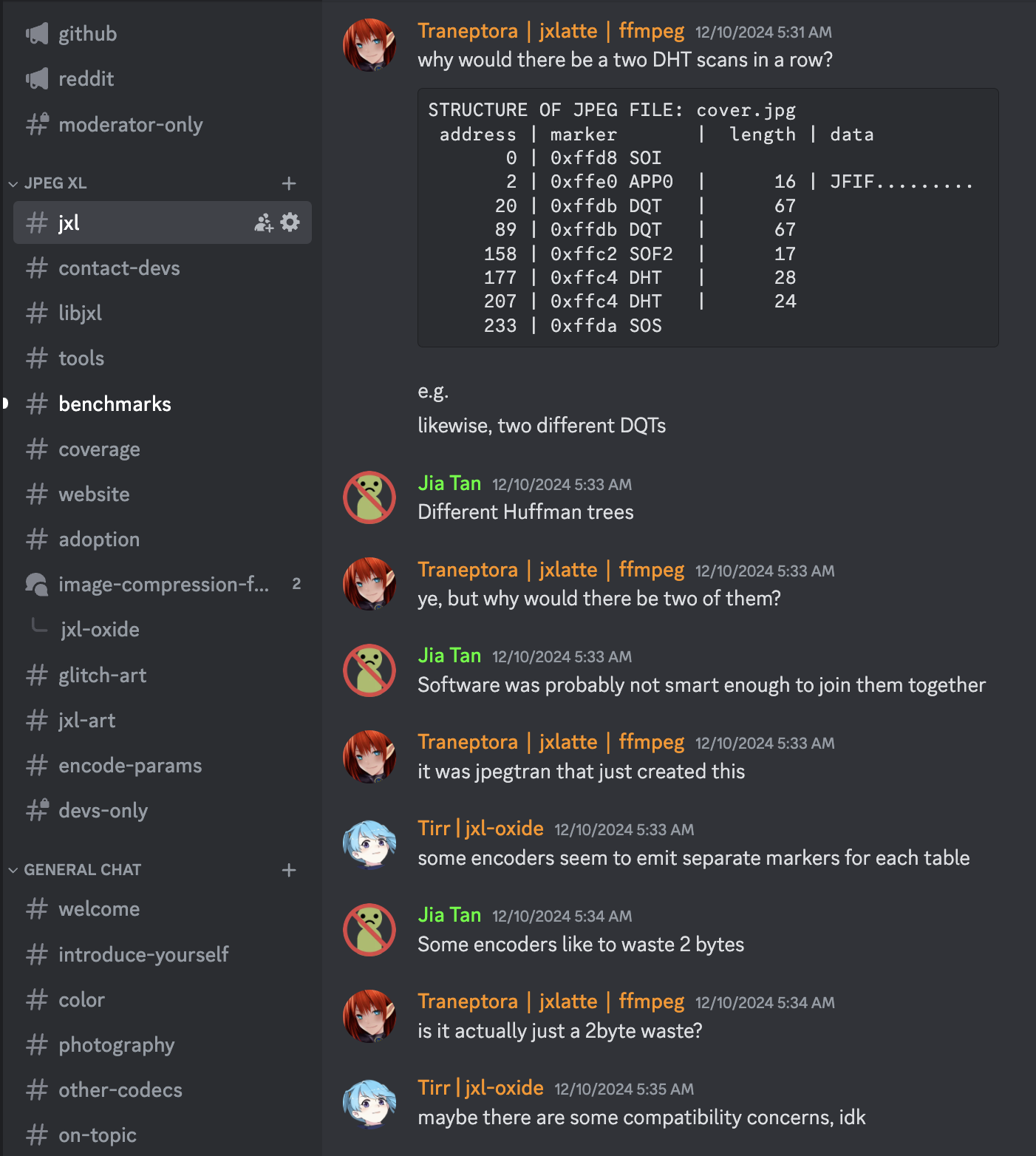} &
\includegraphics[width=0.17\linewidth]{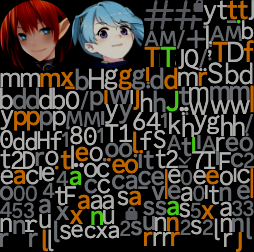} &
\includegraphics[width=0.4\linewidth]{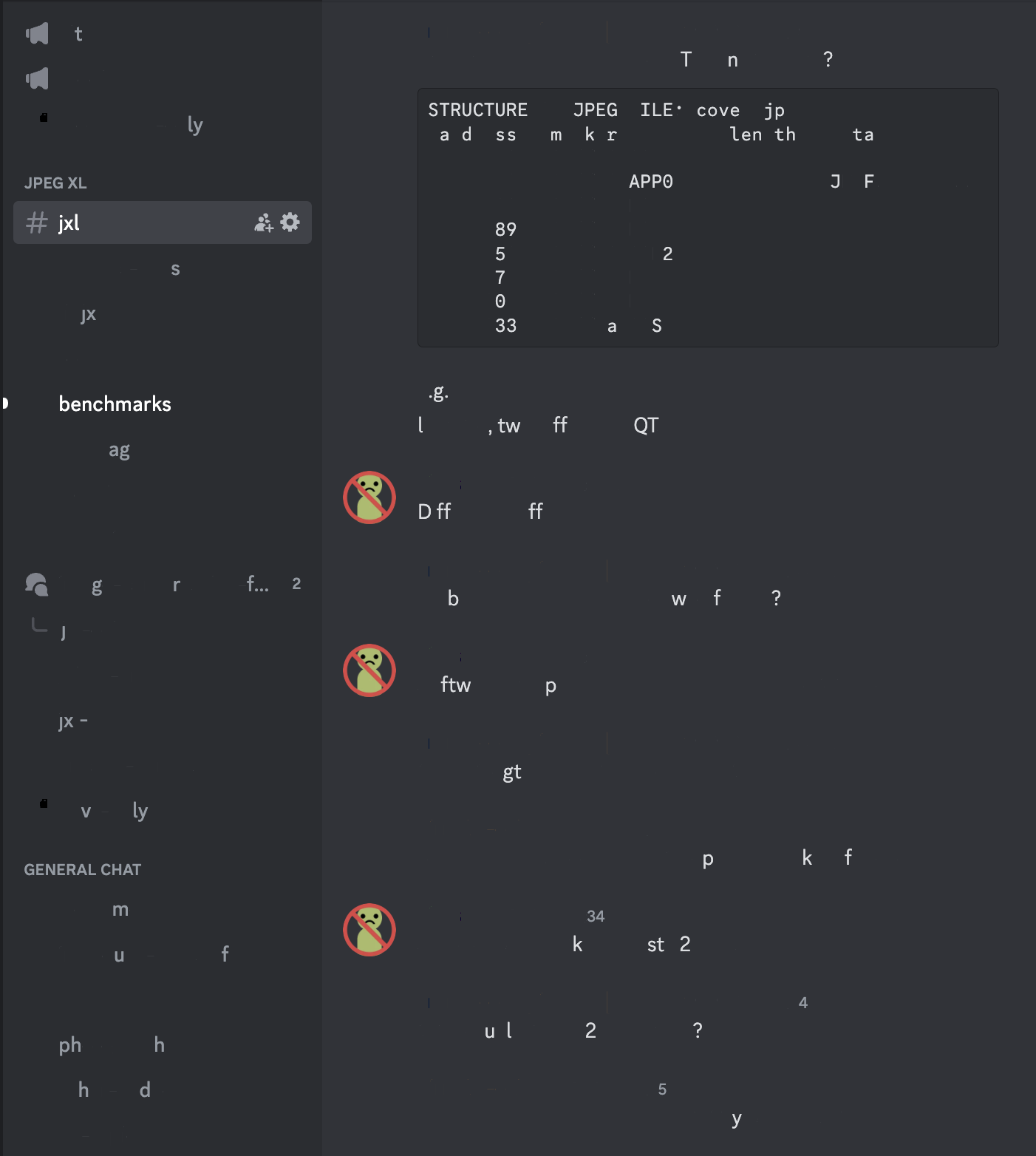} \\
Original image & Reference frame & Residual image \\
\end{tabular}
\caption[Example usage of the Patches coding tool]{Example usage of the Patches coding tool.
Without Patches, the image on the left (297 kB as a PNG file) can be losslessly compressed to a 162 kB JPEG~XL file (0.59 bpp).
With Patches, this can be reduced to 87 kB (0.32 bpp) by extracting repeated image elements, coding them separately and only once in an auxiliary invisible ReferenceOnly frame.
}
\label{fig:patches-example}
\end{figure*}
\begin{figure*}\centering
\includegraphics[width=\linewidth]{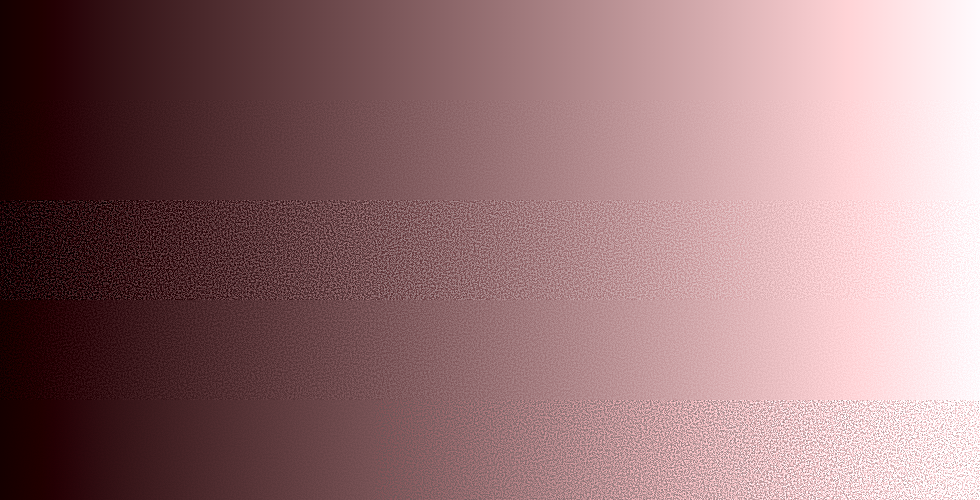}
\caption[Luma-modulated noise generation with various parameters]{Luma-modulated noise generation with various parameter settings. From top to bottom: no noise, photon noise corresponding to ISO 400, ISO 6400, and ISO 1600, and at the bottom noise parameters where the dark regions have no noise while the bright regions do (lookup table goes from 0 in the darks to 0.3 in the brighter range).
\href{https://jxl-art.lucaversari.it/?zcode=rVPLbtswELzrK_ZWIGib5VvsoYccghYo2gJBkeRIS7TNxBYFim2cvy9fBoygFxc-iBqNlsPhcPnweNM9PN7cmiH6sABByoELxDJ0926MW9A9dl-s22wjZLK7voJfi5s2MHm3WJhNMHsbbZpuFhj8fv4d7QirV9i51dNh926Beeujn1q9n6NLH2sfwEBabm83ZnYHu4Mh6QTTfb37ARzxE-BHJMh6mQHtpeoLQ7WoAIXEDFBrKSpDCa9Ac1UA0YpAUSSySXIieKmSTOqqLZSq87RuQHFaBSSllRE9VkZo0ldJ2VwSKYlSBWhCaK4SDDXJgOlUVQATVZthki_LKiUrkIwnyavrroQ7-ZrU-5JRsGsb7DRYSP-_-_jNLBE6t4YBPgN2H-DOpnPJxKER98AbXQ-r5XkyP73zSVws39vcAT_9Apg6CM82J5u7ExmaZI4mLxLvf0R3tJc755_hXaaTTnbNzgzvxbowgpv-pMtn2_V62aZ2gVXI9zXdyFD5VzDTCKMJz0cu1i2VrRTracgPe_ucGuRnGPwL}{(link to jxl-art)}
}
\label{fig:noise-example}
\end{figure*}
\subsubsection{Patches}
\label{patches}
Images may contain elements that are repeated several times. For instance, images containing rasterized typography, icons, or user interface elements can each be seen as compositions of small shapes that are used multiple times.

In multi-frame animations, elements from a previous frame may appear in a subsequent frame, possibly in a different location.
In multi-layer or multi-page images, this may also be the case.

The Patches coding tool exploits such redundancies. It works as follows.
Up to four previously coded frames can be `saved', i.e. the frame header indicates to a decoder that the contents of a decoded frame must be buffered for future reference. There are four `slots' available for this.
The Patches tool allows referencing these frames, selecting a rectangular region in them, and then adding that `patch' to the current frame in one or more arbitrary positions.

Different blend modes are supported, and when there are extra channels, a different blend mode can be selected for the color channels and for each extra channel. The following blend modes are defined, which are different ways to apply a patch to a residual image:
\begin{description}
\item[None.]
Nothing is done; this can be useful to selectively apply patches only to some of the channels.
\item[Replace.]
Overwrites the sample values with the patch.
\item[Add.]
Adds the patch values to the residual image.
\item[Mul.]
Multiplies the sample values with the patch values.
\item[BlendAbove.]
Alpha-blends the patch as if it is a layer that goes on top of the residual image (blend over).
\item[BlendBelow.]
Alpha-blends the patch as if the residual image is a layer on top of the patch (blend under).
\item[MulAddAbove.]
Multiplies the patch sample values with the patch alpha values and adds the result to the underlying image.
\item[MulAddBelow.]
Multiplies the sample values with the patch alpha values and then adds the patch sample values.
\end{description}
The index of the extra channel to use as the `alpha channel' for the purpose of patch blending is signaled. It does not have to be the (first) extra channel of type Alpha. This flexibility allows for example to have a profile picture in a reference frame with two alpha channels: one that corresponds to a rounded rectangle and another that follows the outline of the face. This picture can then be alpha-blended to the underlying image in different ways.
In addition, a choice is signaled regarding whether the alpha values get clamped to the range $[0,1]$ for the purpose of blending or not.

In the current libjxl encoder, the only patch blend mode that is used is Add. In case of lossy compression, this makes it possible to allow imperfect matches without risking fidelity issues, since the imperfect match does not \emph{replace} the original image data but simply gets subtracted from it during encoding. The residual image can then still correct for errors caused by imperfect matching.

\subsubsection{Noise}
From the point of view of entropy coding, noise is challenging since it is high frequency and random (so high entropy).
For (mathematically) lossless compression, this high entropy is unavoidable: noisy images simply cannot be compressed as densely as `clean' images.

For lossy compression, though, it is possible to use synthetic noise that looks similar to the actual noise even though at the pixel level it is completely different.
Typically, a side effect of lossy compression is some amount of denoising: Since high-frequency coefficients are quantized more aggressively, some of the noise will be reduced during encoding, and smoothing filters (like EPF) will further reduce it.
In some cases, this is not a problematic side effect since the noise may not have been intentional or desirable in the first place.

However, in some cases, it may be desirable to preserve noise, e.g. for artistic purposes or to better mask artifacts, i.e. as a form of dithering.

The noise synthesis in JPEG~XL is primarily intended to model photon noise. Photon noise is pixel-sized, caused by the discrete nature of photons. It typically occurs in digital photography at high ISO settings.
JPEG~XL's noise synthesis can also be used to model fine analog film grain, but it is not suitable for modeling coarser grain, whose individual grain diameters spread across multiple pixels.

If enabled, noise is applied to the entire frame, but the amplitude of the noise is modulated based on the luma (Y) component. A signaled lookup table defines the desired noise amplitude at eight brightness points. At intermediate luma values, the amplitude is interpolated.
This allows modeling photography at high ISO settings, which results in more noise in the dark regions than in the bright regions.
The generated noise mostly affects the luminance; chromaticity is affected to a smaller extent. 

Figure~\ref{fig:noise-example} shows a color gradient with four different examples of noise parameters: three corresponding to simulated photon noise (produced by setting the parameter \verb|--photon_noise_iso|  in cjxl) at ISO 400, ISO 6400, and ISO 1600,
and one example of an atypical custom noise lookup table which results in more noise in the bright range and no noise in the dark range.

The generated noise is fully specified and completely deterministic. This ensures that the decoded image is defined exactly, at the pixel level.
It also allows encoders to take the effect of generated noise into account and compensate for it if necessary.

\subsection{Restoration filters}
Restoration filters aim to improve the image quality of the decoded image in a way that is fully part of the codec specification. 
This implies that the encoder can take into account the decoder-side filtering and that the final resulting image is defined exactly.

In JPEG~XL, restoration filters are applied on the entire frame, that is, across group boundaries. This is in contrast with image formats that apply tiling at the container level, such as TIFF, DNG, DICOM, AVIF, and HEIF. If tiling is done at the container level, each tile corresponds to a separate codestream and any filters will only be applied within each tile. This can lead to subtle seams at the tile boundaries.

\begin{figure}\centering
original image:\\
\includegraphics[width=\linewidth]{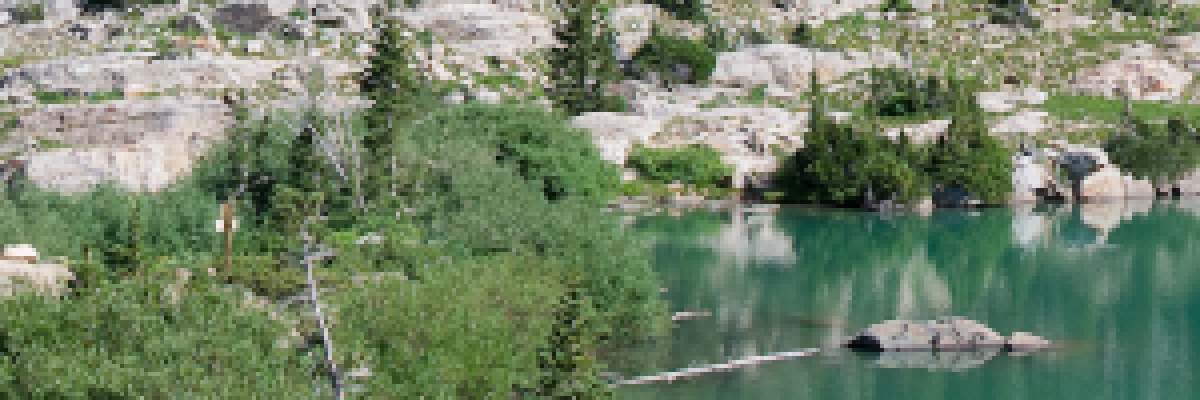}\\
Gabor-like transform disabled:\\
\includegraphics[width=\linewidth]{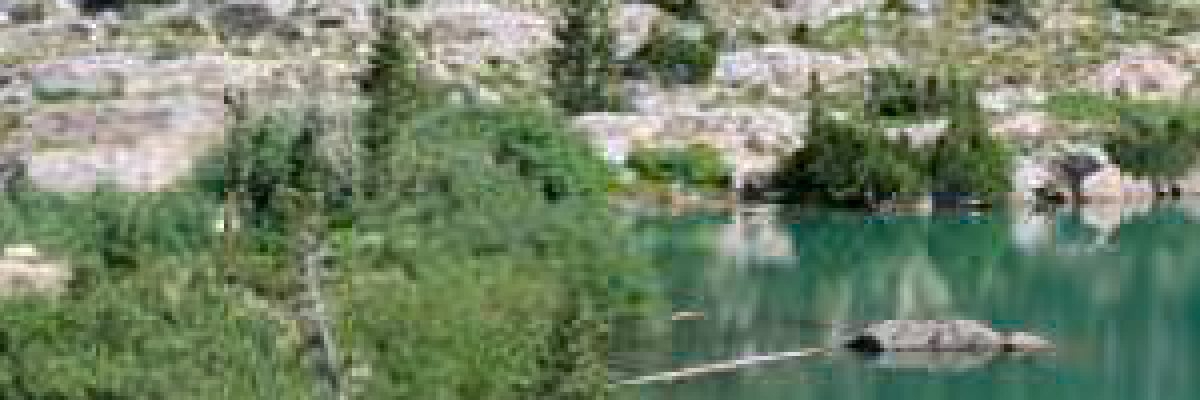} \\
Gabor-like transform enabled:\\
\includegraphics[width=\linewidth]{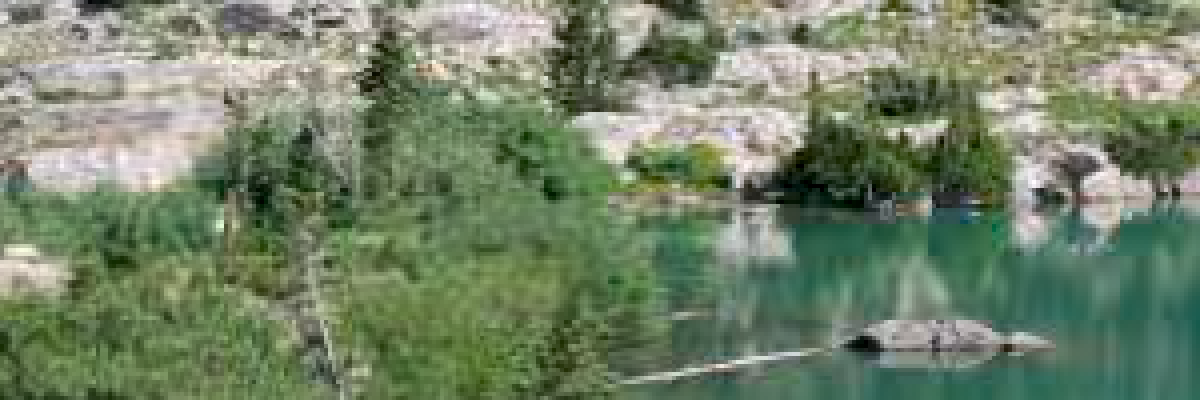}
\caption[Zoomed crop to demonstrate the Gabor-like transform]{Zoomed crop of an image to demonstrate the effect of the Gabor-like transform
Top: original image.
Middle: compressed image at 0.73 bpp, Gaborish disabled. Note the visible DCT macroblocks.
Bottom: compressed image at the same bit rate (0.73 bpp), with Gaborish enabled.
In both examples the EPF was disabled.
}
\label{fig:gaborish}
\end{figure}

\subsubsection{Gabor-like transform}
Also known as ``Gaborish'', the Gabor-like transform corresponds to a slight decode-side blurring convolution, using a symmetric $3 \times 3$ kernel. By default, the following kernel is used:
\begin{equation}
\mathit{gab\_default} =
\begin{bmatrix}
0.0359 & 0.0675 & 0.0359\\
0.0675 & 0.5863 & 0.0675\\
0.0359 & 0.0675 & 0.0359
\end{bmatrix}
\end{equation}

Optionally, different kernel weights can be signaled in the frame header, with the option of having different kernels for each XYB component.

Before encoding, libjxl applies a $5 \times 5$ sharpening convolution that approximately corresponds to the inverse of the decode-time blurring.

The encode-side sharpening and decode-side blurring cancel out. However, the overall effect is a reduction of blocking artifacts and other DCT artifacts, thanks to the smoothing convolution that is applied after the inverse DCT.

This technique brings most of the advantages of lapped transforms, with a lower computational cost, especially in terms of decode complexity.

Figure~\ref{fig:gaborish} shows an example of how the Gabor-like transform helps to reduce noticeable blocking artifacts.

\begin{figure*}\centering
\setlength{\tabcolsep}{1pt}
\begin{tabular}{ccc}
original image: & EPF disabled: & EPF enabled: \\
\includegraphics[width=0.33\linewidth]{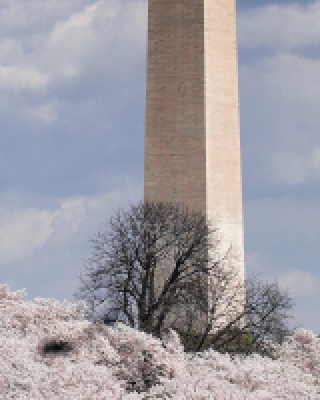} &
\includegraphics[width=0.33\linewidth]{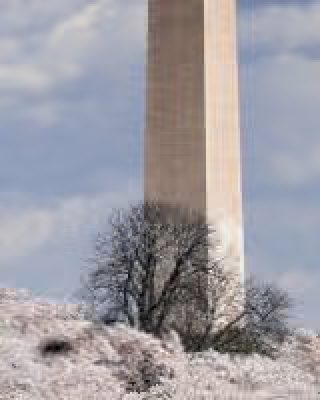} &
\includegraphics[width=0.33\linewidth]{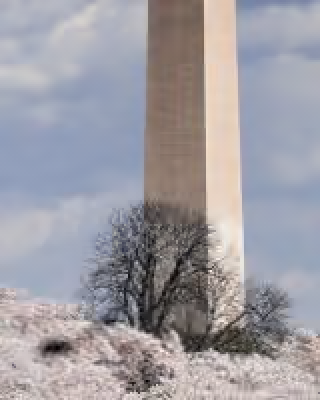}
\end{tabular}
\caption[Zoomed crop to demonstrate the edge-preserving filter (EPF)]{Zoomed crop of an image to demonstrate the effect of the edge-preserving filter (EPF). Left: original image. Middle: compressed image, with a relatively low fidelity setting (distance 4), EPF disabled. Note the DCT ringing artifacts, as well as some DCT base patterns and blocking artifacts.
Right: same compressed image but with EPF enabled.
}
\label{fig:epf}
\end{figure*}

\subsubsection{Edge-preserving filter}
\label{epf}
Compared to the Gabor-like transform, the edge-preserving filter (EPF) is more sophisticated and powerful.
While the Gabor-like transform uses only a $3 \times 3$ kernel, the EPF can apply smoothing over larger distances, using pixels from up to a $9 \times 9$ region around each pixel position.
As the name implies, the EPF is a non-linear filter similar to a bilateral filter, aimed at removing DCT noise and ringing artifacts while avoiding blurred edges.

The overall strength of EPF is configurable with the epf\_iters field in the frame header, which takes an integer value in $[0,3]$. It indicates the number of iterations of the filter to apply,  where 0 means the EPF is disabled.
In every iteration, every pixel is replaced by a weighted sum of reference pixels in a diamond-shaped region around the pixel, where the weights depend on:
\begin{itemize}
\item the color difference between the pixel and the reference pixel (larger weight if the difference is smaller);
\item the color difference between the direct neighbors of the pixel and the corresponding direct neighbors of the reference pixel;
\item the pixel position being at a block border position (assuming $8 \times 8$ blocks) or not, that is, its horizontal or vertical position is 0 or 7 modulo 8;
\item a sigma value that depends on the local amount of HF DCT quantization HfMul and the local EPF Sharpness value (see \SectionName{}s~\ref{hfmul}~and~\ref{sharpness}).
\end{itemize}

The diamond-shaped region is usually just the cross consisting of the pixel itself and its four direct neighbors, except in case there are three iterations: then the first iteration uses a larger reference region forming a $5 \times 5$ diamond around the pixel, that is, the 13 positions within a Manhattan distance ($L^1$) of 2.

Figure~\ref{fig:epf} shows an example image with and without EPF enabled.

\begin{figure*}\centering
\includegraphics[width=\linewidth]{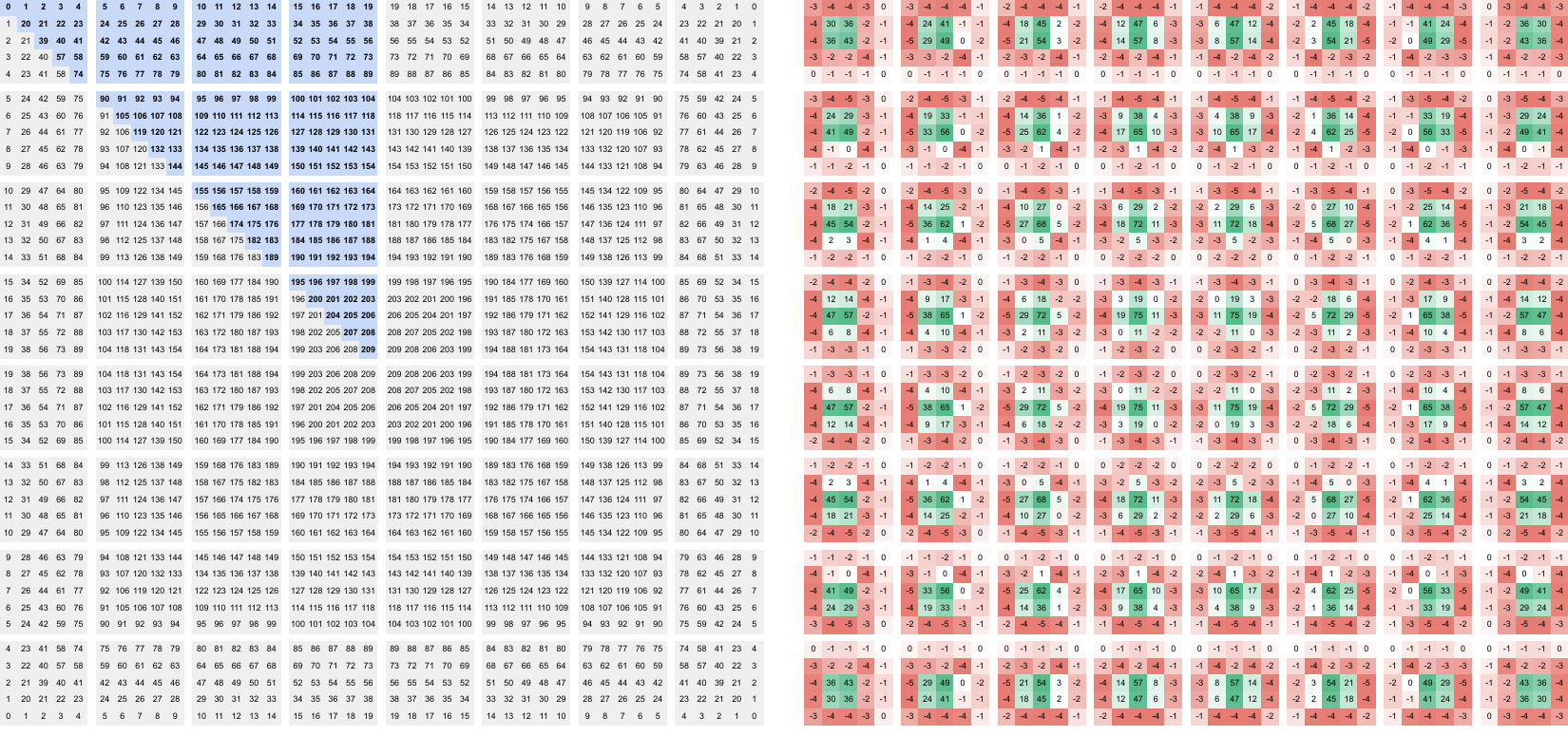}
\caption[Weights for the non-separable 8x upsampling method]{Weights for the non-separable 8x upsampling method. Left: indices of the weights. 
Only 210 of the 1600 weights are different and are optionally signaled.
Right: default weights. Weights in the figure are scaled by a factor of 100.}
\label{fig:upsampling-weights}
\end{figure*}
\begin{figure*}\centering
\setlength{\tabcolsep}{2pt}
\begin{tabular}{cc}
nearest neighbor upsampling: & cubic upsampling:\\
\includegraphics[width=0.49\linewidth]{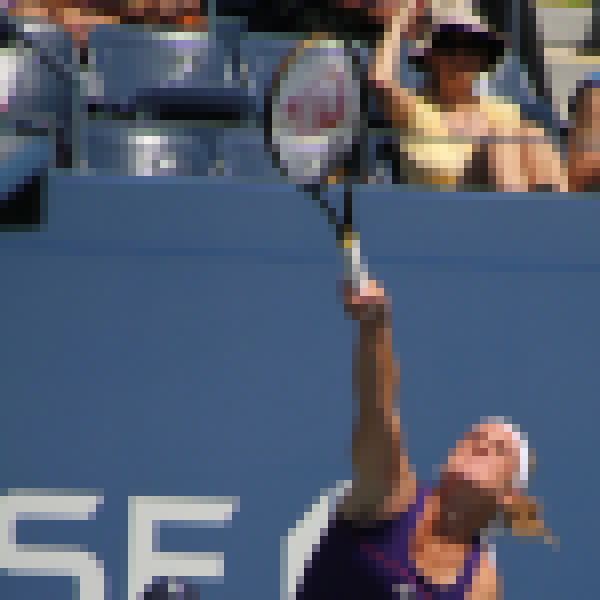} &
\includegraphics[width=0.49\linewidth]{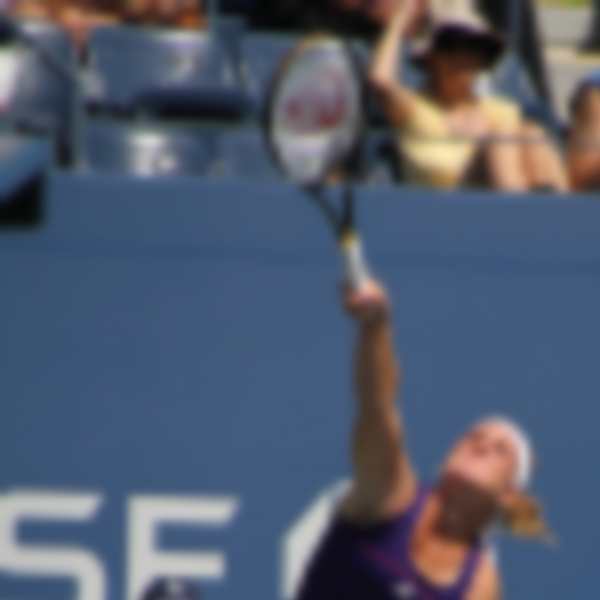} \\
Lanczos upsampling: & default JPEG~XL non-separable upsampling:\\
\includegraphics[width=0.49\linewidth]{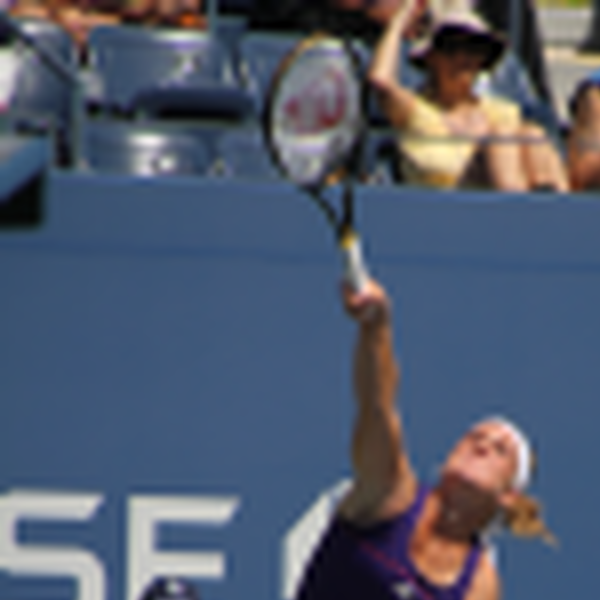} &
\includegraphics[width=0.49\linewidth]{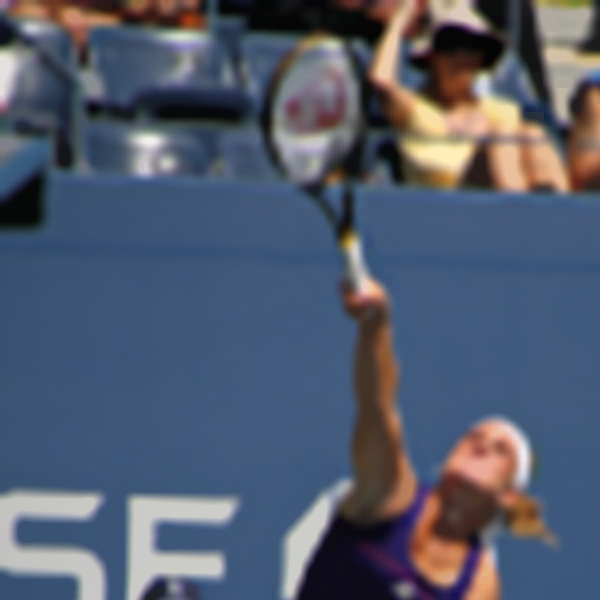}
\end{tabular}
\caption[Result of 8x upsampling using different methods]{Result of 8x upsampling using different methods.
Cubic and Lanczos upsampling are as implemented in ImageMagick {\bf -resize} with respectively {\bf -filter cubic} and {\bf -filter lanczos}.}
\label{fig:upsampling}
\end{figure*}

\begin{figure*}\centering
\setlength{\tabcolsep}{2pt}
\begin{tabular}{cc}
\includegraphics[width=0.49\linewidth]{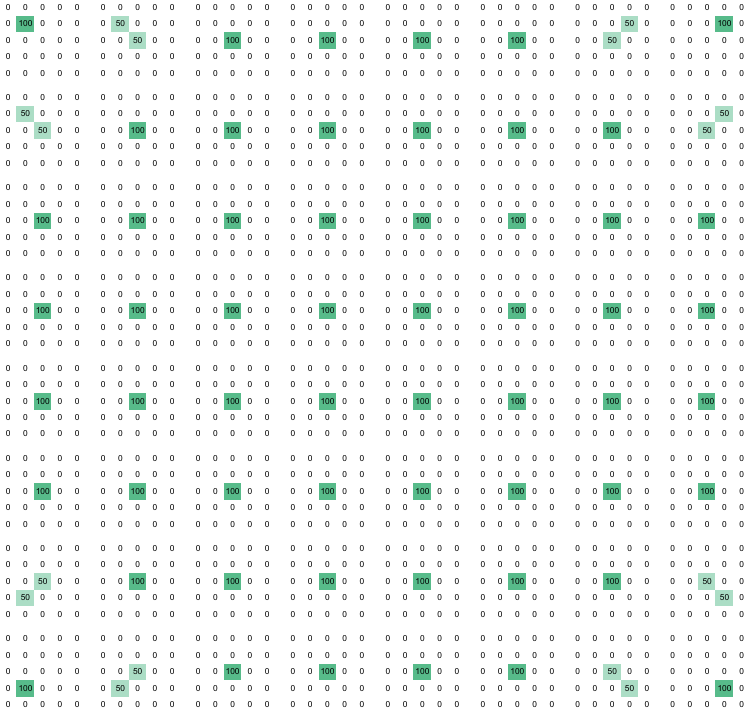} &
\includegraphics[width=0.49\linewidth]{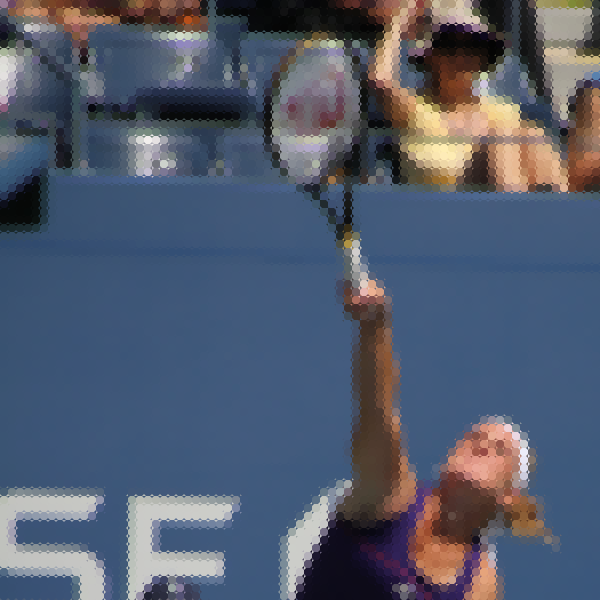} 
\end{tabular}
\caption[Custom weights for 8x upsampling: variation on nearest-neighbor]{Custom weights for 8x upsampling: a variation on nearest-neighbor.
Left: weights. Right: example result.}
\label{fig:upsampling-custom-weights}
\end{figure*}

\subsection{Upsampling}
\label{upsampling}

Subsampling is a crude but effective coding tool. By storing some image components (or the entire image) at a lower resolution and relying on decoder-side upsampling, high-frequency information will obviously be lost, but it may be possible to obtain better image quality at very low bit rates this way.

\paragraph{Simple upsampling.}
For the purpose of lossless JPEG recompression, the simple `triangle' upsampling filter that is commonly implemented in JPEG decoders (e.g. libjpeg-turbo) has been included in the JPEG~XL specification.
This is the upsampling used in case of YCbCr chroma subsampling (see \SectionName{}~\ref{chroma_subsampling}).
Simple upsampling can be applied horizontally (4:2:2), vertically (4:4:0), or both (4:2:0), and it can be applied to individual components (typically Cb and Cr but not Y).

\paragraph{Non-separable upsampling.}
\label{nonsep-upsampling}
Besides the simple upsampling, a more sophisticated upsampling method is defined which can be used to upsample either the color image (all three components at the same time) or one or more of the extra channels.
The upsampling factor is 2x, 4x or 8x.

Figure~\ref{fig:upsampling-weights} illustrates 8x upsampling. In this case, every input sample in the subsampled image becomes a block of $8 \times 8$ samples in the upsampled image. Every output sample is calculated as a weighted sum of a $5 \times 5$ region around the input sample. The method is non-separable since in the general (and default) case, it cannot be performed as a combination of 1D operations.
For each of the 64 output samples, the $5 \times 5$ kernel can be different, although some symmetries must be respected.
Furthermore, the output values are clamped to the range defined by the minimum and maximum sample values in the $5 \times 5$ region around the input sample.

Figure~\ref{fig:upsampling} shows a visual example of 8x upsampling using simple pixel duplication (nearest neighbor upsampling), two typical separable upsampling filters (a cubic filter and Lanczos), and JPEG~XL's non-separable upsampling with default weights.
Obviously none of these methods can produce a highly detailed upsampled image, but arguably the result of the non-separable upsampling method produces the most aesthetically pleasing results: it avoids excessive blur while also avoiding ringing and staircasing, in particular around oblique or curved edges.

The weights can be signaled as part of the image header; the default values (in case no custom weights are signaled) are given in Figure~\ref{fig:upsampling-weights}.
Only about one-eight of the weights are signaled; the remaining weights are defined implicitly through mirroring along horizontal, vertical and diagonal symmetry.

Custom weights can be used to create new kinds of upsampling filters. Custom upsampling filters may be useful for many reasons; they are particularly useful for artistic purposes. For example, for pixel art, nearest-neighbor upsampling can be used to create $2 \times 2$, $4 \times 4$, or $8 \times 8$ `macropixels'. Nearest-neighbor upsampling corresponds to placing all the weight in the center of each $5 \times 5$ kernel.
Figure~\ref{fig:upsampling-custom-weights} shows an example of custom upsampling weights that mostly correspond to nearest-neighbor upsampling, except in each corner of the $8 \times 8$ `macropixel', three corner pixels are defined differently.

\paragraph{LF upsampling.}
In libjxl, the default 8x non-separable upsampling is also used when rendering progressive previews when only the 1:8 LF image is available.

\begin{figure}\centering
\setlength{\tabcolsep}{3pt}
\begin{tabular}{r|cl|cl|cl}
& \multicolumn{2}{c|}{config 2,1,0} & \multicolumn{2}{c|}{config 3,0,1} & \multicolumn{2}{c}{config 3,2,1} \\
$n$ & token & raw bits & token & raw bits & token & raw bits \\
\hline
0  & 0 & ---  & 0  & --- & 0 & --- \\
1  & 1 & ---  & 1  & --- & 1 & --- \\
2  & 2 & ---  & 2  & --- & 2 & --- \\
3  & 3 & ---  & 3  & --- & 3 & --- \\
4  & 4 & 0    & 4  & --- & 4 & --- \\
5  & 4 & 1    & 5  & --- & 5 & --- \\
6  & 5 & 0    & 6  & --- & 6 & --- \\
7  & 5 & 1    & 7  & --- & 7 & --- \\
8  & 6 & 00   & 8  & 00  & 8 & --- \\
9  & 6 & 01   & 9  & 00  & 9 & --- \\
10 & 6 & 10   & 8  & 01  & 10 & --- \\
11 & 6 & 11   & 9  & 01  & 11 & --- \\
12 & 7 & 00   & 8  & 10  & 12 & --- \\
15 & 7 & 11   & 9  & 11  & 15 & --- \\
16 & 8 & 000  & 10 & 000 & 16 & 0 \\
23 & 8 & 111  & 11 & 011 & 19 & 1 \\
24 & 9 & 000  & 10 & 100 & 20 & 0 \\
31 & 9 & 111  & 11 & 111 & 23 & 1 \\
32 & 10 & 0000  & 12 & 0000 & 24 & 00\\
47 & 10 & 1111  & 13 & 0111 & 27 & 11\\
48 & 11 & 0000  & 12 & 1000 & 28 & 00\\
255 & 15 & 111111 & 17 & 111111 & 47 & 1111\\
256 & 16 & 0000000 & 18 & 0000000 & 48 & 00000\\
257 & 16 & 0000001 & 19 & 0000000 & 49 & 00000\\
258 & 16 & 0000010 & 18 & 0000001 & 48 & 00001\\
\end{tabular}

\vspace{1em}

\begin{tabular}{c|c|l}
\multicolumn{2}{r}{$n=7777777$, binary:}      &  
                   \verb|11101101010110111110001| \\
\hline
config & token & raw bits \\
\hline
2,1,0 &  45  &     \verb|  101101010110111110001| \\
3,0,1 &  47  &     \verb| 110110101011011111000 | \\ 
3,2,1 &  167 &     \verb|   0110101011011111000 | \\
3,3,0 &  166 &     \verb|    1101010110111110001| \\
3,0,3 &  161 &     \verb| 1101101010110111110   | \\
7,3,0 &  254 &     \verb|    1101010110111110001| \\
0,0,0 &  23  &     \verb| 1101101010110111110001| \\ 
\end{tabular}
\caption[Hybrid integer coding: example configurations]{Hybrid integer coding: example configurations.
The configurations are specified as a triple of the form
split\_exponent, msb\_in\_token, lsb\_in\_token.}
\label{fig:hybriduint}
\end{figure}

\section{Entropy coding}
\label{entropy}
Besides header data, which is signaled without entropy coding --- relying only on bit packing and conditional signaling --- all data in JPEG~XL is compressed using entropy coding.
The same entropy coding method is used to encode everything: residuals of Modular channel data, the signaling of the MA tree itself, VarDCT HF coefficients, Spline coefficients, Patch reference data, compressed ICC profiles, etc.

At a high level, JPEG~XL's entropy coding method works as follows.
Every entropy coded stream can use either prefix coding (Huffman coding) or ANS coding.
Typically there is one stream per bitstream section (see \SectionName{}~\ref{sections}); the choice between prefix coding and ANS can be made per stream.
There's an option to additionally enable LZ77 codes for distance-length pairs.
The (pre-clustering) number of contexts is implicit.
These contexts are clustered using a context map (see \SectionName{}~\ref{context_map}).
A histogram is signaled for each context cluster.
For prefix coding, compact histogram coding is used as in Brotli \cite{brotli}.
For ANS, histograms are probability distributions with 12-bit precision.

The symbol coding itself is called `hybrid integer coding': every symbol is split into a token, which is entropy coded using ANS or prefix coding, and a number of `raw bits' which are directly written to or read from the bitstream.
The number of raw bits depends on the token and can be between zero (i.e. the symbol is encoded fully through entropy coding) and 31.
Every context cluster has its own configuration that determines how this separation of symbols into tokens and raw bits is done.
The alphabet size for the entropy coded tokens is limited to $2^8$ for ANS and $2^{15}$ for prefix coding.

\subsection{Hybrid integer coding}
The configuration of hybrid integer coding has three parameters:
split\_exponent, msb\_in\_token, and lsb\_in\_token.

Arbitrary unsigned 32-bit integers can be coded using hybrid integer coding; the assumption is that values closer to zero will be more common and will benefit more from entropy coding than high-amplitude symbols.

Tokens smaller than $2^\textrm{split\_exponent}$ correspond directly to a symbol, without any raw bits.
The next tokens represent higher numbers, using an exponent-mantissa representation where the token implies the exponent (the position of the most significant 1 bit), msb\_in\_token of the next most-significant bits, and lsb\_in\_token of the least-significant bits,
and the remaining bits are signaled as raw bits.

Figure~\ref{fig:hybriduint} illustrates this coding scheme.
It is explained and investigated in more detail in Chapter 3 of \cite{luca_phdthesis}, where also theoretical and experimental results are provided that demonstrate the low redundancy of hybrid integer coding compared to other coding schemes such as Rice codes, $\pi$ codes, $\zeta$ codes, and Elias codes.

In a way, the hybrid integer coding method can be seen as a generalization of the DC coding scheme used in JPEG, where the Huffman-coded token only implies the exponent and all the mantissa bits are signaled as raw bits.
The AC coding in JPEG is similar, except that runs of zeros are handled separately, which can be seen as a special case of LZ77.

\subsubsection{Sign bit}
From the point of view of entropy coding, all symbols are unsigned integers.
Whenever signed integers are needed, they are first converted to unsigned integers as follows:
\begin{equation}
k \mapsto
    \begin{cases}
        2k & \text{if } k \geq 0 \\
        -2k-1 & \text{if } k < 0
    \end{cases}
\end{equation}

In other words, the unsigned symbols $0,1,2,3,4,5,6$
correspond to signed integers $0,-1,1,-2,2,-3,3$.
The (encode-side) mapping from signed to unsigned is called PackSigned() and the reverse (decode-side) mapping from unsigned to signed is called UnpackSigned().

As a result, the least significant bit of the entropy coded symbols often corresponds to the sign.
If the data to be entropy coded has a uniformly distributed sign, i.e. (roughly) the same proportion of positive and negative values, then coding the sign bit as a raw bit makes sense.
However, if context modeling is taken into account, there may be a sign bias, in which case it is beneficial to use a hybrid integer coding configuration with lsb\_in\_token $\geq 1$.

\subsubsection{Prefix coding}
In case prefix coding is used, each token corresponds to a unique variable-length sequence of bits such that no token has a code that starts with a subsequence (i.e. has a prefix) that corresponds to the (shorter) code of a different token.
These prefix codes are constructed using Huffman's algorithm \cite{huffman}.
The signaling of the code itself is based on the compact signaling of Brotli \cite{brotli}.
There is a special case when the alphabet size is zero, i.e. there is only one symbol: in this case no code is signaled and every token is equal to 0 without reading or writing any bits.

From an encoder point of view, the main advantage of prefix coding is its relative simplicity.
Given a prefix code (either by using a fixed code, or by computing histograms and constructing an optimized code), the compressed bitstream can be constructed directly from the symbols, writing bits in the same order as they are read at decode time.
This is in contrast to the more advanced ANS coding method, which requires encoding in reverse order and constructing the compressed bitstream starting at the end.

However, a major limitation of prefix coding is that every token requires an integer number of bits.
In the particularly common case that the probability of the zero token is larger than 50\%, this can be far from optimal.

\subsubsection{ANS}
Asymmetric Numeral Systems \cite{ans} is a generic entropy coding method that combines
the fast decode speed of prefix coding and the precise probability distributions of arithmetic coding, leading to better compression.

The specific variant of ANS used in JPEG~XL is rANS with alias mapping, with a 32-bit state (per entropy coded stream, so shared between contexts) and fixed probability distribution per context with a precision of 12-bit, i.e. integer histograms that sum to $2^{12}$.

At encode time, the state is always initialized to the specific value 0x00130000, which means that this will always be the final state after decoding an ANS-coded stream.
This way, the final ANS state acts like a checksum that can be used to detect a corrupted (section of a) bitstream, e.g. due to transmission errors.

\begin{figure}\centering
	\includegraphics[width=1\linewidth]{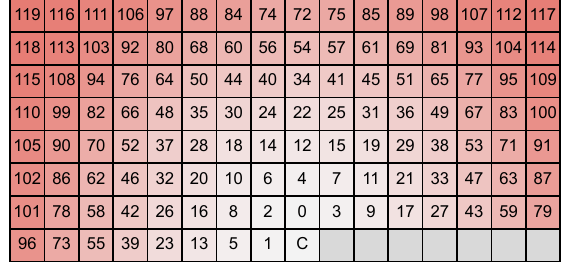}
	\caption{Special LZ77 distance codes for Modular data.}
	\label{fig:special_distances}
\end{figure}
\subsubsection{LZ77}
\label{lz77}
If the option to use LZ77 \cite{lz77} is enabled (this is a choice per entropy coded stream), then the stream header signals two additional numbers:
lz77.min\_symbol and lz77.min\_length.
An implicit extra context is added for coding distances, and an additional hybrid integer coding configuration lz\_len\_conf is signaled for coding lengths.

During entropy coding, a sliding window of $2^{20}$ previously coded symbols is maintained. Symbols are unsigned 32-bit numbers so maintaining this window may require up to 4 megabytes of memory.

If an entropy coded symbol $s$ is at the end of the regular symbol alphabet, that is, $s \geq \textrm{lz77.min\_symbol}$,
then it represents an LZ77 copy instruction.
The value of $s - \textrm{lz77.min\_symbol}$ is then interpreted as a token in hybrid integer coding configuration lz\_len\_conf, and if needed, additional raw bits are coded. The value lz77.min\_length is added to the resulting number and this is then the LZ77 length, that is, the number of symbols to copy from the sliding window.
The LZ77 distance is coded using the implicit extra context and its associated hybrid integer coding configuration.

\subsubsection{Special distances}
In case LZ77 is used in a Modular sub-bitstream, then just like in lossless WebP, the first 120 distance codes are interpreted in a special way to take into account the 2D nature of the data.
For example, distance code 0 does not refer to the previously coded symbol, but to the symbol corresponding to one image row above the current sample position. Figure~\ref{fig:special_distances} gives an overview of the starting positions of these 120 special distances.

There are two caveats:
1) LZ77 refers to previously coded symbols, which in case of Modular coding correspond to prediction residuals, not direclty to sample values.
2) The special distances are defined in terms of the maximum channel width in the current Modular sub-bitstream, which is typically 256 (or rather, the group dimension). For channels that have this width (typically all of them), this means Figure~\ref{fig:special_distances} correctly depicts the positions. However, for channels with a different width (e.g. due to extra channel subsampling), the actual positions will be different.

In JPEG~XL, LZ77 is mostly useful for the special case of run-length encoding (RLE) and for local repetitive patterns such as ordered dithering.
For larger repetitive elements, Patches (see \SectionName{}~\ref{patches}) are a more useful coding tool since entire rectangles can be copied with a single reference, while LZ77 requires multiple references, one per row.

\begin{figure*}\centering
	\includegraphics[width=1\linewidth]{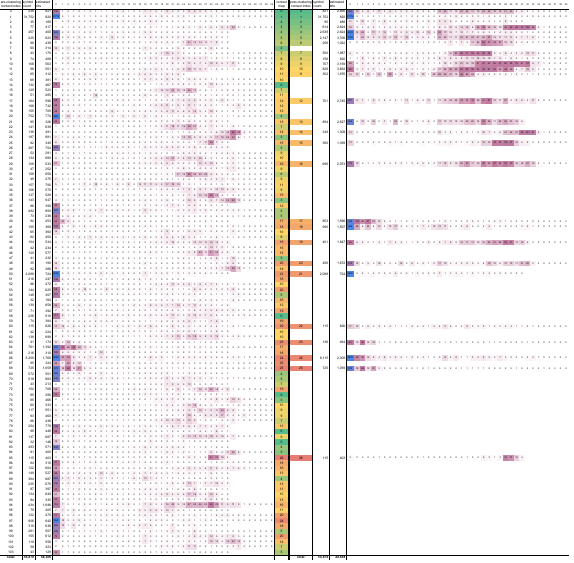}
	\caption[Context clustering example]{Context clustering example. The histograms corresponding to the 104 pre-clustering contexts (left) are clustered and mapped into 27 post-clustering contexts (right), reducing the total signaling cost.}
	\label{fig:context_clustering}
\end{figure*}
\subsection{Context modeling}
The point of context modeling is to achieve better compression by segmenting the entropy-coded symbols into clusters with similar histograms.

For signaling of header data like the Table Of Contents or compressed ICC profile, the number of contexts is small and fixed, and the assignment of symbols to contexts is explicitly specified.
The signaling of image features (\SectionName{}~\ref{image_features}) works in the same way.
For example, for signaling Splines, 6 contexts are used:
one context for the 1D DCT quantization factor;
one for the number of splines;
one for the starting positions, i.e. the $(x,y)$ coordinates of the first control point of each spline;
one for the number of control points of each spline;
one for the double-delta sequences defining the remaining control points;
and one for the quantized DCT coefficients for the color and thickness.
For signaling Patches, a fixed model with 10 contexts is used.

For the HF coefficients of VarDCT data, a larger context model is used, where some of the specifics of the context model can be signaled. This means they can be tuned by an encoder to fit the image data. The context model for HF VarDCT data is explained in \SectionName{}~\ref{context_hf}.

In case of Modular sub-bitstreams, the context model is defined by an MA tree, as explained in \SectionName{}~\ref{modular_trees}.
The MA tree itself is signaled using a fixed context model with 6 contexts.
The size of the resulting context model can be as small as a single context (a trivial tree where the root node is a leaf) or as large as is allowed by the profile and level restrictions.
An encoder has to balance the signaling cost of the MA tree itself with the compression gains brought by a more refined context model.
Another consideration is the impact on decode speed: smaller or more restricted MA trees --- e.g. trees using only some of the properties in their decision nodes, or trees that do not necessitate maintaining the self-correcting predictor --- can be decoded more efficiently.

\subsubsection{Context map}
\label{context_map}
To reduce the signaling overhead, contexts with similar histograms are clustered so that instead of signaling these similar histograms several times, only one histogram per cluster has to be signaled.
This is done by signaling a context map (a concept first introduced in Brotli \cite{brotli}) which is a lookup table mapping pre-clustering context indices to post-clustering context indices.
As a result, the total signaling cost (which includes the cost of signaling the histograms) goes down.
Figure~\ref{fig:context_clustering} illustrates the process, from an encoder point of view.
The decoder only sees the context map and the post-clustering contexts.

The number of post-clustering contexts in any entropy coded stream is limited to 255, which means the context map lookup table can be implemented using 8-bit integers (one byte per pre-clustering context).
Keeping the number of post-clustering contexts low helps to improve memory consumption and in particular memory cache behavior.

In case of VarDCT HF coefficients or large Modular MA trees, the number of pre-clustering contexts can be large (thousands) so the context map itself can be a substantial amount of data that requires some compression itself.
The context map is signaled as follows. First, a single bit indicates whether it is a `simple' clustering or not.

\paragraph{Simple clustering.}
In this case, the context map uses a small fixed number of bits per pre-clustering context index. First, the number of bits to use is signaled using 2 bits, so the number of bits is in the range $[0,3]$ and the maximum index of post-clustering contexts is 0, 1, 3, or 7.

\paragraph{Complex clustering.}
If the context map is not `simple', then an addional bit signals whether the
Move-To-Front transform \cite{MTF} is applied or not.
Then a recursive entropy coded stream with a single pre-clustering context is used to signal the list of post-clustering indices.
Note that this substream can use LZ77 in which case there will be two pre-clustering contexts and it will actually have a small recursive context map; that context map is no longer allowed to use LZ77 in its signaling (which would not be useful anyway).
The Move-To-Front transform (MTF) can be useful to reduce the entropy-coded size of a context map. For example, consider the context map corresponding to the following list of 48 indices mapping 48 pre-clustering contexts to 16 post-clustering contexts:
0, 1, 2, 2, 2, 2, 3, 4, 5, 6, 7, 7, 7, 7, 8, 9, 10, 11, 12, 13, 14, 15, 15, 15, 15, 0, 0, 15, 0, 1, 0, 0, 0, 0, 14, 15, 15, 15, 15, 14, 13, 12, 11, 10, 15, 13, 9, 8.
Complex clustering has to be used since the indices require more than 3 bits.
Without the MTF transform,  204 bits are needed in total to signal this sequence.
With the MTF transform this can be reduced to 173 bits.
The transformed sequence is as follows:
0, 1, 2, 0, 0, 0, 3, 4, 5, 6, 7, 0, 0, 0, 8, 9, 10, 11, 12, 13, 14, 15, 0, 0, 0, 15, 0, 1, 1, 15, 1, 0, 0, 0, 3, 3, 0, 0, 0, 1, 4, 5, 6, 7, 5, 4, 8, 9.

\subsubsection{Histograms}
An entropy-coded stream starts by signaling one bit indicating whether or not LZ77 is used (see \SectionName{}~\ref{lz77}), which has an impact on the number of pre-clustering contexts (an additional context is added in case LZ77 is used), followed by the additional LZ77 information if applicable.
Then the context map is signaled, as described in \SectionName{}~\ref{context_map}.
Then a one-bit flag use\_prefix\_code is signaled indicating whether prefix coding or ANS is used. In case of ANS, an additional 2-bit number $n$ is signaled indicating that the maximum alphabet size is $2^{n+5}$.
In case of prefix coding, the maximum alphabet size is set to $2^{15}$.
Next, for each post-clustering context, a configuration for hybrid integer coding is signaled, using a compact signaling that takes the maximum alphabet size into account.
Then, for each post-clustering context, the distribution of token values is signaled. The way it is signaled depends on whether the entropy-coded stream is based on prefix coding or ANS.

\paragraph{Prefix codes.}
To signal a histogram (in the form of a prefix code description), first a single bit is used that determines whether it is a singleton histogram in which all tokens are zero (and no further bits are spent on coding the tokens). If not, then an additional 4-bit number $n$ is signaled, followed by an $n$-bit number $m$. The alphabet size is $1+2^n+m$, with a maximum of $2^{15}$. A prefix code for this alphabet size is then signaled according to Sections~3.4 and 3.5 of the Brotli specification \cite{brotli}.

\paragraph{ANS distributions.}
An ANS distribution is a table of 12-bit probabilities with the same size as the alphabet, so it has 32, 64, 128 or 256 entries. These probabilities correspond to the frequency of the corresponding tokens in the hybrid integer coding of the entropy-coded symbols.
The signaling of this table has some special cases to allow signaling some specific types of distributions in a very compact way.
Table indices (token values) are signaled using a method called U8(), which codes an 8-bit number using 1 to 11 bits by first using a one-bit boolean is\_zero (if true, it returns zero), then a 3-bit number $n$, and then an $n$-bit number $m$, returning the value $2^n + m$.
The table signaling starts with a single bit $s$ for cases when there are only one or two distinct token values with a nonzero probability. In this case ($s=1$), another bit $c$ signals whether there is one token ($c=0$) or two ($c=1$). If there is only one token, U8() signals the value of the token, which gets a probability of 1. Unlike prefix coding singleton histograms, singleton ANS distributions can correspond to a non-zero token. This implies that in that context, all tokens have that value without requiring any further entropy-coded bits.
If there are two tokens, U8() is used twice to signal their values, and a 12-bit number $k$ signals the probability $p = k / 2^{12}$ of the first token; the other token has probability $1-p$.
In case $s=0$, another bit $u$ is signaled which can be used for uniform distributions. If $u=1$, then U8() is used to signal the largest token value $m$, and all tokens in $[0,m]$ get a probability that is approximately $\frac{1}{m+1}$,
taking into account the limitation of 12-bit precision.
More precisely, for the tokens in $[0, 2^{12} \mod{m+1})$ the probability is $\ceil{\frac{2^{12}}{m+1}} / 2^{12}$ and for the tokens in $[2^{12} \mod{m+1}, m]$ the probability is $\floor{\frac{2^{12}}{m+1}} / 2^{12}$, which ensures that the probabilities add exactly to 1.
In the general case of $u=0$, an arbitrary distribution can be signaled, using a complicated scheme involving a fixed prefix code, run-length encoding, implicitly assigning the remaining probability to the last token, and optional signaling of probabilities with reduced precision while preserving a special case for the minimum probability $1/2^{12}$.

\paragraph{Design rationale.}
Overall, the use of static, explicitly signaled histograms (either as prefix codes or ANS tables) is a design choice aimed at allowing fast decoding and fast encoding.
Entropy coding using dynamic probabilities that are updated at decode time can lead to better compression since the overhead of signaling histograms can be avoided and the probabilities automatically adapt to local behavior. However this is not guaranteed: the static approach  can outperform the dynamic approach since dynamic updates will only converge to optimal probabilities after some time, while static probabilities can be optimal from the start. In any case, the dynamic approach is inherently slower, in particular for decoding.

The use of static histograms allows creating very fast encoders that use a fixed or restricted prefix code, while the option of slower and denser encoding remains open.
Context maps reduce the number of post-clustering contexts. Hybrid integer coding reduces the alphabet sizes needed for the entropy-coded tokens and allows signaling symbols with a very large range. Both reduce the overall memory footprint and signaling overhead of both the histograms and the overall entropy coding, and the compact coding of histograms further reduces the signaling overhead.
Still, most of the compression gains of sophisticated context modeling and near-optimal entropy coding through precise symbol distributions are preserved.

\section{Bitstream ordering}
\label{progressive}
An important aspect of codec design is the order in which the image data, including the metadata and various entropy-coded streams, is serialized into a bitstream.
JPEG~XL aims at balancing flexibility and efficiency. The syntax is expressive enough to satisfy various use cases and encoder constraints, while it also imposes restrictions meant to guarantee efficient decoding.

A key consideration is the ability to implement coding in a parallelizable (e.g., multi-threaded) way, taking advantage of typical modern multi-core processing capabilities in CPUs and GPUs.
This also facilitates bounded-memory coding of large images, as well as region-of-interest decoding where only a specific rectangular region of the image can be rendered without decoding the full image.

Another key consideration is progressive decoding, that is, allowing a decoder implementation to not only start decoding an image while the bitstream is still being transferred but also to render increasingly refined previews of the full image as the transfer progresses.

\subsection{LF and HF groups}
\label{groups}
Every frame of an image is segmented into groups of two types: LF groups, corresponding to $2048 \times 2048$ regions of the frame at 1:8 resolution, and HF groups, corresponding to $256 \times 256$ regions at 1:1 resolution.
Both group types contain image data of dimensions $256 \times 256$, though in the case of LF groups it refers to a 1:8 downsampled version of the image.

In case of VarDCT mode, the group size is fixed to $256 \times 256$; in case of Modular mode there is the option to use a different group size: either $128 \times 128$, $256 \times 256$, $512 \times 512$, or $1024 \times 1024$.
At the right and bottom sides of the image, the group size can be smaller to take into account the actual frame dimensions, although in the case of VarDCT the dimensions of groups are always a multiple of 8 to be aligned with the (smallest) DCT block size; pixels outside the frame dimensions are ignored by a decoder.

Figure~\ref{fig:groups} illustrates the segmentation of a $4032 \times 3024$ image (12 Mpx) into groups, which results in 4 LF groups and 192 HF groups.

In VarDCT mode, the LF groups always contain the data (lowest-frequency coefficients) to reconstruct an 1:8 image, except when an LF frame is used and this data is already available (see \SectionName{}~\ref{lf-frame}).
In Modular mode, the contents of the LF groups depend on the Modular transforms that were used: in the typical case for lossless compression, the LF groups are empty and all the data is stored in HF groups. 
If the Squeeze transform is used, the image data needed to reconstruct a 1:8 image is stored in the LF groups.

\subsubsection{Sections}
\label{sections}
Each group corresponds to an independent section of the bitstream.
Sections are byte-aligned and are padded with zero bits if needed.
All entropy-coded streams are contained within a section, so from the point of view of entropy decoding, sections can be processed independently and in parallel.
There are two additional sections, LfGlobal and HfGlobal, containing global information such as quantization parameters and shared context modeling data which have to be available in order to process LF groups and HF groups.
In Modular mode, LfGlobal also signals the chain of Modular transforms to be applied globally (as opposed to local transforms which only apply to a specific group), as well as image data for channels with dimensions of at most $256 \times 256$, that is, the lowest frequency data after a Squeeze transform.

\begin{figure}\centering
	\includegraphics[width=\linewidth]{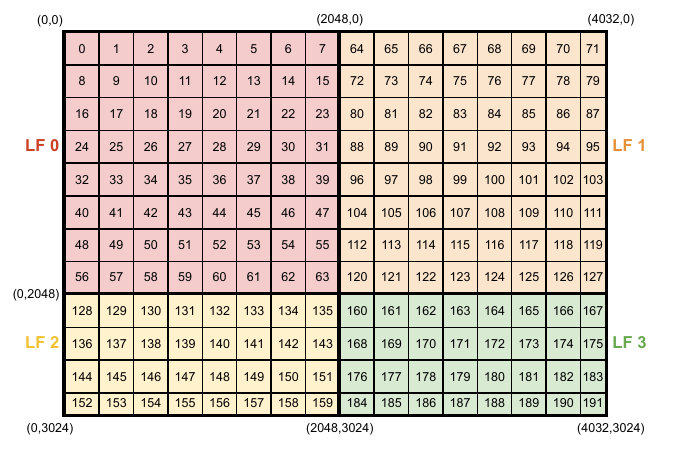}
	\caption{Segmentation of an image frame into groups.}
	\label{fig:groups}
\end{figure}
\begin{figure}\centering
	\includegraphics[height=0.7\textheight]{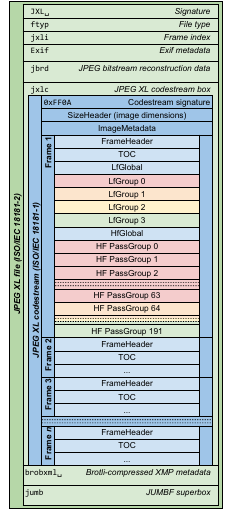}
	\caption[Example bitstream structure of a JPEG~XL file]{Example bitstream structure of a JPEG~XL file and the codestream it contains.}
	\label{fig:bitstream-structure}
\end{figure}

Figure~\ref{fig:bitstream-structure} shows an example file and codestream structure corresponding to the group segmentation shown in Figure~\ref{fig:groups}.
In this example, there is only a single HF pass.

In general, the HF data can be split over multiple refinement passes, in which case the number of HF groups is multiplied by the number of passes.
The number of passes is signaled as part of the FrameHeader and is at most 11. \SectionName{}~\ref{passes} discusses HF passes in more detail.

There is one exception to this general section layout: if the frame is small and there is only a single HF group --- in other words, the frame dimensions are at most $256 \times 256$, in case of VarDCT mode --- then all the data is collapsed into a single section, i.e. without byte-alignment at the start of each element.

\subsubsection{TOC}
After the FrameHeader, a Table Of Contents (TOC) is signaled which is an array of section lengths in bytes. The number of sections and their meaning is implicit: it can be derived from the frame dimensions and the FrameHeader.
The array of section lengths is signaled using the following variable-length encoding:
first two bits indicate a number $m$. Then:
\begin{itemize}
\item if $m=0$, the length is given by a 10-bit number;
\item if $m=1$, the length is 1024 + a 14-bit number;
\item if $m=2$, the length is 17408 + a 22-bit number;
\item if $m=3$, the length is 4211712 + a 30-bit number.
\end{itemize}
This encoding allows for very large section lengths (up to about 1 gigabyte), while in the typical case, there is only 12 or 16 bits of TOC overhead per section.

\begin{figure*}\centering
\setlength{\tabcolsep}{3pt}
\begin{tabular}{l|r}
Default ordering: & Center-first ordering: \\
\includegraphics[width=0.483\linewidth]{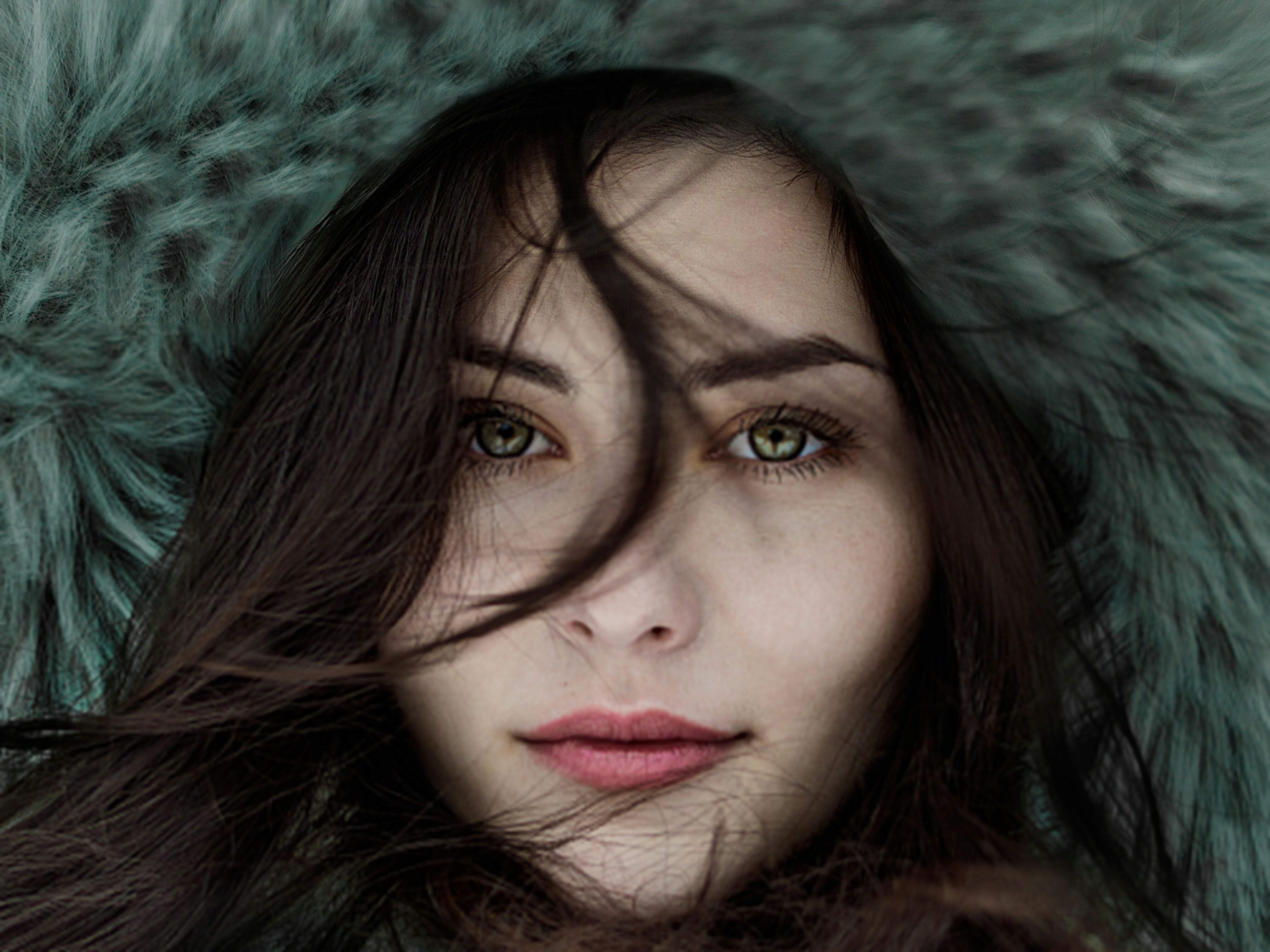}
& \includegraphics[width=0.483\linewidth]{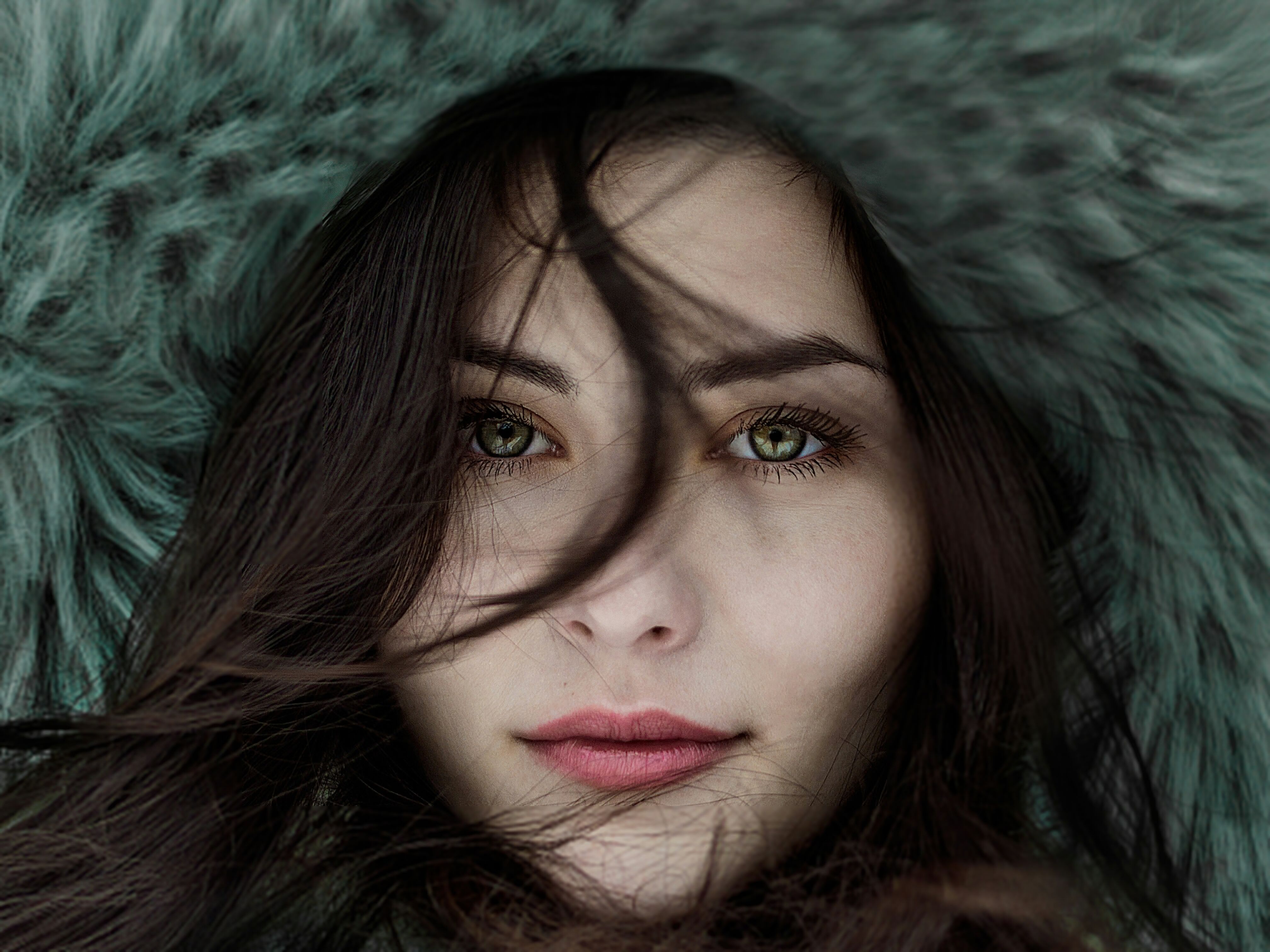} \\
\includegraphics[width=0.24\linewidth]{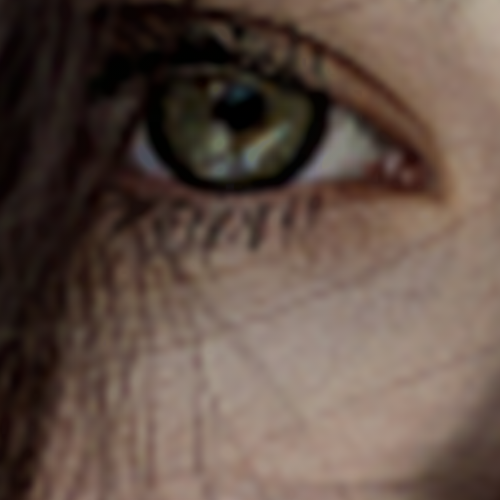}
\includegraphics[width=0.24\linewidth]{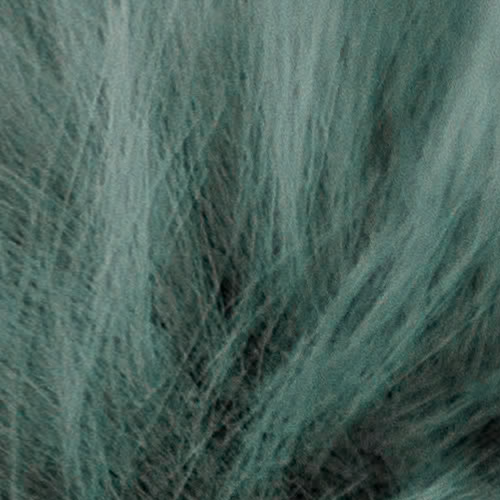} &
\includegraphics[width=0.24\linewidth]{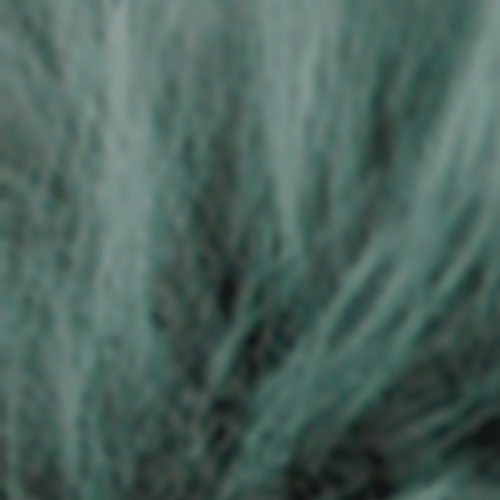} 
\includegraphics[width=0.24\linewidth]{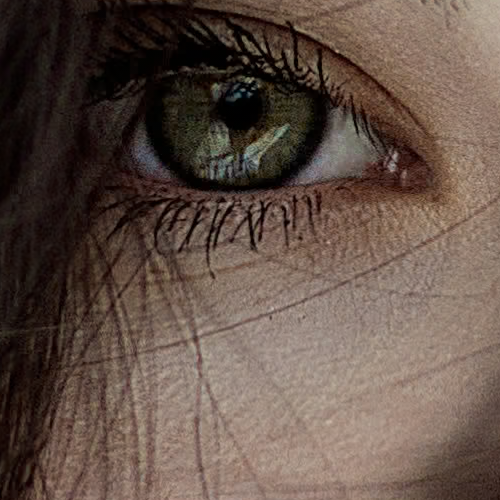}\\
\end{tabular}
\caption[Progressive preview with default and center-first ordering]{Progressive preview after loading 800 kB of a 2,274 kB JPEG~XL file (35\%), with default ordering and with center-first permuted ordering. Photo by Alexandru Zdrobău on Unsplash.}
\label{fig:progressive_ordering}
\end{figure*}

The TOC optionally also includes a permutation (see \SectionName{}~\ref{lehmer}).
This allows the sections to be ordered arbitrarily.
For example, it can be desirable to reorder the HF groups in a center-first way, such that the high-frequency detail of the center of the image is at the beginning of the bitstream and the edges of the image are at the end.
This is useful to move the image data corresponding to more salient or important regions towards the beginning of the bitstream, instead of using the default top-to-bottom ordering. This can result in better progressive previews, as illustrated in Figure~\ref{fig:progressive_ordering}.
Center-first ordering also makes sense when representing 360-degree images, where the initial view typically corresponds to the center of the (projected) image. 

\begin{figure*}\centering
\includegraphics[width=\linewidth]{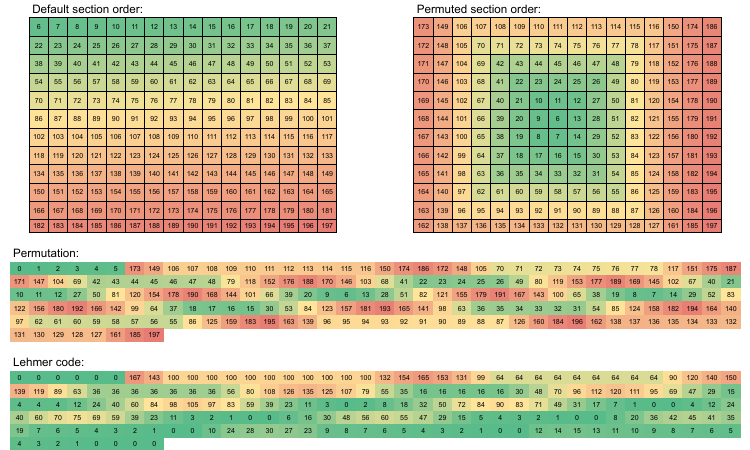}
\caption[Default ordering and center-first permuted section ordering]{Default ordering and center-first permuted section ordering. HF group section numbers start at 6 since indices 0-5 are used for LfGlobal, LF Groups and HfGlobal. Note how the Lehmer code reduces the amplitude of the numbers in the sequence to be entropy coded in permutation signaling.}
\label{fig:permutation}
\end{figure*}

\subsubsection{Permutations}
\label{lehmer}
Permutations are used in both the TOC and in VarDCT HF coefficient reordering (see \SectionName{}~\ref{coef_ordering}). In both cases, there is a default ordering and typically the desired ordering is either the default ordering or a reordering that is relatively similar to the default.
To avoid signaling overhead, permutations are coded in a specific way and entropy-coded.

Permutations are signaled using a Lehmer code \cite{lehmercode}. This means that the identity permutation is represented as a sequence of zeros and, in general, the amplitude of the numbers will be lower than when directly coding a sequence of permuted indices.
Figure~\ref{fig:permutation} shows an example of a TOC permutation that applies the center-first order illustrated in Figure~\ref{fig:progressive_ordering}.
Applying the Lehmer code helps to reduce the entropy-coded size of the permutation data. This is particularly important in the case of HF coefficient reordering when applied to large transforms. For instance, DCT256x256 covers 65536 samples corresponding to 1024 LF coefficients and 64512 HF coefficients which can be permuted.

The entropy coding scheme for the Lehmer sequence first signals the number of elements that will be coded (the remaining elements are implicitly zero).
Then it uses the previous element $L_{i-1}$ of the Lehmer sequence as a context for coding the current element $L_i$ --- or rather, the context index is 0 for the first element and $\ceil{\log_2(L_{i-1} + 1)}$ for the next elements, with the context index clamped to $[0,7]$.

\subsubsection{ROI decoding}
The primary purpose of the TOC is to allow multithreaded decoding: the offsets of every bitstream section can be computed from the TOC (taking into account the optional permutation) and then sections can be processed in parallel, constrained only by the following data dependencies:

\begin{itemize}
\item
Any specific LF group section can be decoded using the section data and the LfGlobal data.
\item
Any specific HF group section can be decoded using the section data, the HfGlobal data, and the LF group data corresponding to the LF group to which the HF group belongs.
\end{itemize}

The TOC also enables region-of-interest (ROI) decoding, also known as cropped decoding. In the best case, the region of interest coincides with an LF group region (a naturally aligned $2048 \times 2048$ square), in which case the region can be decoded without having to decode any data outside the region of interest.
In the general case, it will be necessary to decode all of the LF groups and HF groups that only partially overlap with the region of interest.

As long as the region of interest is aligned with the HF group grid, the worst-case overhead is bounded. Decoding a single HF group requires decoding the entire corresponding LF group, from which only 1/64th of the data is actually used (the LF coefficients and HF metadata for the single HF group), so almost twice as much image data has to be decoded as is needed.
This is the worst case; if the region of interest is larger, the overhead will be smaller.

This can be contrasted with JPEG and PNG where the image data is coded top to bottom, which implies that region-of-interest decoding can only avoid decoding the region below the region of interest; the regions above, to the left and to the right have to be decoded too. So the overhead can be arbitrarily large: decoding a small region at the bottom of the image requires decoding the entire image.

\subsection{Progressive decoding}
Web-based image delivery and other important use cases that involve network transfer of images or accessing image files from relatively slow storage media can benefit from streaming and progressive decoding.

Streaming decoding means the decoder can start processing an input bitstream when only an initial segment is available, not the full bitstream. JPEG~XL was designed for streaming decoding, for example by requiring the TOC to be signaled early (just after the FrameHeader).
For a streaming \emph{encoder}, it would be more convenient to have the TOC at the end, but then a decoder would need to seek to the end of the bitstream to know the position and meaning of the sections, which would make streaming decoding impossible.

Progressive decoding means the decoder can render useful previews of the final decoded image when only partial data is available; gradually these previews or refinement passes get closer to the final image.
In VarDCT mode, some degree of progressive decoding is always possible since the data is always split between LF and HF (assuming the LF data is signalled before the HF data).
This allows rendering an 8x upsampled version of the 1:8 LF image when only a fraction of the full bitstream has been transferred.
For this stage of the preview libjxl uses the non-seperable upsampling described in \SectionName{}~\ref{nonsep-upsampling}.
Then as more image data arrives, HF groups can be decoded one by one, replacing a blurry LF-only image with the detailed final decoded image in the corresponding region.
By using a permutation to reorder the HF groups, more-salient image regions can be prioritized, as shown in Figure~\ref{fig:progressive_ordering}.

This basic two-pass progressive coding (first 1:8, then 1:1) is the default approach that the libjxl encoder takes, as it keeps the overhead of progressive rendering to a minimum (every pixel is painted at most twice) while bringing most of the advantages of progressive decoding.
This is in contrast with progressive JPEG where more passes are needed (at least four, and often around ten are used).

\subsubsection{HF passes}
\label{passes}
It is also possible to split the HF data over multiple passes, to achieve a more gradual refinement from the 1:8 LF image to the final image.
In progressive JPEG, every AC pass consists of a single color component, for which a specific range of AC coefficients is signaled (spectral selection) according to the zig-zag order, and least significant bits can be postponed to a later scan (successive approximation).

In JPEG~XL, there is more flexibility in the progressive passes.
One important difference is that in progressive JPEG, every bit of every AC coefficient corresponds to a specific pass, while in JPEG~XL, every pass can in principle update any HF coefficient.
The HF coefficients of all passes are added together.
This approach allows for example creating saliency-based passes where `important' image regions are updated with more precision than the other regions.
This is impossible in JPEG, which requires passes to update the entire image uniformly.

Explicit successive approximation can be signaled: a pass has a `shift' value and all coefficients are implicitly left-shifted by this amount.
There is no explicit spectral selection; coefficient reordering can be used to put coefficients that are not updated (so their values are all zeroes) at the end.

\subsubsection{Preview frame}
Optionally, the image header can signal that the first frame is a preview frame. To decode the full image, it can be ignored, but for the purpose of quickly showing e.g. a thumbnail image, this preview could be used.

Generally an explicit preview frame is not needed and only harms compression since the preview frame is redundant and does not influence the final image. The LF-HF progression, possibly with additional HF passes, achieves progressive decoding without any overhead.

Still, the preview frame approach can be useful in some cases: for large lossless images, or for animations where the first frame is not representative, it can be useful to add a low-resolution lossy preview frame. In this case, the relative overhead would also be small.

\subsubsection{LF frames}
\label{lf-frame}
The LF coefficients effectively correspond to a 1:8 downsampled image.
The default is to signal them in LF groups.
However it is also possible to signal the LF image as a separate frame that is signaled in advance. In that case a flag in the FrameHeader signals that the LF data is taken from a previously coded LF frame, and the LF groups then no longer contain LF coefficients, only HF metadata.

LF frames can use VarDCT mode, and can recursively use their own LF frame. Up to four levels of recursion are allowed. The first-level LF frame has dimensions that are 1:8 compared to the image dimensions; the second-level LF frame is 1:64, the third-level LF frame is 1:512 and the fourth-level LF frame is 1:4096.

The largest image dimension allowed by the Main Profile, Level 10, is $2^{40}$ pixels, which is a bit over one terapixel ($10^{12}$ pixels).
Even at that huge size, a level-4 LF frame would only be 65 kilopixels (e.g. $400 \times 164$).

This approach makes it possible, at least in principle, to store a huge image in a single JPEG~XL codestream while still being able to efficiently render it, both at the 1:1 scale (using ROI decoding) and at a zoomed-out, `fit to screen' scale --- and at every scale in between.

\begin{figure*}\centering
\setlength{\tabcolsep}{3pt}
\begin{tabular}{cc}
\includegraphics[width=0.49\linewidth]{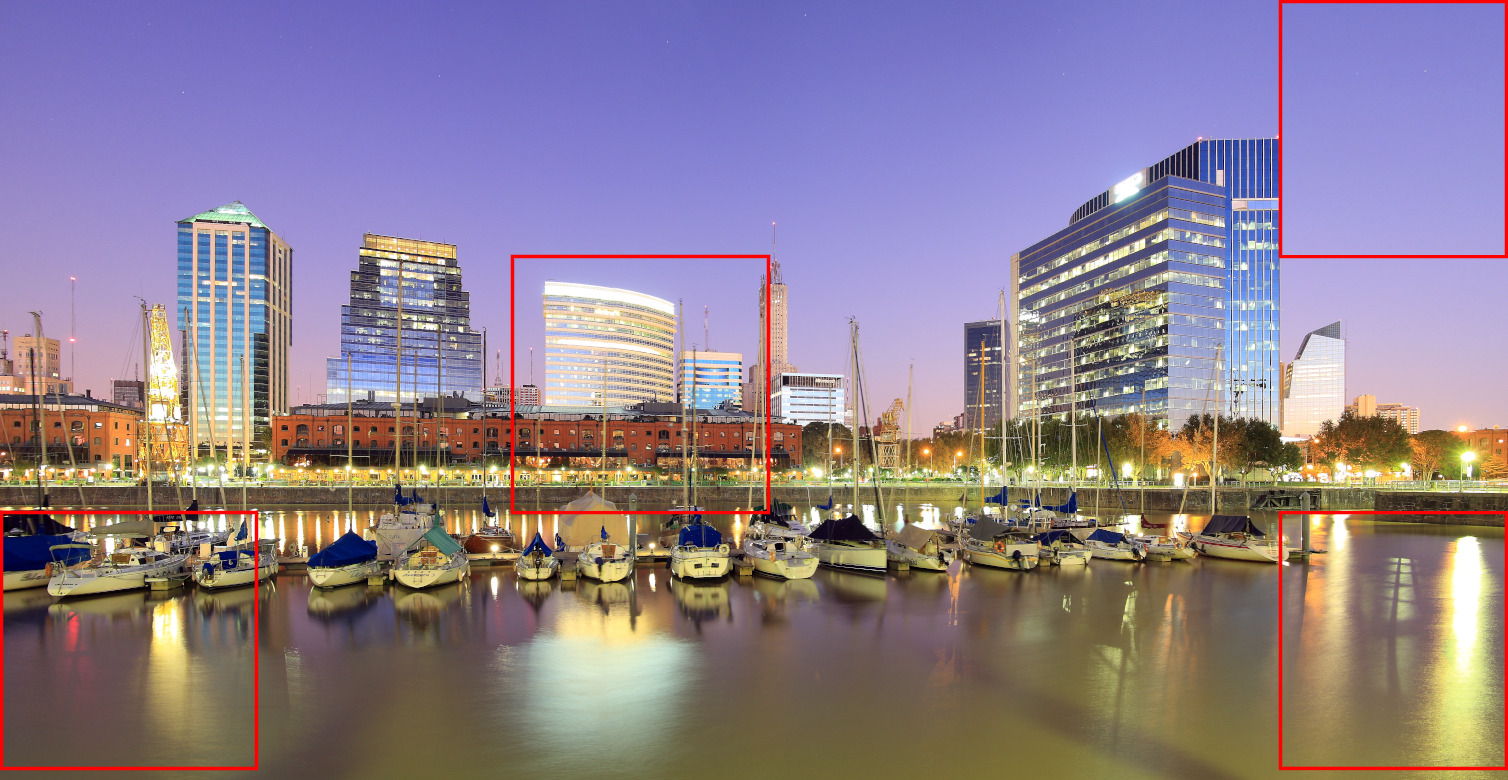} &
\includegraphics[width=0.49\linewidth]{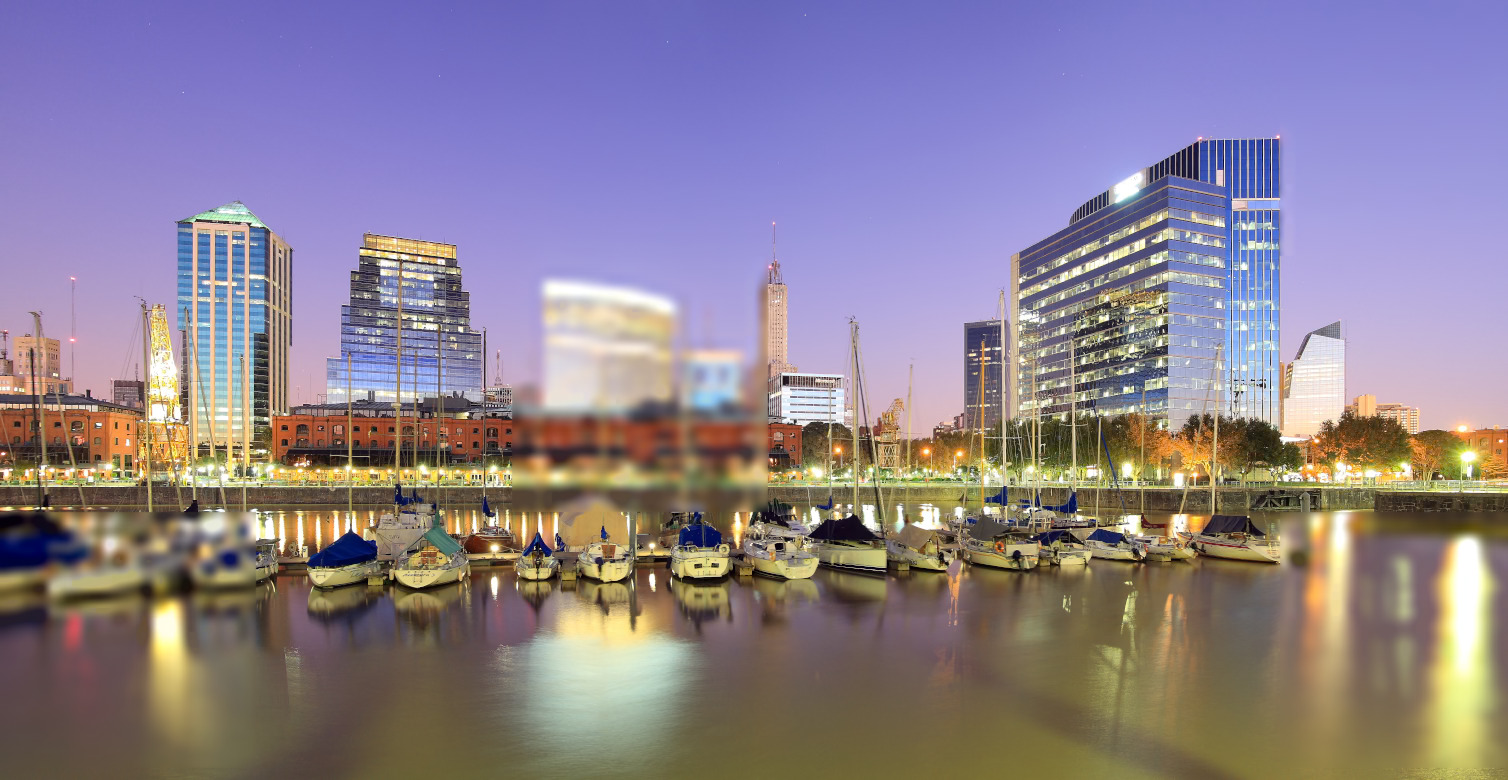}
\end{tabular}
	\caption[Error resilience demonstration]{Error resilience demonstration. Left: original image, with an indication of four HF group regions for which corresponding JPEG~XL bitstream sections are corrupted. Right: image that can be recovered from the corrupted bitstream.}
	\label{fig:error_resilience}
\end{figure*}
\subsubsection{Error resilience}
Although error resilience and data compression are inherently somewhat contradictory goals --- compression is all about eliminating redundant information, while most error recovery approaches require redundancy ---
there are some useful error resilience features in JPEG~XL.
Since groups are coded independently, local bitstream corruption in HF sections will have a bounded impact.
Moreover, errors can be detected (with high likelihood) since entropy-coded streams have a size that is known in advance (so any overrun or underrun while reading indicates corruption), and in the case of ANS, the final ANS state effectively acts like a 32-bit checksum.

Progressive rendering can be used as a partial error recovery mechanism: if corruption is detected in the data of a particular HF group, this group can be rendered using only the LF data --- or an earlier non-corrupt HF group pass, if one is available.
Figure~\ref{fig:error_resilience} shows an example of a bitstream that is corrupted in four different locations; these errors can be detected and partially recovered in this way.

\subsection{Frames and layers}
While JPEG~XL is primarily designed for still images and is not a video codec with sophisticated inter-frame coding tools, a single JPEG~XL codestream can store multiple frames.
Conceptually, there are three types of multi-frame images:
\begin{description}
\item[Layers.]
A composite still image can consist of multiple layers, possibly overlapping one another, which are merged using blend modes.
Layered images are supported in image editors like Photoshop and Gimp.

\item[Pages.]
A multi-page image is a sequence of images that are typically navigated manually. Examples include scanned documents, presentation slides, and comic books.

\item[Animation.]
With the addition of timing information, frames can represent an animation sequence.
Use cases for animation sequences include ``GIFs'' (short looping video fragments), cinemagraphs, and lossless or very high-fidelity intra-only video.
\end{description}

The distinction between these three different types of multi-frame images is made by means of the duration field in the frame header.
Zero-duration frames are layers and non-zero duration frames are part of a layer, unless the duration has the maximum value of $2^{32} - 1$, in which case the frame is interpreted as a page.

Layers, i.e. subsequent zero-duration frames, are composited (blended) by the decoder and only a single merged image is returned.
It is possible to mix the use of layers and animation or pages, e.g. an animation frame can itself have multiple layers.

The number of frames is not signaled explicitly; every frame header has a single-bit field is\_last that indicates whether the frame is the final one or not.

\subsubsection{Animation}
\label{animation}
The single-bit field have\_animation in the image header indicates whether the image is a still image or not.
If this bit is 0, then all frame durations are implicitly zero so there can only be layers.
If the bit is 1, then an additional animation metadata is signaled as explained in \SectionName{}~\ref{header_animation}.
This defines the unit of time the durations in the frame headers refer to.
It also indicates how many times the animation should be played back, with the value 0 meaning it loops forever.

\subsubsection{Blend modes}
\label{blending}
The frame header also signals blending information.
In the simplest case, every frame has the same dimensions as the image and completely replaces all pixels, using a blend mode called Replace.
In this case, all frames are completely independent.

The state of the image canvas after decoding a frame can be recorded by setting the 2-bit save\_as\_reference field in the frame header to a non-zero value. It will then be `saved' in the reference buffer corresponding to that value.
In case the frame has zero duration and is not the last one (is\_last is 0) then it is implicitly saved in reference buffer 0.
Reference buffers can be stored either in the internal XYB or YCbCr color space, or in the RGB color space signaled in the image header. The frame header field save\_before\_ct signals whether to store the canvas state before the inverse color transform (so in the internal color space) or after (so in an RGB color space).

In the general case, a frame is rendered by starting from a reference buffer, called the source, and then blending the frame with the source.
Which of the (up to) four reference buffers to use is signaled in the frame header; if a buffer is referenced that does not contain any previously saved data, it is implicitly zero-filled.

The following frame blend modes are defined:
\begin{description}
\item[Replace.]
Overwrites the source sample values with the frame sample values.
\item[Add.]
Adds the frame sample values to the source sample values.
\item[Blend.]
Alpha-blends the frame as if it is a layer that goes on top of the source image (blend over).
\item[MulAdd.]
Multiplies the frame sample values with the frame alpha values and adds the result to the source sample values.
\item[Mul.]
Multiplies the frame sample values with the source sample values.
\end{description}

In case of Blend and MulAdd, it is signaled which of the extra channels to use as the alpha channel.
It can also be signaled whether or not to clamp the alpha values to the $[0,1]$ range when blending. In case of Mul blending, it can be signaled whether or not the frame sample values should be clamped to that range when blending.

If there are extra channels, the frame header signals the blending information separately for each extra channel, so they do not necessarily use the same blend mode (or even the same source) as the main color channels.

\begin{figure}
\includegraphics[width=0.8\linewidth]{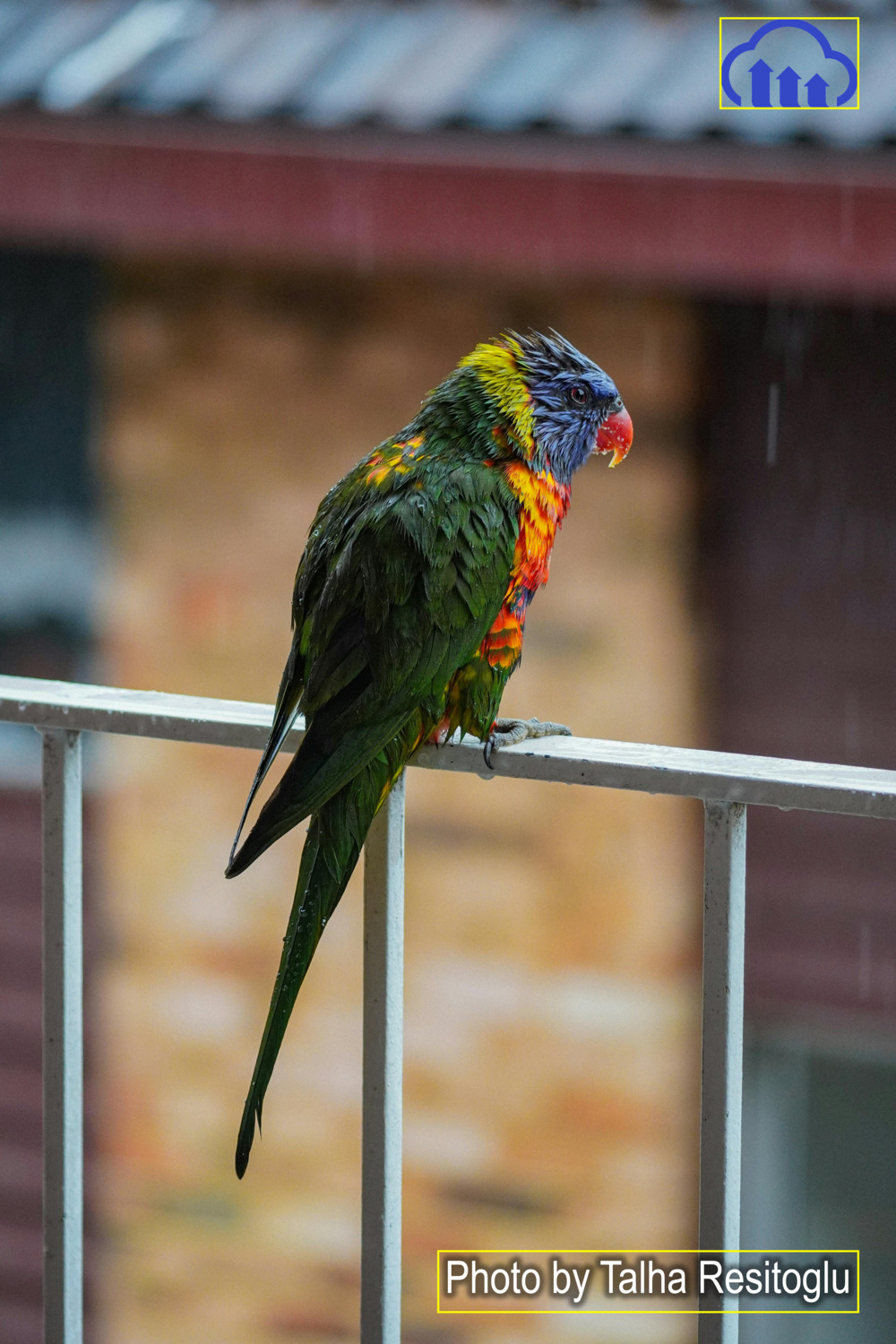}\\
\footnotesize
Image: $4000 \times 6000$\\
--- Frame (``Base''): $4000 \times 6000$ at offset $(0,0)$\\
--- Frame (``Logo''): $615 \times 402$ at offset $(3216,81)$\\
--- Frame (``Text''): $1875 \times 272$ at offset $(1956,5585)$
\caption[Example of a layered image]{Example of a layered image with two\\ smaller layers alpha-blended over a base layer.\\
Original photo by Talha Resitoglu on Pexels.}
\label{fig:layers}
\end{figure}

\subsubsection{Layers}
As explained in \SectionName{}~\ref{header_crop}, frames can have a dimension different from the image dimensions, as well as an offset.
In that case, regardless of the blend mode, any region not covered by the frame is simply copied from the source (one of four reference buffers).

Layers (zero-duration frames) can be used to create a composite still image without the need to merge the layers. The layer blending is part of the codec specification and performed by the decoder, so applications that can only handle a single decoded image frame will correctly render the merged image.
The layer separation is preserved, including an optional name string to identify individual layers. This is useful in authoring workflows, and it can also have compression advantages since e.g. photographic layers can be encoded in a lossy way in VarDCT mode while text or logo overlays may be encoded more effectively in a lossless or near-lossless way in Modular mode.
Figure~\ref{fig:layers} gives an example of a composite still image with three layers, where the regions corresponding to the two overlays are indicated with a yellow border.

\begin{figure}
\includegraphics[width=0.8\linewidth]{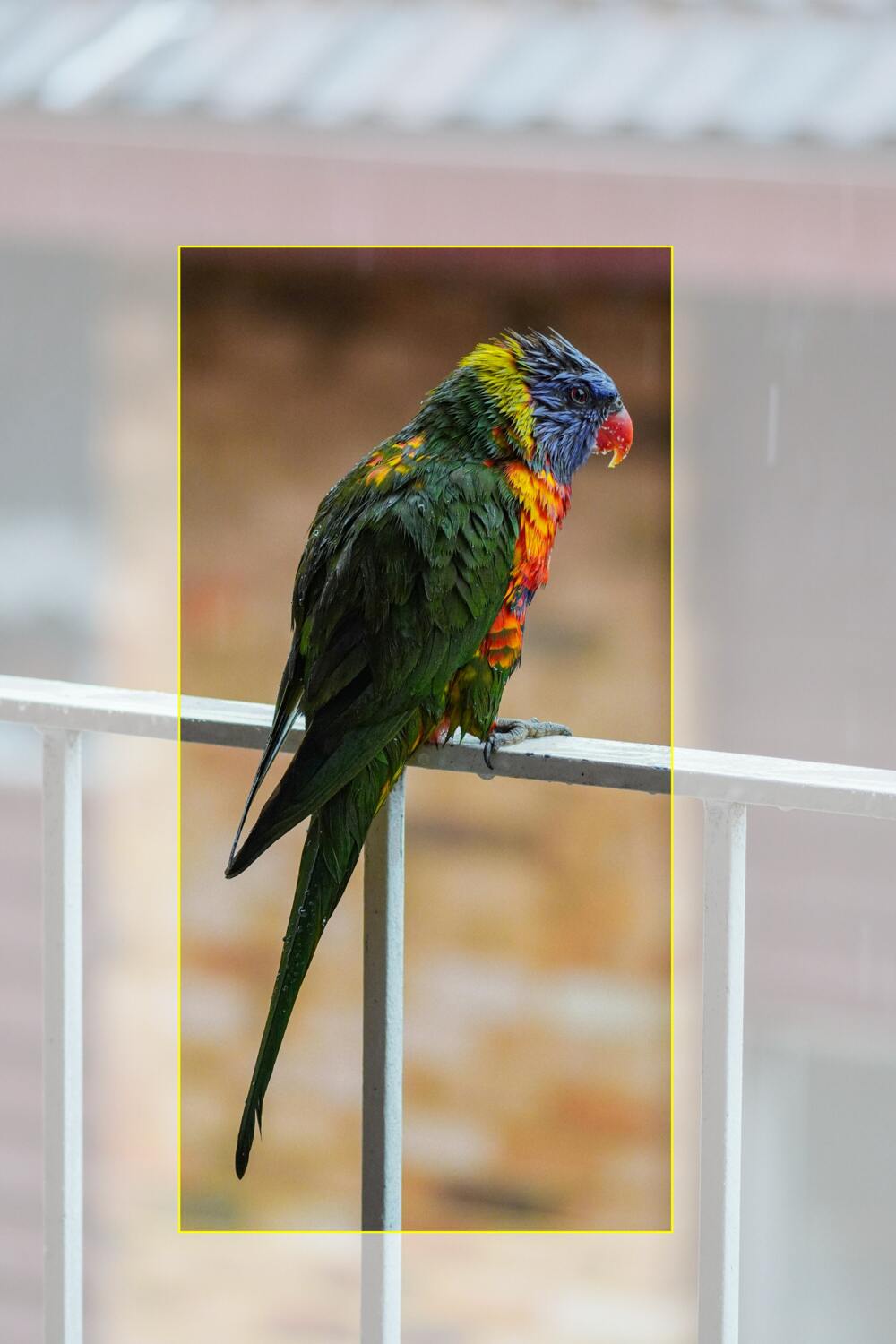}\\
\footnotesize
Image: $2800 \times 4400$\\
--- Frame: $4000 \times 6000$ at offset $(-800, -1100)$
\caption[Example of reversible cropping]{Example of reversible cropping by \\ adjusting the image canvas and frame offset.}
\label{fig:reversible_crop}
\end{figure}
\subsubsection{Reversible cropping}
Since frames can be both smaller and larger than the image dimensions, it is possible to implement a reversible cropping mechanism by adjusting the image and frame headers without modifying the actual image data.
Figure~\ref{fig:reversible_crop} shows an example of a $4000 \times 6000$ image that was cropped to $2800 \times 4400$, that is, the decoded image is only the region indicated by the yellow border.
The frame data still contains the full image, so it is still possible to adjust the crop or recover the full image. For this purpose, libjxl has an option to disable frame coalescing (which is enabled by default). This causes it to return all frames as-is, without applying frame blending and clipping to the image canvas.
Authoring tools can also use this decode option to get access to the layers of a composite still image.

\section{Compression performance}
\label{performance}
In general, it is not possible to evaluate the `inherent' performance of a codec, in terms of speed, compression or quality.
It is only possible to test specific implementations of encoders and decoders. This provides information on what a codec can do, but there is no guarantee that the results are representative for the \emph{best} a codec can do. There might be potential improvements in the encoder or decoder implementation still left untapped.

This is particularly true for codecs like JPEG~XL that leave many degrees of freedom to an encoder implementation. Both for lossy and lossless compression, in many aspects of the encode process choices have to be made, leading to a huge search space. It is not computationally feasible to explore this search space exhaustively. Only in case of lossless compression it is clear what to optimize for (compressed size); in case of lossy compression, an objective metric can be used to optimize for, but that can only be an approximation of the actual visual quality, which is what usually matters. As a result, an encoder typically implements various heuristics and partial search procedures.

For this reason, modern codecs often have encoders with a speed (or effort) setting that allows different trade-offs between encode speed and compression performance.
Even the slowest settings are usually not exhaustive, and even the fastest settings are not necessarily the fastest encode speeds that are possible. Future improvements could always lead to faster or better compression.

Decoders typically have no or almost no degrees of freedom: functionally they are (almost) completely determined by the specifications of the standard. Since there are no choices to made or a search space to be explored, decoders are often an order of magnitude faster than encoders.
The exact decode speed does of course still depend on the implementation though: highly optimized handcrafted SIMD code can be substantially faster than a simple naive implementation in a high-level programming language; a decoder implementing specialized `fast paths' for common special cases can be faster than a decoder only implementing generic methods.
In that sense, the results presented here provide only an indication of the compression performance of current implementations, available at the time of writing. It is possible (and even likely) that in the future, better implementations become available for JPEG~XL and for the codecs it is compared against.

\begin{figure*}\centering
	\includegraphics[width=1\linewidth]{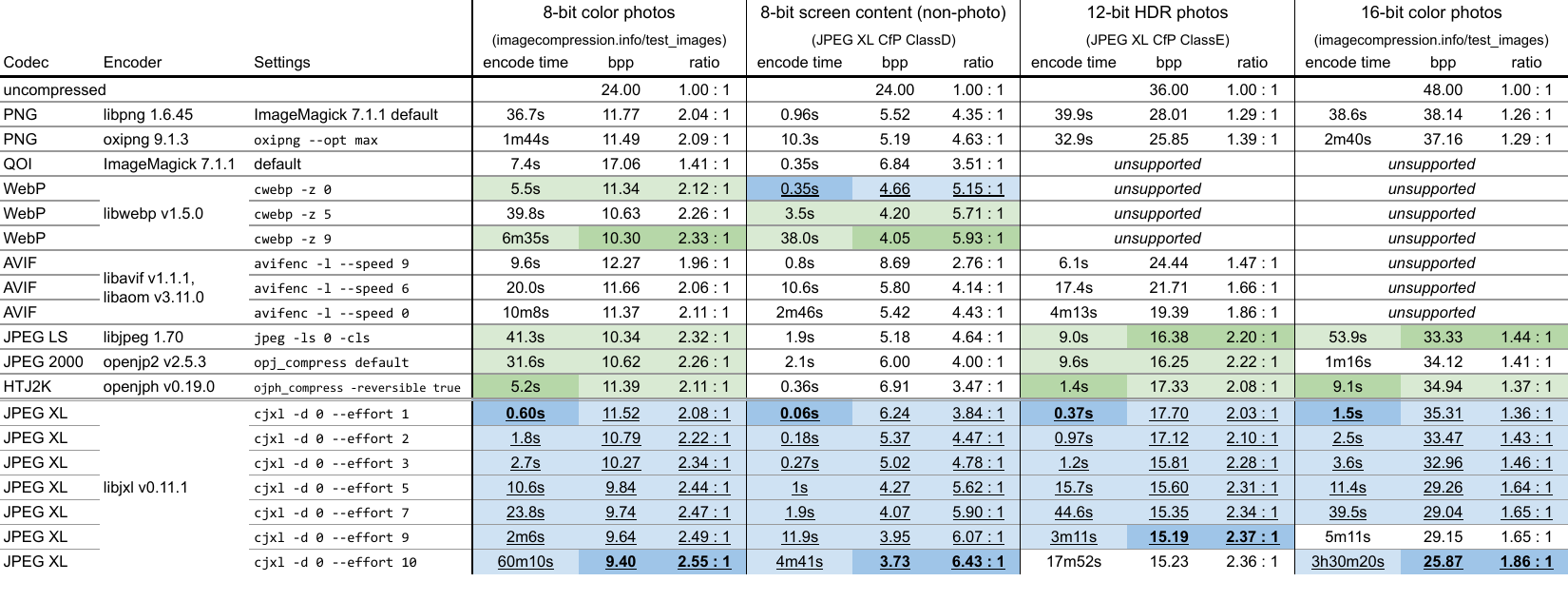}
	\caption[Lossless compression results for four sets of images]{Lossless compression results for four sets of images.
    Pareto-optimal trade-offs between encode speed and compression density are \underline{underlined} and have a blue background; best results are indicated in \textbf{bold}.
    Results that would be Pareto-optimal if not for JPEG~XL are indicated with a green background.}
	\label{fig:lossless_results}
\end{figure*}

\subsection{Lossless compression}
\label{lossless}
Lossless compression --- sometimes called `mathematically lossless' to avoid confusion with `visually lossless' lossy compression ---
is mostly useful in authoring and archival use cases, although there are also cases where it can be useful for end-user image delivery.

The main trade-off is between processing speed (encode and decode time) and compression density, which can be expressed as a bit rate (bits per pixel, or bpp) where lower is better, or as a compression ratio Uncompressed : Compressed where higher is better.
Typically, encoding is computationally more demanding than decoding, but this is not necessarily always the case.

In some codecs, the encoding process is fully or almost fully deterministic, in the sense that there are no substantial choices an encoder has to make. In this case, the codec only offers a single trade-off point; the compression density it achieves is fixed, and the speed only depends on the hardware and software implementation.

Other codecs, especially JPEG~XL, are very expressive in terms of bitstream syntax, with much room for encoders to explore more or less of the search space. This can lead to a range of trade-off points between encode speed and compression density.
In libjxl, this can be configured by setting the ``effort'' option, which has a range from 1 to 10 where higher numbers indicate that more time (and possibly memory) will be used.

To measure the encode time, a November 2023 MacBook Pro was used, with an M3 Pro ARMv8 CPU with 6 performance cores at 4.1~GHz and 6 efficiency cores at 2.7~GHz, and 36~GB of LPDDR5 RAM.
The reported time is the total time taken to encode all images one by one.

Figure~\ref{fig:lossless_results} gives an overview of lossless compression results for JPEG~XL, compared to other codecs.
For each of the sets of images, covering various types of image contents and bit depths, JPEG~XL offers several Pareto optimal trade-offs between encode speed and compression.
These results are consistent with earlier, independent comparisons of lossless compression performance \cite{lossless_eval,mandeel2021comparative,bennett2023benchmarking}.

\begin{figure}\centering
	\includegraphics[width=1\linewidth]{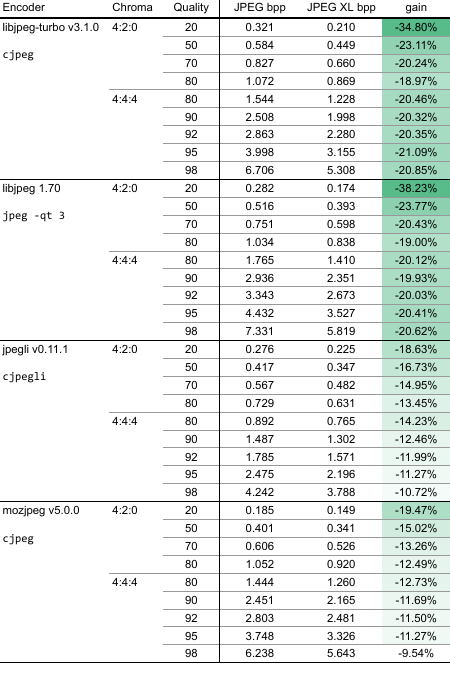}
	\caption[Lossless JPEG recompression results]{Lossless JPEG recompression results. The test image corpus is the set of 8-bit color photos from imagecompression.info/test\_images.
    They are compressed with various JPEG encoders and quality settings, and then recompressed using libjxl v0.11.1 with default settings.
}
	\label{fig:lossless_jpeg_results}
\end{figure}
\paragraph{Lossless JPEG recompression.}
Besides lossless compression of pixels, JPEG~XL can also losslessly recompress JPEG files.
Figure~\ref{fig:lossless_jpeg_results} gives an overview of the compression gains.
The gains that can be expected depend on the JPEG encoder, chroma subsampling, and quality settings. For JPEG files typically produced by a camera (high-quality 4:4:4, fixed Huffman tables), the gains are around 20\%.

\begin{figure*}\centering
	\includegraphics[width=1\linewidth]{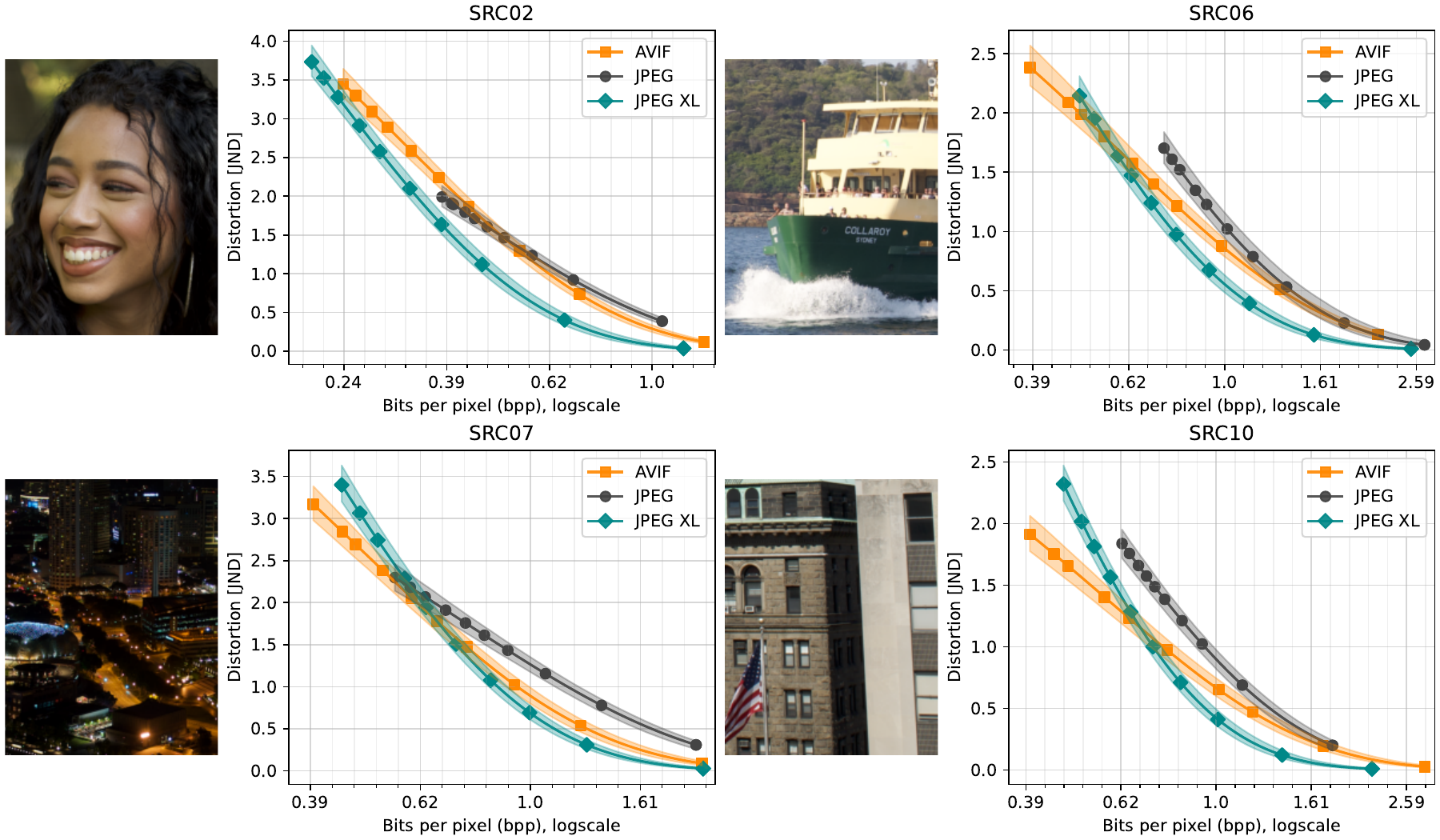}
	\caption[Lossy compression performance of JPEG~XL]{Lossy compression performance of JPEG~XL compared to JPEG and AVIF \cite{dcc_aic3}. The vertical axis indicates distortion in JND units, as estimated using the AIC-3 methodology using boosted and plain triplet comparisons. The shaded regions indicate a 95\% confidence interval. Lower is better.}
	\label{fig:aic3_results}
\end{figure*}
\subsection{Lossy compression}
In lossy image compression, the main trade-off is not between speed and compression, but between image quality and compression. Processing speed is also important, especially the encoding speed. In particular in modern codecs, encoders have many degrees of freedom, or, in other words, a huge search space to consider, which can make encoding very slow.
However, the main parameter that influences the compressed file size is the quality setting.
The goal is to achieve the best possible image quality at a given file size or the smallest possible file size for a given quality point.

\subsubsection{Subjective results}
Figure~\ref{fig:aic3_results} shows plots of bit rate versus distortion for four images.
Distortion is expressed here on a just-noticeable difference (JND) scale, where 1 JND unit is a difference that can be noticed by 50\% of human observers in typical viewing conditions.

The JND values were experimentally determined \cite{dcc_aic3} using the AIC-3 methodology \cite{aic3,testolina2023assessment} for subjective quality assessment. This methodology involves triplet comparisons, where two compressed images A and B are compared to the original and the participants are asked to respond which of the two has the strongest distortion, with a choice between A, B or ``don't know''.
Two types of triplet comparisons are done: plain comparisons (in normal viewing conditions) to establish a meaningful JND scale, and boosted comparisons \cite{men2021subjective} where differences are made more obvious (by flickering, zooming and/or artifact amplification) in order to obtain fine-grained results. 

The CID22 dataset \cite{CID22} is an earlier example of an subjective assessment methodology combining the results obtained through two different experimental protocols.
This dataset has scores for 250 source images with varying image contents.
When comparing the compression performance of codecs (or rather, encoders), the image contents has a large impact on the perceived quality and the relative performance of the codecs.
The results shown in Figure~\ref{fig:aic3_results} are consistent with the results of CID22, which can be summarized in the following general trends:

\begin{description}
\item[Portrait.]
For portrait photography and fashion, JPEG~XL performs better than AVIF, which performs better than WebP, which performs better than JPEG, with a gap of roughly 10-15\% between each of these codecs.

\item[Nature.]
Landscapes, wildlife, cloudy skies, natural materials and textures:
here JPEG~XL outperforms all the other codecs by 20-30\%; in some cases, WebP and AVIF are performing worse than JPEG.

\item[Urban outdoors.]
City streets, buildings and architecture, industrial scenes, cars, sports, nightlife:
JPEG~XL and AVIF have a similar performance, both outperforming WebP and JPEG by 20-30\%. On average JPEG~XL performs slightly better than AVIF but it varies from one image to the other.

\item[Indoors.]
Interior design, indoors events, furniture, decorations:
similar to ``Urban outdoors'' but here on average AVIF performs slightly better than JPEG~XL, again depending on the image.

\item[Illustrations.]
Non-photographic images, diagrams, charts, logos, text:
here AVIF outperforms JPEG~XL and WebP, which perform similarly, by about 20-30\%; JPEG~XL performs 30-40\% better than JPEG.
\end{description}

Overall, for lossy compression of photographs, so far image quality assessment experiments confirm that JPEG~XL performs better than JPEG, JPEG 2000, WebP, HEIF, and AVIF, though the amount of experimental data available is still limited.

\begin{figure*}\centering
\includegraphics[width=0.495\linewidth]{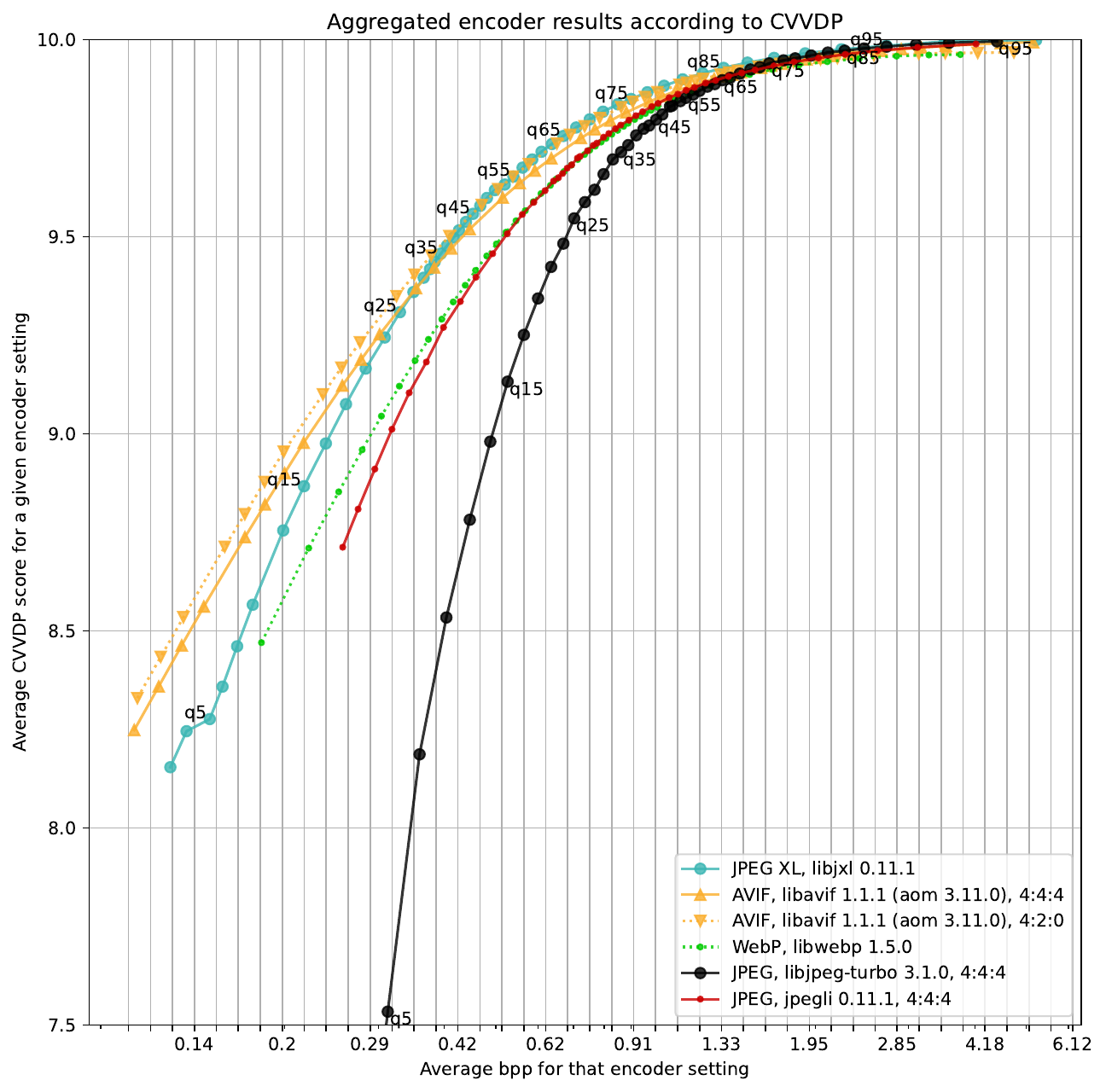}
\includegraphics[width=0.495\linewidth]{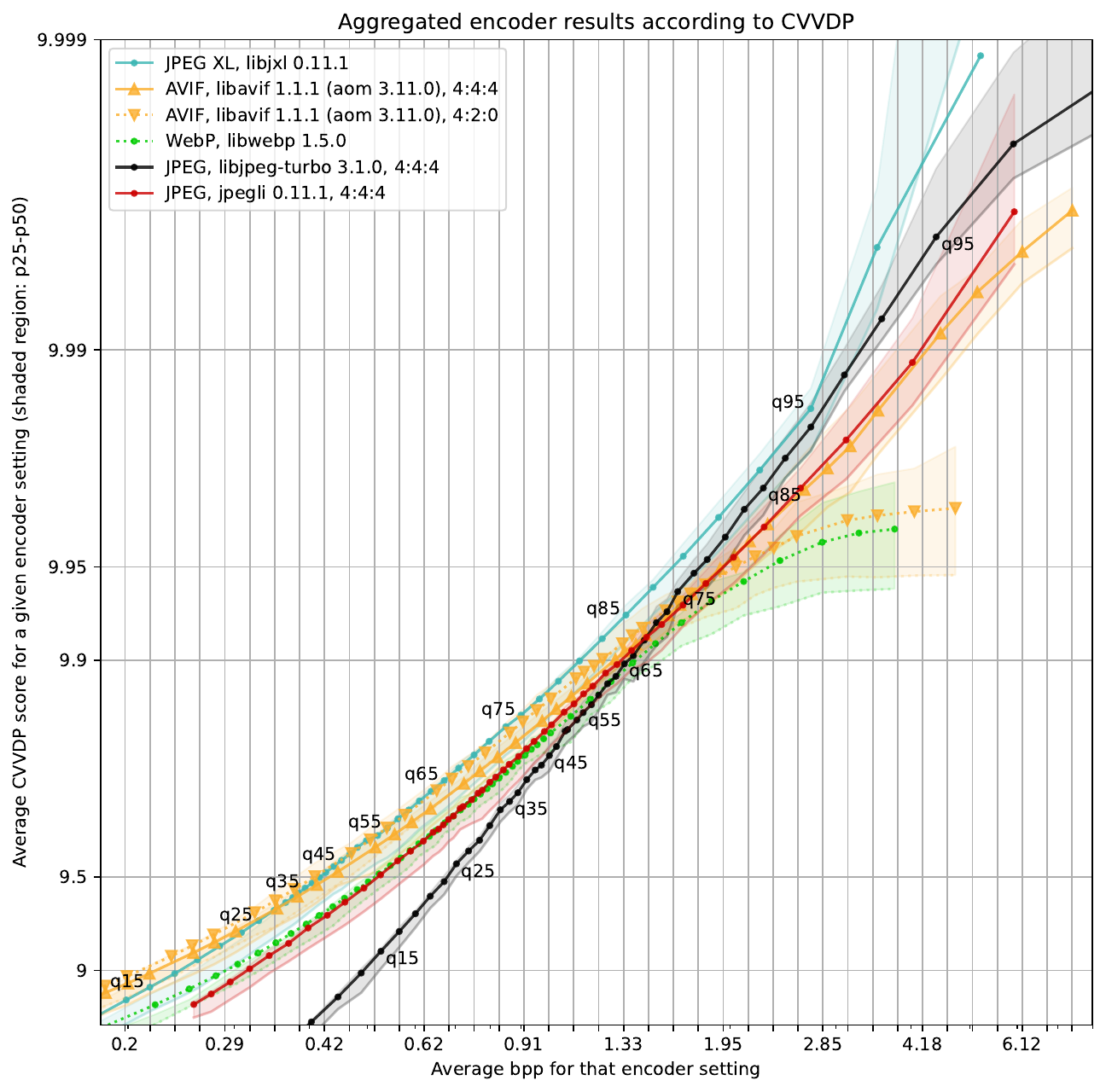}
	\caption[Aggregated CVVDP results for SDR images]{Aggregated bitrate-distortion plots for a set of 49 SDR images (\href{https://github.com/eclipseo/eclipseo.github.io/tree/2f93d184577d273cac729e778d73fdcc8a93d0a8/image-comparison-web/comparisonfiles/subset1/Original}{subset1}) according to CVVDP \cite{cvvdp}.
Every data point corresponds to an encoder quality setting (sampled from 1 to 99 in steps of 2), with the average bit rate on the horizontal axis and the aggregated metric score on the vertical axis.
The vertical gridlines are spaced at intervals corresponding to a 10\% difference.
Left: a commonly used visualization, using a logarithmic scale for the bitrate and averaged metric results on a linear scale.
Right: proposed more suitable visualization, using a logarithmic scale for the metric scores and showing an indication of the encoder consistency across different source images by shading the region between the 25th percentile and the median metric score.}
	\label{fig:cvvdp_sdr_results}
\end{figure*}
\begin{figure}\centering
\includegraphics[width=\linewidth]{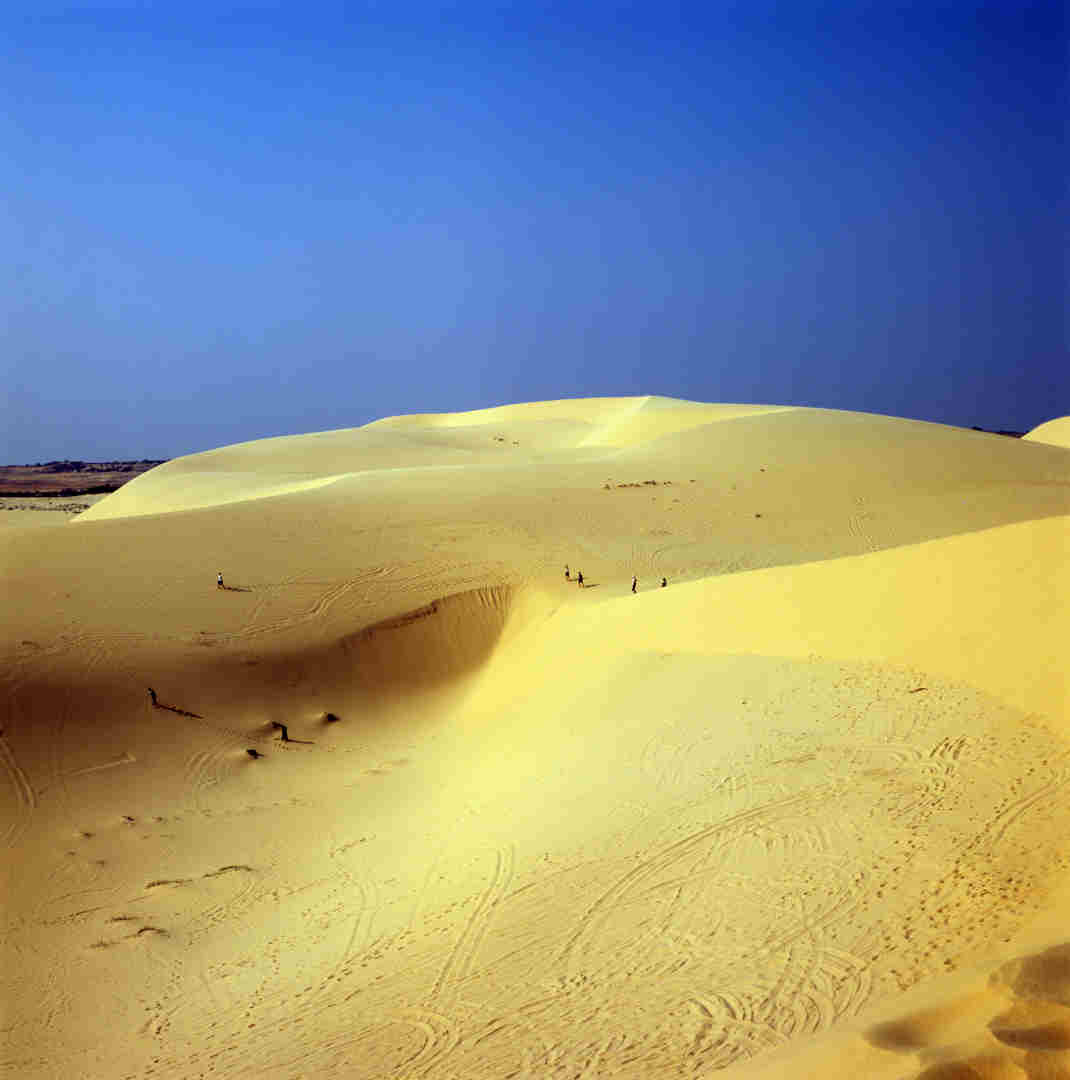}
	\caption[Example of a libjpeg-turbo q19 image]{Example of a libjpeg-turbo q19 image, with CVVDP score 9.0634. The image suffers from obvious color banding and blocking artifacts.}
	\label{fig:low_quality_example}
\end{figure}
\begin{figure}
\includegraphics[width=\linewidth]{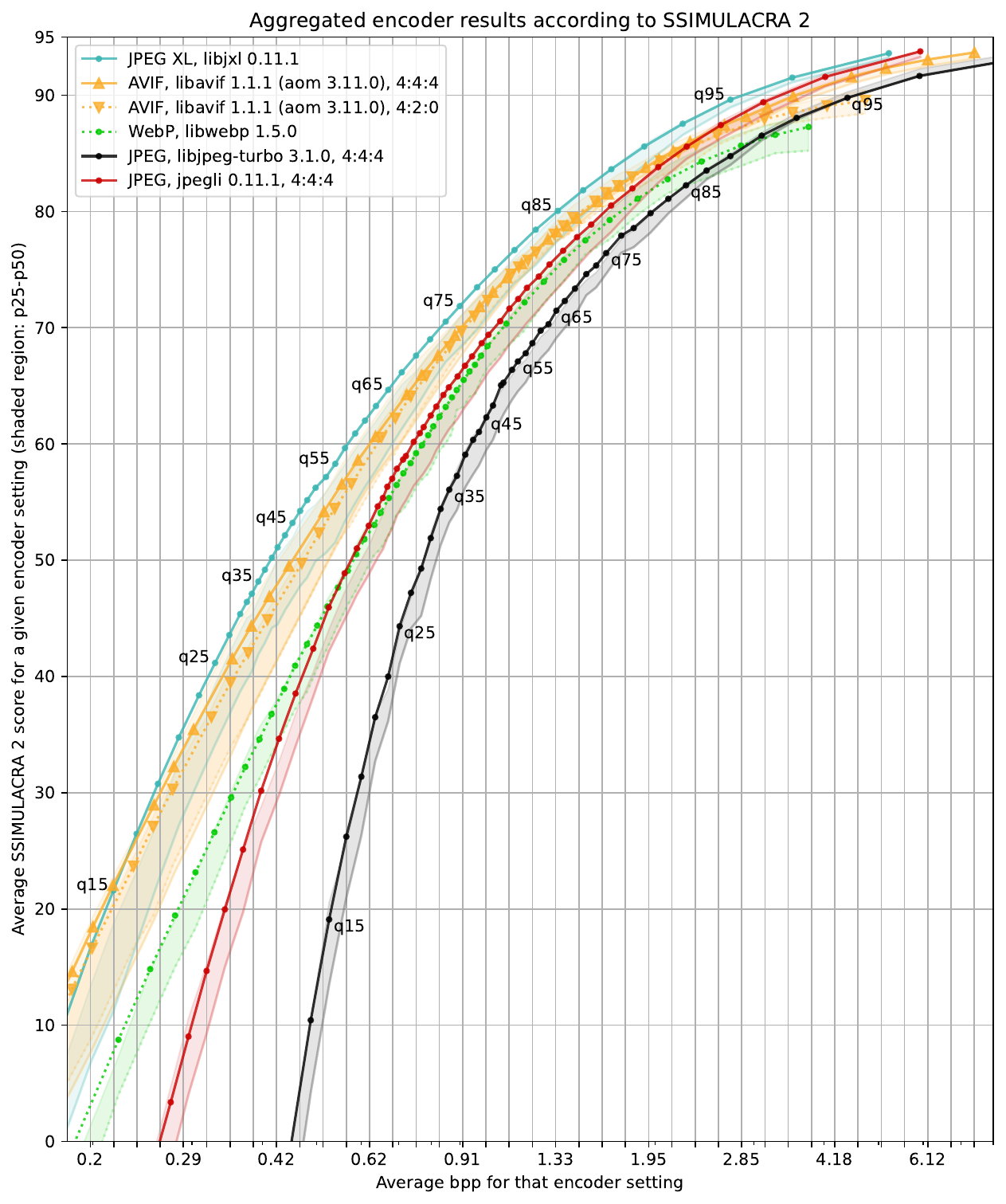}
	\caption[Aggregated SSIMULACRA 2.1 results for SDR images]{Aggregated results for the same set of SDR images as in Figure~\ref{fig:cvvdp_sdr_results}, according to SSIMULACRA 2.1.}
	\label{fig:ssimulacra_sdr_results}
\end{figure}
\subsubsection{Metrics}
Image quality assessment using objective methods (classical algorithmic metrics such as SSIM \cite{ssim}, or AI-based metrics) can be convenient, but the results have to be taken with a grain of salt.
In the end, only subjective quality assessment can provide reliable results that truly reflect the human visual system and human opinions.
While the best metrics achieve a good correlation with subjective results, there is always an `arms race' going on between encoders and metrics: once a metric becomes widely used, encoder developers start tuning encoder choices based on it (or even make encoders that explicitly optimize for it), which makes encoders look more favorable according to the metric but which may over time also undermine the reliability of the metric, i.e. the correlation of a metric with subjective results can deteriorate over time as encoders attempt to `fool the metric'.

The recently proposed quality metric ColorVideoVDP \cite{cvvdp} is based on advanced psychophysical modeling.
It can be used for both SDR and HDR images, and unlike its predecessors HDR-VDP-2 \cite{hdr-vdp-2} and HDR-VDP-3 \cite{hdr-vdp-3}, it takes into account not only luminance but also color.

Figure~\ref{fig:cvvdp_sdr_results} shows results comparing JPEG~XL to AVIF, WebP and JPEG based on a set of 49 SDR images. For AVIF, both 4:4:4 and 4:2:0 encoder configurations are shown, and for JPEG, results for both the libjpeg-turbo and Jpegli \cite{jpegli-study} encoders are shown.
The default configuration of CVVDP was used, corresponding to a 200-nit SDR display at a typical viewing distance:

{\scriptsize
\begin{verbatim}
ColorVideoVDP v0.4.2, 75.4 [pix/deg], Lpeak=200,
Lblack=0.2, Lrefl=0.3979 [cd/m^2], (standard_4k)
\end{verbatim}
}

Both plots in Figure~\ref{fig:cvvdp_sdr_results} show the same data, but the scales used on the axes are different.
The plot on the left shows a commonly used visualization of metric scores, which is aligned with the Bjøntegaard Delta method
\cite{barman2024bjontegaarddeltabdtutorial}, which is very popular especially in video codec assessment.
From the plot on the left, one might be tempted to conclude that AVIF outperforms JPEG~XL across most of the quality range, and both improve upon JPEG by 50\% or more.
However, this way of plotting the data over-emphasizes the low quality range, while making it impossible to compare the codec performance in the higher quality range where the metric scores tend to saturate towards the maximum value.
Figure~\ref{fig:low_quality_example} shows an example of a libjpeg-turbo q19 image, with a CVVDP score around 9. This quality is well below what is typically desired in still-image use cases, yet in the left plot of Figure~\ref{fig:cvvdp_sdr_results}, most of the vertical space is used for qualities around this point or even lower.
In a BD-rate calculation based on these results, the summary result would be mostly based on the low quality region --- say, qualities below the average result of libjpeg-turbo q30 --- while the performance in the higher quality range hardly matters.
This is a problem when using the Bjøntegaard Delta method using any metric that has a saturating behavior, like e.g. SSIM, MS-SSIM, or VMAF.

The plot on the right in Figure~\ref{fig:cvvdp_sdr_results} shows the metric scores in a different way: the vertical axis is logarithmic, plotting CVVDP scores $s$ on a scale proportional to $-\log(10-s)$. That makes it possible to see what happens in the range of qualities that is typically used for still images: the range corresponding to libjpeg-turbo q70-q95.
Looking at this plot, one would arrive at a different conclusion than when looking at the plot on the left: now it looks like for lower qualities, AVIF and JPEG~XL have roughly similar performance, while at higher qualities, JPEG~XL outperforms AVIF by 10-25\%.
This plot also shows not just curves with the average score, but a shaded interval corresponding to the range between the p25 and the median metric scores across the different source images.
This gives some indication of the consistency of the various encoder configurations.

Obviously when doing a codec evaluation using objective metrics, different metrics will lead to different conclusions. For example, Figure~\ref{fig:ssimulacra_sdr_results} shows results according to the SSIMULACRA 2.1 metric, which mostly differ from the CVVDP results in the relative order of WebP and Jpegli compared to libjpeg-turbo.

Nevertheless, most of the perceptual metrics do appear to agree that JPEG~XL performs substantially better than WebP and JPEG across the quality range, and compared to AVIF, JPEG~XL performs similarly or worse in the low-quality region, and better than AVIF starting at medium quality, with a gap that gets larger as the quality increases, from 10-15\% better at `web quality' to 25-40\% better at `camera quality'.

\begin{figure*}\centering
	\includegraphics[width=1\linewidth]{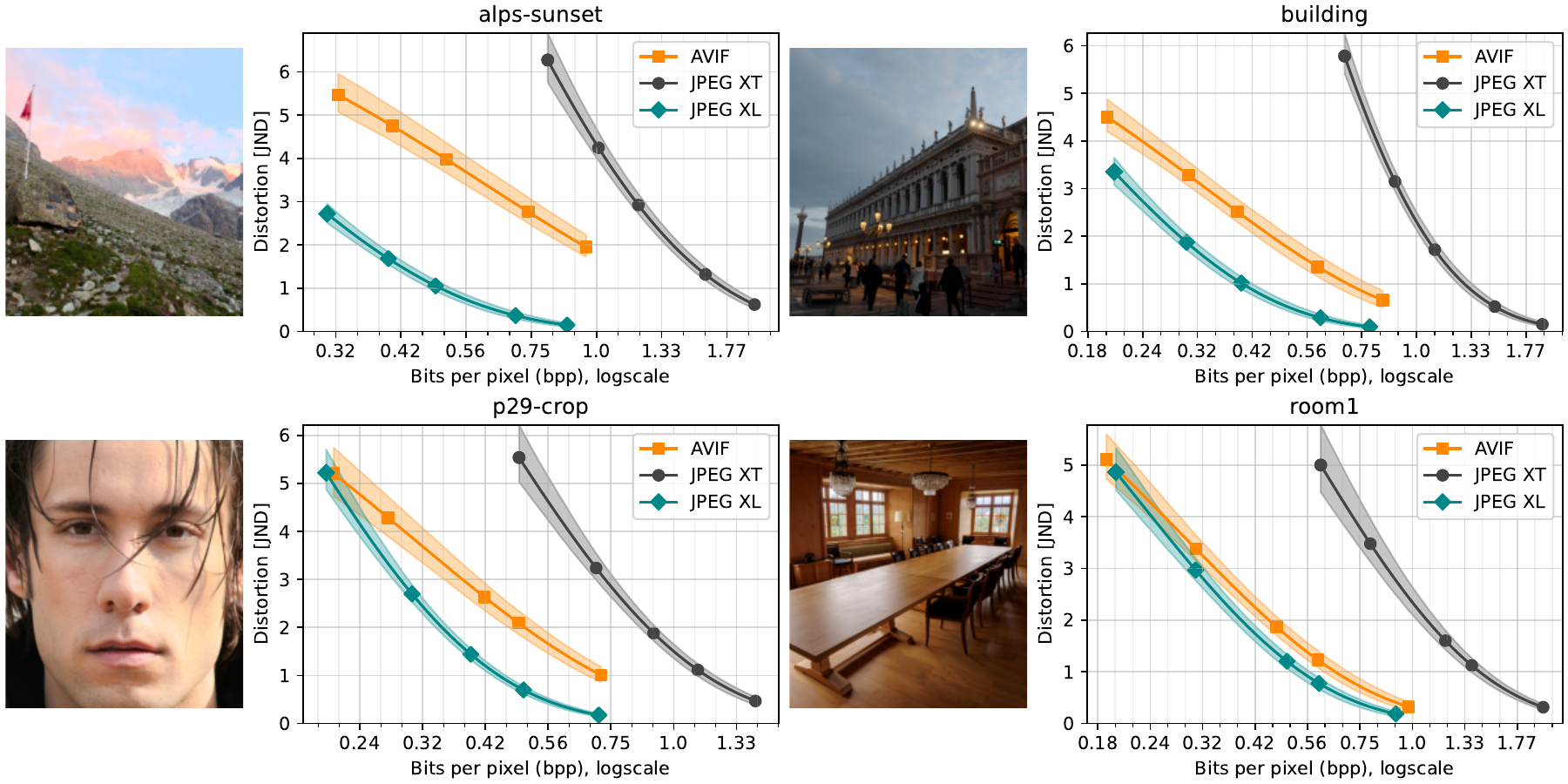}
	\caption[Lossy compression performance on HDR images]{Lossy compression performance of JPEG~XL compared to JPEG XT and AVIF \cite{aic-hdr2025}. The thumbnail images were tone mapped to SDR. The vertical axis indicates distortion in JND units, as estimated using the AIC-3 methodology using boosted and plain triplet comparisons. The shaded regions indicate a 95\% confidence interval. Lower is better.}
	\label{fig:hdr_aic3_results}
\end{figure*}
\begin{figure}
\includegraphics[width=\linewidth]{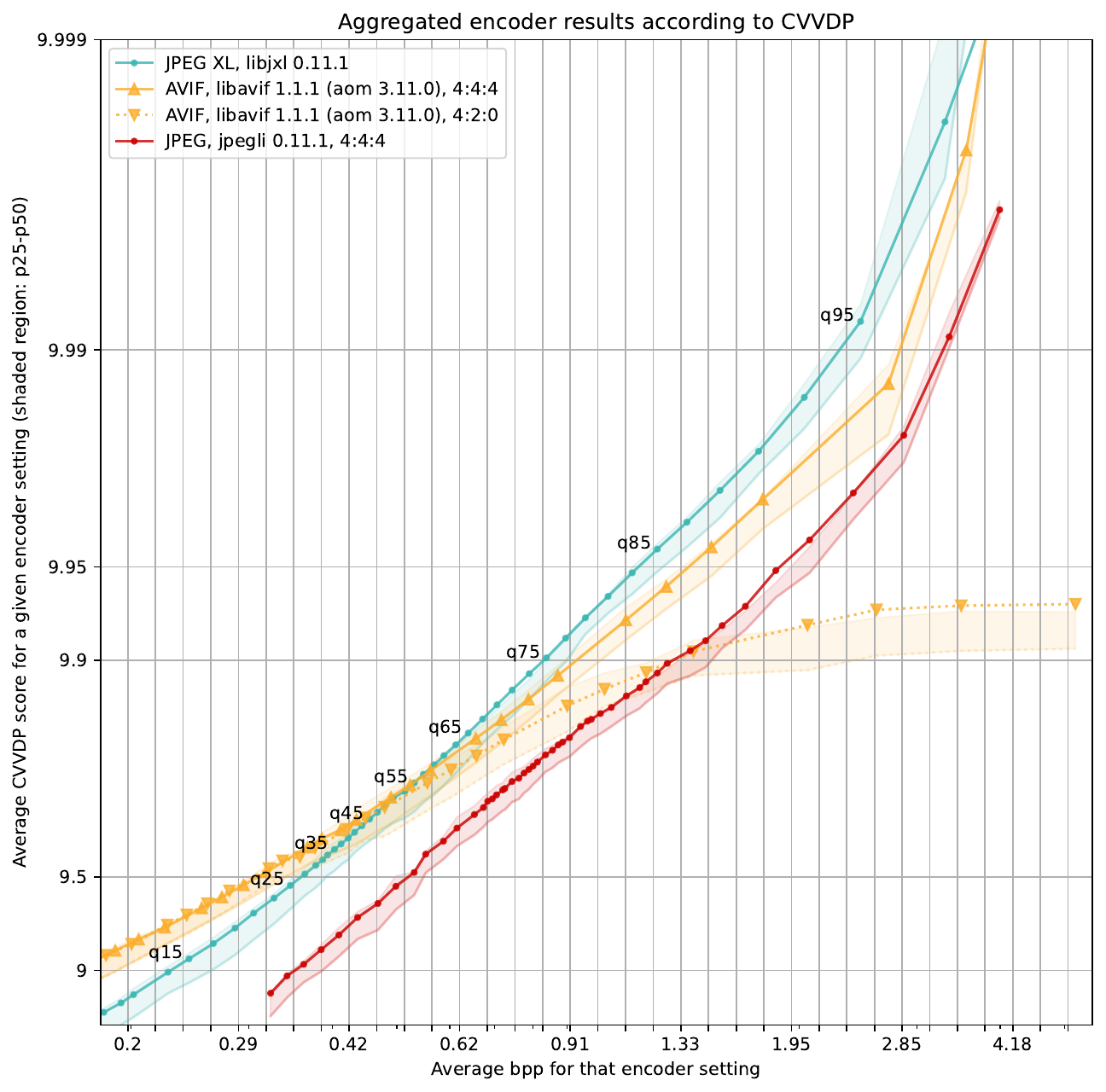}
	\caption[Aggregated CVVDP results for HDR images]{Aggregated results for a set of 24 HDR images (12-bit Rec.2100 PQ, JPEG~XL CfP ClassE) according to CVVDP \cite{cvvdp}.}
	\label{fig:cvvdp_hdr_results}
\end{figure}
\begin{figure}
\includegraphics[width=\linewidth]{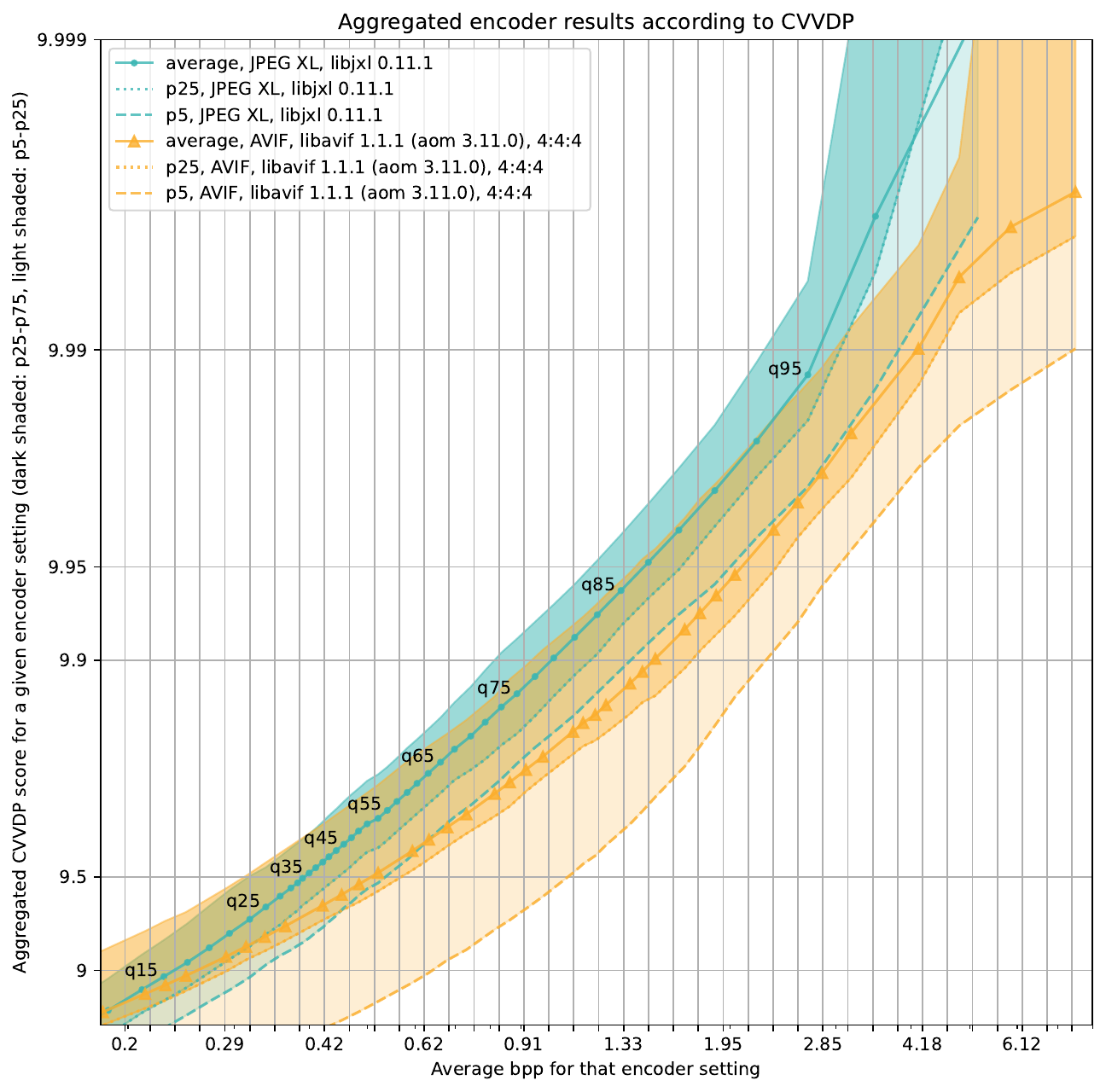}	\caption[Aggregated CVVDP results for SDR and HDR images]{Aggregated results according to CVVDP \cite{cvvdp} for a set of 99 images, of which 24 are the HDR images from Figure~\ref{fig:cvvdp_hdr_results} and the rest are the SDR images from Figure~\ref{fig:cvvdp_sdr_results} plus 26 images from the JPEG~XL CfP ClassA 8-bit set. The dark shaded region corresponds to the quality range per encoder setting of the middle half of the images (p25-p75), while the light shaded region corresponds to 20\% of images towards the ``worst case'' results (p5-p25).}
	\label{fig:cvvdp_hdr_and_sdr_results}
\end{figure}
\subsubsection[HDR]{HDR images}
Figure~\ref{fig:hdr_aic3_results} shows subjective results \cite{aic-hdr2025} for high dynamic range images compressed with JPEG~XL, AVIF and JPEG~XT. At the threshold of visually lossless quality (1 JND), the gap in compression performance between JPEG~XL and the two other codecs is substantial: on several images, JPEG~XL is 70\% smaller than JPEG~XT and around 50\% smaller than AVIF. 
Figure~\ref{fig:cvvdp_hdr_results} shows aggregated objective metric results for a larger set of HDR images.
The standard\_hdr\_pq configuration of CVVDP was used, corresponding to a 1500-nit HDR display at a typical viewing distance:

{\scriptsize
\begin{verbatim}
ColorVideoVDP v0.4.2, 75.4 [pix/deg], Lpeak=1500,
Lblack=0.0015, Lrefl=0.01592 [cd/m^2], (standard_hdr_pq)
\end{verbatim}
}

For HDR images, WebP and the \emph{de facto} JPEG format cannot be used since they are limited to 8-bit.
However, the Jpegli \cite{jpegli-study} implementation of JPEG can still be used. Even though it remains within the bitstream syntax of 8-bit JPEG, it avoids intermediate quantization in the encode and decode process, allowing to make full use of the internal precision of the DCT coefficients in JPEG. By using 16-bit RGB input and output, it can encode HDR images with good fidelity, albeit requiring about twice the bitrate of JPEG~XL.
Other options are to use JPEG~XT (legacy JPEG with additional refinement bits) or JPEG with an embedded gain map according to ISO 21496-1.

According to CVVDP, for HDR images the situation is similar as for SDR images: in the low-quality range (below 0.5 bpp), AVIF performs better, while in the medium to high quality range, JPEG~XL performs about 20\% better than AVIF.
The (limited) subjective data suggest that CVVDP may be somewhat overestimating AVIF quality and underestimating JPEG~XL quality.

\subsubsection{Consistency}
An important aspect of the practical application of codecs is the consistency of the visual quality. It is typically not feasible to manually select an encoder setting on a per-image basis; instead, often a setting is chosen based on the results for a few source images, and then applied throughout the rest of the corpus.
This can lead to unpleasant surprises when the visual quality turns out to be below expectations for some of the images.
In other words, it is not just the bitrate-distortion curve that matters; it is also desirable that the encoder can be configured in a way that produces consistent results.

In \cite{CID22}, the notion of encoder consistency was investigated by considering the standard deviation in mean bias-corrected opinion scores (MCOS) per encoder setting, across a set of 250 source images.
As expected, at the highest quality settings, all encoders consistently produce high-quality images with a score close to that of the original itself and a standard deviation of only 2 on a MCOS scale from 0 to 100. 
However, at lower quality settings, the variation increases in the visual quality obtained.
At intermediate qualities, for JPEG (MozJPEG) and HEIC (x265), the MCOS standard deviation grows to slightly more than 6, indicating a relatively large inconsistency in the visual quality.
For WebP and AVIF (both libaom and aurora), the standard deviation even grows to more than 7.
JPEG~XL (libjxl) has the best consistency, with a standard deviation under 5 at its q70 setting while JPEG, WebP, and AVIF have a standard deviation between 6 and 8 at a similar quality setting.

To some extent, differences in consistency can be explained by choices in the encoder design and implementation: some encoders may be tuned better for perceptual consistency than others.
The codec design itself does play a role too though: e.g. if there are restoration filters available but the bitstream does not allow fine-grained signaling of filtering strengths, it may lead to inconsistent visual quality.

Additionally, when color space information is considered an external, application-level aspect of image coding that is effectively not taken into account by the encoder, it can be expected that results will be less consistent when considering sets of images in different color spaces.
For this reason, the JPEG~XL design philosophy was to consider color space information an essential part of the image data and to use an internal absolute color space (XYB) for lossy encoding.
This makes it possible to maintain perceptual consistency even across images with different color spaces, whether in terms of color gamut or dynamic range.

Figure~\ref{fig:cvvdp_hdr_and_sdr_results} shows aggregated results for a mixed corpus containing both HDR and SDR images. The variation in quality (according to the CVVDP metric) is clearly smaller for JPEG~XL than for AVIF.
This is likely to a large extent caused by the difference in design philosophy between the two codecs. In AVIF the core codestream (AV1) encoder gets a buffer of numerical sample values which it is supposed to compress without knowing their detailed colorimetric interpretation. It will typically get a YCbCr buffer without knowing which RGB space it is relative to. Since color space conversions and signaling are done at the file format level, the encoder is to some extent `blindly' applying lossy compression to numerical sample data.
In JPEG~XL, color transforms are considered not as a file format or application-level external process, but as coding tools within the core codestream. This makes it possible to do more accurate perceptual optimization in an encoder and reach better consistency.

\begin{figure*}\centering
	\includegraphics[width=1\linewidth]{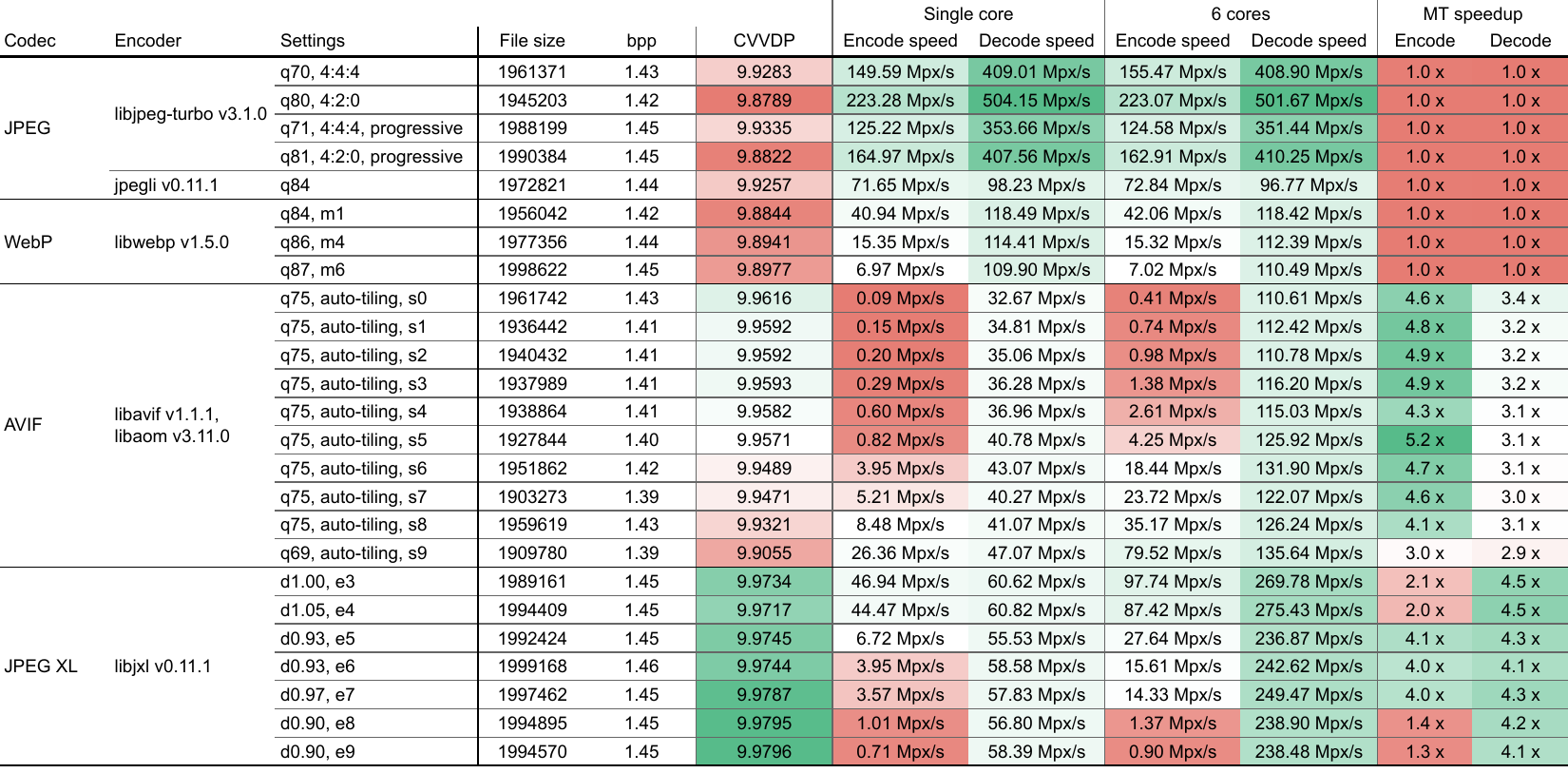}
	\caption[Encode/decode speed measurements]{Encode/decode speed measurements for various encoder settings.}
	\label{fig:lossy_speed}
\end{figure*}
\subsubsection{Speed}
On the same hardware as described in \SectionName{}~\ref{lossless}, the speed of encoding and decoding was tested.
For the purpose of testing, a single 11 megapixel image was used (\verb|p06| from the JPEG~XL CfP test set), and the quality setting was set to the best quality that results in a compressed file size under 2 megabytes (1.45 bpp).

Figure~\ref{fig:lossy_speed} shows the measured speeds for encoding and decoding, expressed in megapixels per second (Mpx/s).
The measurement was done by running the codec multiple times and averaging the measured speed; 10 encodes and 30 decodes were done.
To get an indication of the resulting image quality, the CVVDP metric \cite{cvvdp} with default configuration (standard\_4k) was used.
To understand the parallelization behavior, speeds are reported for both single-threaded (1 core) and multi-threaded processing (6 cores).
The ratio between the single-threaded and multi-threaded speeds is reported in the last two columns. It gives an indication of the speedup from parallelization, where optimal linear speedup would be 6 x.

As expected, slower encode speed settings lead to a higher image quality (at least according to CVVDP) for the same bit rate.
However, the impact of the speed setting on quality is more pronounced in AVIF than it is in WebP or JPEG~XL.
There is also slight impact on decode speed --- higher encode speed leads to higher decode speed --- though the effect is relatively small.

In terms of decode speed, all tested codecs can be considered `fast enough' for most use cases, at least when using multi-threaded decoding.
For example, consider the somewhat extreme use case of an intra-only stream of 4K frames. Then the (multi-threaded) decode speeds in Figure~\ref{fig:lossy_speed} would permit the following maximum frame rates:
for JPEG, 42-60 frames per second (depending on encode settings); for JPEG~XL, around 30 frames per second; for WebP and AVIF, around 15 frames per second.
For more common use cases such as web browsing or photo album viewing, these decode speeds definitely suffice.

\section{Future}
\label{future}
In \SectionName{}~\ref{history}, the history of JPEG~XL and some of the earlier image formats was described.
Obviously, looking back at the past is easier than predicting the future.
In this final section, the current status and recent trends will be sketched, along with some speculation on potential further evolutions related to JPEG~XL.

\subsection{Adoption and application domains}
\label{adoption_applications}
Any standard is useful only to the extent that it gets adopted in practice. For that, firstly software (and devices) dealing with images --- of which there are many! --- have to implement support for the codec.
Secondly, the codec has to actually get deployed in various application domains.

So far, overall prospects for JPEG~XL adoption appear encouraging, but its future on the Web is not yet secured.

\subsubsection[Web]{Web browsers}
From the viewpoint of its proponents, expectations regarding browser adoption have experienced a proverbial roller coaster ride.

Initially, the outlook for adoption of JPEG~XL as a image codec for the Web was very promising:
the Chrome and Firefox browsers added experimental support (behind a flag) for JPEG~XL as early as April 2021.
Engineers at Facebook, Adobe, Intel and various other companies
expressed their enthusiasm for JPEG~XL and asked browsers to enable JPEG~XL support by default. It was looking like universal browser support would be a reality soon.

But then in October 2022, Google Chrome developers decided to remove the experimental support for JPEG~XL, citing lack of ``interest from the entire ecosystem'' and ``not sufficient incremental benefits over existing formats''.
This decision was later additionally motivated by a comparison study performed by the AVIF team \cite{avif_team_comparison}.
While some have raised concerns about the methodology used in these comparison results \cite{contemplating_codec_comparisons}
and Chrome's decision sparked criticism from parts of the community\footnote{See for example the discussion in the Chromium issue tracker following the announcement of the removal of experimental JPEG~XL support: \href{https://issues.chromium.org/issues/40168998\#comment85}
{issues.chromium.org/issues/40168998\#comment85}},
so far it has not been reverted.

The community reaction following Chrome's decision suggests there was still strong interest in JPEG~XL and the desire for widespread adoption had not diminished.
Various Chromium and Firefox forks, such as Pale Moon, Waterfox, and Thorium, decided to enable JPEG~XL support by default.
However, the most popular browsers --- Chrome, Safari, Edge, Firefox --- still lacked support. Edge, being Chromium-based, followed Chrome's decision, while Firefox took a lukewarm `neutral' position. Safari had a reputation of being relatively slow to adopt new web features, so it seemed like Chrome's decision sealed the fate of JPEG~XL, at least as a web codec.

\begin{figure*}\centering
	\includegraphics[width=1\linewidth]{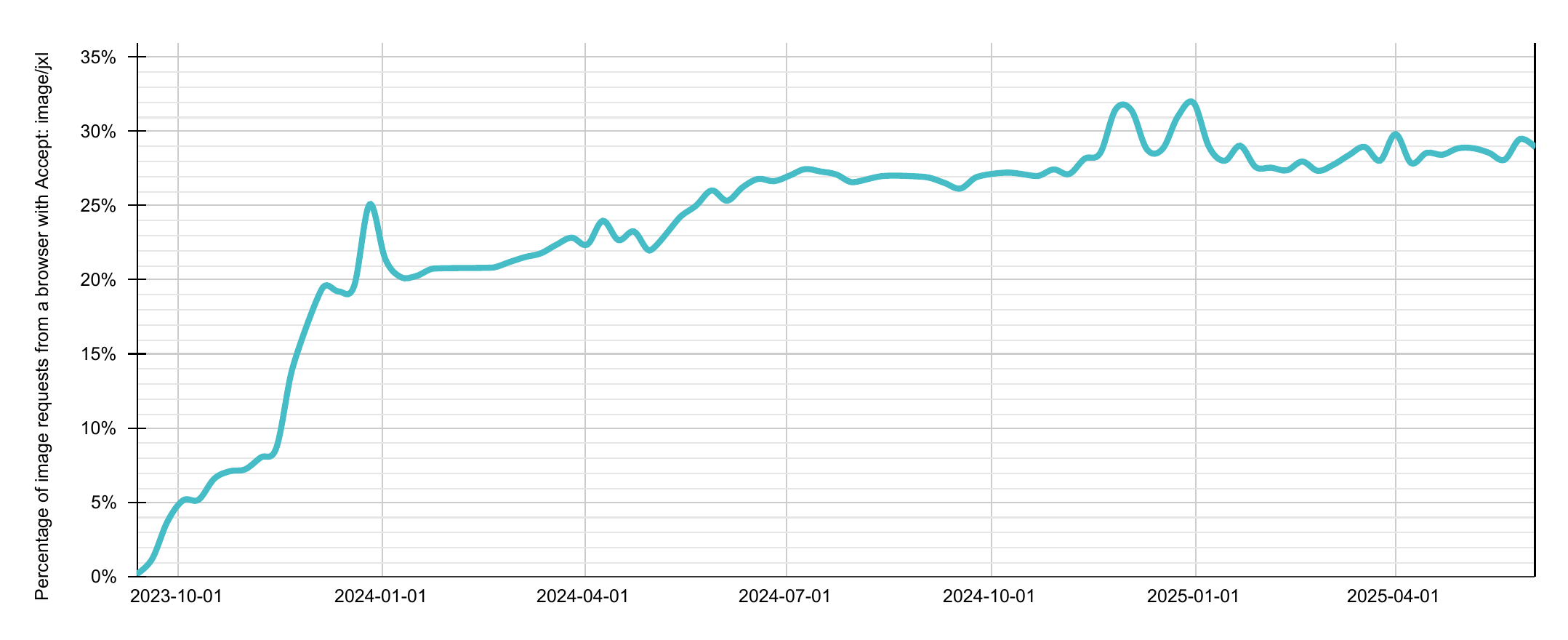}
	\caption[]{Percentage of image requests coming from a browser with JPEG~XL support, as estimated by Cloudinary based on aggregate statistics from approximately 15 billion images served per day.}
	\label{fig:web_support}
\end{figure*}

This changed in September 2023 when Apple unexpectedly announced support for JPEG~XL in Safari.
On iOS devices such as iPhones, the underlying browser rendering engine is always based on Safari --- even when using the iOS version of Chrome, Edge or Firefox.
As shown in Figure~\ref{fig:web_support}, this decision by Apple suddenly changed the percentage of web users with a browser supporting JPEG~XL from close to 0\% to over 20\% in a matter of months, as the iOS and Safari upgrades were rolled out.
Within less than a year, this percentage stabilized to about 27\%, with occasional peaks above 30\% around Thanksgiving and Christmas.

One year later, in September 2024, Firefox adjusted its position to now be ``open to shipping'' JPEG~XL support on the condition that a memory-safe decoder is available that meets their security requirements.
This caused the work to start on a Rust decoder implementation called jxl-rs, which aims to become a high-performance, memory-safe conforming decoder that can be an alternative for the libjxl reference implementation (which is implemented in C++).

\subsubsection[Tools]{Image tools and libraries}
The reference implementation libjxl consists of a software library and command-line encoder/decoder tools. However, usually applications do not directly integrate this library and typical end-users do not use the reference command-line tools. Instead, typically higher-level image libraries are used that offer abstractions for image loading and saving that can deal with multiple codecs.

Various general-purpose image processing libraries including ImageMagick, libvips, Imlib2, FFmpeg, GraphicsMagick, and OpenImageIO (OIIO) added support for JPEG~XL soon after the bitstream syntax was finalized.
Plugins were written for various GUI frameworks such as Qt (KDE), GDK-pixbuf, EFL, and WIC.
Many image viewers based on these libraries or frameworks added support for JPEG~XL.

Open source image editors like GIMP, Krita and darktable also quickly added support.

In October 2022, Adobe Camera Raw 15 was launched with beta support for JPEG~XL.
Soon after, Serif Affinity version 2 was launched with full JPEG~XL support.
In December 2024, Acorn 8.0 was released with JPEG~XL support.

\subsubsection{Operating systems}
Linux distributions were very quick to make libjxl and associated tools and plugins available, in many cases even already in 2021, just after the bitstream was considered frozen and before the standard was published. By adding JPEG~XL support to the backend libraries GdkPixbuf and KImageFormats,
GUI applications in the GNOME and KDE desktop environments can handle JPEG~XL images.
All major Linux distributions currently package JPEG~XL support, including
Arch Linux, Debian, Fedora, Gentoo, openSUSE,
and distributions derived from these such as Manjaro and Ubuntu.
Various independent Linux distributions also maintain libjxl packages, including
ALT Linux, KaOS, NixOS, OpenMandriva, ROSA Linux, Solus, and Void Linux.
Also in other Unix-like operating systems such as FreeBSD and OpenBSD, as well as through
package managers for macOS (Homebrew, MacPorts), Windows (Scoop), and Android (Termux),
libjxl and associated tools can be easily installed.

In September 2023, Apple added JPEG~XL support not just in Safari but also at the OS level across all their operating systems:  macOS 14 (Sonoma), iOS 17, iPadOS 17, tvOS 17, watchOS 10, and visionOS 1.

Microsoft started rolling out official JPEG~XL support for Windows 11 version 24H2 (Hudson Valley)
in March 2025, across all Microsoft platforms: Xbox console, PC, Mobile, Surface Hub, and HoloLens.

\subsubsection[Photography (DNG)]{Digital photography and DNG}
Digital Negative (DNG) is an open format for digital photography.
It can store lossless raw camera data, typically color filter array data such as RGBG in a $2 \times 2$ Bayer pattern.
Another option is to store partly processed image data in what is known as `Linear DNG'; in this case, the data is RGB and can be stored either losslessly or with (very high-fidelity) lossy compression.
Besides the image data itself, a DNG file can store various kinds of metadata, and also a description in the form of `Opcodes' of different processing steps and their parameters, such as lens corrections.

In June 2023, version 1.7 of the DNG specification was released \cite{dng-spec}, which adds JPEG~XL as a payload codec option.
Some popular high-end smartphone models such as the Samsung Galaxy S24 and Apple iPhone 16 Pro added support for DNG 1.7 to their camera apps in 2024.

For a 48~megapixel photo taken on an iPhone Pro, what would typically be a 75~MB ProRAW file (ProRAW is Apple's specific variant of DNG), now thanks to JPEG~XL only requires about 50~MB (losslessly) or about 20~MB (very high fidelity lossy).
Photographers have welcomed these substantial file size reductions \cite{petapixel_iphone16,fstoppers_iphone16pro}.

Whether embedded in DNG or using standard JPEG~XL files, the JPEG~XL codec offers significant advantages for photography.
It can represent wide color gamut and high dynamic range images either losslessly or lossy with sufficient precision for post-production editing.
Development of a hardware encoder is ongoing at the time of writing. This will facilitate deployment of JPEG~XL in digital cameras, where it could very well become the format of choice: it combines the interoperability of JPEG with the fidelity of `raw' formats that is needed for subsequent editing.

\subsubsection[Scanners]{Image scanners}
The TWAIN working group is a not-for-profit organization which represents the imaging industry, in particular manufacturers of scanner devices such as flatbed scanners, sheet-fed scanners, drum scanners, book scanners, and film scanners.
To achieve interoperability between image acquisition devices and software applications, the TWAIN working group promotes the use of the PDF/R format (the R is for Raster). This is a simple, restricted subset of PDF that is used as a scanner output format.
The possibility of adding JPEG~XL to a future PDF (or PDF/R) version is being considered seriously \cite{pdfr}.

In general, JPEG~XL can be applied in various use cases related to image aqcuisition and storage, including digital preservation initiatives, cultural heritage projects, and historical archives.

\subsubsection[DICOM]{Medical Imaging (DICOM)}
DICOM (Digital Imaging and Communications in Medicine) is a widely adopted technical standard for storage and transmission of medical images and associated metadata.
It is used in radiography, ultrasonography, computed tomography (CT), magnetic resonance imaging (MRI), microscopy, and in any field of medicine that involves imaging, including medical and clinical photography.

In September 2024, Supplement 232 (``JPEG~XL Transfer Syntaxes'') was finalized by working group 4 of the DICOM Standards Committee.
The supplement motivation\footnote{DICOM Supplement 232 slideset:\\ \href{https://www.dicomstandard.org/news-dir/progress/docs/sups/sup232-slides.pdf}{dicomstandard.org/news-dir/progress/docs/sups/sup232-slides.pdf}}
included digital pathology workflows where significant compression gains (50-60\%) are reported in whole-slide imaging (WSI),
as well as lossless JPEG recompression which can save costs in existing archives, and DICOMweb where JPEG~XL can replace JPEG, PNG, and GIF.

The supplement was incorporated in the 2024d release of the DICOM standard \cite{dicom}.

\subsubsection[GIS]{Geographic Information Systems (GIS)}
Geospatial data such as geographic information system (GIS) is another application domain where JPEG~XL can bring substantial improvements.
Satellite imagery and aerial photography and their applications in meteorology, oceanography, geography, geology, biodiversity monitoring, etc. require high-resolution and often multi-spectral images.

The Geospatial Data Abstraction Library (GDAL \cite{gdal}), released by the Open Source Geospatial Foundation, is a popular translator library for raster and vector geospatial data formats such as GeoTIFF.
In November 2022, GDAL~3.6 was released, which added support for JPEG~XL.

\subsubsection[more]{Other application domains}
Spectral rendering algorithms require spectral images with a large number of channels. Using a JPEG~XL-based format, file sizes 10 to 60 times smaller than spectral OpenEXR can be achieved \cite{Fichet2025Spectral}.

From digital art or generative AI to legal or insurance evidence,
from observational astronomy to computer-generated imagery (CGI),
the number of application domains for lossless and/or lossy image formats is too large to describe exhaustively.
In many of these domains, JPEG~XL is currently starting to get adopted.

The next chapters of JPEG~XL's history still have to be written, and only time will tell to what extent it will be successful in gaining widespread adoption, let alone in lasting as long as the 1992 JPEG format.
To a large extent, the future success of JPEG~XL will depend on whether or not it achieves universal Web browser support. This is a major precondition for becoming a format that is as ubiquitous and interoperable as JPEG or PNG.
Meanwhile, at least in some specific application domains, JPEG~XL has already become an established image codec.

\subsection{Revisions and extensions}
After the first edition of ISO/IEC 18181-1 was published in March 2022,
work immediately started on a second edition.
No technical changes or additions were made in this second edition:
the main goal was to improve the document structure and to make a large number of technical corrections and clarifications.
It was primarily based on the feedback from independent implementers who discovered minor discrepancies between the reference implementation and the specification, as well as unclear or incomplete descriptions in the first edition of the standard.
For both Part 1 and Part 2, second editions of the standard were published in 2024. The updates to Part 2 (file format) were largely just to synchronize cross-references to Part 1, though also a new Annex was added with the IANA media type registration information for \verb|image/jxl|.

A third edition of ISO/IEC 18181-2 is currently in development, which will include minor technical corrections related to JPEG bitstream reconstruction, as well as introducing a new optional data box in the file format to store ISO 21496‐1 gain maps.

\paragraph{Encoder improvements.}
The JPEG~XL standard mainly specifies the bitstream syntax and the result of a conforming decoder. The encoding process is not specified; any encoder that produces a bitstream conforming to the specification is valid.
The bitstream syntax is very expressive so there are many degrees of freedom from the encoder point of view, both for lossless and lossy compression.
This means there is substantial room for future encoder improvements, in terms of speed, compression density, quality, or the trade-offs between them.
Some coding tools such as Splines and the largest VarDCT block sizes are not yet exploited by the current version of libjxl. Other coding tools such as Patches, noise synthesis, Delta Palette, and frame blending modes are only used in a limited way, or are only exposed through custom options.
To generate the MA trees in Modular mode, currently a greedy search algorithm is used to explore a small subset of the search space; in general, this does not produce optimal solutions.
The block segmentation and block type selection for VarDCT is driven by heuristics that may have room for improvement \cite{cho2021improvement}
.

It could make sense to develop an AI-based encoder that can make better use of the available coding tools than what can be achieved with classical algorithms and heuristics.
The advantage of such an approach compared to a fully AI-based codec, would be that the decode process is still entirely classical, does not require specific hardware, and is already standardized.

\paragraph{New profiles.}
Currently JPEG~XL has only one profile, called the `Main profile'.
Two levels are defined: Level 5, which suffices for most applications including HDR photography and Web images, and Level 10, which is effectively unlimited.
The main reason to avoid having multiple profiles is to avoid encouraging the deployment of decoders that only partially implement the standard, enough to conform to a `simple' profile. That would lead to a similar situation as what happened to JPEG: the \emph{de facto} codec would only consist of the lowest common denominator, making the other coding tools effectively impossible to use if interoperability is desired.
Currently, all deployed implementations are software implementations, and there is no real reason for a software implementation to be incomplete.

In the future, it seems likely that there will also be hardware implementations that can e.g. help improve the battery life of cameras.
The main focus for hardware implementations will likely be encoding, though it may be desirable to also have hardware decoding that can decode the JPEG~XL files produced by the hardware encoder.
For this purpose, at some point a new `Hardware profile' may be defined that would be a subset of the Main profile, with new Levels that would be a subset of Level 5.

\paragraph{Extensions.}
In the ImageMetadata, FrameHeader, and RestorationFilter headers within the codestream, there is a mechanism to create future extensions to the codestream syntax in a backwards-compatible way.
While currently no such extensions are defined, the intention is that if such extensions will be defined, it will still be possible for existing decoders to produce a decoded image, even if they do not know how to interpret the information provided in the extension fields.
This leaves open the possibility to define new standardized types of extra channels (with associated specific metadata), or to embed auxiliary information that can help to further speed up decoding.
While in the current standard the bit depth is limited to 32-bit, the header signaling leaves room for a future extension to a precision of up to 64-bit, should the need arise.

At the file format level, the standard specifies that any boxes with an unrecognized type are to be ignored and skipped by the decoder.
This allows for application-specific additional metadata boxes.
It also leaves room for future extensions to the file format standard while ensuring backwards compatibility.

In general, the JPEG~XL codestream and file format were designed to be as future-proof as possible. At least from the technical perspective, this facilitates a long-term solution for the creation, interchange, and preservation of our digital heritage in terms of still images.

\phantomsection
\section*{Acknowledgments} 
\addcontentsline{toc}{section}{Acknowledgments}
The JPEG~XL project would not have been possible without the JPEG committee (ISO/IEC JTC 1/SC 29/WG 1), including (but not limited to)
the Convener Touradj Ebrahimi;
the Requirements subgroup chair Fernando Pereira
and co-chairs Walt Husak, Takaaki Ishikawa, and Eduardo da Silva;
the Coding, test and quality (CTQ) subgroup chair Peter Schelkens
and co-chair Dale Stolitzka,
later renamed to the Image coding \& quality (ICQ) subgroup with chairs
Thomas Richter and Osamu Watanabe;
and the Systems and integration subgroup chairs and co-chairs
Andy Kuzma, Siegfried Foessel, and Frederik Temmermans.

It would also not have been possible without the participation of the various national bodies of ISO who provided valuable comments at the different stages of the standardization process.

The `Ad Hoc Group on next generation image compression standard' (later renamed to `Ad Hoc Group on JPEG~XL') was established for the first time at the 74th WG1 Meeting that took place in January 2017 in Geneva, Switzerland.
It was chaired by Jan De Cock, David Taubman, and Seungcheol Choi, and produced the Call for Proposals that would launch the project.
At the 82nd WG1 Meeting (Lisbon, Portugal, January 2019) the collaborative phase of the standard development took off, with Jan Wassenberg and Jon Sneyers as chairs of the Ad Hoc Group on JPEG~XL.
At the 89th WG1 Meeting (online, October 2020), the AHG chairs changed to Jon Sneyers and Luca Versari.
The AHG chairs changed again at the 94th WG1 Meeting (online, January 2022) to Jon Sneyers and Jyrki  Alakuijala,
and at the 98th WG1 Meeting (Sydney, Australia, January 2023) to Jon Sneyers and Moritz Firsching.
The AHG on JPEG~XL would be re-established for the last time at the 104th WG1 Meeting (Sapporo, Japan, July 2024).
From the 105th WG1 Meeting (Berlin, Germany, October 2024) onward, it was merged into a broader `Maintenance' Ad Hoc Group.

In terms of the technical development of the core codestream design of JPEG~XL, the main contributors were Jyrki Alakuijala, Jon Sneyers, and Luca Versari. The main focus of Jyrki was VarDCT mode and image quality; the main focus of Jon was Modular mode; the main focus of Luca was the harmonization and efficient implementation of the various coding tools (both encoding and decoding). 
Zoltán Szabadka contributed to various aspects including lossless JPEG recompression.
Jan Wassenberg focused on parallelism (SIMD, multi-threading) and fast decoding.
Lode Vandevenne played a key role in the libjxl library API design.
Sami Boukortt contributed to various aspects including color space signaling (in particular HDR) and Splines.
Alexander Rhatushnyak created the self-correcting predictor.
Alex Deymo, Evgenii  Kliuchnikov, and Moritz Firsching
were key contributors to the libjxl reference software.
Other contributors to aspects of the standard or the reference software include Robert Obryk, Thomas Fischbacher, Iulia-Maria Comșa, 
Krzysztof Potempa, Martin Bruse, Renata Khasanova, 
Ruud van Asseldonk, and Sebastian Gomez.

It is impossible to exhaustively list all the people that were involved in the JPEG~XL project in one way or another, within the JPEG committee or outside.
For example, several research labs played an important supporting role by performing subjective image quality assessment experiments that were instrumental in evaluating codec design decisions.
Also many contributions to libjxl from the broader open source community were very valuable, as well as feedback from a broad community of image compression enthusiasts and independent implementers.


\phantomsection
\bibliographystyle{plain}

\bibliography{refs.bib}

\begin{thebibliography}{10}

\bibitem{dng-spec}
Adobe.
\newblock {Digital Negative (DNG)} specification version 1.7.1.0, September 2023.
\newblock {\small \url{https://helpx.adobe.com/camera-raw/digital-negative.html}}.

\bibitem{akyazi2019assessment}
Pinar Akyazi and Touradj Ebrahimi.
\newblock Assessment of quality of {JPEG~XL} proposals based on subjective methodologies and objective metrics.
\newblock In {\em Applications of Digital Image Processing XLII}, volume 11137, pages 147--163. SPIE, 2019.

\bibitem{brotli}
{IETF RFC 7932}.~Jyrki Alakuijala and Zoltán Szabadka.
\newblock Brotli compressed data format, 2016.

\bibitem{guetzli-study}
Jyrki Alakuijala, Robert Obryk, Zoltán Szabadka, and Jan Wassenberg.
\newblock {Users prefer Guetzli JPEG over same-sized libjpeg}.
\newblock {\em arXiv:1703.04416}, 2017.

\bibitem{jpegxl2019}
Jyrki Alakuijala, Ruud Van~Asseldonk, Sami Boukortt, Martin Bruse, Iulia-Maria Comșa, Moritz Firsching, Thomas Fischbacher, Evgenii Kliuchnikov, Sebastian Gomez, Robert Obryk, et~al.
\newblock {JPEG XL} next-generation image compression architecture and coding tools.
\newblock In {\em Applications of digital image processing XLII}, volume 11137, pages 112--124. SPIE, 2019.

\bibitem{fstoppers_iphone16pro}
Alex Armitage.
\newblock Apple's secret upgrade to the {iPhone~16~Pro} yields incredible results, October 2024.
\newblock \href{https://fstoppers.com/landscapes/apples-secret-upgrade-iphone-16-pro-yields-incredible-results-682936}{Fstoppers website}.

\bibitem{avif_team_comparison}
{AVIF team}.
\newblock Image coding comparisons, December 2022.
\newblock {\small \url{https://storage.googleapis.com/avif-comparison/index.html}}.

\bibitem{barman2024bjontegaarddeltabdtutorial}
Nabajeet Barman, Maria~G. Martini, and Yuriy Reznik.
\newblock {Bj{\o}ntegaard Delta (BD)}: A tutorial overview of the metric, evolution, challenges, and recommendations.
\newblock {\em arXiv:2401.04039}, 2024.

\bibitem{bennett2023benchmarking}
Michael~J Bennett.
\newblock Benchmarking lossless still image codecs: Perspectives on selected compression standards from 1992 through 2022.
\newblock In {\em The Society for Imaging Science and Technology Archiving 2023 Final Program and Proceedings}, pages 165--171, 2023.

\bibitem{MTF}
Jon~Louis Bentley, Daniel~D. Sleator, Robert~E. Tarjan, and Victor~K. Wei.
\newblock A locally adaptive data compression scheme.
\newblock {\em Commun. ACM}, 29(4):320–330, 1986.

\bibitem{jpegli-study}
Martin Bruse, Luca Versari, Zoltán Szabadka, and Jyrki Alakuijala.
\newblock {Users prefer Jpegli over same-sized libjpeg-turbo or MozJPEG}.
\newblock {\em arXiv:2403.18589}, 2024.

\bibitem{exif}
{Camera \& Imaging Products Association}.
\newblock {CIPA DC-008-Translation-2024} --- exchangeable image file format for digital still cameras : Exif version 3.0, 2024.

\bibitem{splines}
Edwin Catmull and Raphael Rom.
\newblock A class of local interpolating splines.
\newblock In Robert~E. Barnhill and Richard~F. Riesenfeld, editors, {\em Computer Aided Geometric Design}, pages 317--326. Academic Press, 1974.

\bibitem{av1}
Yue Chen, Debargha Mukherjee, Jingning Han, Adrian Grange, Yaowu Xu, Sarah Parker, Cheng Chen, Hui Su, Urvang Joshi, et~al.
\newblock An overview of coding tools in {AV1}: the first video codec from the {Alliance for Open Media}.
\newblock {\em APSIPA Transactions on Signal and Information Processing}, 9, 2020.

\bibitem{cho2021improvement}
Joonhyung Cho, Oh-Jin Kwon, and Seungcheol Choi.
\newblock Improvement of {JPEG XL} lossy image coding using region adaptive {DCT} block partitioning structure.
\newblock {\em IEEE Access}, 9:113213--113225, 2021.

\bibitem{ans}
Jarek Duda.
\newblock Asymmetric numeral systems: entropy coding combining speed of {Huffman} coding with compression rate of arithmetic coding.
\newblock {\em arXiv:1311.2540}, 2014.

\bibitem{jpegxr}
Fr{\'e}d{\'e}ric Dufaux, Gary~J Sullivan, and Touradj Ebrahimi.
\newblock The {JPEG XR} image coding standard [standards in a nutshell].
\newblock {\em IEEE Signal Processing Magazine}, 26(6):195--204, 2009.

\bibitem{why_is_color}
Mark~D. Fairchild.
\newblock {\em The Color Curiosity Shop}.
\newblock Rochester Institute of Technology, 2011.

\bibitem{cam}
Mark~D. Fairchild.
\newblock {\em Color Appearance Models}.
\newblock Wiley, 3rd edition, 2013.

\bibitem{Fichet2025Spectral}
Alban Fichet and Christoph Peters.
\newblock Compression of spectral images using spectral {JPEG~XL}.
\newblock {\em Journal of Computer Graphics Techniques (JCGT)}, 14(1):49--69, March 2025.

\bibitem{gdal}
{GDAL/OGR contributors}.
\newblock {\em {GDAL/OGR} Geospatial Data Abstraction software Library}.
\newblock Open Source Geospatial Foundation, 2025.
\newblock {\small \url{https://gdal.org}}.

\bibitem{webp}
Giaime Ginesu, Maurizio Pintus, and Daniele~D Giusto.
\newblock Objective assessment of the {WebP} image coding algorithm.
\newblock {\em Signal processing: image communication}, 27(8):867--874, 2012.

\bibitem{petapixel_iphone16}
Jeremy Gray.
\newblock Why {Apple} uses {JPEG~XL} in the {iPhone~16} and what it means for your photos, September 2024.
\newblock \href{https://petapixel.com/2024/09/18/why-apple-uses-jpeg-xl-in-the-iphone-16-and-what-it-means-for-your-photos/}{PetaPixel website}.

\bibitem{heif}
Miska~M. Hannuksela, Jani Lainema, and Vinod~K. Malamal~Vadakital.
\newblock The {High Efficiency Image File Format} standard {[Standards in a Nutshell]}.
\newblock {\em IEEE Signal Processing Magazine}, 32(4):150--156, 2015.

\bibitem{huffman}
David~A Huffman.
\newblock A method for the construction of minimum-redundancy codes.
\newblock {\em Proceedings of the Institute of Radio Engineers}, 40(9):1098--1101, 1952.

\bibitem{LMS-HPE}
Robert~WG Hunt and MR~Pointer.
\newblock A colour-appearance transform for the {CIE} 1931 standard colorimetric observer.
\newblock {\em Color Research \& Application}, 10(3):165--179, 1985.

\bibitem{tiffspec}
{ISO 12639}.
\newblock {Graphic technology — Prepress digital data exchange — Tag image file format for image technology (TIFF/IT)}.

\bibitem{xmp}
{ISO 16684-1}.
\newblock Graphic technology — extensible metadata platform ({XMP}).

\bibitem{18181-1}
{ISO/IEC 18181-1}.
\newblock {Information technology — JPEG~XL image coding system — Part 1: Core coding system}.

\bibitem{18181-2}
{ISO/IEC 18181-2}.
\newblock {Information technology — JPEG~XL image coding system — Part 2: File format}.

\bibitem{jumbf}
{ISO/IEC 19566-5}.
\newblock Information technology — {JPEG} systems — part 5: {JPEG} universal metadata box format ({JUMBF}).

\bibitem{aic3}
{ISO/IEC DIS 29170-3}.
\newblock {Information technology — Assessment of image coding — Part 3: Subjective quality assessment of high-fidelity images}.

\bibitem{jxl-white-paper}
{ISO/IEC JTC 1 / SC29 / WG1 N100400}.
\newblock {JPEG} white paper: {JPEG~XL} image coding system.
\newblock 98th {JPEG} meeting, Sydney, Australia, January 2023.
\newblock {\small \url{https://ds.jpeg.org/whitepapers/jpeg-xl-whitepaper.pdf}}.

\bibitem{jxl-use-cases}
{ISO/IEC JTC1 / SC29 / WG1 N83043}.
\newblock {JPEG~XL} use cases and requirements.
\newblock 83rd {JPEG} Meeting, Geneva, Switzerland, March 2019.
\newblock {\small \url{https://ds.jpeg.org/documents/wg1n83043-REQ-JPEG_XL_Use_Cases_and_Requirements.pdf}}.

\bibitem{h273}
{ITU-T recommendation H.273}.
\newblock Coding-independent code points for video signal type identification.

\bibitem{aic-hdr2025}
Mohsen Jenadeleh, Jon Sneyers, Davi~N. Lazzarotto, Shima Mohammad, Dominik Keller, Atanas Boev, António Pinheiro, Thomas Richter, Alexander Raake, Touradj Ebrahimi, João Ascenso, and Dietmar Saupe.
\newblock Fine-grained {HDR} image quality assessment from noticeably distorted to very high fidelity.
\newblock In {\em 17th International Conference on Quality of Multimedia Experience (QoMEX)}, 2025.

\bibitem{lehmercode}
C.-A. Laisant.
\newblock Sur la num\'eration factorielle, application aux permutations.
\newblock {\em Bulletin de la Soci\'et\'e Math\'ematique de France}, 16:176--183, 1888.

\bibitem{mozjpeg}
Kornel Lesiński.
\newblock {MozJPEG 3.0}.
\newblock Web Performance Calendar, December 28th, 2014.
\newblock {\small \url{https://calendar.perfplanet.com/2014/mozjpeg-3-0/}}.

\bibitem{CIEDE2000}
Ming Luo, Guihua Cui, and B.~Rigg.
\newblock The development of the {CIE} 2000 colour‐difference formula: {CIEDE2000}.
\newblock {\em Color Research \& Application}, 26:340 -- 350, 10 2001.

\bibitem{amiga}
J.~Maher.
\newblock {\em The Future Was Here: The {Commodore Amiga}}.
\newblock Platform Studies. MIT Press, 2018.

\bibitem{mandeel2021comparative}
Thulfiqar~H Mandeel, Muhammad~Imran Ahmad, Noor Aldeen~A Khalid, and Mohd Nazrin~Md Isa.
\newblock A comparative study on lossless compression mode in {WebP}, better portable graphics ({BPG}), and {JPEG~XL} image compression algorithms.
\newblock In {\em 8th Intl. Conf. on Comp. and Commun. Eng. (ICCCE)}. IEEE, 2021.

\bibitem{hdr-vdp-3}
Rafa{\l} Mantiuk, Dounia Hammou, and Param Hanji.
\newblock {HDR-VDP-3}: A multi-metric for predicting image differences, quality and contrast distortions in high dynamic range and regular content.
\newblock {\em arXiv:2304.13625}, 2023.

\bibitem{cvvdp}
Rafa{\l} Mantiuk, Param Hanji, Maliha Ashraf, Yuta Asano, and Alexandre Chapiro.
\newblock {ColorVideoVDP}: A visual difference predictor for image, video and display distortions.
\newblock {\em ACM Trans. Graph.}, 43(4), July 2024.

\bibitem{hdr-vdp-2}
Rafa{\l} Mantiuk, Kil~Joong Kim, Allan~G Rempel, and Wolfgang Heidrich.
\newblock {HDR-VDP-2}: A calibrated visual metric for visibility and quality predictions in all luminance conditions.
\newblock {\em ACM Transactions on graphics (TOG)}, 30(4):1--14, 2011.

\bibitem{j2k-overview}
Michael~W. Marcellin, Michael~J. Gormish, Ali Bilgin, and Martin~P. Boliek.
\newblock An overview of {JPEG-2000}.
\newblock In {\em Proceedings {DCC} 2000. Data compression conference}, pages 523--541. IEEE, 2000.

\bibitem{men2021subjective}
Hui Men, Hanhe Lin, Mohsen Jenadeleh, and Dietmar Saupe.
\newblock Subjective image quality assessment with boosted triplet comparisons.
\newblock {\em IEEE Access}, 9:138939--138975, 2021.

\bibitem{gif_etc}
John Miano.
\newblock {\em Compressed image file formats: {JPEG}, {PNG}, {GIF}, {XBM}, {BMP}}.
\newblock Addison-Wesley, 1999.

\bibitem{dicom}
{NEMA PS3 / ISO 12052}.
\newblock {\em {Digital Imaging and Communications in Medicine (DICOM)} Standard}.
\newblock National Electrical Manufacturers Association, Rosslyn, VA, USA, 2024.
\newblock {\small \url{http://www.dicomstandard.org/}}.

\bibitem{ffv1}
{IETF RFC 9043}.~Michael Niedermayer, Dave Rice, and Jérôme Martinez.
\newblock {FFV1} video coding format versions 0, 1, and 3, 2021.

\bibitem{tiff}
Charles~A Poynton.
\newblock Overview of {TIFF} 5.0.
\newblock In {\em Image Processing and Interchange: Implementation and Systems}, volume 1659, pages 152--158. SPIE, 1992.

\bibitem{png}
Greg Roelofs.
\newblock {\em {PNG}: the definitive guide}.
\newblock O'Reilly, 1999.

\bibitem{j2k-suite}
Peter Schelkens, Athanassios Skodras, and Touradj Ebrahimi, editors.
\newblock {\em The {JPEG 2000} suite}.
\newblock Wiley, 2009.

\bibitem{contemplating_codec_comparisons}
Jon Sneyers.
\newblock Contemplating codec comparisons, December 2022.
\newblock {\small \url{https://cloudinary.com/blog/contemplating-codec-comparisons}}.

\bibitem{CID22}
Jon Sneyers, Elad Ben~Baruch, and Yaron Vaxman.
\newblock {CID22}: Large-scale subjective quality assessment for high fidelity image compression.
\newblock {\em TechRxiv}, 2023.

\bibitem{flif}
Jon Sneyers and Pieter Wuille.
\newblock {FLIF}: Free lossless image format based on {MANIAC} compression.
\newblock In {\em {2016 IEEE international conference on image processing (ICIP)}}, pages 66--70. IEEE, 2016.

\bibitem{packjpg}
Matthias Stirner and Gerhard Seelmann.
\newblock Improved redundancy reduction for {JPEG} files.
\newblock In {\em {Proc. of Picture Coding Symposium (PCS 2007)}}, pages 7--9, 2007.

\bibitem{Stockman-Sharpe}
Andrew Stockman and Lindsay~T. Sharpe.
\newblock The spectral sensitivities of the middle- and long-wavelength-sensitive cones derived from measurements in observers of known genotype.
\newblock {\em Vision Research}, 40(13):1711--1737, 2000.

\bibitem{hevc}
Vivienne Sze, Madhukar Budagavi, and Gary~J Sullivan.
\newblock High efficiency video coding ({HEVC}).
\newblock {\em Integrated circuit and systems, algorithms and architectures}, 39:40, 2014.

\bibitem{jpegtrust}
Frederik Temmermans, Sabrina Caldwell, Deepayan Bhowmik, and Touradj Ebrahimi.
\newblock {JPEG Trust: an international standard facilitating the assessment of trustworthiness of digital media assets}.
\newblock In {\em Applications of Digital Image Processing XLVII}, volume 13137, page 131370B. International Society for Optics and Photonics, SPIE, 2024.

\bibitem{jumbf_trust}
Frederik Temmermans and Leonard Rosenthol.
\newblock Adopting the {JPEG} universal metadata box format for media authenticity annotations.
\newblock In {\em Applications of Digital Image Processing XLIV}, volume 11842, pages 165--170. SPIE, 2021.

\bibitem{dcc_aic3}
Michela Testolina, Mohsen Jenadeleh, Shima Mohammadi, Shaolin Su, Joao Ascenso, Touradj Ebrahimi, Jon Sneyers, and Dietmar Saupe.
\newblock Fine-grained subjective visual quality assessment for high-fidelity compressed images.
\newblock In {\em Data Compression Conference (DCC)}. IEEE, 2025.

\bibitem{testolina2023assessment}
Michela Testolina, Evgeniy Upenik, and Touradj Ebrahimi.
\newblock {On the assessment of high-quality images: advances on the JPEG AIC-3 activity}.
\newblock In {\em Applications of Digital Image Processing XLVI}, volume 12674, pages 180--190. SPIE, 2023.

\bibitem{av1cfl}
Luc~N. Trudeau, Nathan~E. Egge, and David Barr.
\newblock Predicting chroma from luma in {AV1}.
\newblock {\em arXiv:1711.03951}, 2018.

\bibitem{pdfr}
{TWAIN Working Group}.
\newblock White paper: The benefits of adding {JPEG-XL} to the {ISO PDF} standard and {PDF-Raster}, 2025.
\newblock {\small \url{https://twain.org/}}.

\bibitem{luca_phdthesis}
Luca Versari.
\newblock {\em Compression Techniques for Large Graphs: Theory and Practice}.
\newblock PhD thesis, Università di Pisa, 2021.

\bibitem{css}
{W3C Recommendation 07 June 2011}.
\newblock Cascading style sheets level 2 revision 1 ({CSS} 2.1) specification.
\newblock {\small \url{http://www.w3.org/TR/2011/REC-CSS2-20110607}}.

\bibitem{jpeg}
Gregory~K. Wallace.
\newblock The {JPEG} still picture compression standard.
\newblock {\em Commun. ACM}, 34(4):30--44, 1991.

\bibitem{ssim}
Zhou Wang, A.C. Bovik, H.R. Sheikh, and E.P. Simoncelli.
\newblock Image quality assessment: from error visibility to structural similarity.
\newblock {\em IEEE Transactions on Image Processing}, 13(4):600--612, 2004.

\bibitem{jpegls}
Marcelo~J Weinberger, Gadiel Seroussi, and Guillermo Sapiro.
\newblock The {LOCO-I} lossless image compression algorithm: Principles and standardization into {JPEG-LS}.
\newblock {\em IEEE Transactions on Image Processing}, 9(8):1309--1324, 2000.

\bibitem{lz77}
Jacob Ziv and Abraham Lempel.
\newblock A universal algorithm for sequential data compression.
\newblock {\em IEEE Transactions on information theory}, 23(3):337--343, 1977.

\bibitem{lossless_eval}
Emir Öztürk and Altan Mesut.
\newblock Performance evaluation of {JPEG} standards, {WebP} and {PNG} in terms of compression ratio and time for lossless encoding.
\newblock In {\em 2021 6th International Conference on Computer Science and Engineering (UBMK)}, pages 15--20, 2021.

\end{thebibliography}


\end{document}